\documentclass[preprint,prd,superscriptaddress,amsmath,nofootinbib,preprintnumbers]{revtex4}

\usepackage{graphicx}
\usepackage{epsfig}
\usepackage{color}
\usepackage{amsmath}
\usepackage{amssymb}

\newcommand{\bea}{\begin{eqnarray}}
\newcommand{\eea}{\end{eqnarray}}
\newcommand{\nn}{\nonumber}

\newcommand{\eq}[1]{Eq.~(\ref{#1})}
\newcommand{\real}{\mathrm{Re}\,}
\newcommand{\imag}{\mathrm{Im}\,}
\newcommand{\bdtau}{\ensuremath{B\to D\tau\nu\;}}
\newcommand{\bdstau}{\ensuremath{B\to D^*\tau\nu\;}}
\newcommand{\btau}{\ensuremath{B\to \tau\nu\;}}
\newcommand{\rd}{\ensuremath{{\cal R}(D)}\;}
\newcommand{\rds}{\ensuremath{{\cal R}(D^*)}\;}

\newcommand{\dd}{\ensuremath{D\!-\!\Dbar{}\,}}
\newcommand{\kk}{\ensuremath{K\!-\!\Kbar{}\,}}
\newcommand{\bbd}{\ensuremath{B_d\!-\!\Bbar{}_d\,}}
\newcommand{\bbs}{\ensuremath{B_s\!-\!\Bbar{}_s\,}}

\newcommand{\Bbar}{\,\overline{\!B}}
\newcommand{\Dbar}{\,\overline{\!D}}
\newcommand{\Kbar}{\,\overline{\!K}}

\newcommand{\be}{\begin{equation}}
\newcommand{\ee}{\end{equation}}

\begin{document}

\title{Flavor-phenomenology of two-Higgs-doublet models with generic Yukawa structure}

\author{Andreas Crivellin}
\author{Christoph Greub}
\author{Ahmet Kokulu}
\affiliation{Albert Einstein Center for Fundamental Physics, Institute
  for Theoretical Physics,\\ University of Bern, CH-3012 Bern,
  Switzerland}

\begin{abstract}
In this article, we perform an extensive study of flavor observables in a two-Higgs-doublet model (2HDM) with generic Yukawa structure (of type III). This model is interesting not only because it is the decoupling limit of the Minimal Supersymmetric Standard Model (MSSM) but also because of its rich flavor phenomenology which also allows for sizable effects not only in FCNC processes but also in tauonic $B$ decays. We examine the possible effects in flavor physics and constrain the model both from tree-level processes and from loop-observables.
The free parameters of the model are the heavy Higgs mass, $\tan\beta$ (the ratio of vacuum expectation values) and the ''non-holomorphic'' Yukawa couplings $\epsilon^f_{ij} \,(f=u,d,\ell)$. In our analysis we constrain the elements $\epsilon^f_{ij}$ in various ways: In a first step we give order of magnitude constraints on $\epsilon^f_{ij}$ from 't Hooft's naturalness criterion, finding that all $\epsilon^f_{ij}$ must be rather small unless the third generation is involved. In a second step, we constrain the Yukawa structure of the type-III 2HDM from tree-level FCNC processes ($B_{s,d}\to\mu^+\mu^-$, $K_L\to\mu^+\mu^-$, ${\bar D}^0 \to\mu^+\mu^-$, $\Delta F=2$ processes, $\tau^-\to\mu^-\mu^+\mu^-$, $\tau^-\to e^-\mu^+\mu^-$ and $\mu^-\to e^- e^+ e^-$) and observe that all flavor off-diagonal elements of these couplings, except $\epsilon^u_{32,31}$ and $\epsilon^u_{23,13}$ must be very small in order to satisfy the current experimental bounds. In a third step, we consider Higgs mediated loop contributions to FCNC processes ($b\to s(d) \gamma$, $B_{s,d}$ mixing, \kk mixing and $\mu\to e\gamma$) finding that also $\epsilon^u_{13}$ and $\epsilon^u_{23}$ must be very small, while the bounds on $\epsilon^u_{31}$ and $\epsilon^u_{32}$ are especially weak. Furthermore, considering the constraints from electric dipole moments (EDMs) we obtain constrains on some parameters $\epsilon^{u,\ell}_{ij}$.
Taking into account the constraints from FCNC processes we study the size of possible effects in the tauonic $B$ decays (\btau, \bdtau and \bdstau) as well as in $D_{(s)} \to\tau  \nu$, $D_{(s)} \to\mu  \nu$, $K(\pi) \to e \nu$, $K(\pi) \to \mu \nu$ and $\tau \to K(\pi) \nu$ which are all sensitive to tree-level charged Higgs exchange. Interestingly, the unconstrained $\epsilon^u_{32,31}$ are just the elements which directly enter the branching ratios for \btau, \bdtau and \bdstau. We show that they can explain the deviations from the SM predictions in these processes without fine tuning. Furthermore, \btau, \bdtau and \bdstau can even be explained simultaneously. 
Finally, we give upper limits on the branching ratios of the lepton flavor-violating neutral $B$ meson decays ($ B_{s,d} \to \mu e$, $ B_{s,d} \to \tau e$ and $ B_{s,d} \to \tau \mu$) and correlate the radiative lepton decays ($\tau\to\mu\gamma$, $\tau\to e\gamma$ and $\mu\to e\gamma$) to the corresponding neutral current lepton decays ($\tau^-\to \mu^-\mu^+\mu^-$, $\tau^-\to e^-\mu^+\mu^-$ and $\mu^-\to e^-e^+e^-$).
A detailed appendix contains all relevant information for the considered processes for general scalar-fermion-fermion couplings. 

\end{abstract}


\maketitle

\tableofcontents

\newpage
\section{Introduction}

Two-Higgs-doublet models (2HDMs) \cite{Lee:1973iz} have been under intensive investigation for a long time (see for example Ref.~\cite{Gunion:1989we} for an introduction or Ref.~\cite{Branco:2011iw} for a recent review article). There are several reasons for this great interest in 2HDMs: Firstly, 2HDMs are very simple extensions of the Standard Model (SM) obtained by just adding an additional scalar $SU(2)_L$ doublet to the SM particle content. This limits the number of new degrees of freedom and makes the model rather predictive. Secondly, motivation for 2HDMs comes from axion models \cite{Kim:1986ax} because a possible CP-violating term in the QCD Lagrangian can be rotated away \cite{Peccei:1977hh} if the Lagrangian has a global $U(1)$ symmetry which is only possible if there are two Higgs doublets. Also the generation of the baryon asymmetry of the Universe motivates the introduction of a second Higgs doublet because in this way the amount of CP violation can be large enough to accommodate for this asymmetry, while the CP violation in the SM is too small \cite{Trodden:1998qg}. Finally, probably the best motivation for studying 2HDMs is the Minimal Supersymmetric Standard Model (MSSM) where supersymmetry enforces the introduction of a second Higgs doublet \cite{Haber:1984rc} due to the holomorphic superpotential. Furthermore, the 2HDM of type III is also the effective theory obtained by integrating out all super-partners of the SM-like particles (the SM fermion, the gauge boson and the Higgs particles of the 2HDM) from MSSM.
\medskip

2HDMs are not only interesting for direct searches for additional Higgs bosons at colliders. In addition to these high energy searches at the LHC also low-energy precision flavor observables provide a complementary window to physics beyond the SM, i.e. to the 2HDMs. In this respect, FCNC processes, e.g. neutral meson decays to muon pairs ($B_{s(d)}\to\mu^+\mu^-$, $D\to\mu^+\mu^-$ and $K_L\to\mu^+\mu^-$) are especially interesting because they are very sensitive to flavor changing neutral Higgs couplings. However, also charged current processes like tauonic $B$-meson decays are affected by the charged Higgs boson and $b\to s\gamma$ provides currently the best lower limit on the charged Higgs mass in the 2HDM of type~II.
\medskip

Recently, tauonic $B$ decays received special attention because the BABAR collaboration performed an analysis of the semileptonic $B$ decays \bdtau and \bdstau reporting a discrepancy of 2.0\,$\sigma$ and 2.7\,$\sigma$ from the SM expectation, respectively. The measurements of both decays exceed the SM predictions, and combining them gives a $3.4\, \sigma$ deviation from the SM~\cite{BaBar:2012xj,Lees:2013qea} expectation, which constitutes first evidence for new physics in semileptonic $B$ decays to tau leptons. This evidence for the violation of lepton flavor universality is further supported by the measurement of \btau by BABAR \cite{Lees:2012ju,Aubert:2008zzb} and BELLE \cite{Hara:2010dk,Adachi:2012mm} which exceeds the SM prediction by $1.6\, \sigma$ using $V_{ub}$ from the global fit \cite{Charles:2004jd}.
Assuming that these deviations from the SM are not statistical fluctuations or underestimated theoretical or systematic uncertainties, it is interesting to ask which model of new physics can explain the measured values. Since, a 2HDM of type II cannot explain \btau, \bdtau and \bdstau simultaneously \cite{BaBar:2012xj}, one must look at 2HDMs with more general Yukawa structures. Also 2HDMs of type III with Minimal Flavor Violation (MFV) \cite{MFV} cannot explain these deviations from the SM but a 2HDM of type III (where both Higgs doublets couple to up quarks and down quarks as well) with flavor-violation in the up sector, is capable of explaining \btau, \bdtau and \bdstau without fine tuning~\cite{Crivellin:2012ye}. 
\medskip

These points motivate us to perform a complete analysis of flavor-violation in 2HDMs of type III in this article. For this purpose we take into account all relevant constraints from FCNC processes (both from tree-level contributions and from loop-induced effects) and consider afterwards the possible effects in charged current processes.
\medskip

This article is structured as follows: In Sec.~\ref{setup}, we review the Yukawa Lagrangian of the 2HDM of type III. In Sec.~\ref{sec:general} we give a general overview on the constraints on 2HDMs and update the bounds on the 2HDM of type~II. The following sections discuss in detail the constraints on the 2HDM of type III parameter space from 't Hooft's naturalness argument (Sec.~\ref{sec:naturalness}), from tree-level FCNC processes (Sec.~\ref{tree-level-constraints}) and from loop-induced charged and neutral Higgs mediated contributions to the flavor observables (Sec.~\ref{sec:loop-contributions}). Sec.~\ref{sec:charged-current} studies the possible effects in charged current decays (\btau, \bdtau, \bdstau, $D_{(s)} \to\tau\nu$, $D_{(s)} \to\mu\nu$, $ K(\pi) \to e \nu$, $K(\pi) \to \mu \nu$, $\tau \to K(\pi) \nu$) and Sec.~\ref{LFV-b-decays} is denoted to the study of the upper limits on the branching ratios $B_{s,d} \to \tau \mu$, $B_{s,d} \to \tau e$, $B_{s,d} \to \mu e$ and the correlations among $\tau^-\to \mu^-\mu^+\mu^-$, $\tau^-\to e^-\mu^+\mu^-$, $\mu^-\to e^-e^+e^-$ and $\tau\to \mu \gamma$, $\tau\to e \gamma$, $\mu\to e \gamma$. Finally, we conclude. A detailed appendix contains some of the input parameters used in our analysis, general expressions for some branching ratios as well as all the relevant Wilson coefficients for $b\to s(d) \gamma$, $\Delta F=2$ processes, leptonic neutral meson decays ($\Delta F=1$),  LFV transitions, EDMs, anamolous magnetic moment (AMM) of muon and (semi-) leptonic charged meson decays for general charged and/or neutral scalar-fermion-fermion couplings.

\section{Setup}
\label{setup}

The SM contains only one scalar weak-isospin doublet, the Higgs doublet. After electroweak symmetry breaking its vacuum expectation value (''vev'') gives masses to up quarks, down quarks and charged leptons. The charged (CP-odd neutral) component of this doublet becomes the longitudinal component of the $W$ ($Z$) boson, and thus we have only one physical CP-even neutral Higgs particle in the SM. In a 2HDM we introduce a second Higgs doublet and obtain four additional physical Higgs particles (in the case of a CP conserving Higgs potential): the neutral heavy CP-even Higgs $H^0$, a neutral CP-odd Higgs $A^0$ and the two charged Higgses $H^{\pm}$.
\medskip

As outlined in the introduction we consider a 2HDM with generic Yukawa structure (2HDM of type III). One motivation is that a 2HDM with natural flavor-conservation (like type I or type II) cannot explain \bdtau, \bdstau and \btau simultaneously, while the type III model is capable of doing this \cite{Crivellin:2012ye}. Beside this, our calculations in the 2HDM III are the most general ones in the sense that they can be applied to models with specific flavor-structures like 2HDMs with MFV\cite{MFV,Buras:2010mh,Blankenburg:2011ca}. In this sense also our bounds are model independent, because they apply to any 2HDM with specific Yukawa structures as well (in the absence of large cancellations which are unlikely). Finally the type-III 2HDM is the decoupling limit of the MSSM and the calculated bounds can be translated to limits on the MSSM parameter space. 
\medskip

The fact that the 2HDM III is the decoupling limit of the MSSM also motivates us to choose for definiteness a MSSM like Higgs potential\footnote{If we would require that the Higgs potential possesses a $Z_2$ symmetry the results would be very similar. The heavy Higgs masses squared would still differ by terms of the order of $v^2$ and only Higgs self-couplings would be different, but they do not enter the flavor-processes at the loop-level under consideration.} which automatically avoids dangerous CP violation. The matching of the MSSM on the 2HDM Yukawa sector has been considered in detail. For the MSSM with MFV it was calculated in Ref.~\cite{Hamzaoui:1998nu,Babu:1999hn,Carena:1999py,Isidori:2001fv,Buras:2002vd,Hofer:2009xb} and for the MSSM with generic flavor structure in Ref.~\cite{Isidori:2002qe} (neglecting the effects of the $A$-terms) and in Ref.~\cite{Crivellin:2010er} (including the $A$-terms). Even the next-to-leading order corrections were calculated for the flavor-conserving case in \cite{Noth:2010jy} and for the flavor-changing one in the general MSSM in Ref.~\cite{Crivellin:2012zz}. Also the one-loop corrections to the Higgs potential have been considered \cite{Okada:1990vk,Haber:1990aw,Ellis:1990nz,Brignole:1991wp,Chankowski:1992er,Dabelstein:1994hb,Pilaftsis:1999qt,Carena:2000yi,Freitas:2007dp}, but their effects on flavor-observables were found to be small \cite{Gorbahn:2009pp}.
\medskip

Following the notation of Ref.~\cite{Crivellin:2010er,Crivellin:2011jt,Crivellin:2012zz} we have the following Yukawa Lagrangian in the 2HDM of type III starting in an electroweak basis:
\begin{equation}
\renewcommand{\arraystretch}{1.8}
\begin{array}{l}
\mathcal{L}_Y = \bar{Q}^a_{f\,L} \left[
  Y^{d\,{\rm ew}}_{fi}\epsilon_{ba}H^{b\star}_d\,-\,\epsilon^{d\,{\rm ew}}_{fi} H^{a}_u \right]d_{i\,R}
+ \, \bar{Q}^a_{f\,L} \left[ Y^{u\,{\rm ew}}_{fi} 
 \epsilon_{ab} H^{b\star}_u \, - \, \epsilon^{ u\,{\rm ew}}_{fi} H^{a}_d
  \right]u_{i\,R}\,+ \, {h.c.} \,\,\, .
  \end{array}
  \label{Leff}
\end{equation}
Here $a$, $b$ denote $SU(2)_L$\,-\,indices, $\epsilon_{ab}$ is the
two-dimensional antisymmetric tensor with $\epsilon_{12}=-1$ and the Higgs doublets are defined as :

\begin{equation}
\renewcommand{\arraystretch}{1.4}
\begin{array}{l}
H_{d}  =    \left(~ 
\begin{matrix}      
H^{1}_{d}   \\    
H^{2}_{d} \\    
 \end{matrix} ~  \right)    \, =   \,  \left( ~
\begin{matrix}      
H^{0}_{d}   \\    
H^{-}_{d} \\    
 \end{matrix}   ~ \right) \, \,   {\rm with}  \, \,   \left\langle H_{d} \right\rangle   =    \left(~ 
\begin{matrix}      
v_{d}   \\    
0 \\    
 \end{matrix} ~  \right)    \,   ,  \\ [0.35cm]
 H_{u}  =    \left( ~
\begin{matrix}      
H^{1}_{u}   \\    
H^{2}_{u} \\    
 \end{matrix} ~  \right)    \, =   \,  \left( ~
\begin{matrix}      
H^{+}_{u}   \\    
H^{0}_{u} \\    
 \end{matrix}  ~ \right)  \, \,   {\rm with}  \, \,   \left\langle H_{u} \right\rangle   =    \left(~ 
\begin{matrix}      
0   \\    
v_{u} \\    
 \end{matrix} ~  \right)    \, .
  \end{array}
 \label{doublets}
 \end{equation}
Apart from the holomorphic Yukawa-couplings $Y^{u\,{\rm ew}}_{fi}$ and $Y^{d\,{\rm ew}}_{fi}$, we included the non-holomorphic couplings $\epsilon^{q\,{\rm ew}}_{fi}$ ($q=u,d$) as well.
\medskip

As a next step we decompose the $SU(2)$ doublets into their components and switch to a basis in which the holomorphic Yukawa couplings are diagonal:
\begin{equation}
\renewcommand{\arraystretch}{1.4}
\begin{array}{l}
\mathcal{L}_Y = - \bar d_{f\,L} \left[Y^{d_i }\delta_{fi} H_d^{0\star}\,+\,\tilde\epsilon_{fi}^{ d}\,H_u^0 \right]d_{i\,R}
 - \bar u_{f\,L} \left[Y^{u_i } \delta_{fi} H_u^{0\star}\,+\,\tilde\epsilon_{fi}^{u}\,H_d^{0} \right] u_{i\,R} \\
\qquad ~ + \bar u_{f\,L} V_{fj} \left[ {Y^{d_i }\delta_{ji}-\cot \beta \tilde\epsilon_{ji}^{ d}  } \right]H^{2\star}_d d_{i\,R}  \\
\qquad ~ + \bar d_{f\,L} V_{jf}^{\star} \left[  Y^{u_i }\delta_{ji}-\tan \beta\tilde\epsilon_{ji}^{ u}   \right]H^{1\star}_u u_{i\,R} +\,{h.c.}    \,\,\, . \\ 
  \end{array}
\label{LeffY}
\end{equation}
where, $\tan\beta=v_{u}/v_{d}$ is the ratio of the vacuum expectation values $v_{u}$ and $v_{d}$ acquired by $H_{u}$ and $H_{d}$, respectively. 
We perform this intermediate step, because this is the basis which corresponds to the super-CKM basis of the MSSM and the couplings $\tilde\epsilon^{d}_{ij}$ can be directly related to loop-induced non-holomorphic Higgs coupling. The wave-function rotations $U^{q\,L,R}_{fi}$ necessary to arrive at the physical basis with diagonal quark mass matrices are defined by

\begin{equation}
	U^{q\,L\star}_{jf} m^q_{jk\,{}}U^{q\,R}_{ki}=m_{q_i}\delta_{fi}\, .
\end{equation}
They modify the Yukawa Lagrangian as follows:

\bea
\mathcal{L}_Y = &-& \bar d_{f\,L} \left[\left(\dfrac{{m_{d_i }
  }}{{v_d }}\delta_{fi} - \epsilon_{fi}^{ d}\tan\beta
  \right)H_d^{0\star}\,+\,\epsilon_{fi}^{ d}\,H_u^0 \right]d_{i\,R}
\nn \\
&-& \bar u_{f\,L} \left[\left(\dfrac{{m_{u_i } }}{{v_u }}\delta_{fi} -
  \epsilon_{fi}^{ u}\cot\beta \right)H_u^{0\star}\,+\,\epsilon_{fi}^{ u}\,H_d^{0} \right] u_{i\,R} \nn\\
&+& \bar u_{f\,L} V_{fj} \left[ {\dfrac{{m_{d_i } }}{{v_d
    }}\delta_{ji}-\left( {\cot \beta + \tan \beta }
    \right) \epsilon_{ji}^{ d}  } \right]H^{2\star}_d\ d_{i\,R} \nn \\
&+& \bar d_{f\,L} V_{jf}^{\star} \left[ { \dfrac{{m_{u_i
    } }}{{v_u }}\delta_{ji}-\left( {\tan \beta +
      \cot\beta } \right)\epsilon_{ji}^{ u}  } \right] H^{1\star}_u u_{i\,R}\,+\,{h.c.}  \,\,\, .
\label{L-Y-FCNC}
\eea
Here, $m_{q_i}$ are the physical running quark masses and 

\begin{equation}
	V_{fi}=U^{u\,L*}_{jf}U^{d\,L}_{ji} \, ,
\end{equation}
is the CKM matrix. The Higgs doublets $H_u$ and $H_d$ project onto the physical mass eigenstates $H^0$ (heavy CP-even Higgs), $h^0$ (light CP-even Higgs), $A^0$ (CP-odd Higgs) and $H^{\pm}$ in the following way:
\begin{eqnarray}
H_u^0&=&\frac{1}{\sqrt{2}}\left(H^0\sin\alpha + h^0\cos\alpha +
iA^0\cos\beta\right)\,, \nonumber\\
H_d^0&=&\frac{1}{\sqrt{2}}\left(H^0\cos\alpha - h^0\sin\alpha +
iA^0\sin\beta\right)\,,  \nonumber\\
H^{1}_{u}\, &=&\, \cos\beta \, H^{+}  \nn \, , \\
H^{2}_{d}\, &=&\, \sin\beta \, H^{-}  \, ,
\label{Htilde}
\end{eqnarray}
\medskip
where, $\alpha$ is the mixing angle necessary to diagonalize the neutral CP-even Higgs mass matrix (see e.g. \cite{Rosiek:1995kg}). Since we assume a MSSM-like Higgs potential\footnote{MSSM-like Higgs potential implies that in the large $\tan\beta$ limit and for $v\ll m_H$ the charged Higgs mass $m_{H^\pm}$, the heavy CP even Higgs mass $m_{H^0}$ and the CP odd Higgs mass $m_{A^0}$ are equal.} we have

\begin{equation}
\begin{array}{l}
	\tan\beta=\dfrac{v_u}{v_d}\,,\\
	\tan 2\alpha=\tan 2\beta \; \dfrac{m^{2}_{A^{0}}+M_Z^2}{m^{2}_{A^{0}}-M_Z^2}\,,\\
	m_{H^\pm}^2=m^{2}_{A^{0}}+M_W^2 \,, \hspace{0.3cm}
	m_{H^0}^2=m^{2}_{A^{0}}+M_Z^2-m_{h^0}^2 \, ,
\end{array}
\end{equation}
with $\dfrac{-\pi}{2}<\alpha<0$ and $0<\beta<\dfrac{\pi}{2}$. \\
\\
This means that in the phenomenologically interesting and viable limit of large values of $\tan\beta$ and $v \ll m_{A^{0}}$ we have to a good approximation\footnote{For the SM-like Higgs boson $h^0$ we use $m_{h^0}\approx125$~GeV in our numerical analysis.}:
\begin{equation}
\begin{array}{l}
\tan\beta\approx-\cot\alpha\,,\\
m_{H^0}\approx m_{H^\pm}\approx m_{A^{0}} \, \equiv  m_{H}\, .
\end{array}
\end{equation}

Without the non-holomorphic corrections $\epsilon^{q}_{ij}$, the
rotation matrices $U^{q\,L,R}_{fi}$ would simultaneously diagonalize the mass terms and the neutral Higgs couplings in \eq{L-Y-FCNC}.
However, in the presence of non-holomorphic corrections, this is no
longer the case and flavor changing neutral Higgs couplings are present in the basis in which the physical quark mass matrices are diagonal.
\medskip

The Yukawa Lagrangian in
\eq{L-Y-FCNC} leads to the following Feynman rules\footnote{Hermiticity of the Lagrangian implies the relation $\Gamma_{q_f q_i }^{RL\, H}\, =\, \Gamma_{q_i q_f }^{LR\, H\,\star}$. } for Higgs-quark-quark couplings
 
\begin{equation}
 	i \left( \Gamma_{q_f q_i }^{LR\, H} P_R + \Gamma_{q_f q_i }^{RL\, H} P_L \right) 
\end{equation}
with
\begin{eqnarray}
{\Gamma_{u_f u_i }^{LR\, H_k^0} } &=& x_u^k\left( \frac{m_{u_i }}{v_u}
\delta_{fi} - \epsilon_{fi}^{u}\cot\beta \right) + x_d^{k\star}
\epsilon_{fi}^{u}\,, \nonumber\\[0.1cm]
{\Gamma_{d_f d_i }^{LR\, H_k^0 } } &=& x_d^k \left( \frac{m_{d_i
}}{v_d} \delta_{fi} - \epsilon_{fi}^{d}\tan\beta \right) +
x_u^{k\star}\epsilon_{fi}^{ d} \,,\nonumber \\[0.1cm]
{\Gamma_{u_f d_i }^{LR\, H^\pm } } &=& \sum\limits_{j = 1}^3
{\sin\beta\, V_{fj} \left( \frac{m_{d_i }}{v_d} \delta_{ji}-
  \epsilon^{d}_{ji}\tan\beta \right)\, ,}
\nonumber\\[0.1cm]
{\Gamma_{d_f u_i }^{LR\,H^ \pm } } &=& \sum\limits_{j = 1}^3
{\cos\beta\, V_{jf}^{\star} \left( \frac{m_{u_i }}{v_u} \delta_{ji}-
  \epsilon^{u}_{ji}\tan\beta \right)\, }\, .
 \label{Higgs-vertices-decoupling}
\end{eqnarray}
\medskip
Similarly, for the lepton case, the non-vanishing effective Higgs vertices are
\begin{equation}
\renewcommand{\arraystretch}{1.8}
\begin{array}{l}
{\Gamma_{\ell_f \ell_i }^{LR\,H_k^0} } = x_d^k\left( \dfrac{m_{\ell_i }}{v_d}
\delta_{fi} - \epsilon_{fi}^{\ell}\tan\beta \right) + x_u^{k\star} \epsilon_{fi}^{\ell}\,, \\[0.25cm]
{\Gamma_{\nu_f \ell_i }^{LR\, H^\pm } } = \sum\limits_{j = 1}^3
{\sin\beta\, V^{\rm PMNS}_{fj} \left( \dfrac{m_{\ell_i }}{v_d} \delta_{ji}-
  \epsilon^{\ell}_{ji}\tan\beta \right)\, .}
 \end{array}
  \label{Higgs-leptons-vertices-decoupling}
\end{equation}
Here, $H^0_k=(H^0,h^0,A^0)$ and the coefficients $x_q^{k}$ are given
by
\begin{equation}
\renewcommand{\arraystretch}{1.4}
\begin{array}{l}
x_u^k \, = \, \left(-\dfrac{1}{\sqrt{2}}\sin\alpha,\,-\dfrac{1}{\sqrt{2}}\cos\alpha,
\,\dfrac{i}{\sqrt{2}}\cos\beta\right) \,,\\[0.3cm]
x_d^k \, = \,\left(-\dfrac{1}{\sqrt{2}}\cos\alpha,\,\dfrac{1}{\sqrt{2}}\sin\alpha,
\,\dfrac{i}{\sqrt{2}}\sin\beta\right) \, .
 \end{array}
\end{equation}
\smallskip

This means that flavor-violation (beyond the one already present in the 2HDM of type~II) is entirely governed by the couplings $\epsilon^{q,\ell}_{ij}$. If one wants to make the connection to the MSSM, the parameters $\epsilon^{q,\ell}_{ij}$ will depend only on SUSY breaking parameters and $\tan\beta$.

\section{Constraints on the 2HDM parameter space \\-- general discussion and overview}
\label{sec:general}

In this section we give an overview on flavor observables sensitive to charged Higgs contributions. We review the constraints on the 2HDM of type II and discuss to which extent these bounds will hold in the 2HDM of type III. A detailed analysis of flavor constraints on the type-III 2HDM parameter space will be given in the following sections. 
\medskip

The most common version of 2HDMs, concerning its Yukawa sector, is the 2HDM of type II which respects natural flavor conservation \cite{Glashow:1976nt} by requiring that one Higgs doublet couples only to up-quarks while the other one gives masses to down-type quarks and charged leptons (like the MSSM at tree-level). Flavor-observables in 2HDMs of type II have been studied in detail \cite{Miki:2002nz,WahabElKaffas:2007xd,Deschamps:2009rh}. In the type II model there are no tree-level flavor-changing neutral currents and all flavor violation is induced by the CKM matrix entering the charged Higgs vertex. In this way the constraints from FCNC processes can be partially avoided. This is true for $\Delta F=2$ processes where the charged Higgs contribution is small, for $K_L\to\mu^{+}\mu^{-}$, $D^0\to\mu^{+}\mu^{-}$ (due to the tiny Higgs couplings to light quarks) and all flavor observables in the lepton sector. However, the FCNC processes $b\to s \gamma$ (also to less extent $b\to d \gamma$) and $B_s\to\mu^+\mu^-$ are sensitive the charged Higgs contributions. In addition, direct searches at the LHC and charged current processes restrict the type-II 2HDM parameter space.
\medskip

Among the FCNC processes, the constraints from $b\to s \gamma$ are most stringent due to the necessarily constructive interference with the SM contribution \cite{Bertolini:1990if,Ciuchini:1997xe,Borzumati:1998tg,Misiak:2006zs}. The most recent lower bound on the charged Higgs mass obtained in Ref.~\cite{Hermann:2012fc} is $m_{H^\pm}\geq 360\,{\rm GeV}$ which includes NNLO QCD corrections and is rather independent of $\tan\beta$. In the type-III 2HDM  this lower bound on the charged Higgs mass can be weakened due to destructive interference with contributions involving $\epsilon^q_{ij}$. Also in $B_s\to\mu^+\mu^-$ (and $B_d\to\mu^+\mu^-$) a sizable loop-induced effect is possible in the 2HDM~II, but the constrains are still not very stringent even if the new LHCb measurement are used. The reason for this is that, taking into account the constraints from $b\to s \gamma$ on the charged Higgs mass, the branching ratio for $B_s\to\mu^+\mu^-$ in the 2HDM II is even below the SM expectation for larger values of $\tan\beta$ \cite{He:1988tf,Skiba:1992mg,Logan:2000iv} due to the destructive interference between the charged Higgs and the SM contribution. 
\medskip

Regarding charged current processes, tauonic $B$ decays are currently most sensitive to charged Higgs effects. Here, the charged-Higgs contribution in the type-II 2HDM to \btau interferes destructively with the SM contribution \cite{Hou:1992sy,Akeroyd:2003zr}. The same is true for \bdstau \cite{Fajfer:2012vx} and \bdtau \cite{Tanaka:1994ay,Miki:2002nz,Nierste:2008qe}. As outlined in the introduction this leads to the fact that the 2HDM II cannot explain  \btau, \bdtau and \bdstau simultaneously \cite{BaBar:2012xj}. Other charged current observables sensitive to charged Higgses are $D_{(s)}\to\mu\nu$, $D_{(s)}\to\tau\nu$~\cite{Akeroyd:2003jb,Akeroyd:2007eh,Akeroyd:2009tn}, $\tau\to K(\pi) \nu$ and $K\to\mu\nu/\pi\to\mu\nu$ \cite{Antonelli:2008jg} (see \cite{Deschamps:2009rh} for a global analysis).
\medskip
\begin{figure}[htbp]
\begin{center}
\includegraphics[width=0.58\textwidth]{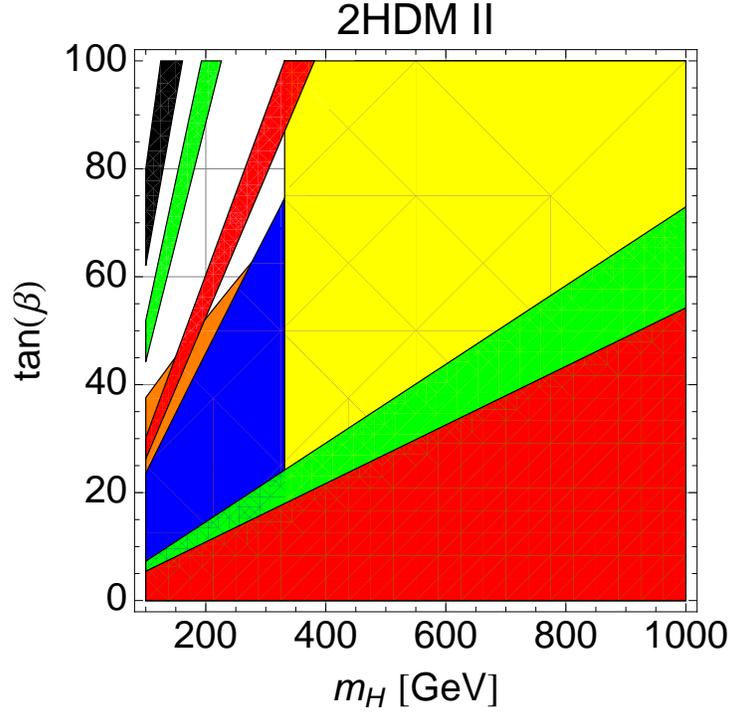}
\end{center}
\caption{Updated constraints on the 2HDM of type II parameter space. The regions compatible with experiment are shown: $b\to s\gamma$ (yellow), $B\to D\tau\nu$ (green), $B\to \tau \nu$ (red), $B_{s}\to \mu^{+} \mu^{-}$ (orange), $K\to \mu \nu/\pi\to \mu \nu$ (blue) and $B\to D^*\tau \nu$ (black). Note that no region in parameter space is compatible with all processes. Explaining $B\to D^*\tau \nu$ would require very small Higgs masses and large values of $\tan\beta$ which is not compatible with the other observables. To obtain this plot, we added the theoretical uncertainty linear on the top of the $2 \, \sigma$ experimental error.}
\label{fig:2HDMII}
\end{figure}

Fig.~\ref{fig:2HDMII} shows our updated constraints on the 2HDM II parameters space from $b\to s\gamma$, \btau, \bdtau, \bdstau, $B_{s}\to \mu^{+} \mu^{-}$ and $K\to \mu \nu/\pi\to \mu \nu$. We see that in order to get agreement within $2\,\sigma$ between the theory prediction and the measurement of \bdstau, large values of $\tan\beta$ and light Higgs masses would be required which is in conflict with all other processes under consideration. 
\medskip

\begin{figure}[htbp]
\begin{center}
\includegraphics[width=0.6\textwidth]{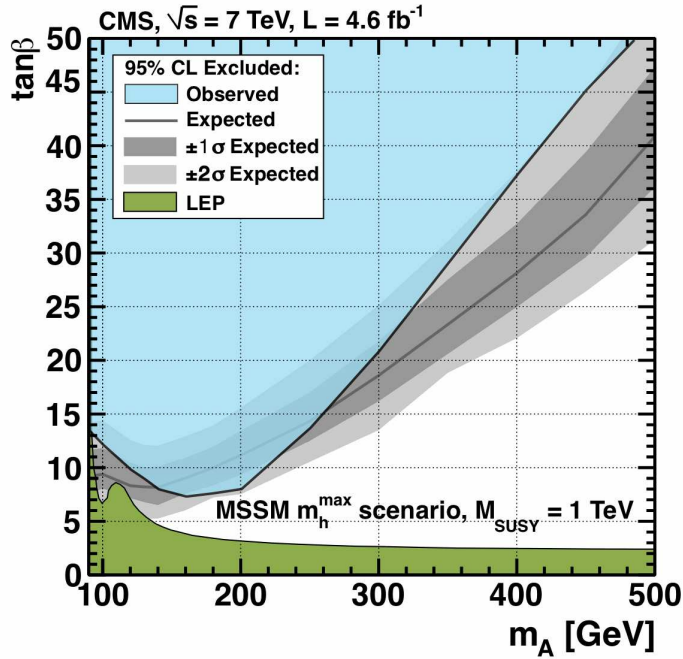}
\end{center}
\caption{Plot from the CMS collaboration taken from Ref.~\cite{CMS}: allowed regions in the $m_{A^0}$--$\tan\beta$ plane from $A^{0}\to \tau^{+}\tau^{-}$. The analysis was done in the MSSM, but since we consider a 2HDM with MSSM-like Higgs potential and the MSSM corrections to the $A^0\tau\tau$ vertex are small, we can apply this bound to our model. However, a large value of $\epsilon^\ell_{33}$ in the 2HDM of type~III could affect the conclusions.}
\label{cmsplot}
\end{figure}

Concerning direct searches the bounds on the charged Higgs mass are rather weak due to the large background from $W$ events. The search for neutral Higgs bosons is easier and the CMS bounds\footnote{Note that we did not use the bounds from unpublished CMS update of the $A^0\to\tau^{+}\tau^{-}$ analysis.} on $m_{A^0}$ from $A^0\to\tau^{+} \tau^{-}$ are shown in Fig.~\ref{cmsplot}. These bounds were obtained in the MSSM, but since the MSSM corrections to $A^0\to\tau^{+} \tau^{-}$ are rather small and since we consider a MSSM-like Higgs potential, these bounds also hold in the 2HDM III as long as the Peccei-Quinn symmetry breaking in the lepton sector is small.
\medskip

Going beyond the simple Yukawa structure of the 2HDM of type II, also 2HDMs of type III with MFV \cite{MFV,Buras:2010mh,Blankenburg:2011ca}, alignment \cite{Pich:2009sp,Jung:2010ik} or natural flavor conservation \cite{Glashow:1976nt,Buras:2010mh} have been analyzed in detail. However, flavor-observables in type III models with generic flavor-structure have received much less attention. Ref.~\cite{Mahmoudi:2009zx} considered the possible effects of the flavor-diagonal terms and Ref.~\cite{Iltan:2001rp} considers leptonic observables. 
As outlined in the introduction, 2HDMs of type II (or type III with MFV) cannot explain \bdtau and \bdstau simultaneously \cite{BaBar:2012xj} (and for \btau fine tuning is needed \cite{Blankenburg:2011ca}). 
\medskip

In the following sections we will study in detail the flavor-observables in the 2HDM with generic flavor-structure \cite{Cheng:1987rs}, but for definiteness, with MSSM-like Higgs potential. For this purpose, all processes described above are relevant. In addition, $\Delta F=2$ processes, lepton flavor violating observables (LFV), EDMs, $\tau \to K(\pi)  \nu/K(\pi)\to\mu \nu$ and $K \to \mu (e) \nu/\pi\to\mu (e) \nu$ will turn out to give information on the flavor structure of the 2HDM of type III. Furthermore, we will investigate to which extent contributions to $B_{s,d}\to \tau \mu$, $B_{s,d}\to\tau e$, $B_{s,d}\to\mu e$ and muon anomalous magnetic moment are possible.
\medskip

\section{Constraints from 't Hooft's naturalness criterion}
\label{sec:naturalness}
The naturalness criterion of 't Hooft states that the smallness of a quantity is only natural if a symmetry is gained in the limit in which this quantity is zero. This means on the other hand that large accidental cancellations, which are not enforced by a symmetry, are unnatural and thus not desirable. Let us apply this reasoning to the fermion mass matrices in the 2HDM. We recall from the last section the expressions for the fermion mass matrices in the electroweak basis:
\be
\renewcommand{\arraystretch}{1.2}
\begin{array}{l}
m^{d}_{ij}=v_{d} Y^{d\,{\rm ew}}_{ij} + v_{u} \epsilon^{d\,{\rm ew}}_{ij}\, ,\\ [0.1cm]
 m^{u}_{ij}=v_{u} Y^{u\,{\rm ew}}_{ij} + v_{d} \epsilon^{u\,{\rm ew}}_{ij}\, , \\[0.1cm]
 m^{\ell}_{ij}=v_{d} Y^{\ell\,{\rm ew}}_{ij} + v_{u} \epsilon^{\ell\,{\rm ew}}_{ij}\,.
\end{array}
\label{meffew}
\ee

Diagonalizing these fermion mass matrices gives the physical fermion masses and the CKM matrix. Using 't Hooft's naturalness criterion we can demand the absence of fine-tuned cancellations between $v_d Y^{d,\ell}_{ij}$ ($v_u Y^{u}_{ij}$) and $v_u \epsilon^{d,\ell}_{ij}$ ($v_d \epsilon^u_{ij}$). Thus, we require that the contributions of $v_u \epsilon^{d,\ell}_{ij}$ and $v_d \epsilon^u_{ij}$ to the fermion masses and CKM matrix do not exceed the physical measured quantities.
\medskip

In first order of a perturbative diagonalization of the fermion mass matrices, the diagonal elements $m^f_{ii}$ give rise to the fermion masses, while (in our conventions) the elements $m^f_{ij}$ with $i<j$ ($i>j$) affect the left-handed (right-handed) rotations necessary to diagonalize the fermion mass matrices. The left-handed rotations of the quark fields are linked to the CKM matrix and can therefore be constrained by demanding that the physical CKM matrix is generated without a significant degree of fine-tuning. However, the right-handed rotations of the quarks are not known and the mixing angles of the PMNS matrix are big so that for these two cases we can only demand that the fermion masses are generated without too large accidental cancellations. Note, that in \eq{meffew} the elements $\epsilon^{f\,{\rm ew}}_{ij}$ enter, while the elements $\epsilon^{f}_{ij}$ which we want to constrain from flavor observables are given in the physical basis with diagonal fermion masses. This means that in order to constrain $\epsilon^{f}_{ij}$ from 't Hooft's naturalness criterion we have to assume in addition that no accidental cancellation occur by switching between the electroweak basis and the physical basis. In conclusion this leads to the following upper bounds

\begin{equation}
\begin{array}{l}
\left| v_{u(d)} \epsilon^{d(u)}_{ij} \right| \leq  \left|V^{\rm CKM}_{ij}\right| \times{ \rm max }\left[m_{d_i(u_i)},m_{d_j(u_j)}\right]\;{\rm for}\;i<j \,,\\ [0.25cm]
\left| v_{u(d)} \epsilon^{d(u)}_{ij} \right| \leq { \rm max }\left[m_{d_i(u_i)},m_{d_j(u_j)}\right]\, \;{\rm for}\;i\geq j \,,\\  [0.25cm]
\left| v_{u} \epsilon^{\ell}_{ij} \right| \leq  { \rm max }\left[m_{\ell_i},m_{\ell_j}\right]\,.
\end{array}
\label{Naturalness}
\end{equation}
\smallskip

In the large $ \tan\beta$ limit, inserting the quark masses $m_{q}(\mu)$ at the Higgs scale (which we choose here to be $\mu_{\rm Higgs}=500\,{\rm GeV}$), we can immediately read off the upper bounds on $\epsilon^{u,d,\ell}_{ij}$ from \eq{Naturalness}:

\begin{equation}
\renewcommand{\arraystretch}{1.2}
\begin{array}{l}
\left| \epsilon^{d}_{ij}\right|  \le   \left( 
\begin{matrix}      
1.3 \times10^{-4} ~&~5.8 \times10^{-5} ~&~ 5.1\times10^{-5}    \\    
2.6 \times10^{-4}  ~&~2.6 \times10^{-4} ~&~5.9\times10^{-4} \\    
1.4\times10^{-2}~&~1.4 \times10^{-2}~&~ 1.4\times10^{-2}
 \end{matrix} \right)_{ij} \,,   \\ [0.25cm]
  \left| \epsilon^{u}_{ij} \right| \le  ({\tan\beta}/50) \, \left( \begin{matrix}
 3.4 \times10^{-4} ~&~3.2\times10^{-2} ~&~ 1.6\times10^{-1}    \\    
 1.4\times10^{-1} ~&~1.4\times10^{-1} ~&~1.9 \\    
-~&~-~&~ -
 \end{matrix} \right)_{ij} \,, 
 \\ [0.25cm]
\left| \epsilon^{\ell}_{ij}\right|  \le   \left( \begin{matrix}      
2.9\times10^{-6} ~&~6.1\times10^{-4} ~&~ 1.0\times10^{-2}    \\    
6.1\times10^{-4} ~&~6.1\times10^{-4} ~&~1.0\times10^{-2} \\    
1.0\times10^{-2}~&~1.0\times10^{-2}~&~ 1.0\times10^{-2}
 \end{matrix} \right)_{ij} \,.  
  \end{array}
 \label{Natlimits}
 \end{equation}
\medskip

Of course, these constraints are not strict bounds in the sense that they must be respected in any viable model. Anyway, big violation of naturalness is not desirable and \eq{Natlimits} gives us a first glance on the possible structure of the elements $\epsilon^f_{ij}$. As we will see later, it is possible to explain \btau, \bdtau and \bdstau using $\epsilon^u_{31,32}$ without violating \eq{Natlimits}, while if one wants to explain \btau with $\epsilon^d_{33}$ 't Hooft's naturalness criterion is violated.

\section{Constraints from tree-level neutral-current processes}
\label{tree-level-constraints}

The flavor off-diagonal elements $\epsilon^f_{ij}$ (with $i\neq j$) give rise to flavor-changing neutral currents (FCNCs) already at the tree-level. Comparing the Higgs contributions to the loop-suppressed SM contributions, large effects are in principle possible. However, all experimental results are in very good agreement with SM predictions, which put extremely stringent constraints on the non-holomorphic terms $\epsilon^f_{ij}$.  
\medskip

In this section we consider three different kinds of processes: 
\begin{itemize}
	\item Muonic decays of neutral mesons ($B_{s,d}\to\mu^+\mu^-$, $K_L\to\mu^+\mu^-$ and ${\bar D}^0\to\mu^+\mu^-$).
	\item $\Delta F=2$ processes (\dd, \kk, \bbs and \bbd mixing).
	\item Flavor changing lepton decays ($\tau^-\to \mu^-\mu^+\mu^-$, $\tau^-\to e^-\mu^+\mu^-$ and $\mu^-\to e^-e^+e^-$).
\end{itemize}

As we will see in detail in Sec.~\ref{NeutralMesonDecays}, the leptonic neutral meson decays $B_{s,d}\to\mu^+\mu^-$, $K_L\to\mu^+\mu^-$ and ${\bar D}^0\to\mu^+\mu^-$ put constraints on the elements $\epsilon^d_{ij}$ (with $i\neq j$) and $\epsilon^u_{12,21}$ already if one of these elements is non-zero, while \bbd, \bbs, \kk and \dd mixing only provide constraints on the products $\epsilon^d_{ij}\epsilon^{d\star}_{ji}$ and $\epsilon^u_{12}\epsilon^{u\star}_{21}$ (Sec.~\ref{DeltaF2-processes}). This means that the constraints on $\Delta F=2$ processes can be avoided if one element of the product $\epsilon^q_{ij}\epsilon^{q\star}_{ji}$ is zero, while the constraints from the leptonic neutral meson decays can only be avoided if the Peccei Quinn symmetry breaking for the leptons is large such that $\epsilon^\ell_{22}\approx m_{\mu}/v_u$ is possible. 
\medskip

In Sec.~\ref{taumumumu} we will consider the flavor changing lepton decays $\tau^-\to \mu^-\mu^+\mu^-$, $\tau^-\to e^-\mu^+\mu^-$ and $\mu^-\to e^-e^+e^-$ which constrain the off-diagonal elements $\epsilon^\ell_{23,32}$, $\epsilon^\ell_{13,31}$ and $\epsilon^\ell_{12,21}$, respectively.
\medskip

\subsection{Leptonic neutral meson decays: 
$B_{s,d}\to\mu^+\mu^-$, $K_L\to\mu^+\mu^-$ and ${\bar D}^0\to\mu^+\mu^-$}
\label{NeutralMesonDecays}

Muonic decays of neutral mesons ($B_s\to\mu^+\mu^-$, $B_d\to\mu^+\mu^-$, $K_L\to\mu^+\mu^-$ and ${\bar D}^0\to\mu^+\mu^-$) are strongly suppressed in the SM for three reasons: they are loop-induced, helicity suppressed and they involve small CKM elements. Therefore, their branching ratios (in the SM) are very small and in fact only $K_L\to \mu^+\mu^-$ and recently also $B_s\to\mu^+\mu^-$ \cite{Aaij:2012ct} have been measured, while for the other decays only upper limits on the branching ratios exist (see Table \ref{expbounds}). We do not consider decays to electrons (which are even stronger helicity suppressed) nor $B_{d,s}\to\tau^+\tau^-$ (where the tau leptons are difficult to reconstruct) because the experimental limits are even weaker. The study of meson decays to lepton flavor-violating final states is postponed to Sec.~\ref{LFV-b-decays}.
\begin{table}[h]
\centering \vspace{0.9cm}
\renewcommand{\arraystretch}{1.5}
\begin{tabular}{|c|c|c|}
\hline
  Process & Experimental value & SM prediction
\\   \hline \hline
${\cal B}\left[\ensuremath{B_{s}\to \mu^{+} \mu^{-}}\right]$  
& $ \, 3.2^{+1.5}_{-1.2} \times 10^{-9}$ \cite{Aaij:2012ct}
& $(3.23\pm 0.27)\times 10^{-9}$  \cite{Buras:2012ru}   \\ \hline 
${\cal B}\left[\ensuremath{B_{d}\to \mu^{+} \mu^{-}}\right]$  
& $\leq \, 9.4\times 10^{-10}$ { (95\% CL)} \cite{Aaij:2012ct}
& $(1.07 \pm 0.10)\times 10^{-10}$ \cite{Buras:2012ru}  \\ \hline
${\cal B}\left[\ensuremath{K_{L}\to \mu^{+} \mu^{-}}\right]_{\rm short}$  
& $\leq 2.5 \times 10^{-9}$ \cite{Isidori:2003ts}
&  $\approx 0.9 \times 10^{-9}$ \cite{Isidori:2003ts} \\ \hline
${\cal B}\left[\ensuremath{D^{0}\to \mu^{+} \mu^{-}}\right]$  
& $\leq \,1.4\times 10^{-7}$ {(90\% CL)}\,\cite{Beringer:1900zz}
& --  \\ \hline 
\end{tabular}
  \caption{Experimental values and SM predictions for the branching ratios of neutral meson decays to muon pairs. For $K_L\to \mu^+\mu^-$ we only give the upper limit on the computable short distance contribution \cite{Isidori:2003ts} extracted from the experimental value $ (6.84\; \pm\; 0.11) \times 10^{-9}$ {(90\% CL)} \cite{Beringer:1900zz}. The SM prediction for $D^{0}\to\mu^+\mu^-$ cannot be reliably calculated due to hadronic uncertainties. } 
\label{expbounds}
\end{table}

\begin{figure*}[t]
\centering
\includegraphics[width=0.34\textwidth]{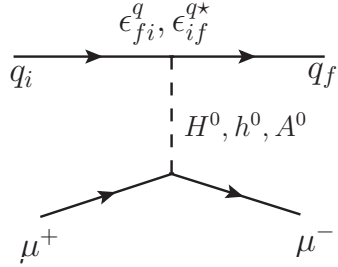}
\caption{Feynman diagram showing the neutral Higgs contribution to  $B_{s,d}\to\mu^+\mu^-$, $K_L\to\mu^+\mu^-$ and ${\bar D}^{0}\to\mu^+\mu^-$.}
\label{feynPSmumuH0}
\end{figure*}

We see from Fig.~\ref{feynPSmumuH0} that the off-diagonal elements of $\epsilon^{d}_{13,31}$, $\epsilon^{d}_{23,32}$, $\epsilon^{d}_{12,21}$ and $\epsilon^{u}_{12,21}$ directly give rise to tree-level neutral Higgs contributions to $B_d\to\mu^+\mu^-$, $B_s\to\mu^+\mu^-$, $K_L\to\mu^+\mu^-$ and ${\bar D}^{0}\to\mu^+\mu^-$, respectively. 
\medskip

In principle, the constraints from these processes could be weakened, or even avoided, if $\epsilon^\ell_{22}\approx m_{\ell_{2}}/v_u$. Anyway, in this section we will assume that the Peccei Quinn breaking for the leptons is small and neglect the effect of $\epsilon^\ell_{22}$ in our numerical analysis for setting limits on $\epsilon^q_{ij}$.

\subsubsection{$B_{s,d}\to \mu^+\mu^-$}

For definiteness, consider the decay of a neutral $B_s  \left( \bar{b}s \right)$ meson (the corresponding decay of a $B_d$ meson follow trivially by replacing $s$ with $d$ and $2$ with $1$) to a moun pair. The effective Hamiltonian governing this transition is\footnote{The complete expression for the Hamiltonian and the branching ratio including lepton flavor-violating final states is given in the appendix.}
\begin{equation}
{\cal H}^{B_s\to\mu^+\mu^-}_{\rm eff}=-\frac{G_{F}^{2} M_{W}^{2}}{\pi^{2}} \left[C^{bs}_{A}O^{bs}_{A}+C^{bs}_{S}O^{bs}_{S}+C^{bs}_{P}O^{bs}_{P}+C_{A}^{\prime bs}O_{A}^{\prime bs}+C^{\prime bs}_{S} O^{\prime bs}_{S}+C^{\prime bs}_{P} O_{P}^{\prime bs}\right]+{h.c.} \,,
\label{Heffsemilep2}
\end{equation}
where the operators are defined as
\be
\renewcommand{\arraystretch}{1.2}
\begin{array}{l}
 {O}^{bs}_A \,= \!
 \left(\bar{b} \gamma_\mu P_{L} s\right)
  \left(\bar{\mu} \gamma^\mu \gamma_{5}  \mu \right)\,,   \\
{O}^{bs}_S \,= \!
  \left(\bar{b}  P_{L} s\right)
  \left(\bar{\mu} \mu \right)\,, \\ 
{O}^{bs}_P \,= \!
  \left(\bar{b}  P_{L} s\right) 
  \left(\bar{\mu} \gamma_{5}  \mu \right)\,,
\end{array} 
\label{opbasis2}
\ee
and the primed operators are obtained replacing $P_L$ with $P_R$. The corresponding expression for the branching ratio in terms of the Wilson coefficients reads
%
\begin{equation}
\begin{array}{l}
\renewcommand{\arraystretch}{2.2}
{\cal{B}}\left[B_{s} \to \mu^{+} \mu^{-} \right]=

\dfrac{G_{F}^{4} M_{W}^{4} }{8 \pi^{5}}  \,  \sqrt{1-4\dfrac{m^{2}_\mu}{M^{2}_{B_s}}} \,  M_{B_{s}} \,  f_{B_{s}}^{2} \, m_\mu^{2}   \,   \tau_{B_{s}}   \\
 ~~~~~~~~~~~ \times \left[      \left|  \dfrac{M_{B_{s}}^{2} \left(C_{P}^{bs}-C_{P}^{\prime bs}\right)}{2\left(m_{b}+m_{s}\right)m_\mu} 
-(C_{A}^{bs}-C_{A}^{\prime bs})  \right|^{2}   + \left|  \dfrac{M_{B_{s}}^{2} (C_{S}^{bs}-C_{S}^{\prime bs})}{2\left(m_{b}+m_{s}\right)m_\mu}   \right| ^{2} \times \left(1-4\dfrac{m_{\mu}^2}{m_{B_s}^2}\right)    \right] \,.
\label{BRBsmumu}
\end{array}
\end{equation}
Concerning the running of the Wilson coefficients due to the strong interaction, the operators ${O}^{bs}_A$ and $O^{\prime bs}_A$ correspond to conserved vector currents with vanishing anomalous dimensions. This means that their Wilson coefficients are scale independent. The scalar and pseudo-scalar Wilson coefficients $C_S^{bs}$ and $C_P^{bs}$ ($C_S^{\prime bs}$ and $C^{\prime bs}_P$) have the same anomalous dimension as quark masses in the SM which means that their scale dependence is given by:
\begin{eqnarray}
C_{S,P}^{(\prime)bs}(\mu_{{low}})=\frac{m_{q}(\mu_{{low}})}{m_{q}(\mu_{{high}})} \, C_{S,P}^{(\prime)bs}(\mu_{{high}})\,,
\end{eqnarray}
where $m_q$ is the running quark mass with the appropriate number of active flavors. In the SM, $C_{A}$ is the only non-vanishing Wilson coefficient
\be
{ C^{bs}_{A}  = -V^{\star}_{tb}V_{ts} Y\left(\dfrac{ m^{2}_{t}}{ M^{2}_{W}}\right)-V^{\star}_{cb}V_{cs} Y\left(\dfrac{ m^{2}_{c}}{ M^{2}_{W}}\right)\,,   }
\label{CASM}
 \ee
where, the function $Y$ is defined as $Y=\eta_{Y} Y_{0}$ such that the NLO QCD effects are included in $\eta_{Y}=1.0113$ \cite{Buras:2012ru} and the one loop Inami-Lim function $Y_{0}$ reads \cite{InamiLim81}
\be
Y_{0}(x)= \dfrac{x}{8} \, \left[  \dfrac{4-x}{1-x}+ \dfrac{3\,x}{(1-x)^{2}}\,{\ln (x)}  \right] \,.
 \ee
The complete Wilson coefficients for general quark-quark-scalar couplings are given in the appendix. In the 2HDM of type III, in the case of large $\tan\beta$ and $v\ll m_H$, the terms involving $\epsilon^q_{ij}$ simplify to
\begin{equation}
\renewcommand{\arraystretch}{2.0}
\begin{array}{l}
 C_S^{bs}  = C_P^{bs}  = -\dfrac{{\pi ^2 }}{{G_F^2 M_W^2 }}\dfrac{1}{{2m_H^2 }}\dfrac{{m_{\ell _2 } -v_u \epsilon^\ell_{22} }}{v}\epsilon _{23}^{d\star} \tan^{2} \beta  \, ,\\ 
 C^{\prime bs}_S  =  - C^{\prime bs}_P  =   -\dfrac{{\pi ^2 }}{{G_F^2 M_W^2 }}\dfrac{1}{{2m_H^2 }}\dfrac{{m_{\ell _2 } -v_u \epsilon^\ell_{22}}}{v}\epsilon _{32}^d \tan^{2}  \beta \, .\\ 
 \end{array}
 \label{CSCPBsmumu}
\end{equation}
To these Wilson coefficients the well known loop-induced type II 2HDM contributions\footnote{Since we want to put constraints on the elements $\epsilon^d_{13,23}$ we assume that the loop-induced 2HDM II contribution is not changed by elements $\epsilon^u_{i3}$ or $\epsilon^d_{33}$.} 
\begin{equation}
\label{eqn:type2Cis}
  C^{bs}_S = C^{bs}_P = -\dfrac{m_{b}\, V^{\star}_{tb}V_{ts}}{2} \dfrac{m_{\mu}}{2 M^{2}_{W}}  \, \tan^{2}\beta \, \dfrac{\log\left({{m_H^2}/{m_t^2} }\right)}{ { {m_H^2}/{m_t^2}-1}}  \, ,
\end{equation}
have to be added as well \cite{Logan:2000iv}. Note that since we give the Wilson coefficients at the matching scale, also $m_b$ and $m_t$ must be evaluated at this scale.
\medskip

\begin{figure*}[t]
\centering
\includegraphics[width=0.3\textwidth]{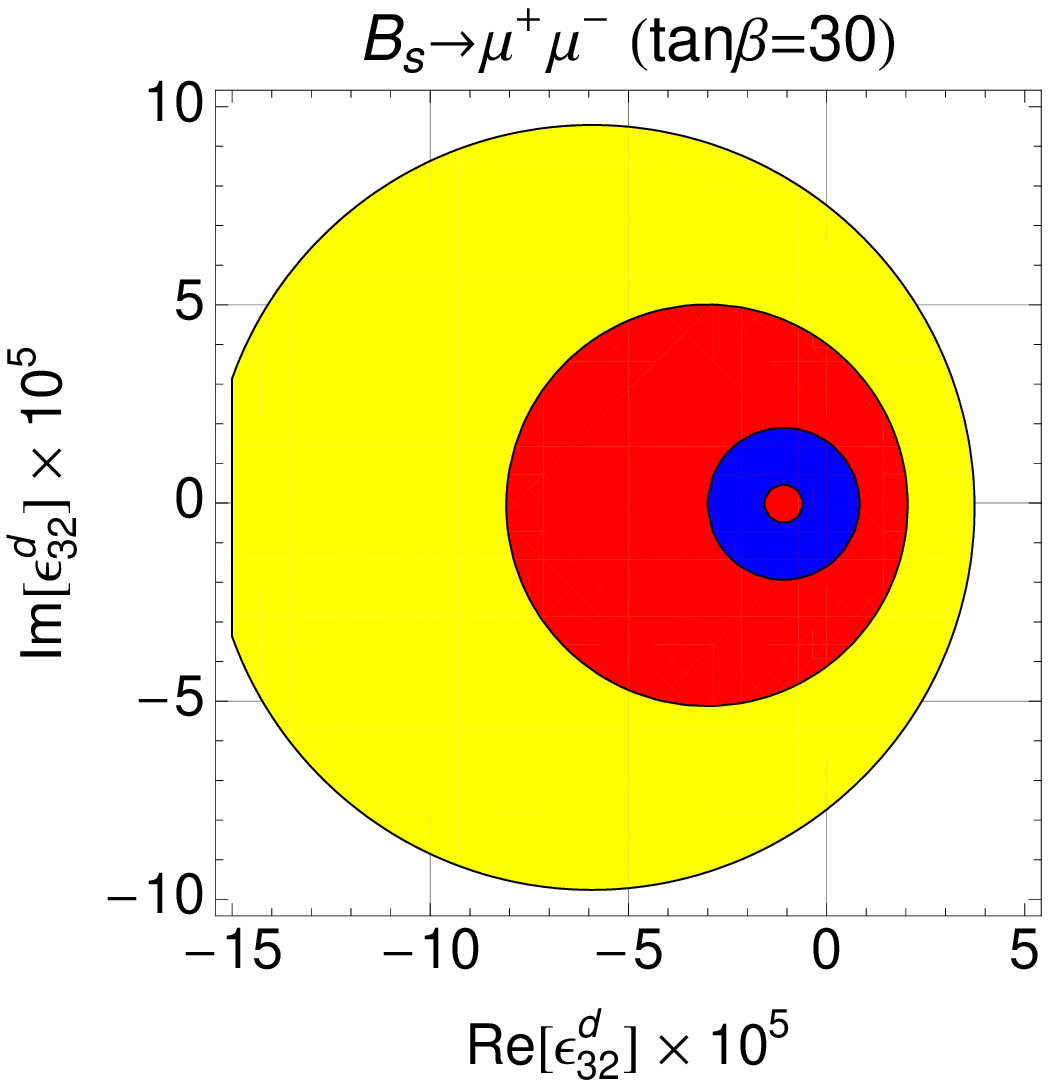}~~~~~
\includegraphics[width=0.3\textwidth]{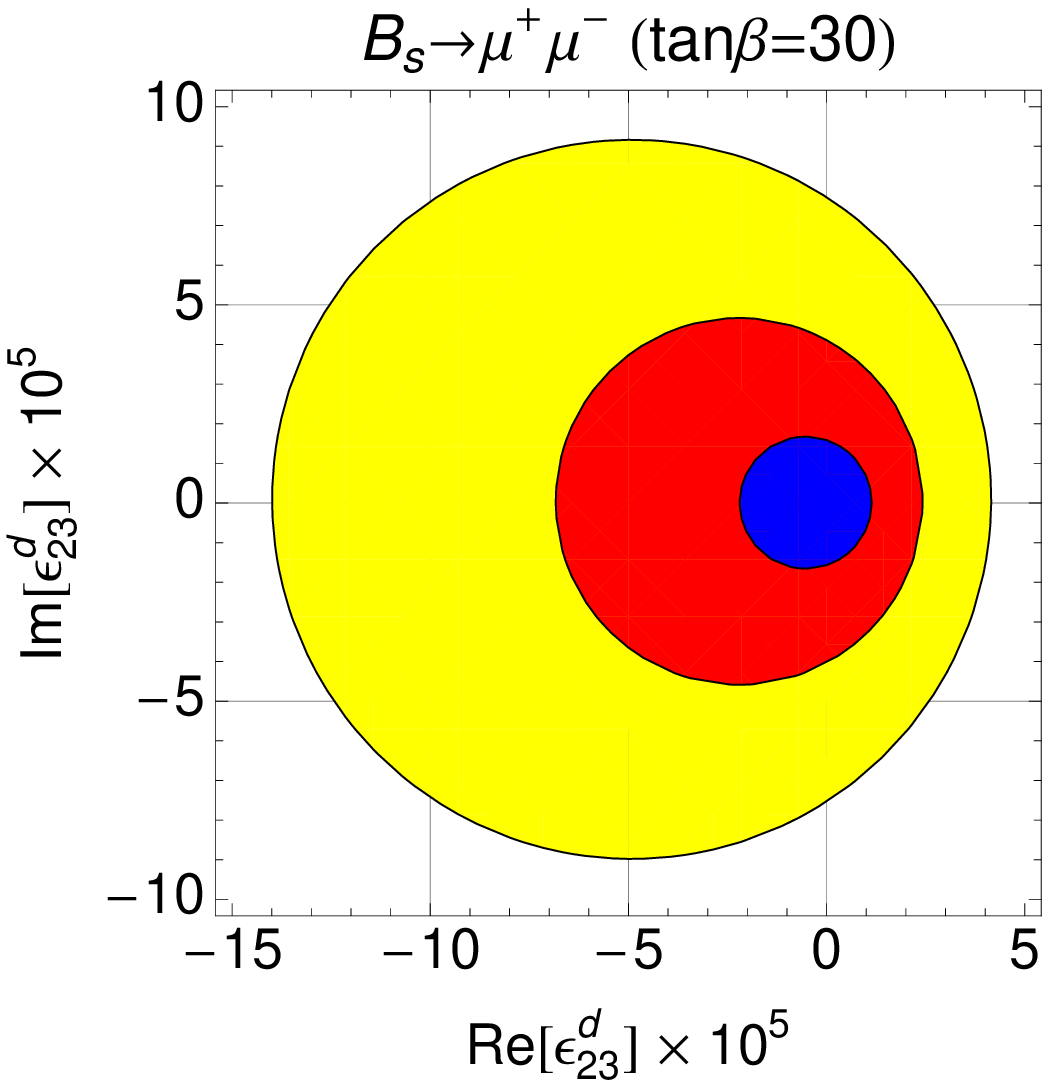} \\ [0.2cm]
\includegraphics[width=0.3\textwidth]{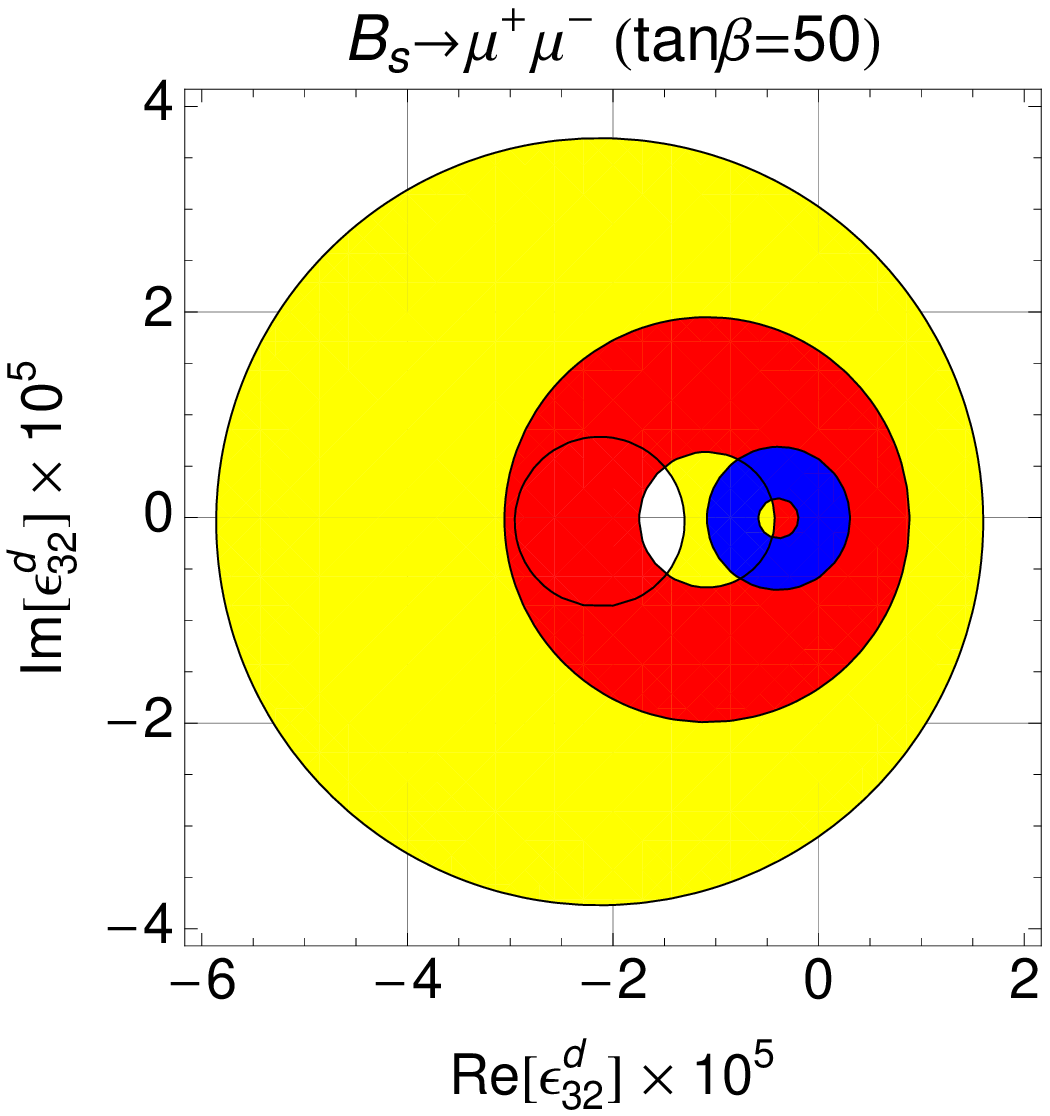}~~~~~
\includegraphics[width=0.3\textwidth]{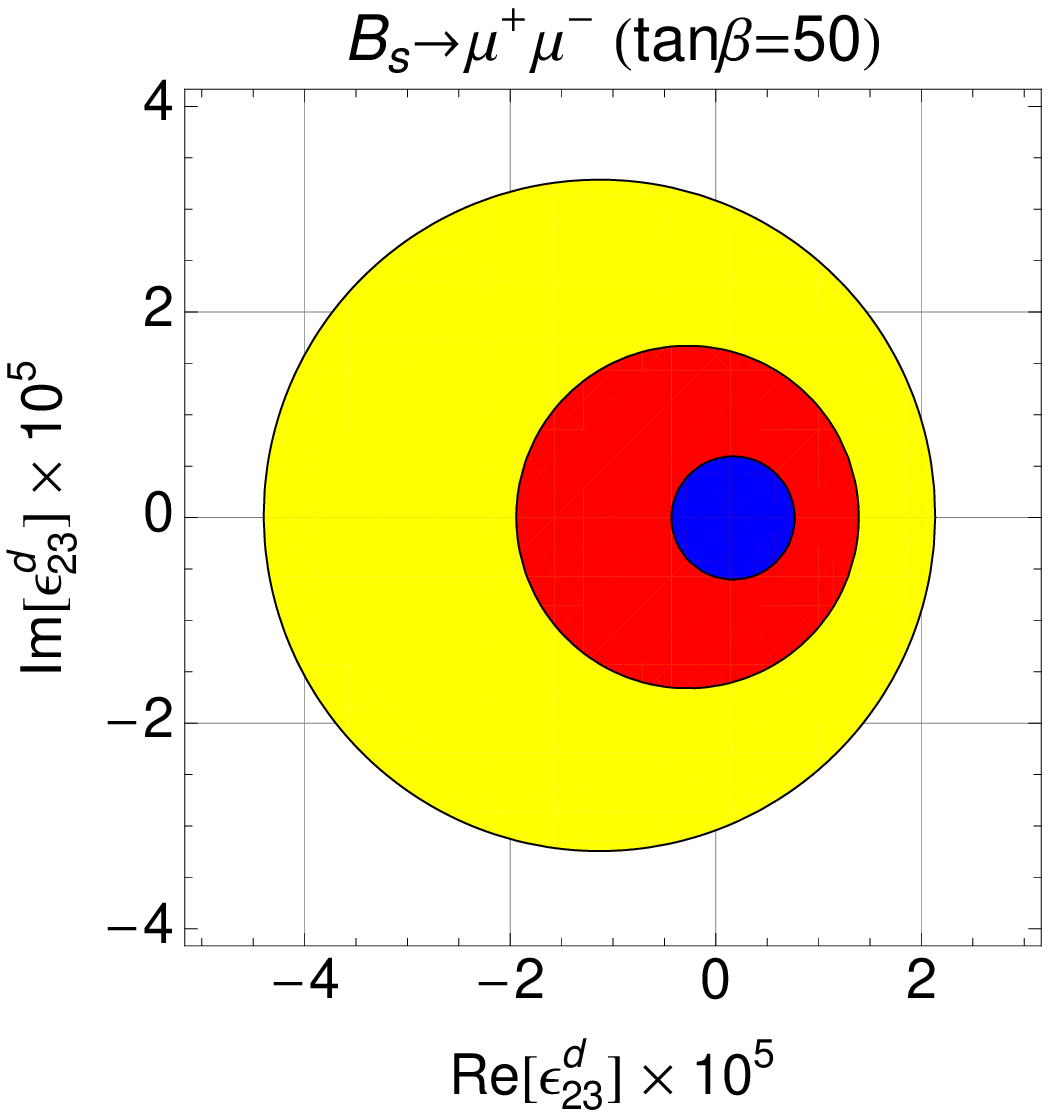}
\caption{Allowed regions in the complex $\epsilon^{d}_{23,32}$--plane from $B_s\to\mu^+\mu^-$ for $\tan\beta=30$, $\tan\beta=50$ and $m_{H}=700\mathrm{~GeV}$ (yellow), $m_{H}=500\mathrm{~GeV}$ (red) and $m_{H}=300\mathrm{~GeV}$ (blue). Note that the allowed regions for $\epsilon^{d}_{32}$--plane are not full circles because in this case a suppression of ${\cal B}\left[B_{s}\to\mu^+\mu^-\right]$ below the experimental lower bound is possible.}
\label{fig:Bstomumu}
\end{figure*}

\begin{figure*}[t]
\centering
\includegraphics[width=0.3\textwidth]{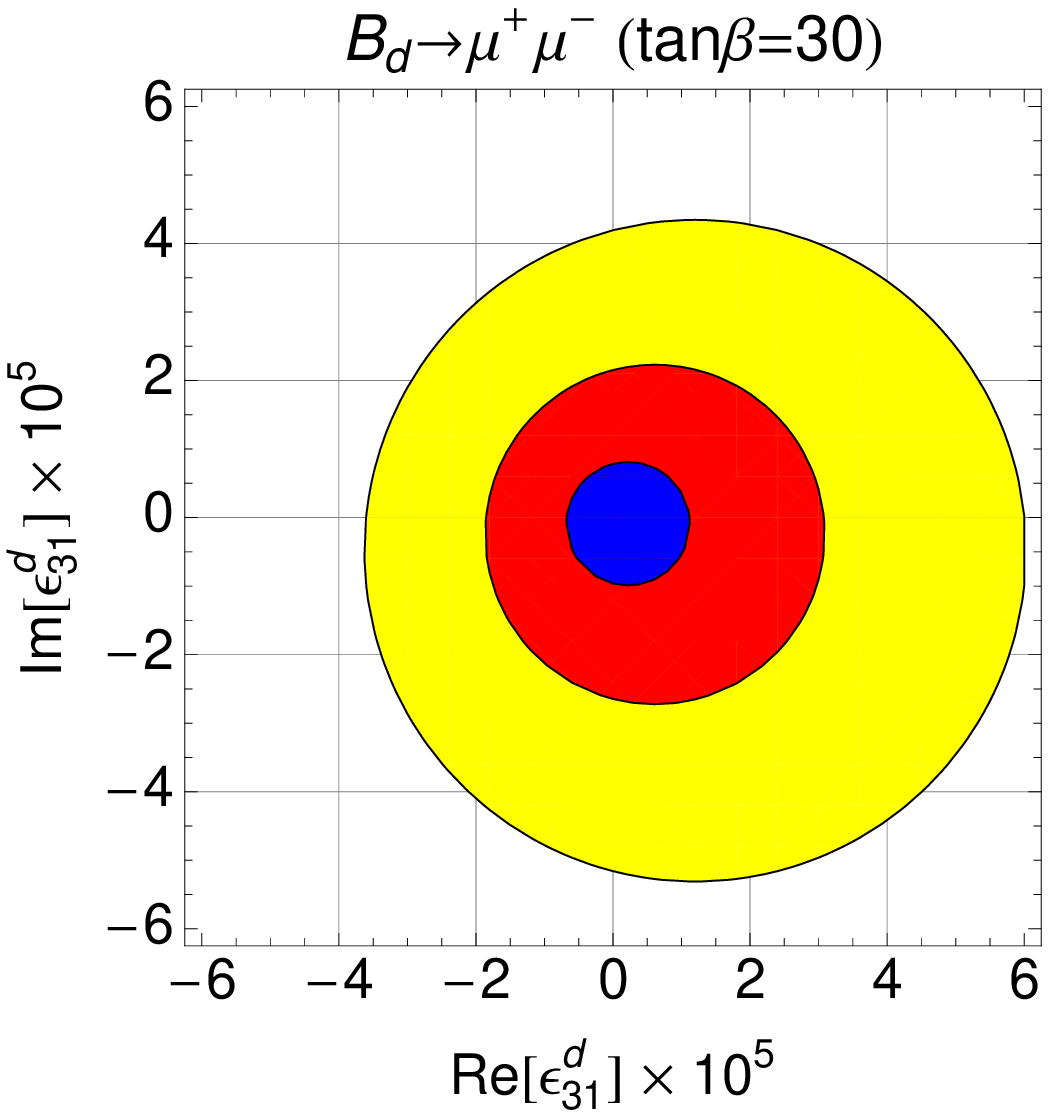}~~~~~
\includegraphics[width=0.3\textwidth]{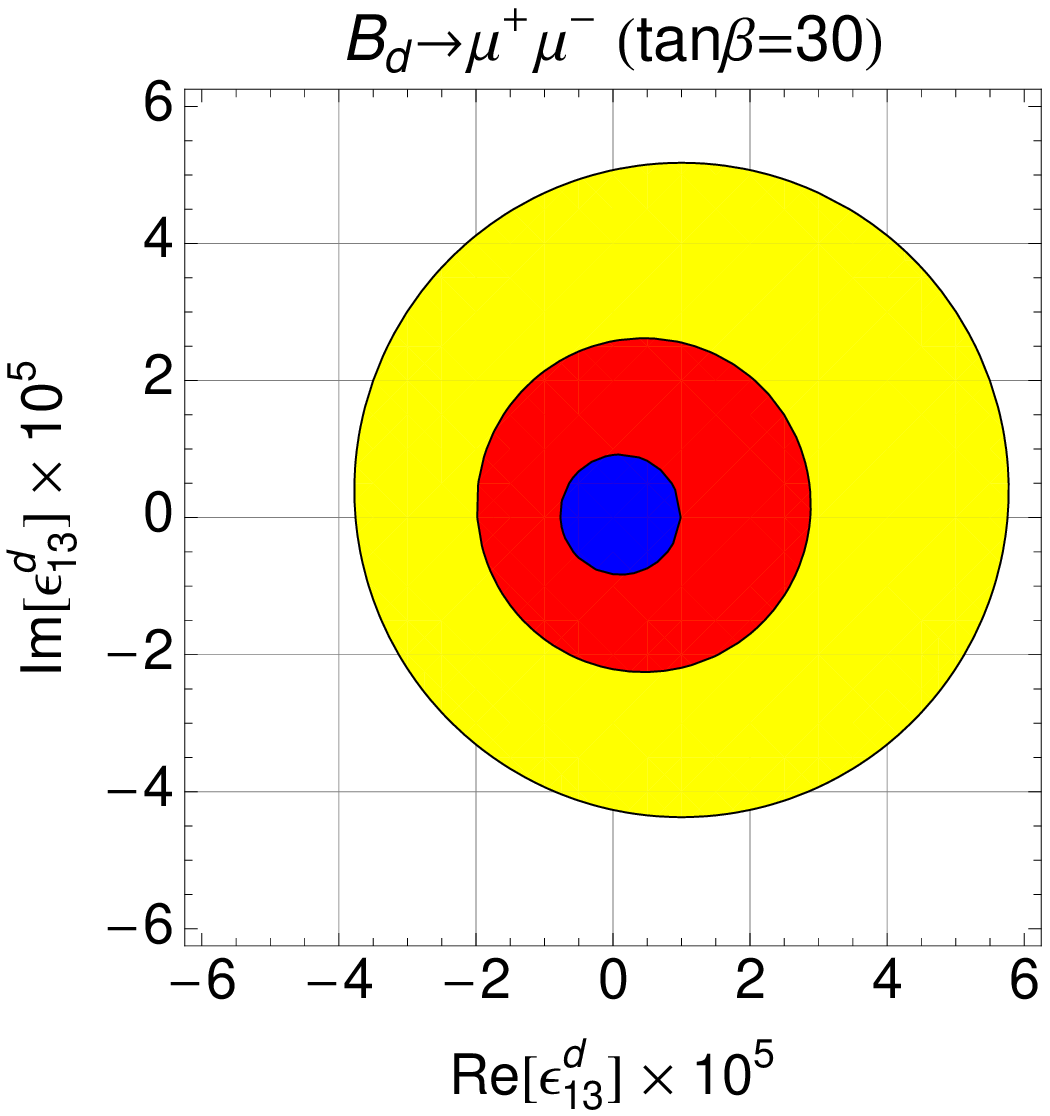} \\ [0.2cm]
\includegraphics[width=0.3\textwidth]{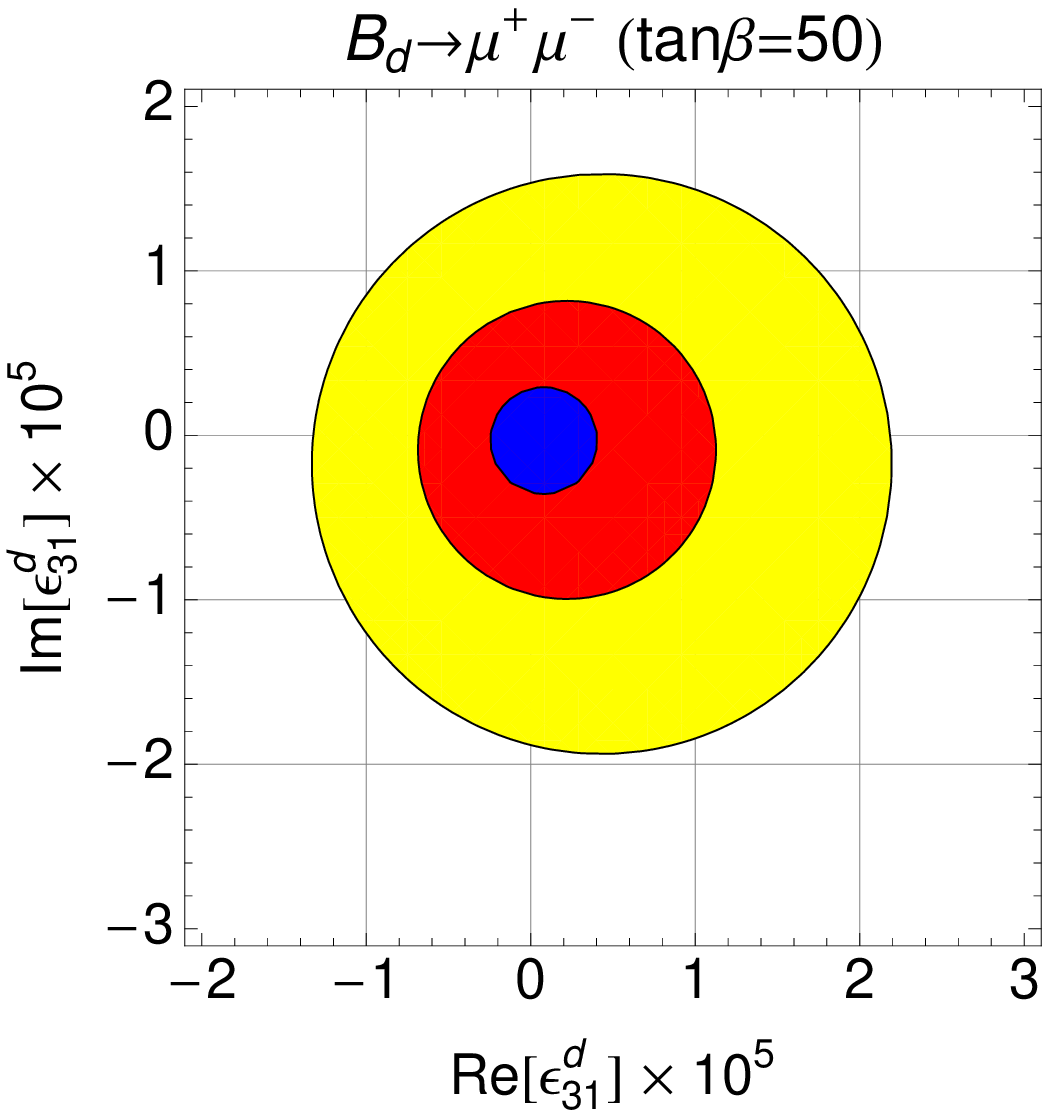}~~~~~
\includegraphics[width=0.3\textwidth]{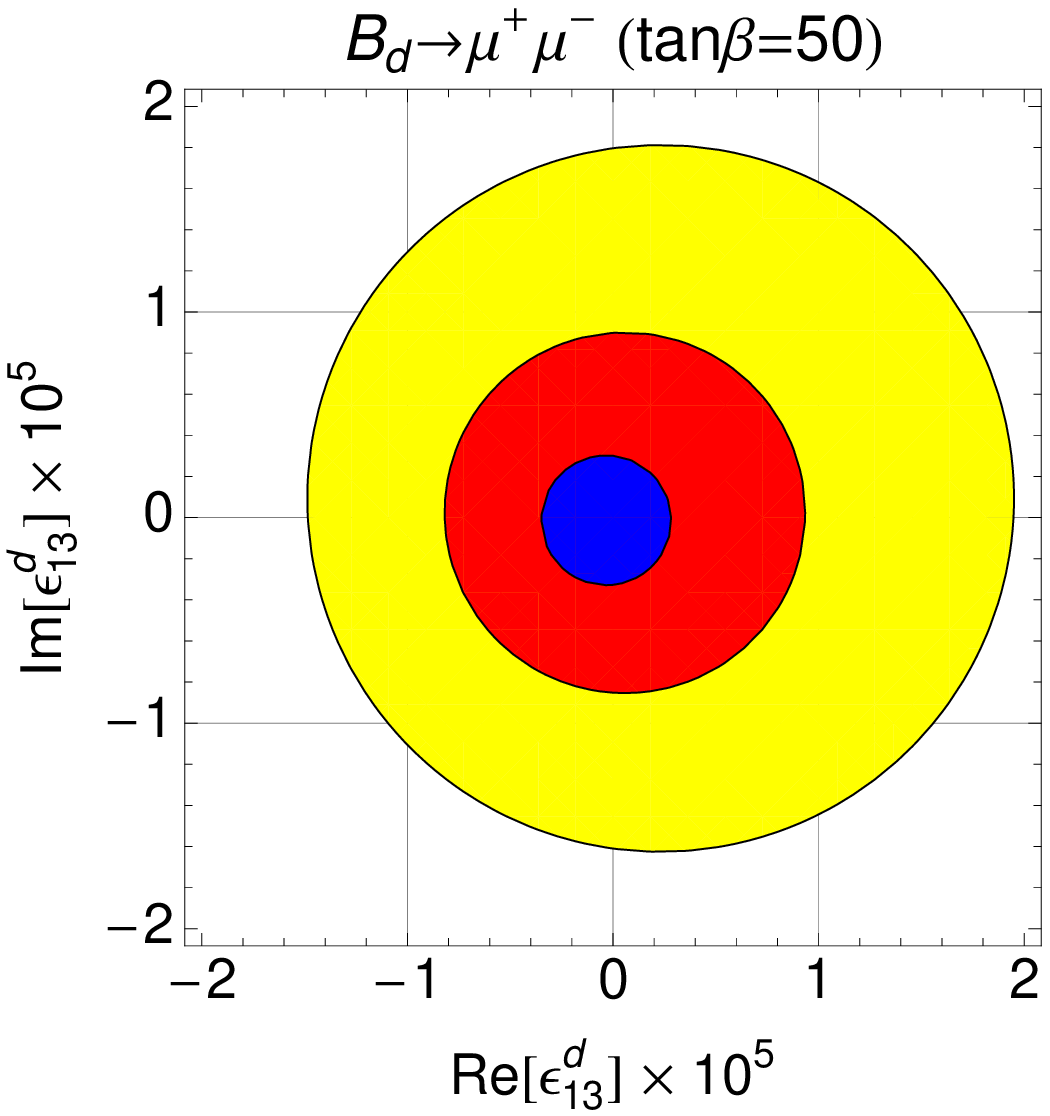}
\caption{Allowed regions in the complex $\epsilon^{d}_{13,31}$--plane  from $ B_d\to\mu^+\mu^-$ for $ \tan\beta=30$, $ \tan\beta=50$ and $m_{H}=700\mathrm{~GeV}$(yellow), $m_{H}=500\mathrm{~GeV}$(red) and $m_{H}=300\mathrm{~GeV}$(blue).}
\label{fig:Bdtomumu}
\end{figure*}

We can now constrain the elements $\epsilon^d_{23,32}$ and $\epsilon^d_{13,31}$ by demanding that the experimental bounds are satisfied within two standard deviations for $B_s\to\mu^+\mu^-$ or equivalently at the 95\% CL concerning $B_d\to\mu^+\mu^-$. The results for the constraints on $\epsilon^d_{23}$ and $\epsilon^d_{32}$ ($\epsilon^d_{13}$ and $\epsilon^d_{31}$) from $B_s\to\mu^+\mu^-$ ($B_d\to\mu^+\mu^-$) are shown in Fig.~\ref{fig:Bstomumu} (Fig.~\ref{fig:Bdtomumu}).
\medskip

All constraints on $\epsilon^d_{13,31}$ and $\epsilon^d_{23,32}$ are very stringent; of the order of $10^{-5}$. Both an enhancement or a suppression of ${\cal B}\left[ B_{d,s}\to\mu^+\mu^-\right]$ compared to the SM prediction is possible. While in the 2HDM II the minimal value for ${\cal B}\left[ B_{d,s}\to\mu^+\mu^-\right]$ is half the SM prediction, in the 2HDM III also a bigger suppression of $B_{d,s}\to\mu^+\mu^-$ is possible if $\epsilon^d_{13,23}\neq0$. In principle, the constraints on $\epsilon^d_{23}$ ($ \epsilon^d_{13}$) from $B_{s(d)}\to\mu^+\mu^-$ are not independent of $\epsilon^d_{32}$ ($ \epsilon^d_{31}$). Anyway, in the next section it will turn out that the constraints from $\Delta F=2$ processes are more stringent if both $\epsilon^d_{32}$ and $ \epsilon^d_{23}$ are different from zero (the same conclusions hold for $\epsilon^d_{31,13}$, $\epsilon^d_{21,12}$ and $\epsilon^u_{21,12}$).
\smallskip

$B_s \to \mu^+\mu^-$ and $B_d \to \mu^+\mu^-$ can also be used to constrain the leptonic parameter $\epsilon^\ell_{22}$. We will discuss the corresponding subject in Sec.\ref{sec:loop-contributions}.

\subsubsection{$K_L\to \mu^{+} \mu^{-}$}

\begin{figure*}[t]
\centering
\includegraphics[width=0.3\textwidth]{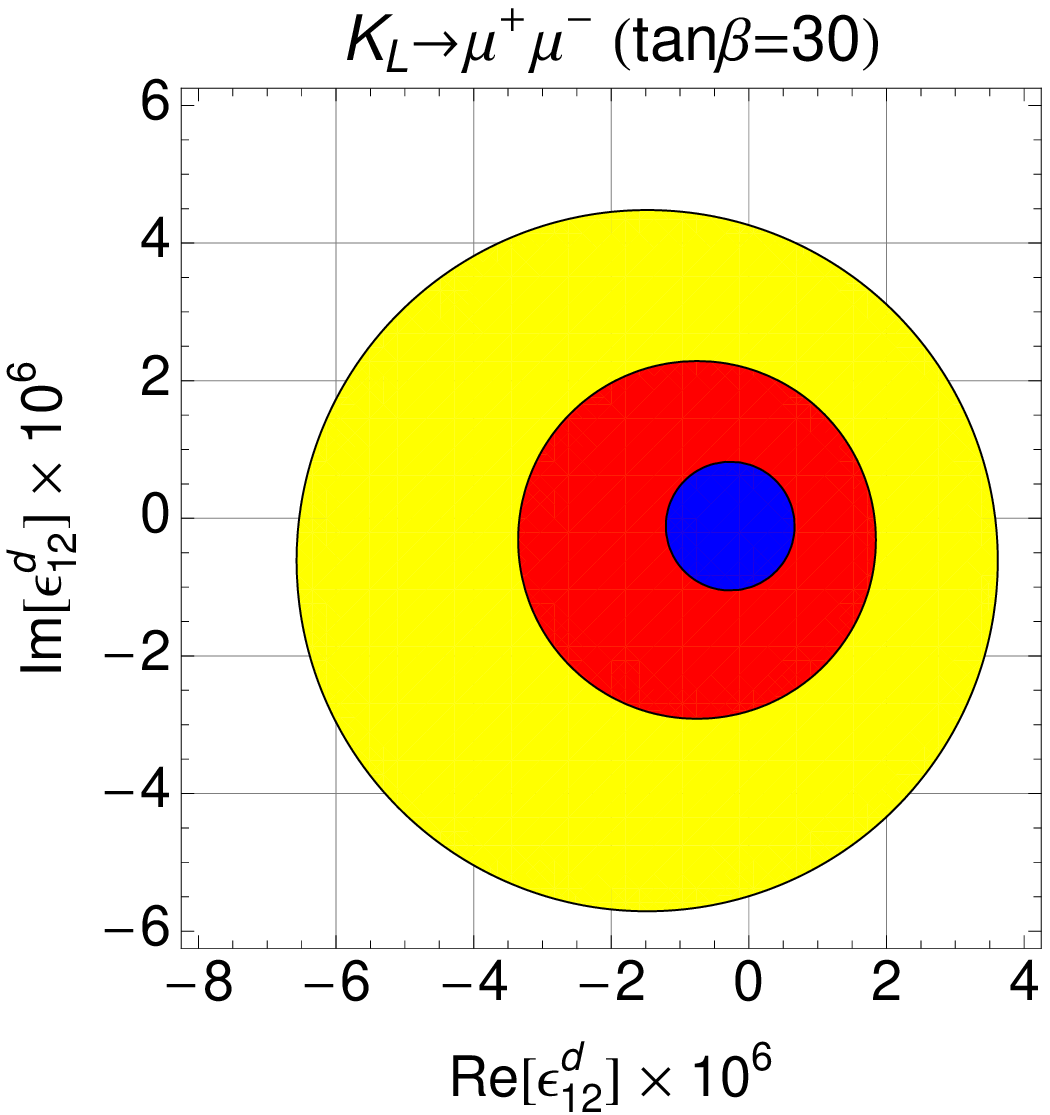}~~~~~
\includegraphics[width=0.3\textwidth]{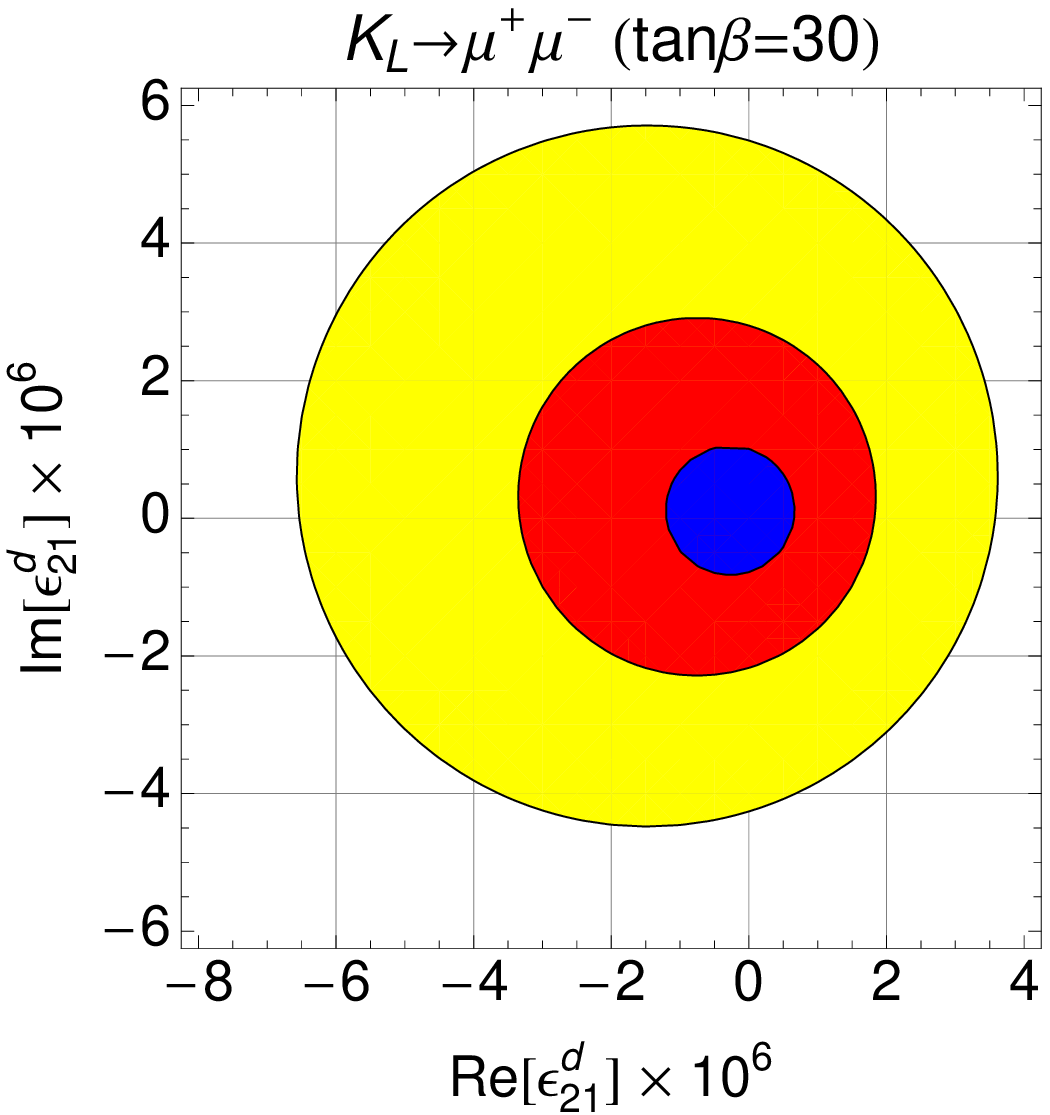} \\ [0.2cm]
\includegraphics[width=0.3\textwidth]{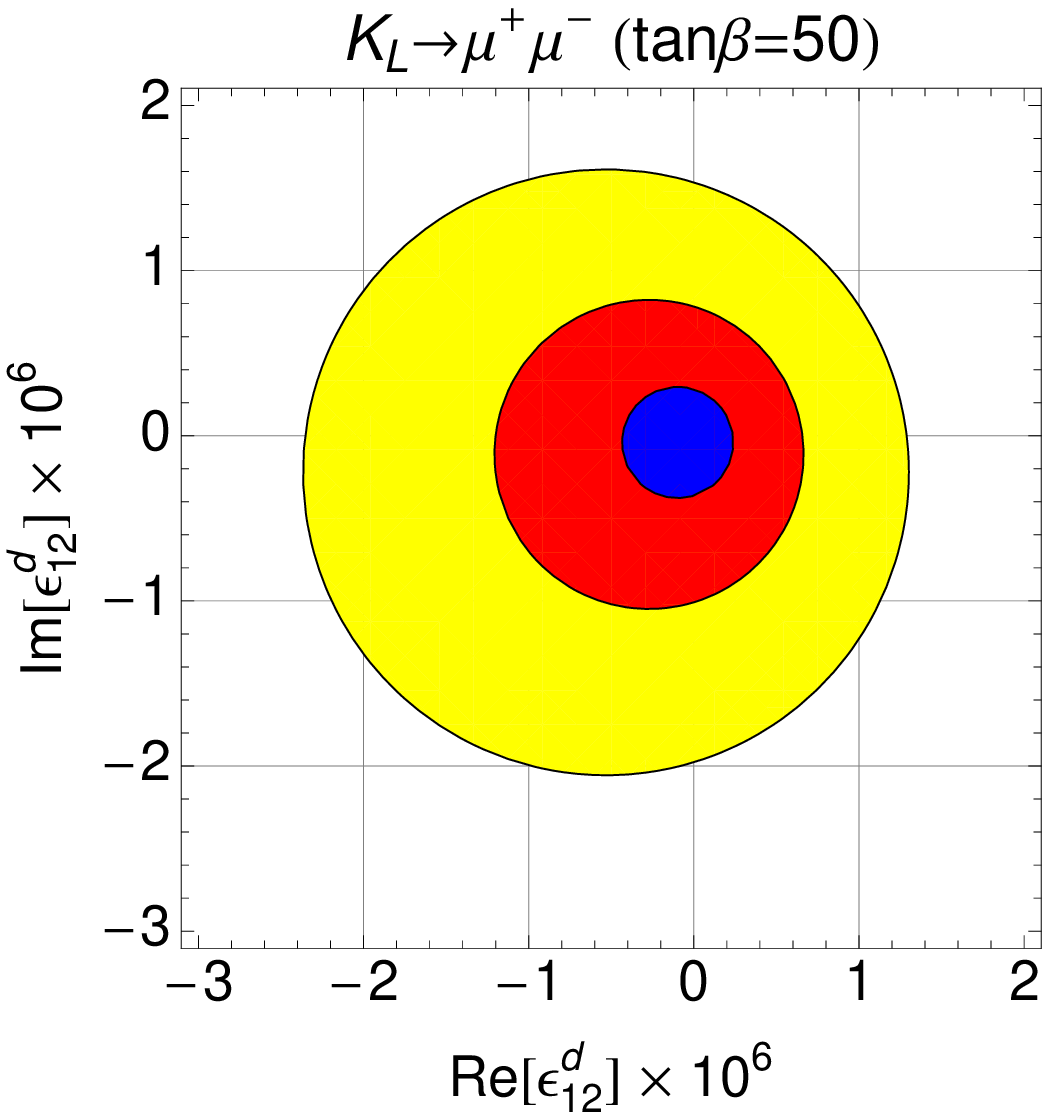}~~~~~
\includegraphics[width=0.3\textwidth]{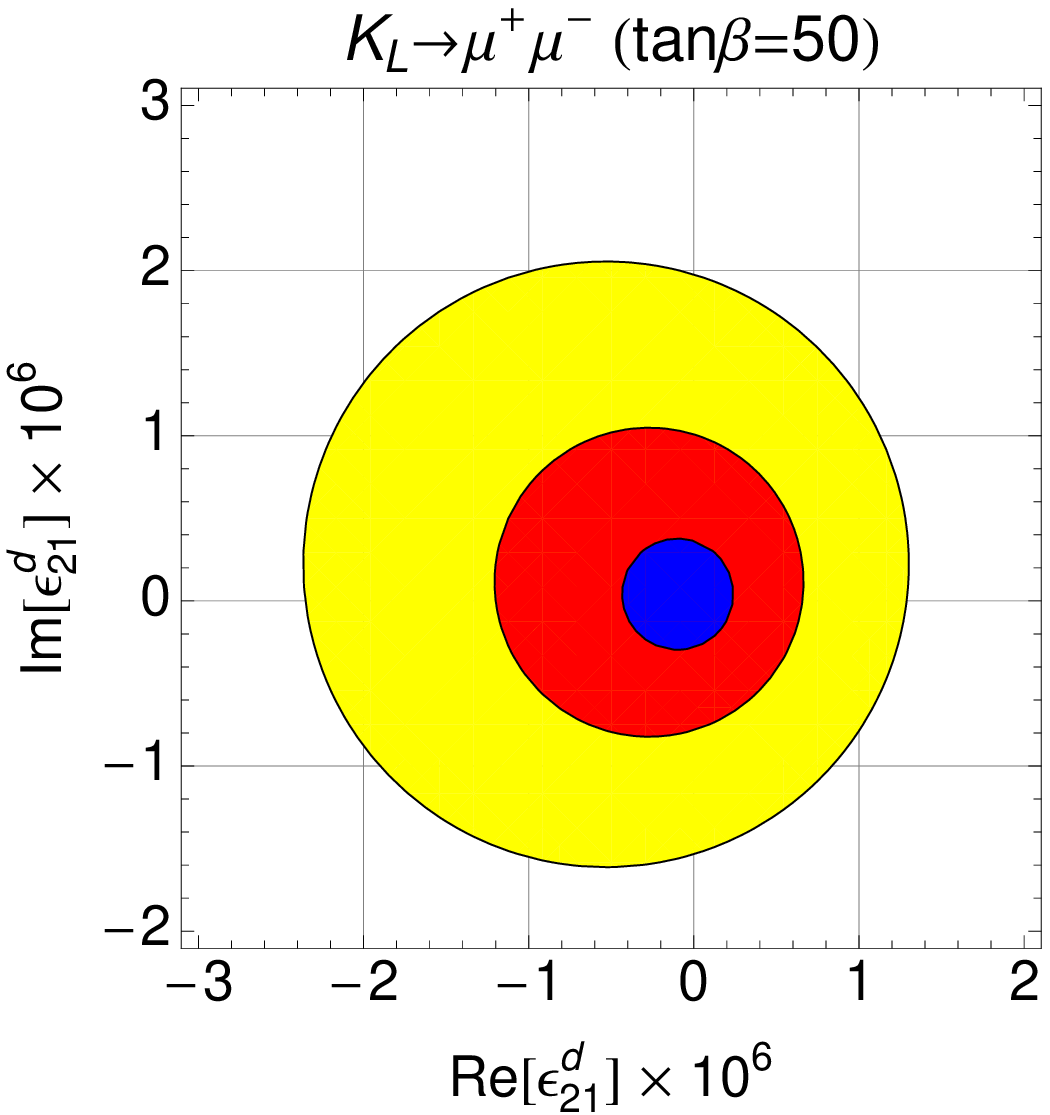}
\caption{Allowed regions in the complex $\epsilon^{d}_{12,21}$-plane from $K_L \to \mu^+\mu^-$ for $ \tan\beta=30$, $ \tan\beta=50$ and $m_{H}=700\mathrm{~GeV}$(yellow), $m_{H}=500\mathrm{~GeV}$(red) and $m_{H}=300\mathrm{~GeV}$(blue).}
\label{fig:KLmumu}
\end{figure*}

Concerning $K_L\to\mu^+\mu^-$, the branching ratio and the Wilson coefficients can be obtained by a simple replacement of indices from \eq{BRBsmumu}, \eq{CASM} and \eq{CSCPBsmumu}. Due to the presence of large non-perturbative QCD effects, we require that the 2HDM III contribution together with the short distance piece of the SM contribution does not exceed the upper limit on the short distance contribution to the branching ratio calculated in Ref.~\cite{Isidori:2003ts}. The resulting constraints on $\epsilon^d_{12,21}$ are shown in Fig.~\ref{fig:KLmumu}. They are found to be extremely stringent (of the order of $10^{-6}$).

\subsubsection{${\bar D}^0\to\mu^+\mu^-$}

\begin{figure*}[t]
\centering
\includegraphics[width=0.4\textwidth]{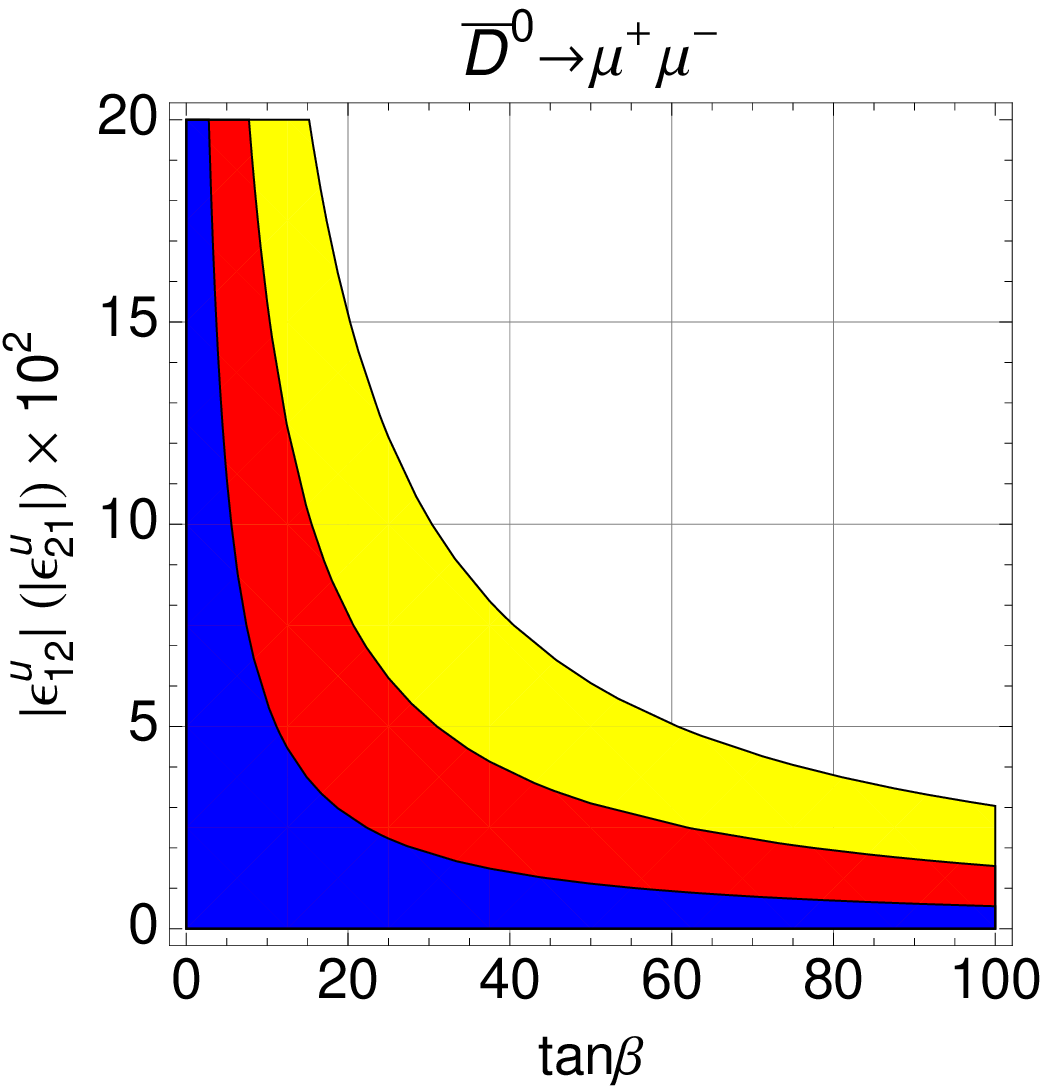}
\caption{Allowed regions in the complex $\epsilon^{u}_{12,21}$--$\tan\beta$ plane from $ {\bar D}^{0} \to \mu^+\mu^-$ for $m_{H}=700\mathrm{~GeV}$ (yellow), $m_{H}=500\mathrm{~GeV}$ (red) and $m_{H}=300\mathrm{~GeV}$ (blue).}
\label{fig:Dmumu}
\end{figure*}

The analogous expressions for the branching ratio for ${\bar D}^0\to\mu^+\mu^-$ (${\bar D}^0(\bar{c}u)$) follow by a straightforward replacement of indices in \eq{BRBsmumu} but the Wilson coefficients in the type-III 2HDM for ${\bar D}^{0}\to\mu^+\mu^-$ have a different dependence on $\tan\beta$:
\begin{equation}
\renewcommand{\arraystretch}{2.0}
\begin{array}{l}
C^{cu}_S  = -C^{cu}_P  = \dfrac{{\pi ^2 }}{{G_F^2 M_W^2 }}\dfrac{1}{{2m_H^2 }}\dfrac{{m_{\ell _2 }-v_u \epsilon^\ell_{22} }}{v}\epsilon _{12}^{u\star} \tan \beta \,, \\ 
\
C^{\prime cu}_S  = C^{\prime cu}_P  = \dfrac{{\pi ^2 }}{{G_F^2 M_W^2 }}\dfrac{1}{{2m_H^2 }}\dfrac{{m_{\ell _2 }-v_u \epsilon^\ell_{22} }}{v}\epsilon _{21}^{u} \tan \beta  \, . \\ 
 \end{array}
\end{equation}
Differently than for $B_{d,s}\to\mu^+\mu^-$ the SM contribution cannot be calculated due to non-perturbative effects and the 2HDM II contribution is numerically irrelevant. Since we do not know the SM contribution, we require that the 2HDM III contribution alone does not generate more than the experimental upper limit on this branching ratio. 
\medskip

It is then easy to express the constraints on $\epsilon^{u}_{12,21} $ in terms of the parameters $m_{H}$ and $\tan\beta$:
\begin{equation}
\left|   \epsilon^u_{12,21}  \right | \; \le \;  3.0 \times 10^{-2}  \dfrac{\left( m_H/500\,  \mathrm{GeV} \right)^2 }{ \tan\beta/50}  \,.
\end{equation}
The resulting bounds on $\epsilon^{u}_{12,21}$ (setting one of these elements to zero) are shown in Fig.~\ref{fig:Dmumu}.

\subsection{Tree-level contributions to $\Delta F=2$ processes}
\label{DeltaF2-processes}

\begin{figure*}[t]
\centering
~~~~~~~~~~~~\includegraphics[width=0.38\textwidth]{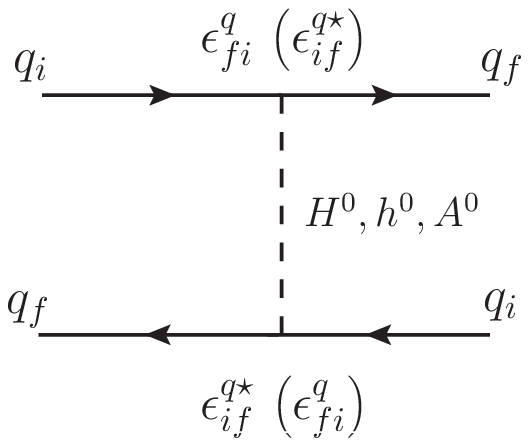}
\caption{Feynman diagram contributing to $B_{d,s}$--$\overline{B}_{d,s}$, $K$--$\overline{K}$ and $D$--$\overline{D}$ mixing.}
\label{mixingH0}
\end{figure*}

In the presence of non-zero elements $\epsilon_{ij}^q$ neutral Higgs mediated contributions to neutral meson mixing ($B_{d,s}$--$\overline{B}_{d,s}$, $K$--$\overline{K}$ and $D$--$\overline{D}$ mixing) arise (see Fig.~\ref{mixingH0}). In these processes, the 2HDM contribution vanishes if the $U(1)_{\rm{PQ}}$ symmetry is conserved. This has the consequence that the leading  $\tan\beta$-enhanced tree-level contribution to the $\Delta F=2$ processes (shown in Fig.~\ref{mixingH0}) is only non-vanishing if $\epsilon_{ij}^q$ and $\epsilon_{ji}^q$ are simultaneously different from zero (in the approximation $m_{A^0}=m_{H^0}$ and $\cot\beta=0$). Making use of the effective Hamiltonian defined in \eq{Heff_DeltaF2} of the appendix we get the following contributions to $B_{s}$--$\overline{B}_{s}$ mixing (the expressions for $B_{d}$--$\overline{B}_{d}$ and $K$--$\overline{K}$ mixing again follow by a simple replacement of indices):
\begin{equation}
C_4  = -\frac{{ \epsilon _{23}^d \epsilon _{32}^{d\star} }}{{m_{H}^2 }}\tan^{2}\beta   \,.
\end{equation}
All other Wilson coefficients are sub-leading in $\tan\beta$. For $D$ mixing, again only $C_4$ is non-zero and given by
\begin{equation}
C_4  = -\frac{{ \epsilon _{12}^u \epsilon _{21}^{u\star} }}{{m_{H}^2 }} \,.
\end{equation}

After performing the renormalization group evolution \cite{Ciuchini:1997bw,Buras:2000if,Bona:2007vi,Becirevic:2001jj,Ciuchini:1998ix} (here we used $\mu_{H}=500~$GeV at the high scale) it turns out that the dominant contribution to the hadronic matrix elements stems from $O_{4}$. Inserting the bag factors \cite{Lubicz:2008am,Becirevic:2001yv} and decay constants from lattice QCD (see Table.~\ref{tab:input3-4}), we get for the 2HDM of type III contribution
\begin{equation}
\renewcommand{\arraystretch}{1.6}
\begin{array}{l}
 \left\langle {B_d^0 } \right|C_4 O_4  \left| {\bar B_d^0 } \right\rangle  \approx 0.26\;C_4\,{\rm GeV}^3 \,,\\ 
 \left\langle {B_s^0 } \right|C_4  O_4  \left| {\bar B_s^0 } \right\rangle  \approx 0.37\;C_4\,{\rm GeV}^3 \,,\\ 
 \left\langle {K^0 } \right|C_4 O_4 \left| {\bar K^0 } \right\rangle  \approx 0.30\;C_4\,{\rm GeV}^3   \,,\\ 
 \left\langle {D^0 } \right|C_4  O_4 \left| {\bar D^0 } \right\rangle  \approx 0.18\;C_4\,{\rm GeV}^3  \,, \\ 
 \end{array}
\label{DeltaF2C4}
\end{equation}
where, we used the normalization of the meson states as defined for example in~\cite{Becirevic:2001jj}. In \eq{DeltaF2C4} the Wilson coefficients within the matrix elements are at the corresponding meson scale while $C_4$ on the right-handed side is given at the matching scale $m_H$. For computing the constraints on $\epsilon_{13}^d \epsilon_{31}^{d\star}$, $\epsilon_{23}^d \epsilon _{32}^{d\star}$ and $\epsilon_{12}^d  \epsilon _{21}^{d\star}$ we use the online update of the analysis of the UTfit collaboration \cite{Ciuchini:2000de}\footnote{See also the online update of the CKMfitter group for an analogous analysis~\cite{Charles:2004jd}.}. For this purpose we define
\begin{equation}
C_{B_q } e^{2i\varphi _{B_q } }  = 1 + \frac{{\left\langle {B_q^0 } \right| {\cal H}_{eff}^{NP} \left| {\bar B_q^0 } \right\rangle }}{{\left\langle {B_q^0 } \right| {\cal H}_{eff}^{SM} \left| {\bar B_q^0 } \right\rangle }}\,,
\end{equation}
for \bbd and \bbs mixing and 
\begin{equation}
\renewcommand{\arraystretch}{1.8}
\begin{array}{l}
C_{\epsilon_K} = 1 + \dfrac{{\imag \left[ \left\langle {K^0 } \right| {\cal H}_{eff}^{NP} \left| {\bar K^0 } \right\rangle \right]}}{{\imag\left[\left\langle {K^0 } \right| {\cal H}_{eff}^{SM} \left| {\bar K^0 } \right\rangle \right] }}\,,\\ [0.25cm]
C_{\Delta M_K} = 1 + \dfrac{{\real \left[ \left\langle {K^0 } \right| {\cal H}_{eff}^{NP} \left| {\bar K^0 } \right\rangle \right]}}{{\real\left[\left\langle {K^0 } \right| {\cal H}_{eff}^{SM} \left| {\bar K^0 } \right\rangle \right] }}\,,
 \end{array}
\end{equation}
for \kk mixing. Using for the matrix elements of the SM Hamiltonian\footnote{To obtain a value consistent with the NP analysis of the UTfit collaboration, we also used their input for computing the matrix elements of the SM $\Delta F=2$ Hamiltonian in \eq{DeltaF2SM}.}~\cite{Lenz:2010gu}
\begin{equation}
\renewcommand{\arraystretch}{1.6}
\begin{array}{l}
 \left\langle {B_d^0 } \right| {\cal H}^{\Delta F=2}_{{SM}} \left| {\bar B_d^0 } \right\rangle  \approx \left( {1.08 + 1.25 i} \right) \times 10^{ - 13}  \, {\rm GeV} \,,\\ 
 \left\langle {B_s^0 } \right| {\cal H}^{\Delta F=2}_{{SM}} \left| {\bar B_s^0 } \right\rangle  \approx  \left( 59 -2.2 i \right) \times 10^{ - 13}  \, {\rm GeV}  \,,\\ 
  \left\langle {K^0 } \right| {\cal H}^{\Delta F=2}_{{SM}} \left| {\bar K^0 } \right\rangle  \approx  \left( 115+1.16 i \right) \times 10^{-17}    \, {\rm GeV}  \,,
 \end{array}
 \label{DeltaF2SM}
\end{equation}
we can directly read off the bounds on $C_4$ and thus on $\epsilon_{12}^d  \epsilon _{21}^{d\star}$,  $\epsilon_{13}^d \epsilon _{31}^{d\star}$ and $\epsilon_{23}^d  \epsilon _{32}^{d\star}$:
\begin{eqnarray}
-2.0 \times 10^{-10}  \, \leq && {\rm Re} \left[ \epsilon^d_{23}  \epsilon^{d\star}_{32} \right]   \left(\dfrac{ \tan\beta/50}{  m_H/500 \,   \mathrm{GeV}} \right)^2 \,    \leq \, 6.0 \times 10^{-10}  \,, \\
-3.0\times 10^{-10} \,  \,  \leq  && {\rm Im } \left[  \epsilon^d_{23}  \epsilon^{d\star}_{32} \right]  \, \left(\dfrac{ \tan\beta/50}{  m_H/500  \,  \mathrm{GeV}} \right)^2 \,   \leq  7.0\times 10^{-10} \,   \,,  \\
-3.0 \times 10^{-11} \,  \leq &&  \rm Re \left[  \epsilon^d_{13}  \epsilon^{d\star}_{31} \right]  \left(\dfrac{ \tan\beta/50}{  m_H/500  \,  \mathrm{GeV}} \right)^2 \,   \leq 1.5 \times 10^{-11}   \,, \\
-1.5 \times 10^{-11}  \,  \leq && \rm Im \left[  \epsilon^d_{13}  \epsilon^{d\star}_{31} \right]  \left(\dfrac{ \tan\beta/50}{  m_H/500  \,  \mathrm{GeV}} \right)^2 \,   \leq 2.5 \times 10^{-11} \,,  \\
-1.0\times 10^{-12} \,  \,  \leq && \rm Re \left[  \epsilon^d_{12}  \epsilon^{d\star}_{21} \right]   \left(\dfrac{ \tan\beta/50}{  m_H/500  \,  \mathrm{GeV}} \right)^2 \,    \leq 3.0\times 10^{-13}  \,,  \\
-4.0\times 10^{-15} \,  \leq && \rm Im \left[  \epsilon^d_{12}  \epsilon^{d\star}_{21} \right]   \left(\dfrac{ \tan\beta/50}{  m_H/500  \,  \mathrm{GeV}} \right)^2 \,   \leq  2.5\times 10^{-15}   \, .
\end{eqnarray}
We see that if $\epsilon^d_{ij}$ is of the same order as $\epsilon^d_{ji}$ these bound are even more stringent than the ones from $B_{d,s}\to \mu^+\mu^-$ and $K_L\to \mu^+\mu^-$ computed in the last subsection. 
\medskip

\begin{figure*}[t]
\centering
\includegraphics[width=0.34\textwidth]{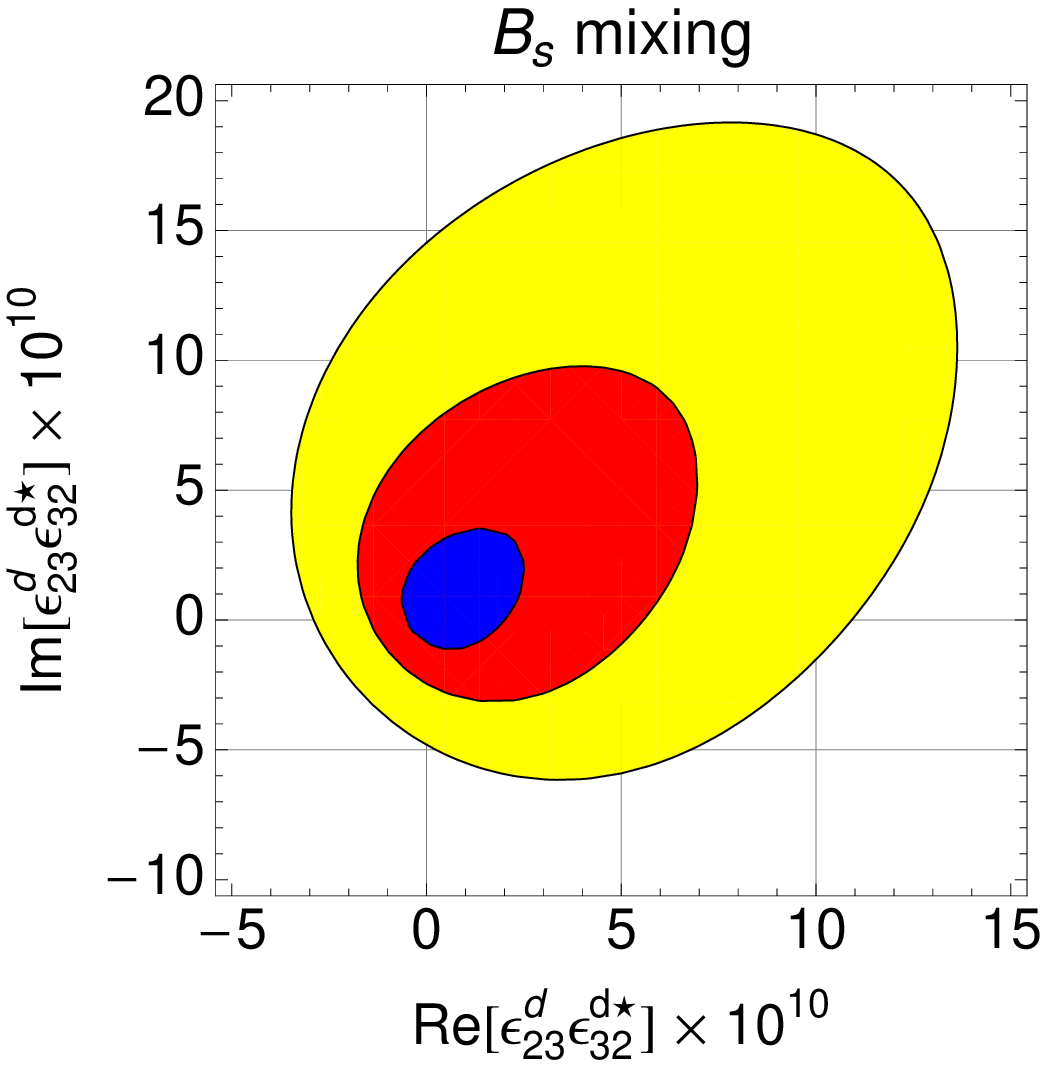}~~~~~
\includegraphics[width=0.35\textwidth]{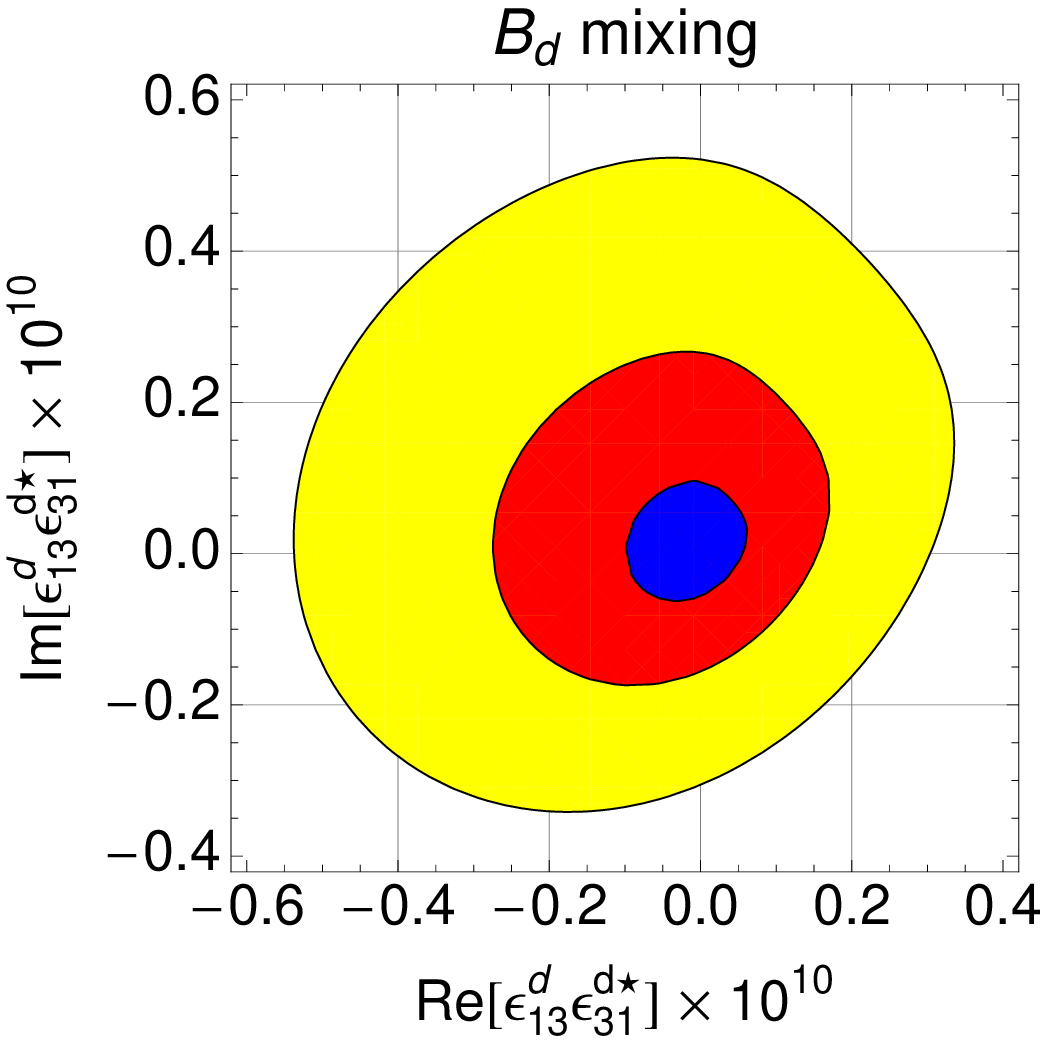}
\caption{ Allowed regions in the complex $\epsilon^{d}_{ij}$-plane from $B_{d,s}$-$\overline{B}_{d,s}$ mixing for $\tan\beta=50$  and $m_{H}=700 $ GeV (yellow), $m_{H}=500$ GeV (red) and  $m_{H}=300$ GeV (blue). 
\label{mixingplots}}
\end{figure*}

\begin{figure*}[t]
\centering
\includegraphics[width=0.373\textwidth]{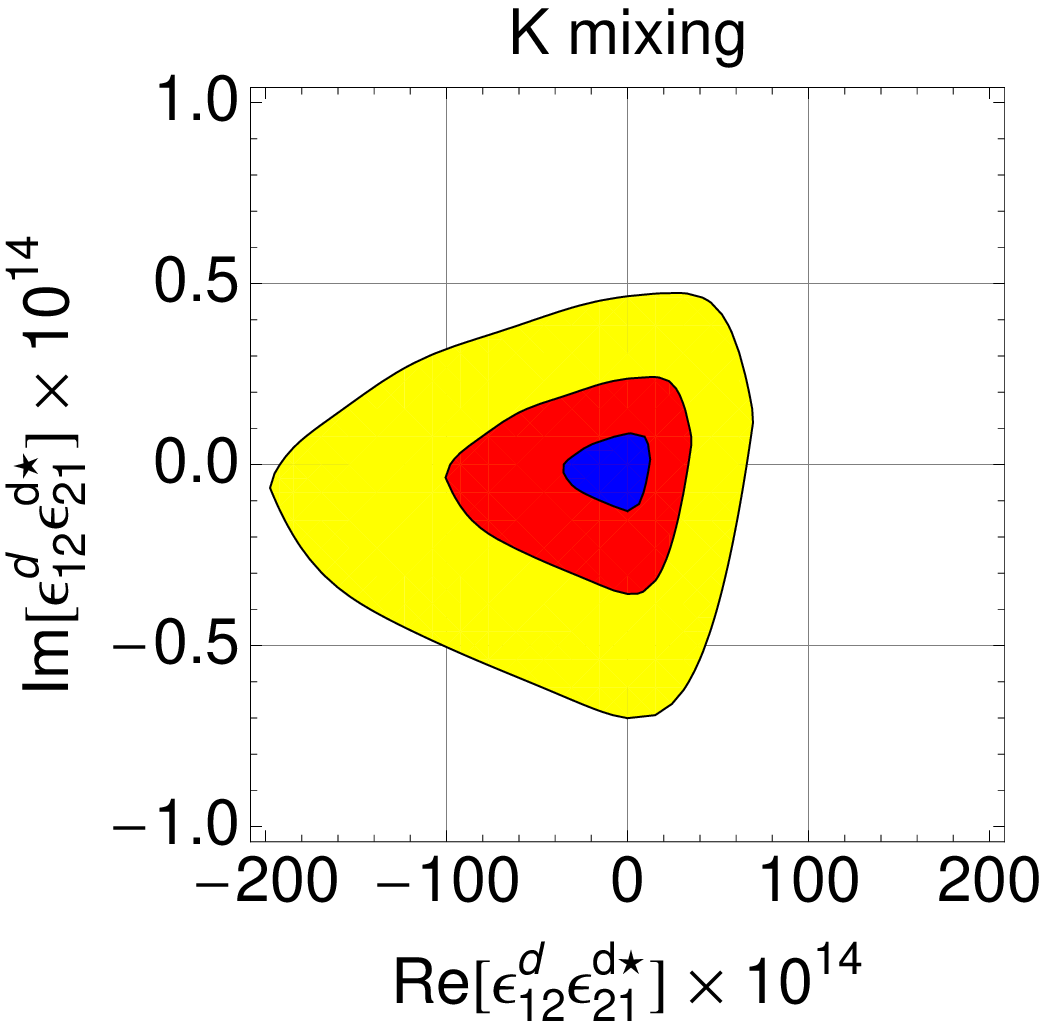}~~~~
\includegraphics[width=0.354\textwidth]{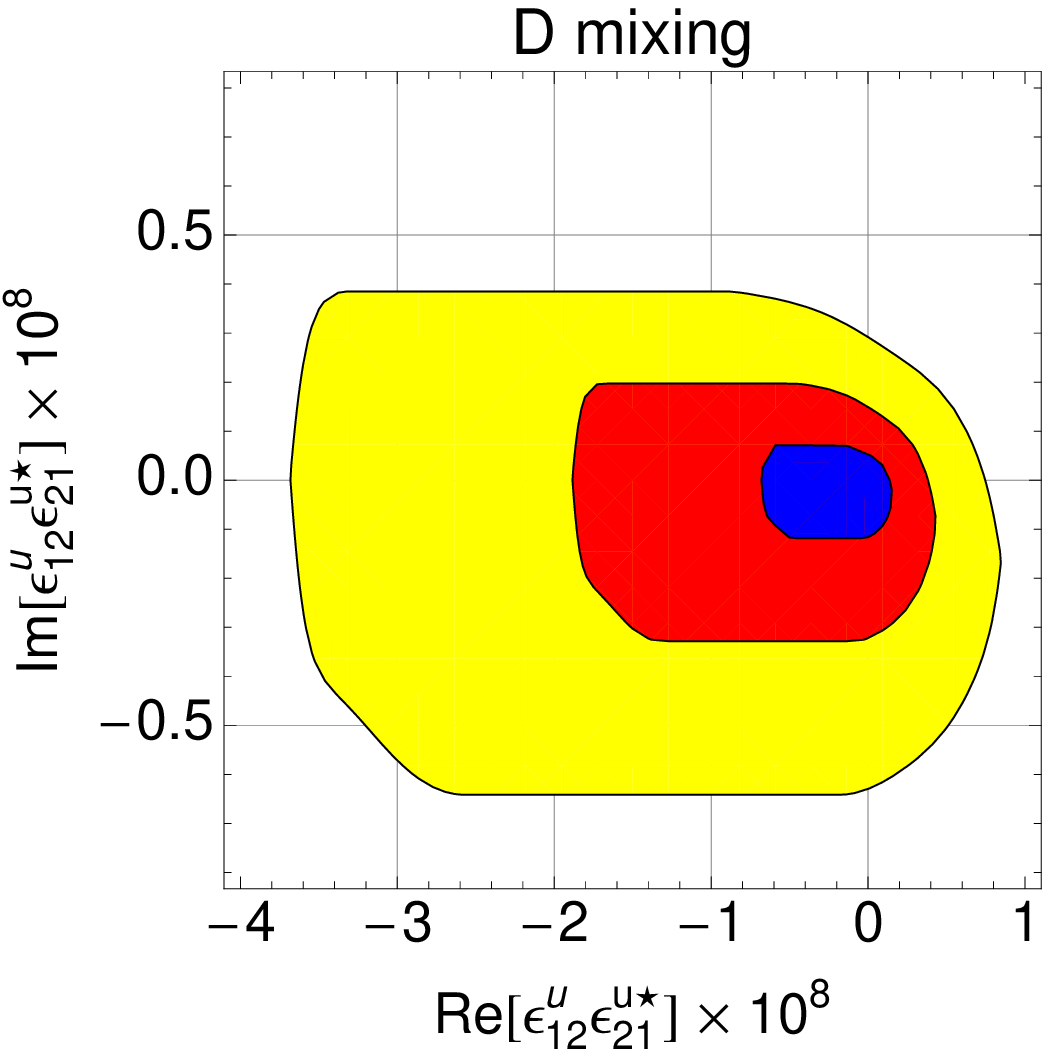}
\caption{Allowed regions in the complex $\epsilon^{q}_{12}\epsilon^{q\star}_{21}$--plane from \kk and \dd mixing for $\tan\beta=50$ and $m_{H}=700$~GeV (yellow), $m_{H}=500$~GeV (red) and  $m_{H}=300$~GeV (blue).}
\label{mixingplotsKD}
\end{figure*}

For \dd mixing, the SM predictions is not known due to very large hadronic uncertainties. In order to constrain the NP effects we demand the absence of fine tuning, which means that the NP contribution, which are calculable short distance contributions, should not exceed the measured values. Concerning the 2HDM III contribution, there is no ${\tan\beta}$ enhancement and taking into account the recent analysis of UTfit collaboration \cite{UTfit:2012zm} we arrive at the following constraints (for $m_{H}=500$ GeV):

\be
\left| \epsilon^u_{12} \epsilon^{u\star}_{21} \right| <  2.0 \times 10^{-8} \,.
\ee
Note that although these bounds look more stringent than the corresponding $\Delta F=1$ constraints, they scale differently with $\tan\beta$ and also involve products of pairs of $\epsilon^{u}_{ij}$. Therefore, contrary to the $\Delta F=1$ case, in principle all of these limits can be evaded for one of the couplings by suppressing the other one. Fig.~\ref{mixingplots} and Fig.~\ref{mixingplotsKD} show the allowed regions for these parameters obtained from neutral Higgs contribution to $B_{d,s}$--$\overline{B}_{d,s}$, $K$--$\overline{K}$ and $D$--$\overline{D}$ mixing (see the Feynman diagram in Fig.~\ref{mixingH0}).
\medskip

\subsection{Lepton-flavor-violating decays: $\tau^- \to \mu^-\mu^+\mu^-$, $\tau^- \to e^- \mu^+\mu^-$ and $\mu \to e^-e^+e^-$ }
\label{taumumumu}

In this section, we investigate the constraints that $\tau^- \to \mu^-\mu^+\mu^-$, $\tau^- \to e^- \mu^+\mu^-$ and $\mu \to e^-e^+e^-$ place on the flavor changing couplings $\epsilon^{\ell}_{32,23}$, $\epsilon^{\ell}_{31,13}$ and $\epsilon^{\ell}_{21,12}$, respectively.
\begin{figure}[t]
\centering
\includegraphics[width=0.42\textwidth]{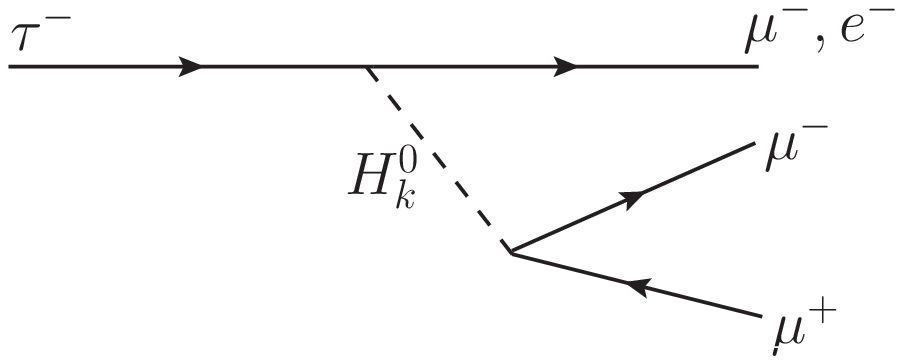}~~~~~
\includegraphics[width=0.42\textwidth]{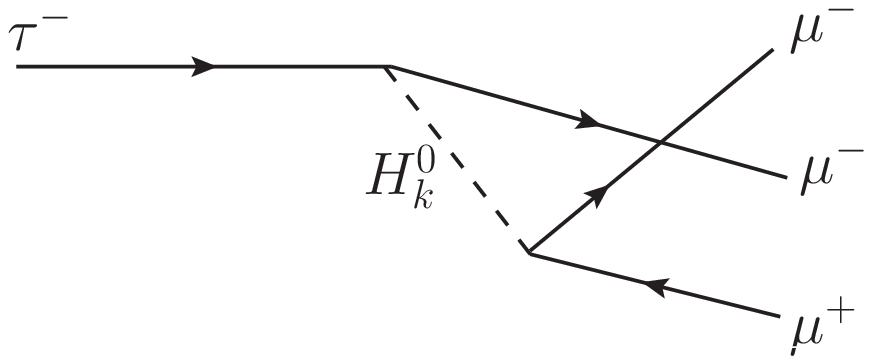}
\caption{Feynman diagrams contributing to $\tau^- \to \mu^-\mu^+ \mu^-$ and $\tau^- \to e^{-} \mu^+ \mu^-$ via neutral Higgs exchange. Note that for $\tau^- \to \mu^-\mu^+\mu^-$ (or $\mu \to e^-e^+e^-$) two distinct diagrams exist which come with a relative minus sign due to the exchange of the two fermion lines.}
\label{tau3mudiag}
\end{figure}

\begin{figure}[t]
\centering
\includegraphics[width=0.32\textwidth]{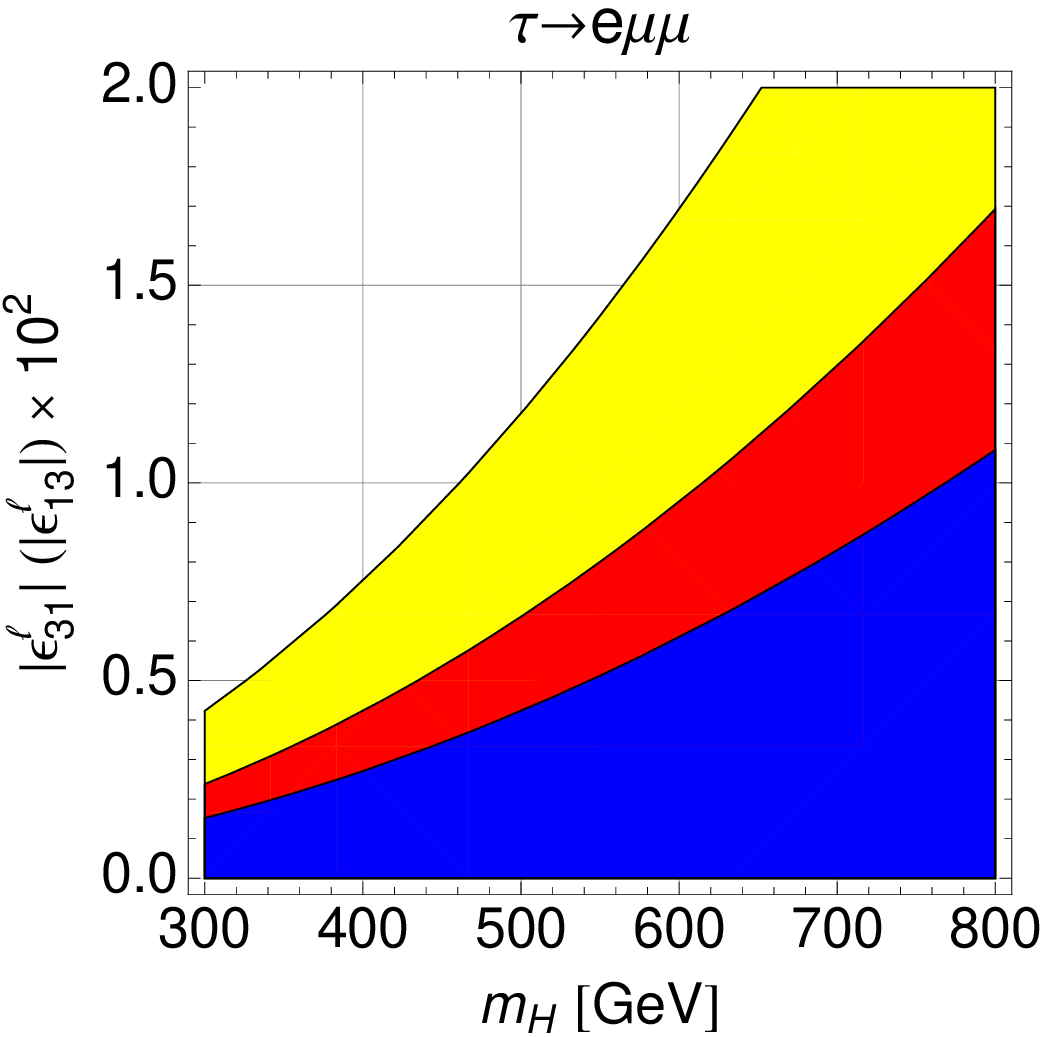}~
\includegraphics[width=0.32\textwidth]{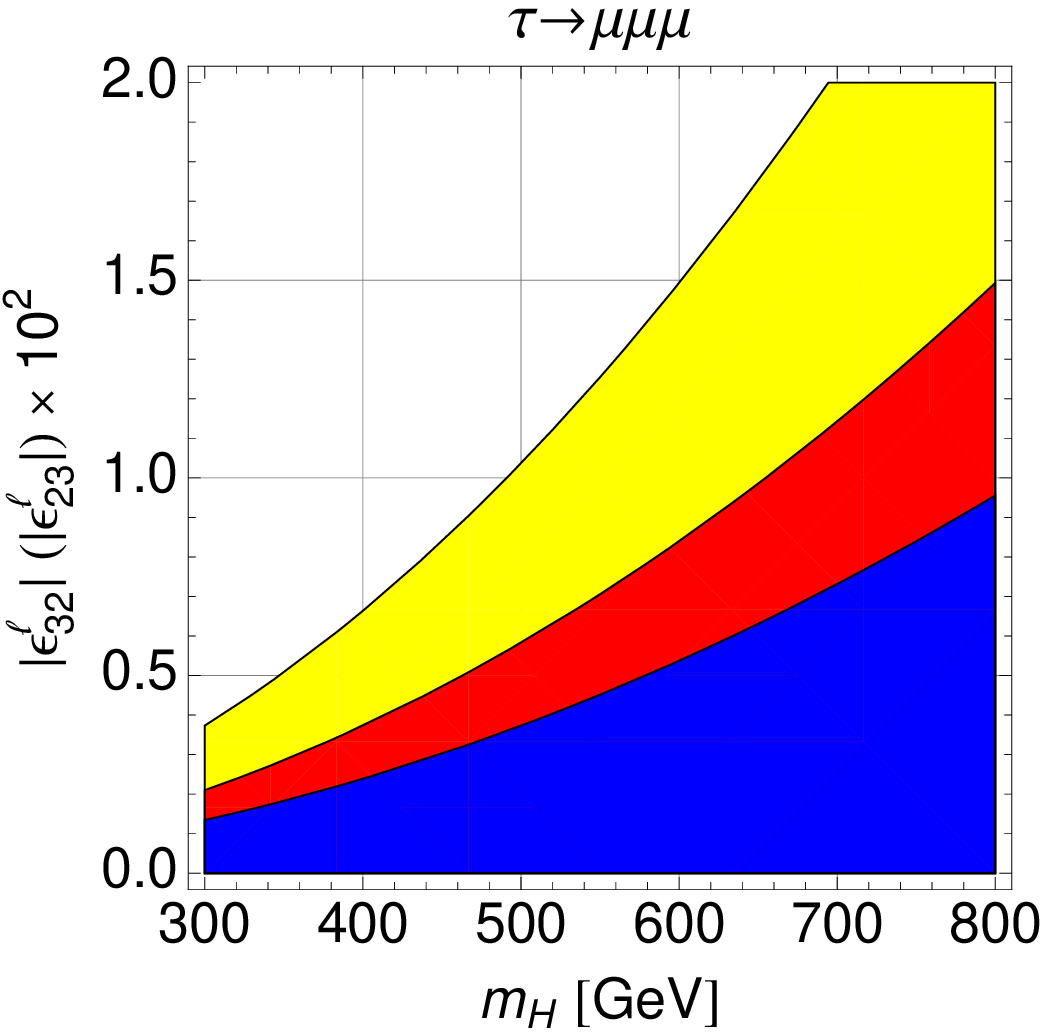}~
\includegraphics[width=0.318\textwidth]{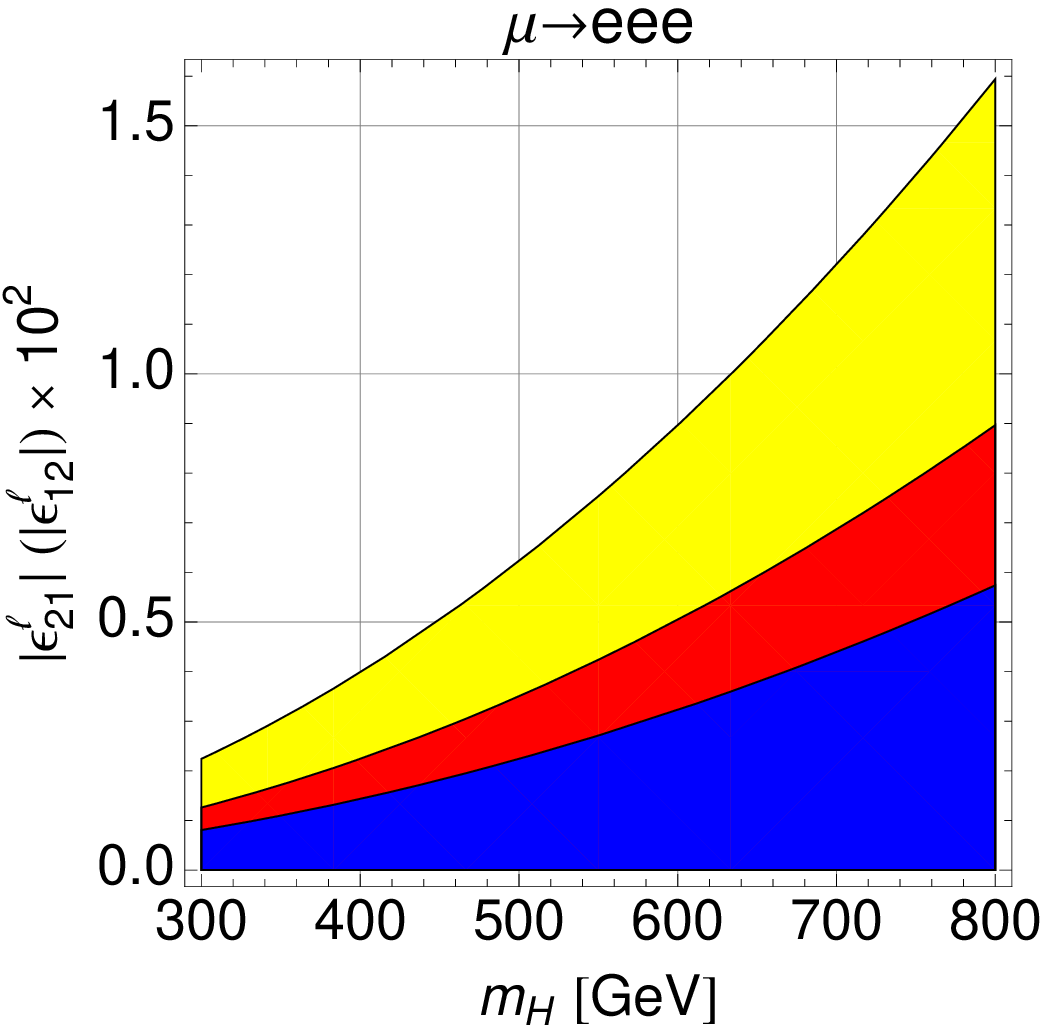}
\caption{Allowed regions for the absolute value of $\epsilon^{\ell}_{13,31}$, $\epsilon^{\ell}_{23,32}$ and $\epsilon^{\ell}_{12,21}$ for $\tan\beta=30$ (yellow), $\tan\beta=40$ (red) and $\tan\beta=50$ (blue) from $\tau^-\to e^-\mu^+\mu^-$, $\tau^-\to \mu^-\mu^+\mu^-$ and $\mu^-\to e^-e^+e^-$, respectively. In each plot only one of the elements $\epsilon^{\ell}_{if}$ or $\epsilon^{\ell}_{fi}$ is assumed to be different from zero.}
\label{fig:TauToMuMuMu}
\end{figure}
For these decays, the experimental upper limits \cite{Hayasaka:2010np,Bellgardt:1987du} are
\begin{equation}
\renewcommand{\arraystretch}{1.6}
\begin{array}{l}
	{\cal B}\left[\tau^-\to\mu^-\mu^+\mu^-\right] \leq 2.1\times 10^{-8}\,,  \\
		{\cal B}\left[\tau^-\to e^- \mu^+\mu^- \right] \leq 2.7\times 10^{-8}\,,   \\
		{\cal B}\left[\mu^-\to e^-e^+e^- \right] \leq 1.0\times 10^{-12}\,,
\end{array}
\label{taumumumuExpBounds}
\end{equation}
at 90\% CL. Let us consider the processes $\tau^-\to \mu^-\mu^+\mu^-$ and $\tau^-\to e^-\mu^+\mu^-$ which are shown in Fig.~\ref{tau3mudiag}. The expressions for the branching-ratio for $\tau^-  \to e^-\mu^+ \mu^-$ can be written as

\begin{equation}
{\cal B} \left[ {\tau^-  \to e^-\mu^+ \mu^- } \right] = \frac{{m_\tau ^5}}{{12{{\left( {8\pi } \right)}^3}{\Gamma _\tau }}}\frac{{{{\tan }^4}\beta }}{{m_H^4}}{\left| {\left( {\frac{{{m_\mu }}}{v} - \varepsilon _{22}^\ell } \right)} \right|^2}\left( {{{\left| {\varepsilon _{31}^\ell } \right|}^2} + {{\left| {\varepsilon _{13}^\ell } \right|}^2}} \right)	
\label{Br:leptondecays}
\end{equation}
\\
where, $\Gamma_{\tau}$ is the total decay width of the $\tau$-lepton. The branching ratios for $\tau^-\to e^-e^+e^-$ and $\mu^-\to e^-e^+e^-$ can be obtained by an obvious replacement of masses, indices and total decays widths. Note that the full expression for general scalar couplings given in \eq{taumumumuFull} of the appendix is different for $\tau^-  \to e^-\mu^+ \mu^-$ than for $\tau^-  \to \mu^-\mu^+ \mu^-$ and only approaches a common expression in the limit of large $\tan\beta$ and large Higgs masses. 
\medskip

Comparing the type-III 2HDM expression with experiment we obtain the following constraints on $\epsilon^{\ell}_{fi}$ (assuming $\epsilon^{\ell}_{jj} = 0$)   
\begin{equation}
\renewcommand{\arraystretch}{2}
\begin{array}{l}
\left |\epsilon^{\ell}_{12} \right|^2+\left |\epsilon^{\ell}_{21} \right|^2   \le   \,\left( 2.3 \times 10^{-3}\right)^2 \, \left(\dfrac{m_{H}/500\,{\rm GeV}}{\tan\beta/50} \right )^{4}  \dfrac{{\cal B}\left[\mu^-\to e^-e^+e^- \right] }{1.0\times 10^{-12}}\, ,  \\
\left |\epsilon^{\ell}_{13}  \right|^2+\left |\epsilon^{\ell}_{31}  \right|^2  \le   \, \left(4.2 \times 10^{-3} \right)^2\, \left(\dfrac{m_{H}/500\,{\rm GeV}}{\tan\beta/50} \right )^{4}  \dfrac{{\cal B}\left[\tau^-\to e^- \mu^+\mu^- \right] }{2.7\times 10^{-8}}\, ,  \\
\left |\epsilon^{\ell}_{23} \right|^2+\left |\epsilon^{\ell}_{32} \right|^2  \le   \, \left(3.7 \times 10^{-3}\right)^2 \, \left(\dfrac{m_{H}/500\,{\rm GeV}}{\tan\beta/50} \right )^{4} \dfrac{{\cal B}\left[\tau^-\to\mu^-\mu^+\mu^-\right]}{2.1\times 10^{-8}}\, .
\end{array}
\end{equation}
These constraints are also illustrated in Fig.~\ref{fig:TauToMuMuMu} for the experimental limits given in \eq{taumumumuExpBounds}.

\section{Loop-contributions to FCNC processes}
\label{sec:loop-contributions}

We observed in the previous section that all elements $\epsilon^d_{ij}$, $\epsilon^\ell_{ij}$ (with $i \neq j$) and $\epsilon^u_{12,21}$ must be extremely small due to the constraints from tree-level neutral Higgs contributions to FCNC processes. Furthermore, the constraints on $\epsilon^q_{ij}$ and $\epsilon^q_{ji}$ get even more stringent if both of them are non-zero at the same time due to the bounds from $\Delta F=2$ processes. Nevertheless, the elements $\epsilon^u_{13,23}$ and $\epsilon^u_{31,32}$ are still unconstrained because we have no data from neutral current top decays. In addition, also the flavor-conserving elements $\epsilon^f_{ii}$ are not constrained from neutral Higgs contributions to FCNC processes. 
\medskip

In this section, we study the constraints from Higgs mediated loop contributions to FCNC observables. First, in Sec.~\ref{Charged-Higgs-DeltaF2} we consider the $\Delta F=2$ processes, $B_{s}$--$\overline{B}_{s}$, $B_{d}$--$\overline{B}_{d}$ and \kk mixing and then examine the constraints on  $\epsilon^u_{13,23}$ and $\epsilon^u_{31,32}$ from $b\to s(d) \gamma$. Also $\epsilon^u_{22}$ ($\epsilon^u_{33}$) can be constrained from these processes due to the relative $\tan\beta$ enhancement compared to $m_{c}$ ($m_{t}$) in the quark-quark-Higgs vertices. In this analysis, we neglect the effects of the elements $\epsilon^d_{ij}$, which means that we assume the absence of large accidental cancellations between different contributions.
\medskip

Also $\Delta F=0$ processes (electric dipole moments) place relevant constraints on the type-III 2HDM parameter space, as we will see in Sec~\ref{DeltaF0}.

\subsection{\bbs, \bbd and \kk mixing}
\label{Charged-Higgs-DeltaF2}

For the charged Higgs contributions to $\Delta F=2$ processes we calculated the complete set of Wilson Coefficients in a general $R_\xi$-gauge. The result is given, together with our conventions for the Hamiltonian, in the appendix. For the QCD evolution we used the NLO running of the Wilson coefficients of Ref.~\cite{Ciuchini:1997bw,Buras:2000if}. 
\medskip

\begin{figure}[t]
\centering
\includegraphics[width=0.32\textwidth]{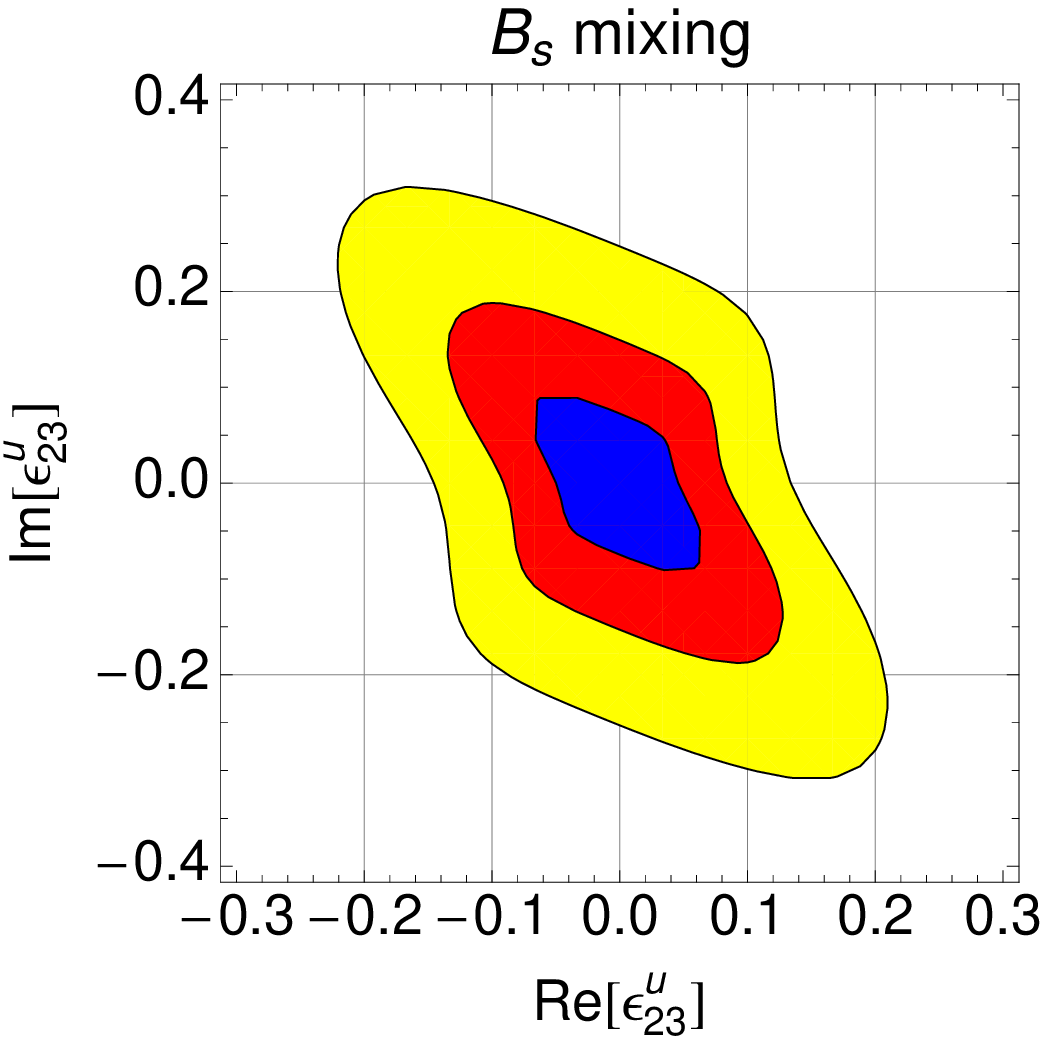}~~
\includegraphics[width=0.32\textwidth]{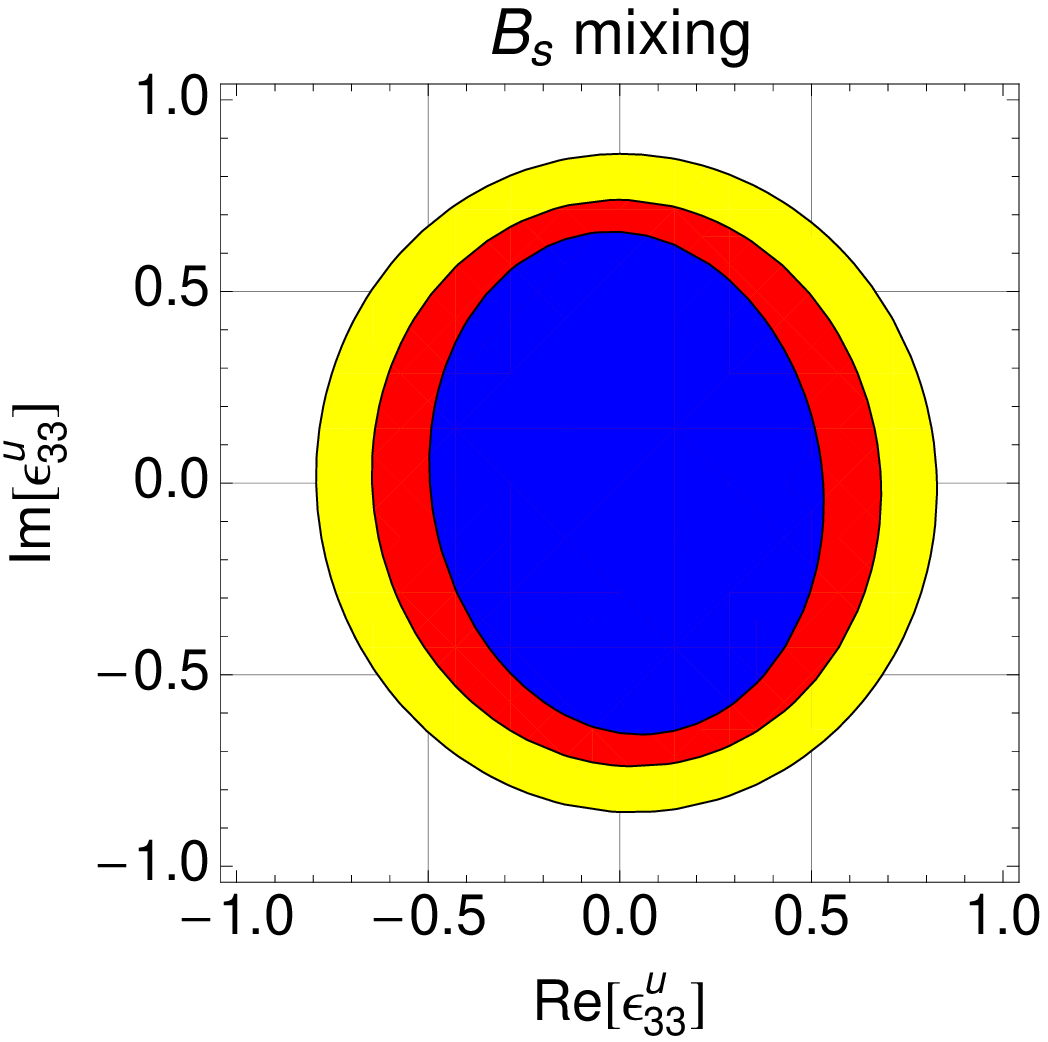}~~
\includegraphics[width=0.32\textwidth]{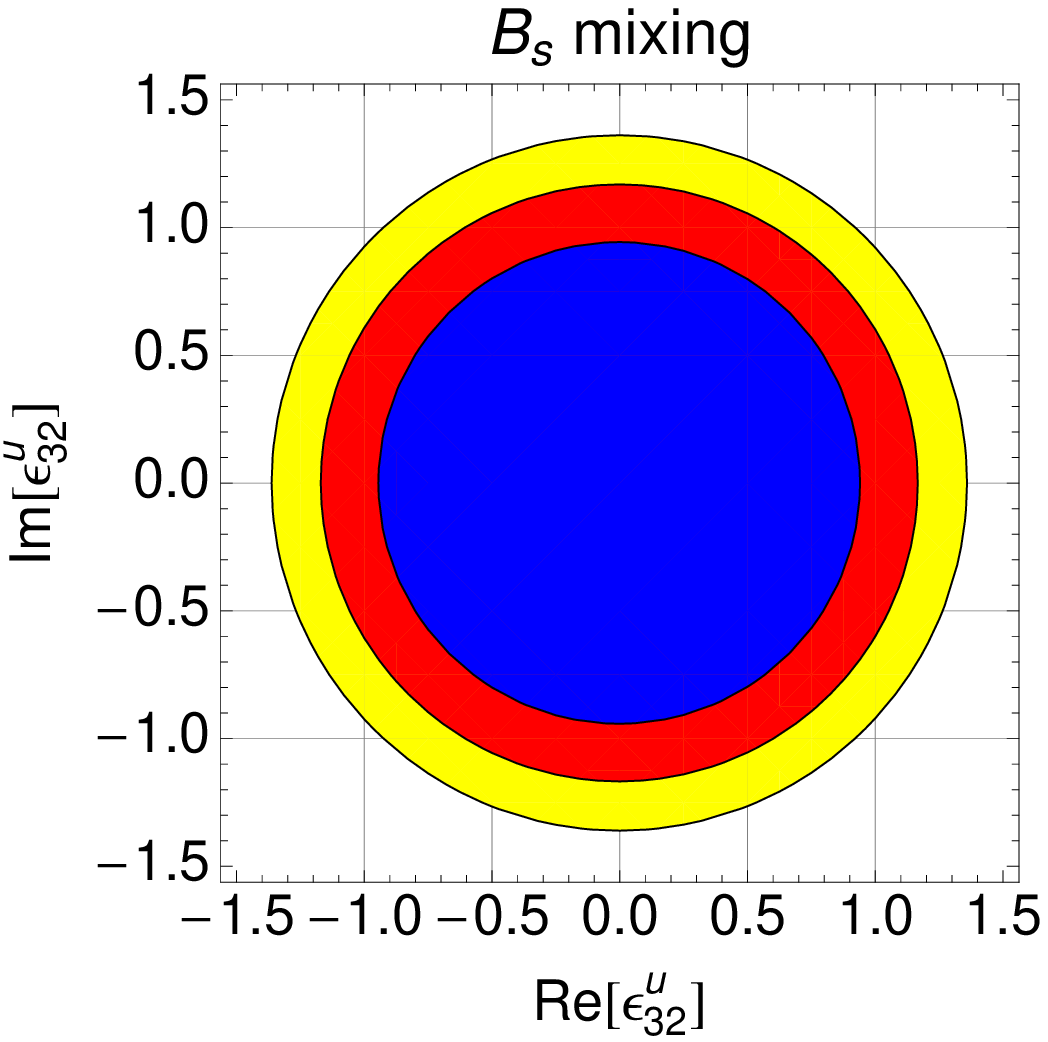}
\caption{Allowed regions in the complex $\epsilon^{u}_{ij}$-plane from $B_s$ mixing for  $\tan\beta=50$ and  $m_{H}=700 \,  \mathrm{ GeV}$ (yellow), $m_{H}=500\,  \mathrm{ GeV}$ (red) and  $m_{H}=300\, \mathrm{ GeV}$ (blue).}
\label{BsmixingBoxCont}
\end{figure}

\begin{figure}[t]
\centering
\includegraphics[width=0.33\textwidth]{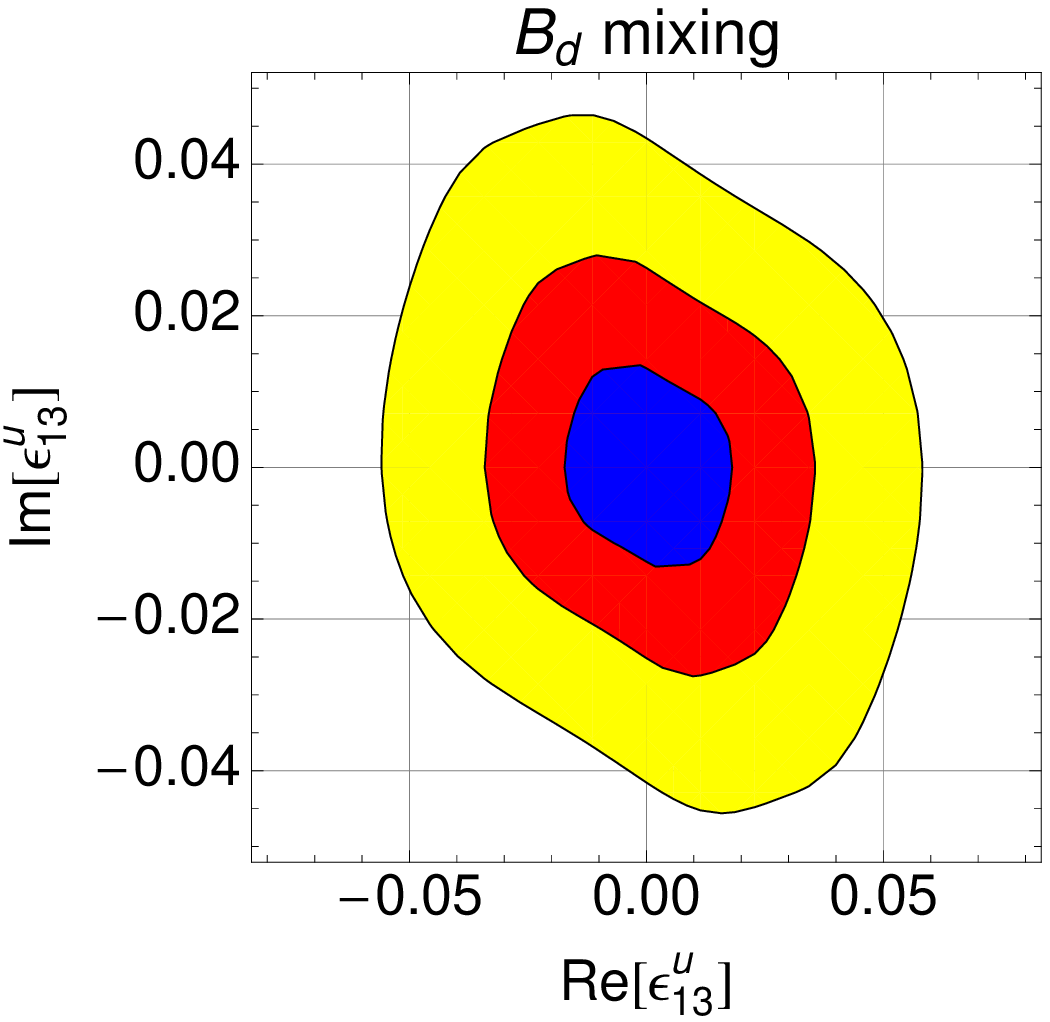}~~
\includegraphics[width=0.328\textwidth]{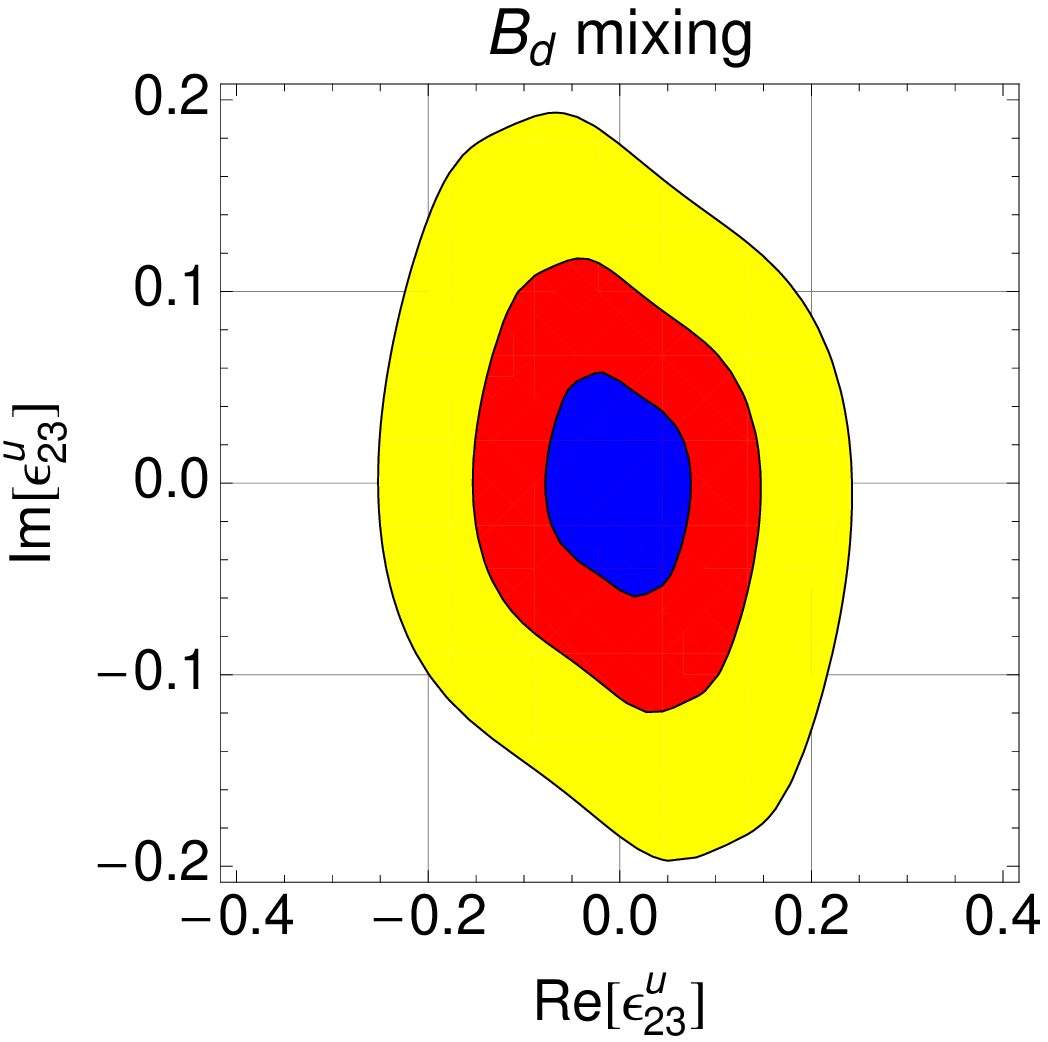}~~
\includegraphics[width=0.31\textwidth]{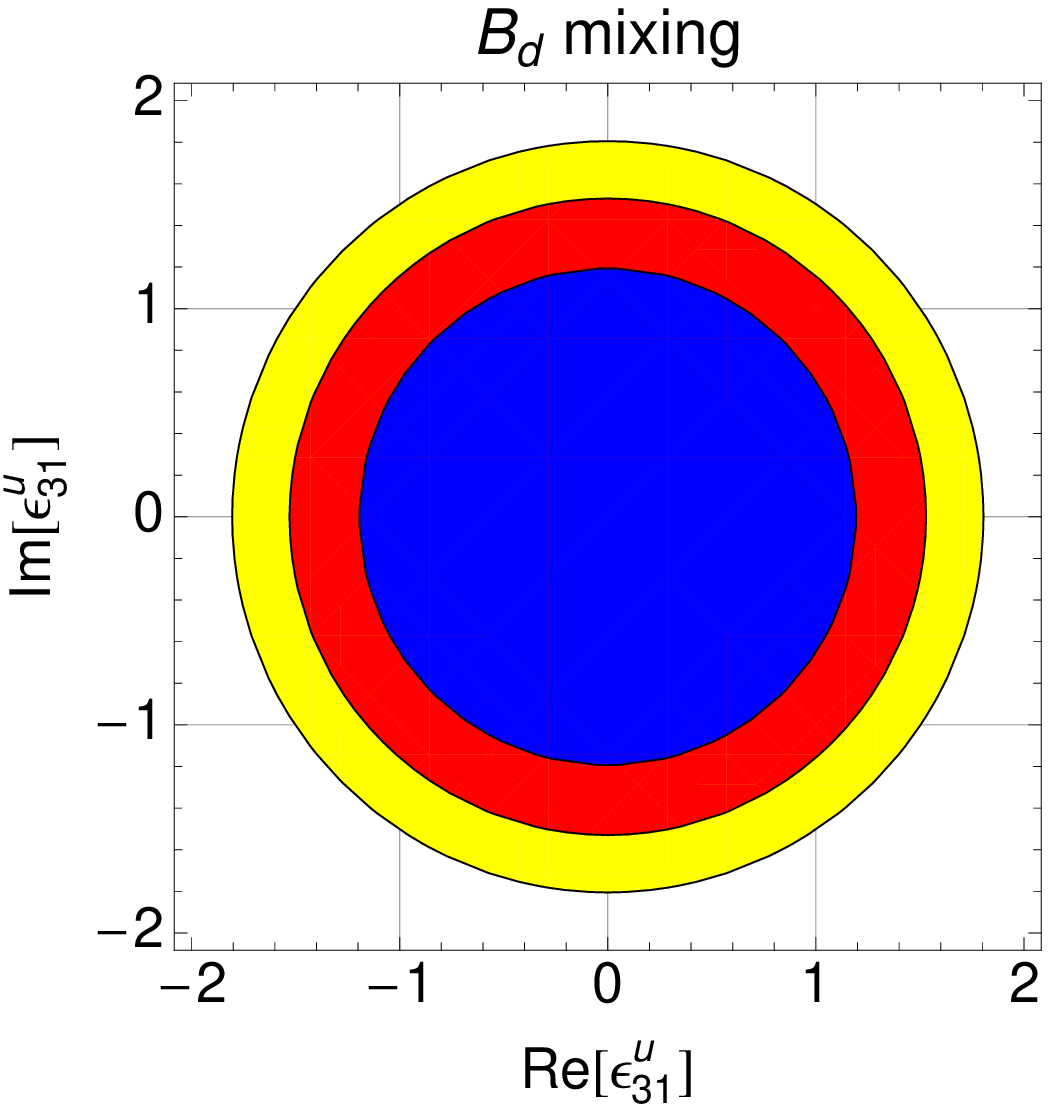}
\caption{Allowed regions in the complex $\epsilon^{u}_{ij}$-plane from $ B_d$ mixing for  $\tan\beta=50$ and  $m_{H}=700\, \mathrm{ GeV}$ (yellow), $m_{H}=500\, \mathrm{ GeV}$ (red) and  $m_{H}=300\, \mathrm{ GeV}$ (blue).}
\label{BdmixingBoxCont}
\end{figure}

\begin{figure}[t]
\centering
\includegraphics[width=0.4\textwidth]{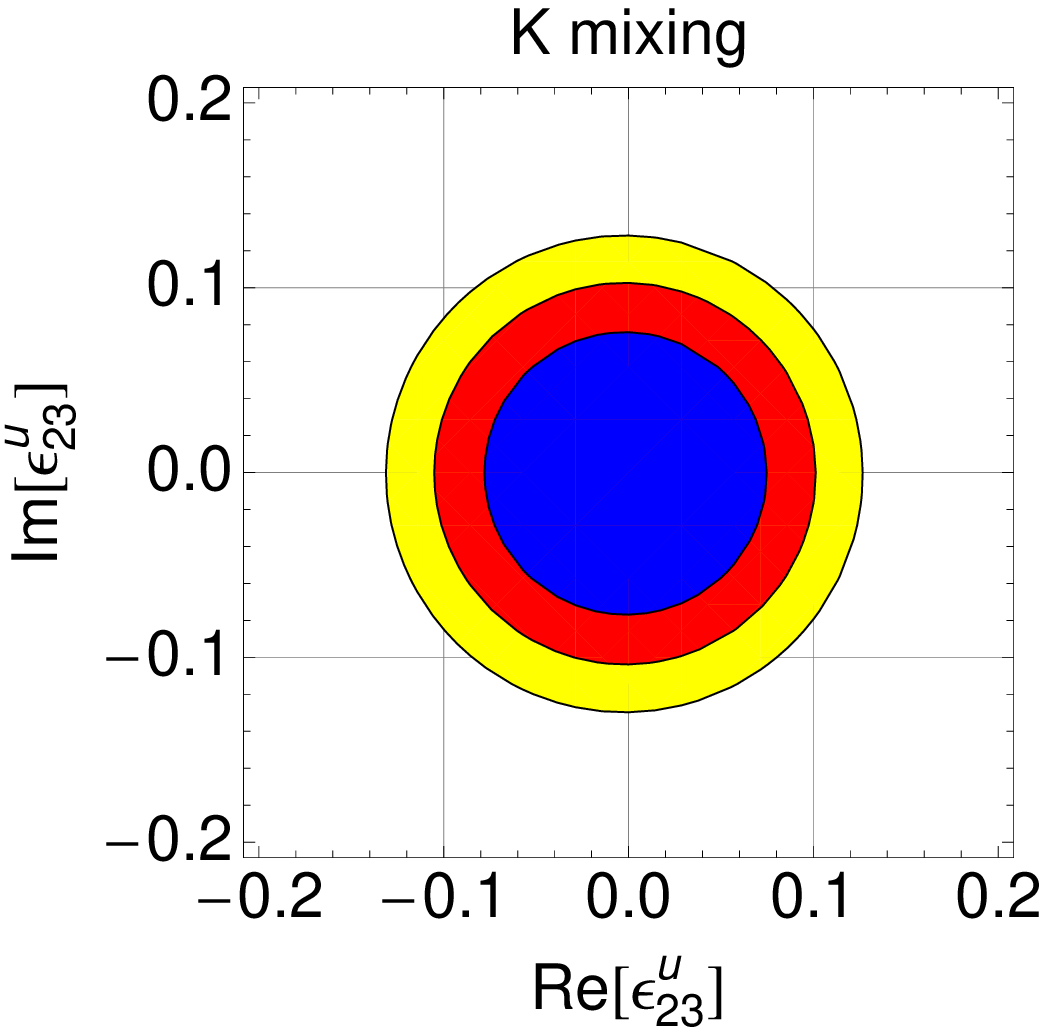}~~~
\includegraphics[width=0.4\textwidth]{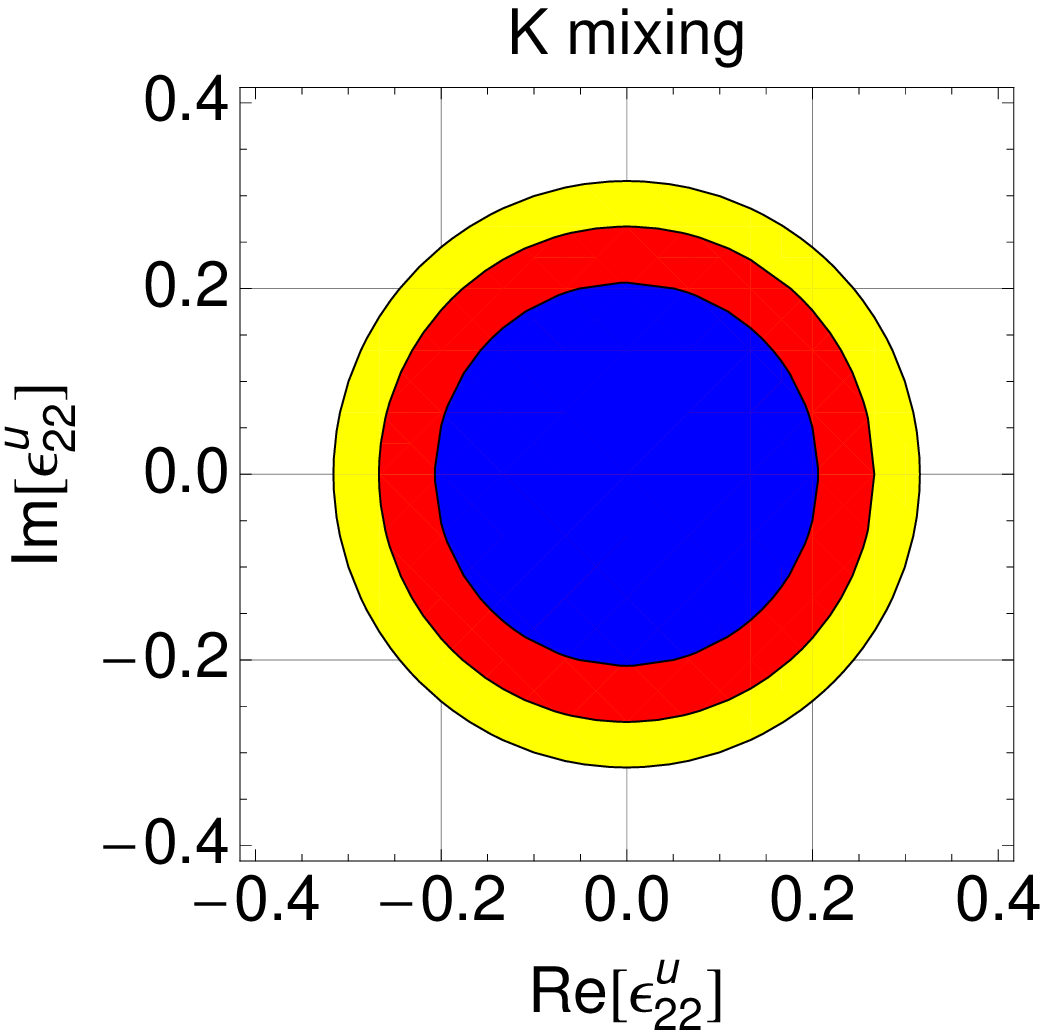}
\caption{ Allowed regions in the complex $\epsilon^{u}_{ij}$-plane from \kk mixing for  $\tan\beta=50$ and  $m_{H}=700 \,\mathrm{ GeV}$ (yellow), $m_{H}=500\, \mathrm{ GeV}$ (red) and  $m_{H}=300 \, \mathrm{ GeV}$ (blue). The constraints are practically independent of $\tan\beta$. }
\label{KmixingBoxCont}
\end{figure}

For computing the allowed regions in parameter space we used the same procedure as explained in the last section. The results are shown in Fig.~\ref{BsmixingBoxCont}, \ref{BdmixingBoxCont} and \ref{KmixingBoxCont} and can be summarized as follows: $B_{s}$--$\overline{B}_{s}$ ($B_{d}$--$\overline{B}_{d}$) mixing gives constraints on $\epsilon^u_{23}$ ($\epsilon^u_{13}$) which are of the order of $10^{-1}$ ($10^{-2}$) for our typical values of $\tan\beta$ and $m_H$. In addition, $B_{d}$--$\overline{B}_{d}$ mixing also constrains $\epsilon^u_{23}$ to a similar extent as $B_{s}$--$\overline{B}_{s}$ mixing. The constraints on $\epsilon^u_{33}$, $\epsilon^u_{32}$ and $\epsilon^u_{31}$ are all very weak (of order one). Also Kaon mixing gives comparable bounds on $\rm{Abs}\left[\epsilon^u_{23}\right]$ and the bounds on $\rm{Abs}\left[\epsilon^u_{22}\right]$ are of the order $10^{-1}$.
\medskip


\subsection{Radiative $B$ meson decays: $b\to s\gamma$ and $b\to d\gamma$}
\label{Charged-Higgs-b-s-gamma}

\begin{figure}[t]
\centering
\includegraphics[width=0.42\textwidth]{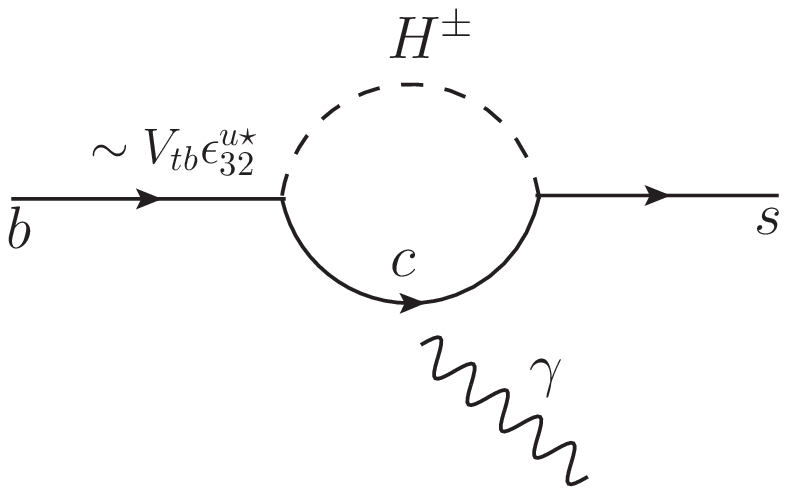}~~~
\includegraphics[width=0.44\textwidth]{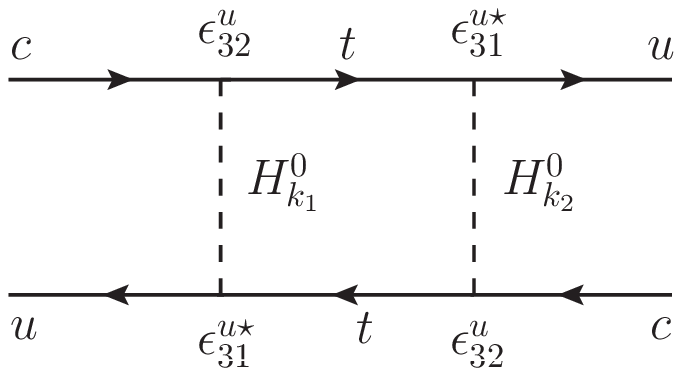}
\caption{Left: Feynman diagram contributing to $ b\to s \gamma$ via a charm-loop containing $\epsilon^{u\star}_{32}$. The contribution is suppressed, since the small charm mass enters either form the propagator or from the charged Higgs coupling to the charm and strange quark.\\
 Right: Feynman diagram showing a neutral Higgs box contribution to \dd mixing arising if $\epsilon^{u}_{31}$ and $\epsilon^{u}_{32}$ are simultaneously different from zero.}
\label{diagsx}
\end{figure}
\begin{figure}[t]
\centering
\includegraphics[width=0.32\textwidth]{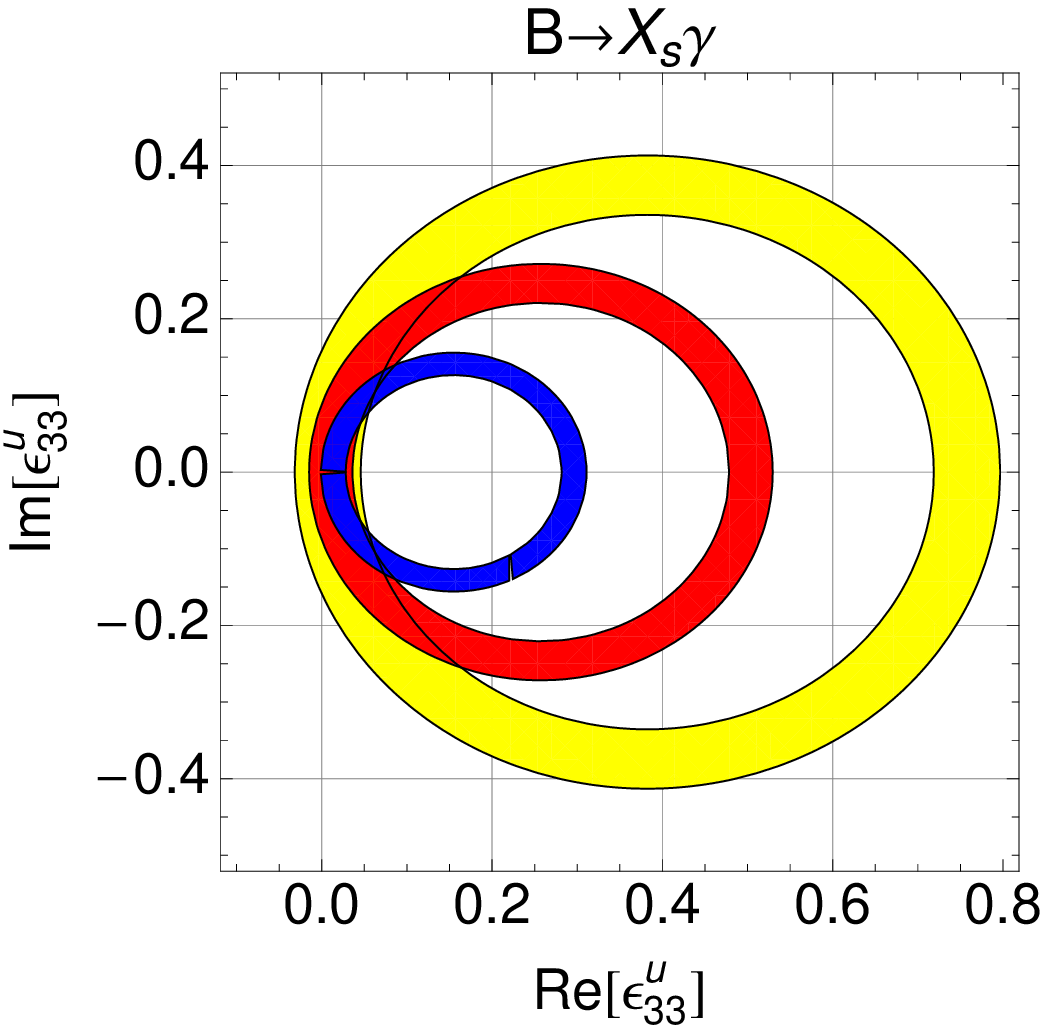}~
\includegraphics[width=0.334\textwidth]{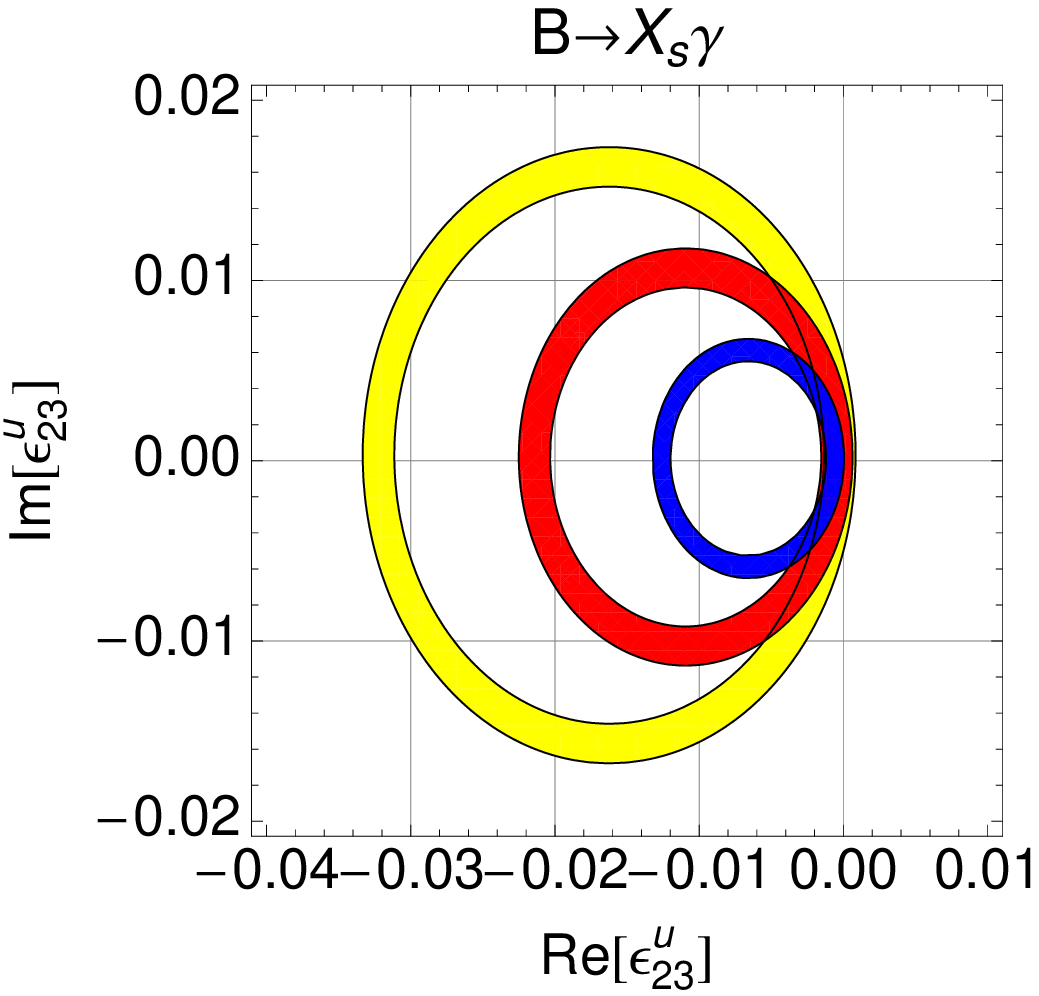}~
\includegraphics[width=0.33\textwidth]{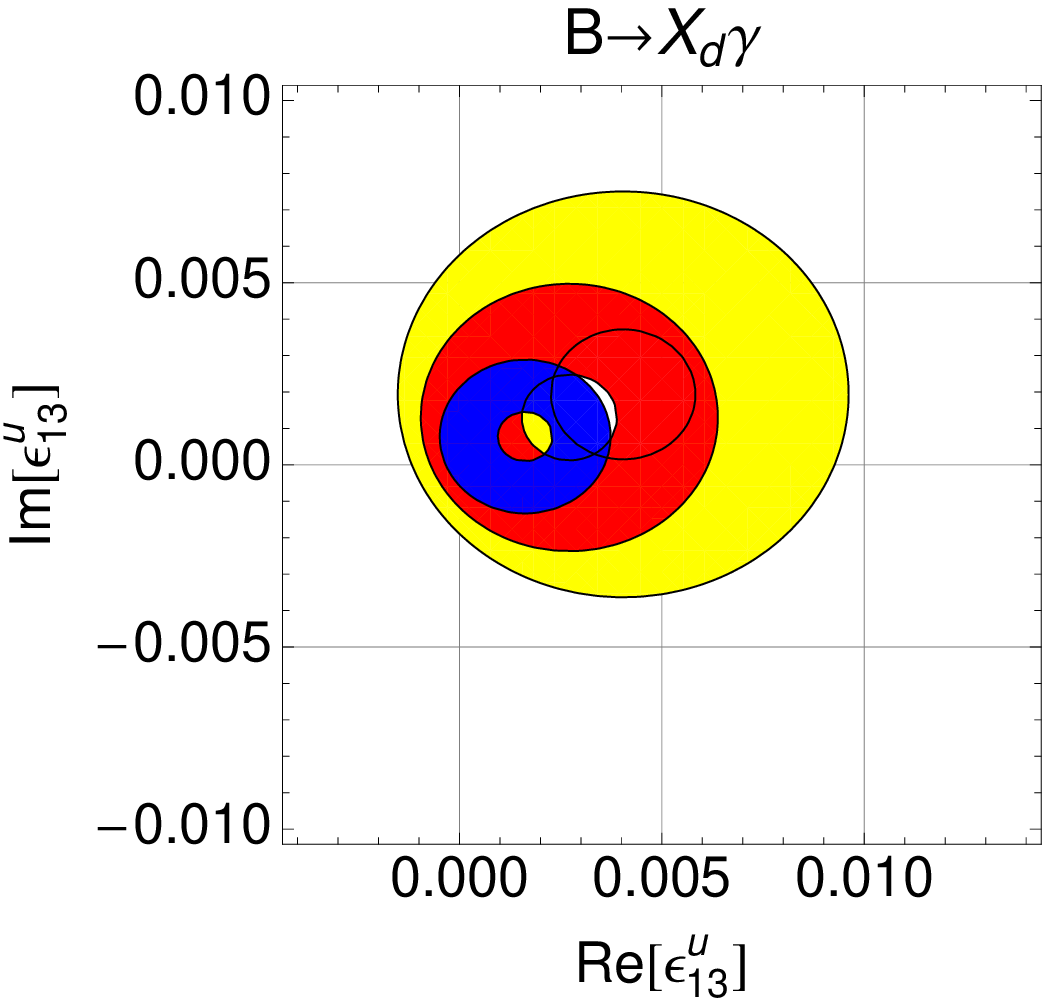}
\caption{Allowed regions for $\epsilon^{u}_{ij}$ from $ B \to X_{s(d)} \gamma$, obtained by adding the $2\,\sigma$ experimental error and theoretical uncertainty linear for $\tan\beta=50$ and $m_{H}=700 \, \mathrm{ GeV}$ (yellow), $m_{H}=500\, \mathrm{ GeV}$ (red) and  $m_{H}=300 \,\mathrm{ GeV}$ (blue). }
\label{BXsdgamma}
\end{figure}
\begin{figure}[t]
\centering
\includegraphics[width=0.32\textwidth]{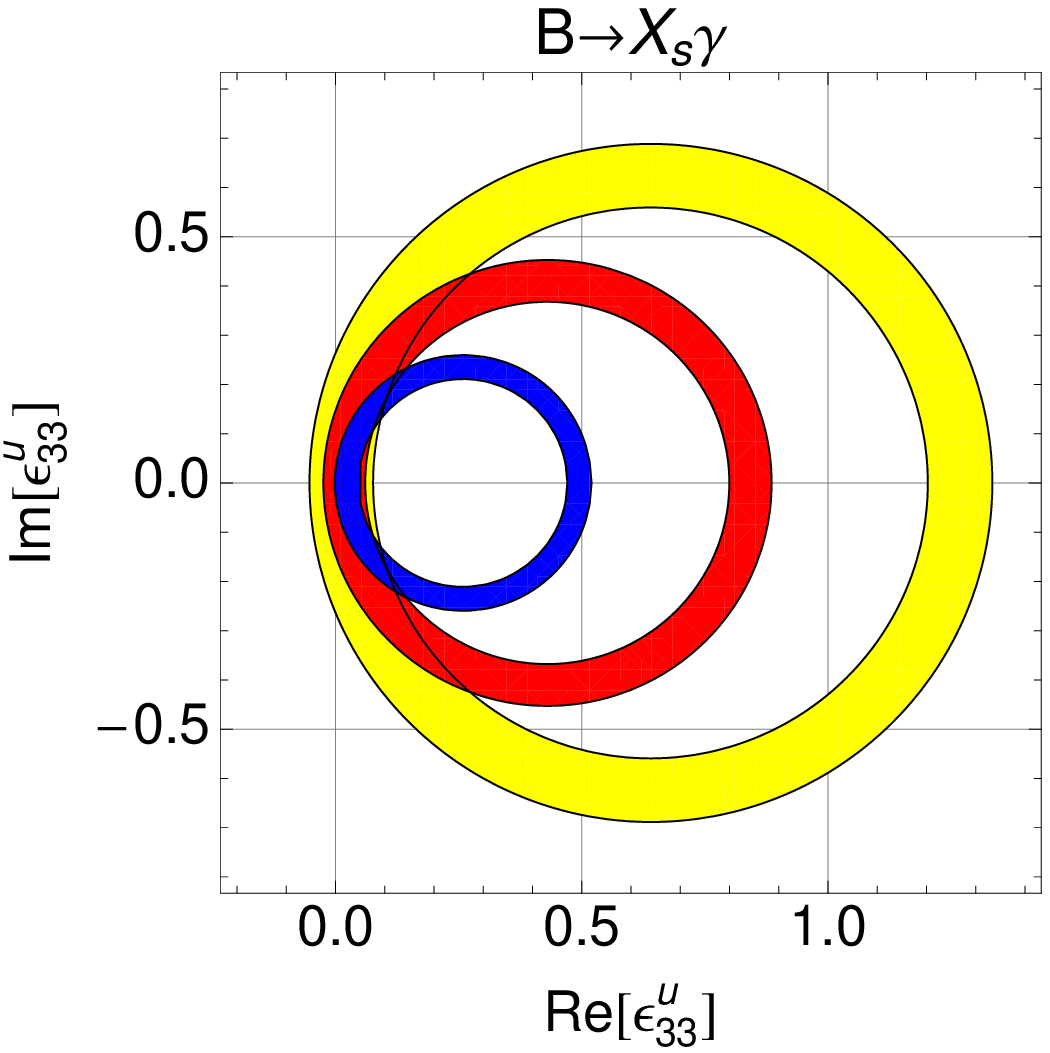}~
\includegraphics[width=0.33\textwidth]{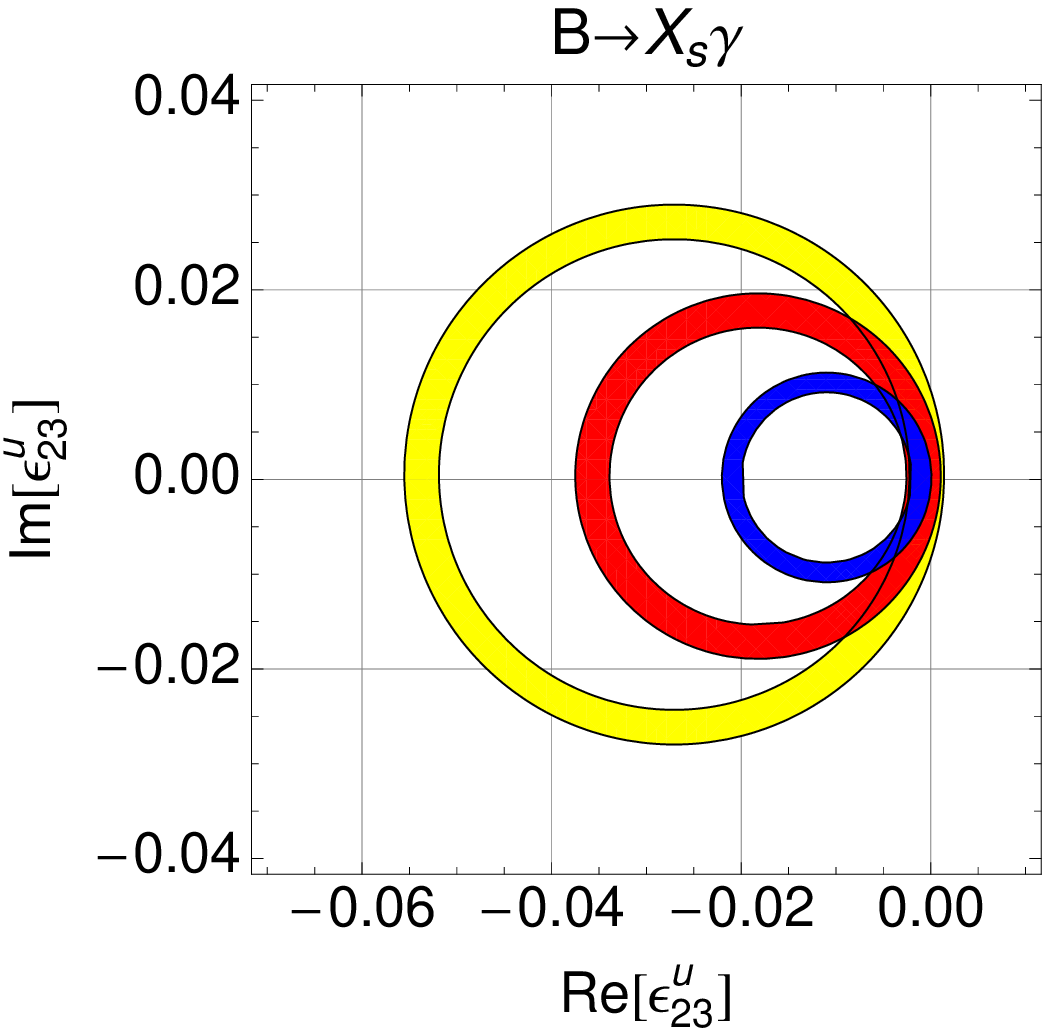}~
\includegraphics[width=0.34\textwidth]{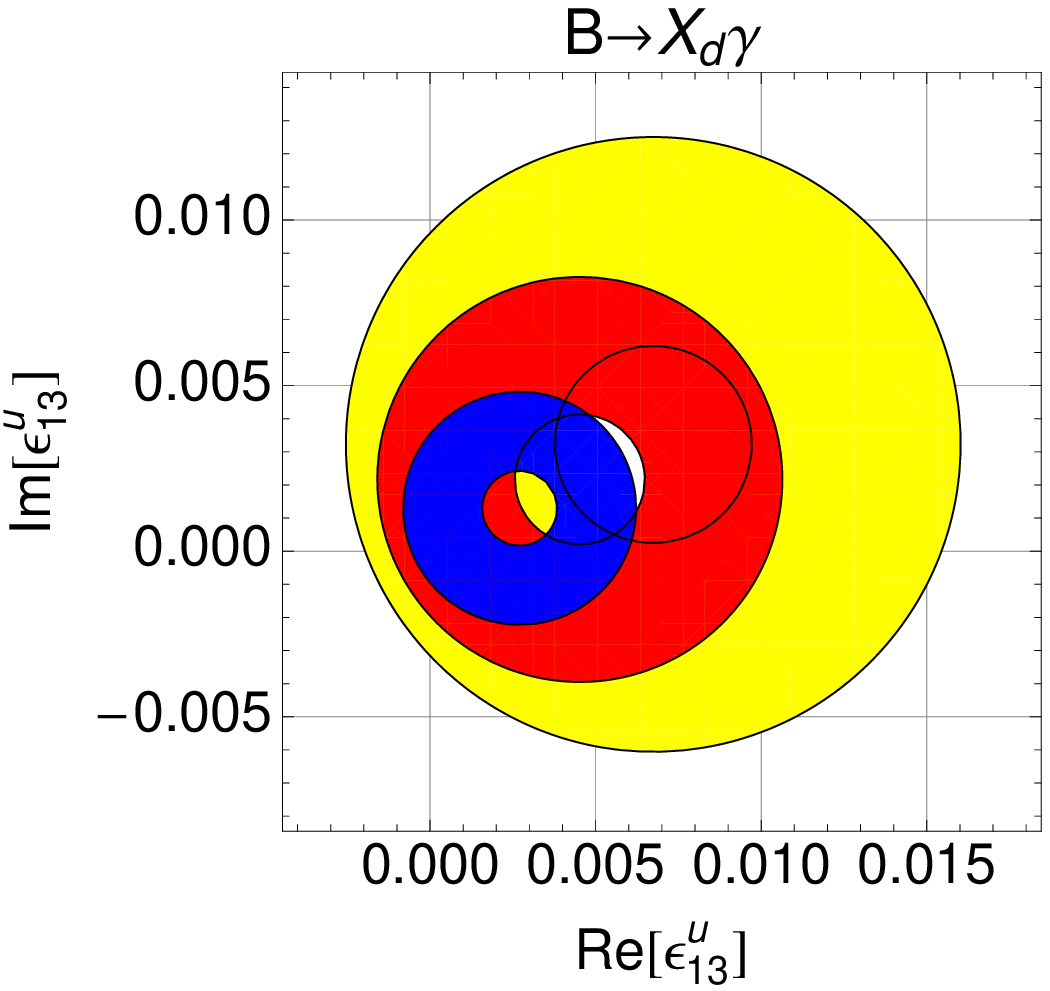}
\caption{Allowed regions for $\epsilon^{u}_{ij}$ from $B \to X_{s(d)} \gamma$, obtained by adding the $2\,\sigma$ experimental error and theoretical uncertainty linear for $\tan\beta=30$ and $m_{H}=700\, \mathrm{ GeV}$ (yellow), $m_{H}=500\, \mathrm{ GeV}$ (red) and  $m_{H}=300\, \mathrm{ GeV}$ (blue). }
\label{fig:BXsdgammatanb30}
\end{figure}

The radiative $B$ decay $ b\to s \gamma$ ($ b\to d \gamma$) imposes stringent constraints on the element $\epsilon^u_{23}$ ($\epsilon^u_{13}$) while also in this case the constraints on $\epsilon^u_{32}$ ($\epsilon^u_{31}$) are very weak due to the light charm (up) quark involved (see left diagram in Fig.~\ref{diagsx}). For these processes both a neutral and a charged Higgs contribution occur. Since the flavor off-diagonal elements $\epsilon^d_{13,23}$ and $\epsilon^d_{31,32}$ are already stringently constrained from tree-level decays we neglect the neutral Higgs contribution here. We give the explicit results for the Higgs contributions to the Wilson coefficients governing $b\to s(d) \gamma$ in the appendix.
\medskip

For $ B \to X_{s} \gamma$, we obtain the constraints on the 2HDM of type III parameters $\epsilon^u_{ij}$ by using $ {\cal B} \left[B \to X_{s} \gamma\right]$ from Ref.~\cite{StoneICHEP12} (BABAR) and Ref.~\cite{Collaboration:2012fk,Lees:2012wg} (BELLE). Combined and extrapolated to a photon energy cut of ${1.6}$ GeV, the HFAG value is \cite{Amhis:2012bh}
\begin{equation}
 {\cal B}\left.\left[B \to X_{s} \gamma\right]\right|^{\rm exp}_{E_{\gamma}>1.6\, {\rm GeV}} = \left(3.43 \pm 0.21 \pm 0.07\right)\times 10^{-4} \, .
\end{equation}
In order to estimate the possible size of NP we use the NNLO SM calculation of Ref.~\cite{Misiak:2006zs} (again for a photon energy cut of ${1.6}$\, GeV) 
\begin{equation}
 {\cal B} \left[B \to X_{s} \gamma\right]^{\rm SM} = \left(3.15 \pm 0.23 \right) \times 10^{-4} \, ,
\end{equation}
and calculate the ratio
\begin{equation}
R^{b\to s \gamma}_{\rm exp}=\dfrac{ {\cal B}\left.\left[B \to X_{s} \gamma\right]\right|^{exp}}{ {\cal B}\left.\left[B \to X_{s} \gamma\right]\right|^{SM}} \,.
\end{equation}
This leads to a certain range for $R^{b\to s \gamma}_{\rm exp}$. Now, we require that in our leading-order calculation the ratio

\begin{equation}
R^{b\to s \gamma}_{\rm theory}=\dfrac{ {\cal B}\left.\left[B \to X_{s} \gamma\right]\right|^{\rm 2HDM}}{ {\cal B}\left.\left[B \to X_{s} \gamma\right]\right|^{\rm SM}} 
\end{equation}
lies within this range. In this way, we obtain the constraints on our model parameters $\epsilon^u_{ij}$ as illustrated in Fig.~\ref{BXsdgamma} and Fig.~\ref{fig:BXsdgammatanb30}. 
\medskip

The analysis for $b \to d \gamma$ is performed in an analogous way. In addition we use here the fact that most of the hadronic uncertainities cancel in the CP-averaged branching ratio for $ B \to X_{d} \gamma$ \cite{Benzke:2010js,Hurth:2010tk}. The current experimental value of the BABAR collaboration \cite{delAmoSanchez:2010ae,Wang:2011sn} for the CP averaged branching ratio reads
\begin{equation}
 {\cal B}\left.\left[B \to X_{d} \gamma\right]\right|^{\rm exp}_{E_{\gamma}>1.6\, {\rm GeV}}\,=\,(1.41\pm0.57) \times 10^{-5}\,.
 \end{equation}
Here we take into account a conservative estimate of the uncertainty coming from the extrapolation in the photon energy cut~\cite{Crivellin:2011ba}. For the theory prediction we use the NLL SM predictions of the CP-averaged branching ratio ${\cal B}(B \to X_{d} \gamma)|_{E_{\gamma}>1.6\, {\rm GeV}}$ of Ref.~\cite{Ali:1998rr,Hurth:2003dk}, which was recently updated in Ref.~\cite{Crivellin:2011ba} and reads
\begin{equation}
 {\cal B}\left.\left[B \to X_{d} \gamma\right]\right|^{\rm SM}_{E_{\gamma}>1.6\, {\rm GeV}}\,=\,(1.54^{+0.26}_{-0.31}) \times 10^{-5}\, .
 \end{equation}
After defining the ratios $R^{b\to d \gamma}_{\rm exp}$ and $R^{b\to d \gamma}_{\rm theory}$ we continue as in the case of ${\cal B}\left[B \to X_{s} \gamma\right]$ in order to constrain $\epsilon^u_{13}$.
\medskip

As can be seen from Fig.\ref{BXsdgamma} and Fig.~\ref{fig:BXsdgammatanb30}, the constraints that $ B \to X_{s(d)} \gamma$ enforces on $\epsilon^u_{23(13)}$ are  stronger than the ones from $ B_{s(d)}$ mixing. Even $\epsilon^u_{33}$ can be restricted to a rather small range. 
\medskip

While in the 2HDM of type II $b\to s\gamma$ enforces a lower limit on the charged Higgs mass of $360$~GeV \cite{Hermann:2012fc} this constraint can get weakened in the 2HDM of type III: The off-diagonal element $\epsilon^u_{23}$ can lead to a destructive interference with the SM (depending on its phase) and thus reduce the 2HDM contribution. Lighter charged Higgs masses are also constrained from $b\to d\gamma$ but also this constraint can be avoided by $\epsilon^u_{13}$.
\medskip

\subsection{Neutral Higgs box contributions to \dd mixing}
\begin{figure}[t]
\centering
\includegraphics[width=0.45\textwidth]{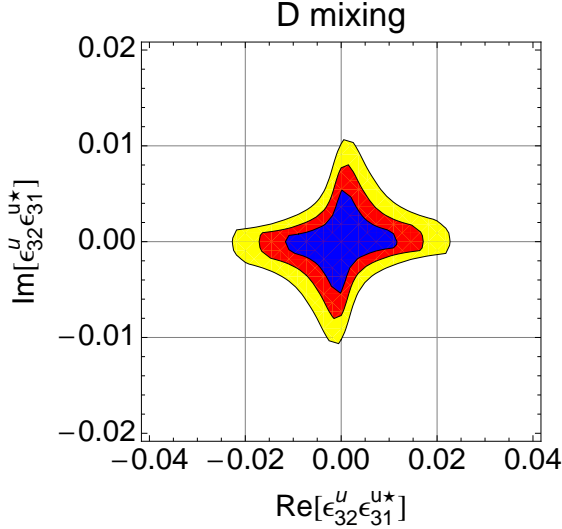}
\caption{ Allowed region in the complex $\epsilon^{u}_{32}\epsilon^{u\star}_{31}$--plane obtained from neutral Higgs box contributions to \dd mixing for $\tan\beta=50$ and $m_{H}=700\, \mathrm{ GeV}$ (yellow), $m_{H}=500 \,\mathrm{ GeV}$ (red) and  $m_{H} = 300\, \mathrm{ GeV}$ (blue).}
\label{DmixingH0boxplot}
\end{figure}

Nearly all the loop-induced neutral Higgs contributions to FCNC processes can be neglected because the elements involved are already stringently constrained from tree-level processes. However, there is one exception: since the constraints on $\epsilon^u_{31,32}$ are particularly weak (because of the light charm or up quark entering the loop) this can give a sizable effect in \dd mixing via a neutral Higgs box\footnote{In principle, one can also get contribution to ${\bar D}^{0}\to\mu^{+}\mu^{-}$ through $H^{0}_{k}$ box and penguin contributions if the elements $\epsilon^{u}_{32}$ and $\epsilon^{u}_{31}$ are simultaneously non-zero. However, we observe that they are negligible.} (see Fig.~\ref{diagsx}). As we will use $\epsilon^u_{31}$ and $\epsilon^u_{32}$ in Sec.~\ref{sec:charged-current} for explaining the mentioned deviations from the SM prediction in \btau, \bdtau and \bdstau it is interesting to ask if all processes can be explained simultaneously without violating \dd mixing. In principle also charged Higgs contributions to \dd mixing arise but we find that they are very small compared to the $H^{0}_{k}$ contributions. The explicit expression for the Wilson coefficients can be found in the appendix.
\medskip

Fig.~\ref{DmixingH0boxplot} shows the allowed regions in the complex $\epsilon^{u}_{32}\epsilon^{u\star}_{31}$--plane. The constraints are again obtained by using the recent UTFit \cite{UTfit:2012zm} analysis for the $D$--$\overline{D}$ system.


\subsection{Radiative lepton decays :  $\mu \to e \gamma$, $\tau\to e \gamma$ and $\tau\to \mu \gamma$ }
\label{sec:muegamma}

\begin{figure}[t]
\centering
\includegraphics[width=0.4\textwidth]{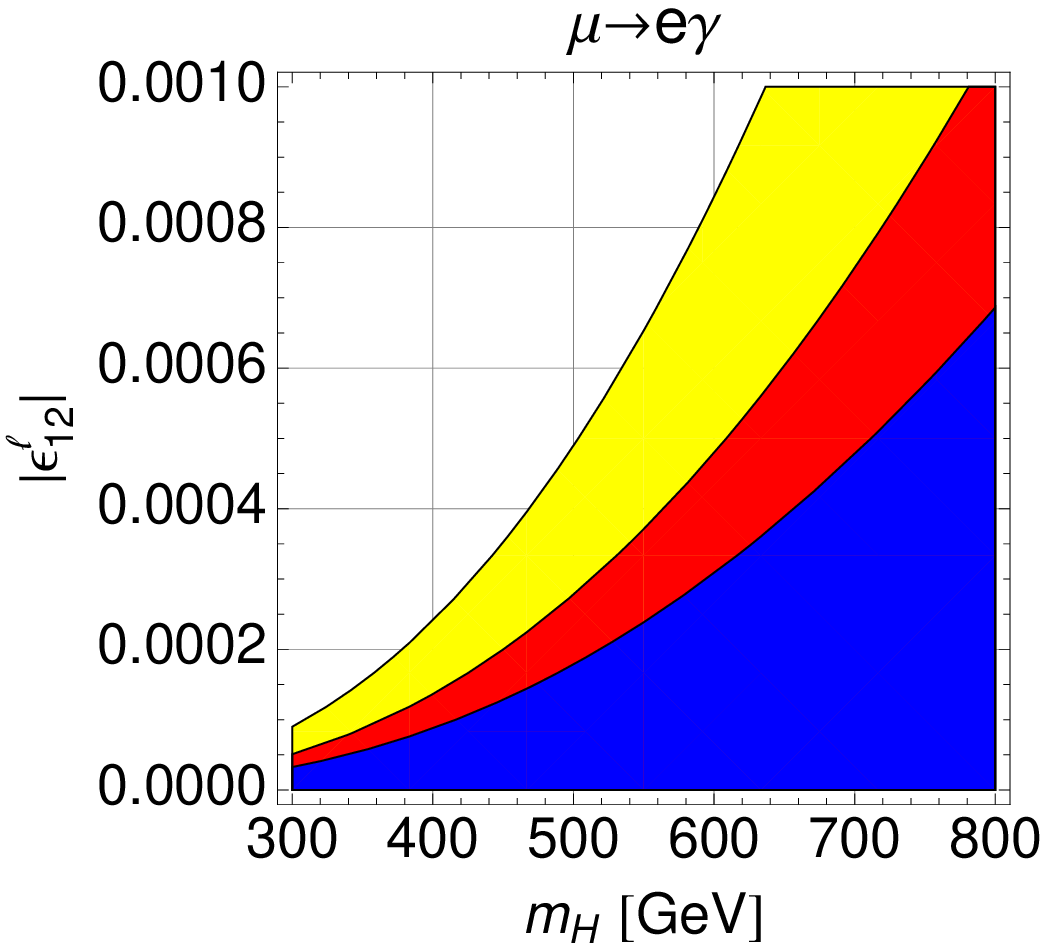}~~~
\includegraphics[width=0.4\textwidth]{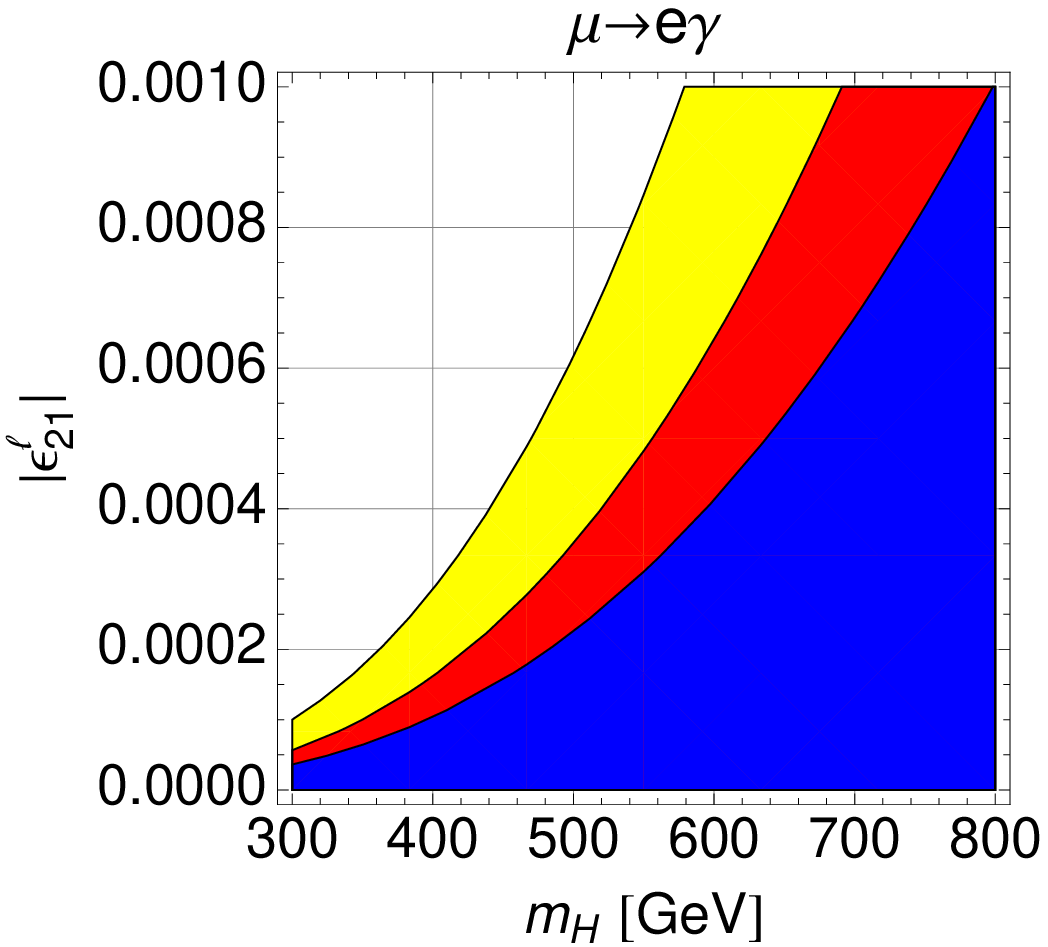}
\caption{ Allowed region for $\epsilon^{\ell}_{12}$ (left plot) and $\epsilon^{\ell}_{21}$ (right plot) from $\mu\to e\gamma$ for $\tan\beta=30$ (yellow), $\tan\beta=40$ (red) and $\tan\beta=50$ (blue).}
\label{Fig:muegamma}
\end{figure}

The bounds on $\epsilon^\ell_{13,31}$ and $\epsilon^\ell_{23,32}$ from the radiative lepton decays $\tau\to e \gamma$ and $\tau\to \mu \gamma$ (using the experimental values given in Table~\ref{table:LFVdecays}) turn out to be significantly weaker than the ones from $\tau^- \to \mu^-\mu^+\mu^-$ and $\tau^- \to e^- \mu^+\mu^-$. Concerning $\mu\to e\gamma$ we expect constraints which are at least comparable to the ones from $\mu^-\to e^-e^+e^-$ since $\mu\to e\gamma$ does not involve the small electron Yukawa coupling entering $\mu^-\to e^-e^+e^-$. In fact, using the new MEG results \cite{Adam:2013mnn} the constraints from $\mu\to e\gamma$ turn out to be stronger than the ones from $\mu^-\to e^-e^+e^-$ (see Fig.~\ref{Fig:muegamma}). Note that the constraints from $\mu^-\to e^-e^+e^-$ can be avoided if $v_u \epsilon^\ell_{11}\approx m_e$ while the leading contribution to $\mu\to e\gamma$ vanishes for $v_u \epsilon^\ell_{22}\approx m_\mu$. 

\begin{table}[htdp]
\begin{minipage}{2in}
\centering \vspace{0.8cm}
\renewcommand{\arraystretch}{1.2}
  \begin{tabular}{@{}|c|c|}
 \hline
 Process  & Experimental bounds
\\   \hline \hline
${\cal B}\left[ \tau \to \mu \gamma  \right] $   & ~$\leq \, 4.5\times 10^{-8}$ ~\cite{Aubert:2009ag,Hayasaka:2007vc}  \\ \hline 
${\cal B}\left[ \tau \to e \gamma \right] $     & ~$\leq \, 1.1 \times 10^{-7}$~ \cite{Aubert:2009ag}  \\ \hline
$ {\cal B}\left[ \mu \to e \gamma \right] $      & ~$\leq \, 5.7 \times 10^{-13}$~\cite{Adam:2013mnn} 
\\ \hline \hline
\end{tabular}
 \end{minipage} 
 \caption{Experimental upper limits on the branching ratios of lepton-flavor violating decays.}
\label{table:LFVdecays}
\end{table}   

In principle, for $\mu\to e\gamma$ a simplified expression for the branching ratio in the large $\tan\beta$ limit and $v\ll m_H$ could also be given. However, due to the large logarithm with a relative big prefactor (last term of \eq{CiH0klitolfgamma}) this is only a good approximation for very heavy Higgses and we therefore use the full expression in our numerical analysis.
\medskip

We will return to the radiative lepton decays in Sec.~\ref{LFV-b-decays} and correlate them to the decays $\tau^- \to \mu^-\mu^+\mu^-$, $\tau^- \to e^- \mu^+\mu^-$ and $\mu^-\to e^-e^+e^-$.

\subsection{$B_s\to\mu^+\mu^-$}

\begin{figure*}[t]
\centering
\includegraphics[width=0.4\textwidth]{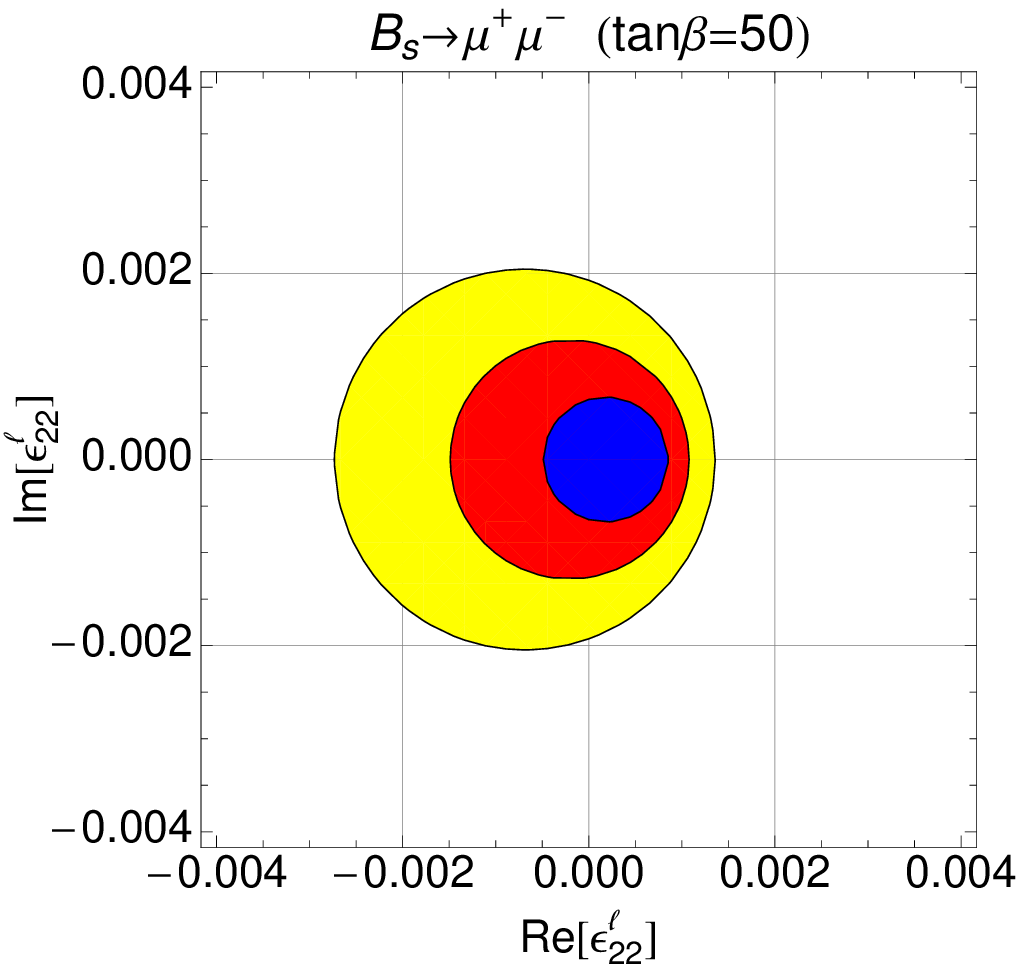}~~~
\includegraphics[width=0.42\textwidth]{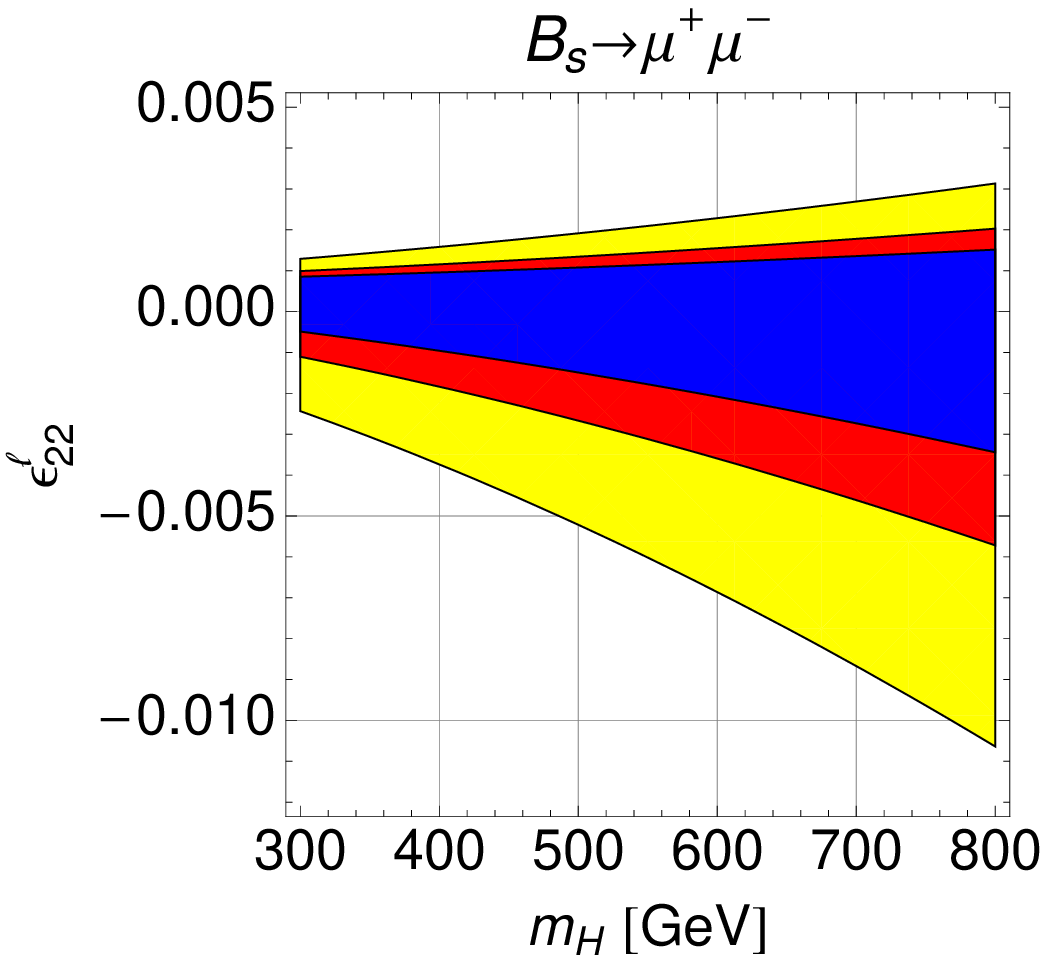}
\caption{Left: Allowed regions in the complex $\epsilon^\ell_{22}$--plane from $B_s\to\mu^+\mu^-$ for $\tan\beta=50$ and $m_H=700$ GeV (yellow), $m_H=500$ GeV (red) and $m_H=300$ GeV (blue). Right: Allowed regions in the $\epsilon^{\ell}_{22}$--$m_{H}$ plane from $B_s\to\mu^+\mu^-$ for $ \tan\beta=30$ (yellow), $ \tan\beta=40$ (red), $ \tan\beta=50$ (blue) and real values of $\epsilon^\ell_{22}$.}
\label{fig:BstomumuEl22}
\end{figure*}

Setting $\epsilon^q_{ij}=0$ only the loop induced charged Higgs contribution to $B_s\to\mu^+\mu^-$ (and $B_d\to\mu^+\mu^-$) exist. This contribution (see \eq{eqn:type2Cis}) gets altered in the presence of non-zero elements $\epsilon^\ell_{ij}$, e.g. $\epsilon^\ell_{22}$. In the large $\tan\beta$ limit, the loop induced result in \eq{eqn:type2Cis} is modified to 
\begin{equation}
  C^{bs}_S = C^{bs}_P = -\dfrac{m_{b}\, V^{\star}_{tb}V_{ts}}{2} \dfrac{m_{\mu}-v_{u}\epsilon^{\ell}_{22}}{2 M^{2}_{W}}  \, \tan^{2}\beta \, \dfrac{\log\left({{m_H^2}/{m_t^2} }\right)}{ { {m_H^2}/{m_t^2}-1}}\,.  
\end{equation}
\medskip
The resulting constraints on $\epsilon^\ell_{22}$ from $B_s \to \mu^+\mu^-$ are shown in Fig.~\ref{fig:BstomumuEl22} and the ones from $B_d \to \mu^+\mu^-$ are found to be weaker.
%

\subsection{Electric dipole moments and anomalous magnetic moments}
\label{DeltaF0}
\subsubsection{Charged leptons}

\begin{figure*}[t]
\centering
\includegraphics[width=0.4\textwidth]{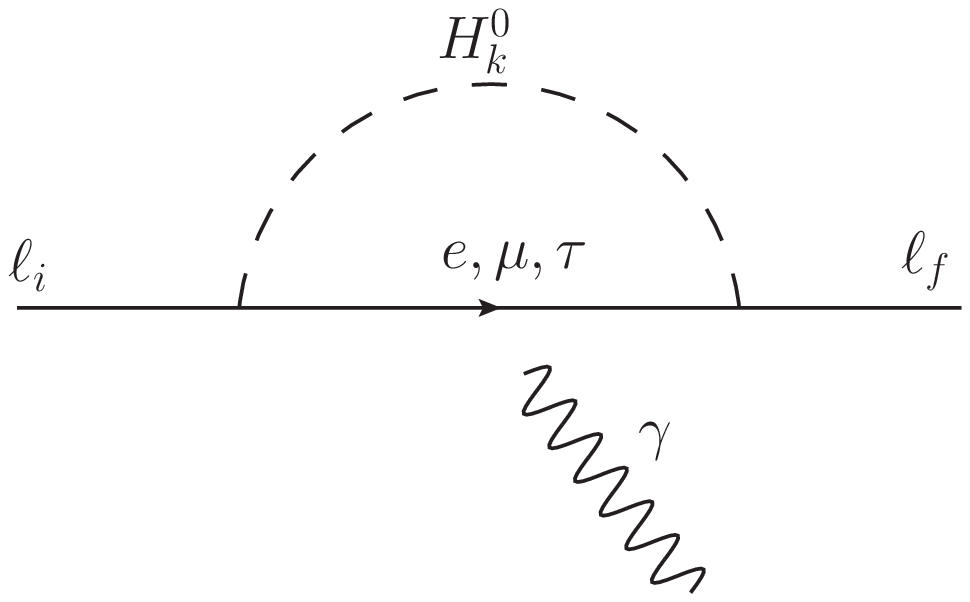}~~~
\includegraphics[width=0.4\textwidth]{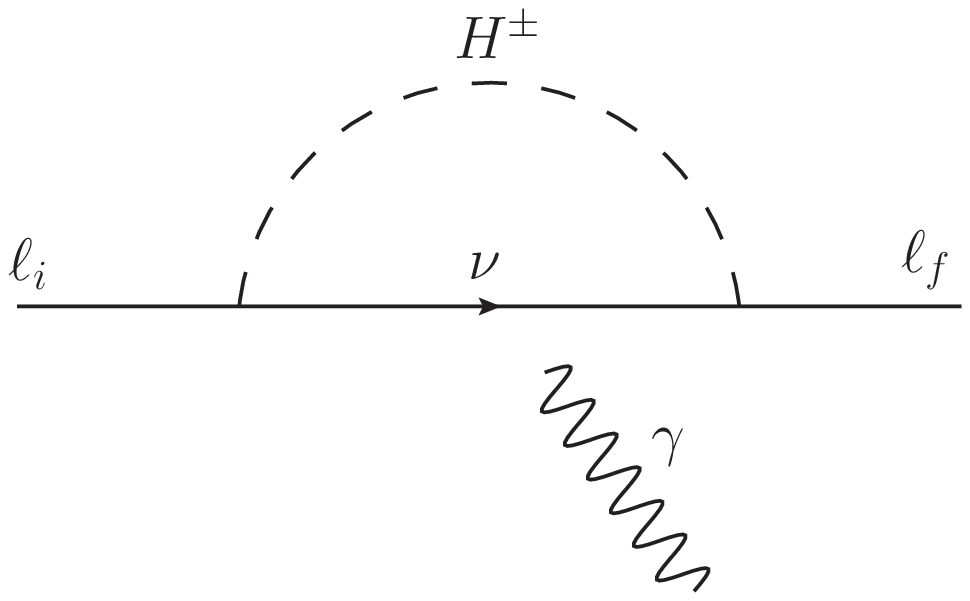}
\caption{Left: Feynman diagram contributing to EDMs (for $i=f$) or LFV decays (for $i \ne f$) involving a neutral-Higgs boson. Right: Feynman diagram contributing to EDMs (for $i=f$) or LFV decays (for $i \ne f$) involving a charged-Higgs boson.}
\label{EMMAMMfeyndiag}
\end{figure*}

The same diagrams which contribute to the radiative lepton decays for $\ell_{i}\neq\ell_{f}$ also affect the electric dipole moments and the anomalous magnetic moments of leptons for $\ell_{i}=\ell_{f}$. For this reason we use the same conventions as in \eq{HeffLFV} and express the EDMs of leptons in terms of the coefficients $c^{\ell_{f}\ell_{i}}_{L,R}$ of the magnetic dipole operators $ {{O}}^{\ell_{f}\ell_{i}}_{L,R} $ in the following way (using that for flavor conserving transitions $c^{\ell_{i}\ell_{i}}_{L}=c^{\ell_{i}\ell_{i}\star}_{R}$)
\begin{equation}
 d_{\ell_i} =\, {2 m_{\ell_{i}}}\, {\rm Im}\left[ {c^{\ell_{i}\ell_{i}}_{R} } \right]   \, . 
\end{equation}

In SM there is no contribution to the EDMs of leptons at the one-loop level. This is also true in the 2HDM of type II, because the Wilson coefficients are purely real since the phases of the PMNS matrix drop out in the charged Higgs contributions after summing over the massless neutrinos. However, in a 2HDM of type III, one can have neutral Higgs mediated contributions to EDMs. Note that there is no charged Higgs contribution to the charged lepton EMDs also in the 2HDM of type III because the Wilson coefficients are purely real in this case. Comparing the expression for the EDMs in the 2HDM of type III with the experimental upper bounds on $d_{e}$, $d_{\mu}$ and $d_{\tau}$ (see Table~\ref{tableEDMs}), one can constrain the parameters $\epsilon^\ell_{ij}$ (or combination of them) if they are complex.
\begin{table}[htdp]
\centering \vspace{0.8cm}
\renewcommand{\arraystretch}{1.2}
\begin{tabular}{|c| c| c| c| c|}
\hline \hline
 EDMs & ${\left|d_{e}\right|}$ &  ${|d_{\mu}|}$   &  ${d_{\tau}}$ & ${|d_{n}|}$ 
\\   \hline \hline
Bounds (${\rm e \, cm}$) & $10.5 \times 10^{-28}$ \cite{Hudson:2011zz} & $1.9 \times 10^{-19} $ \cite{Bennett:2008dy} & $\in \left[  -2.5, \,0.8 \right]  \times 10^{-17}$ \cite{Inami:2002ah}   &  $2.9 \times 10^{-26}$ \cite{Baker:2006ts}  \\ \hline \hline
\end{tabular}
  \caption{Experimental (upper) bounds on electric dipole moments.} 
\label{tableEDMs}
\end{table}
\medskip

\begin{figure}[t]
\centering
\includegraphics[width=0.4\textwidth]{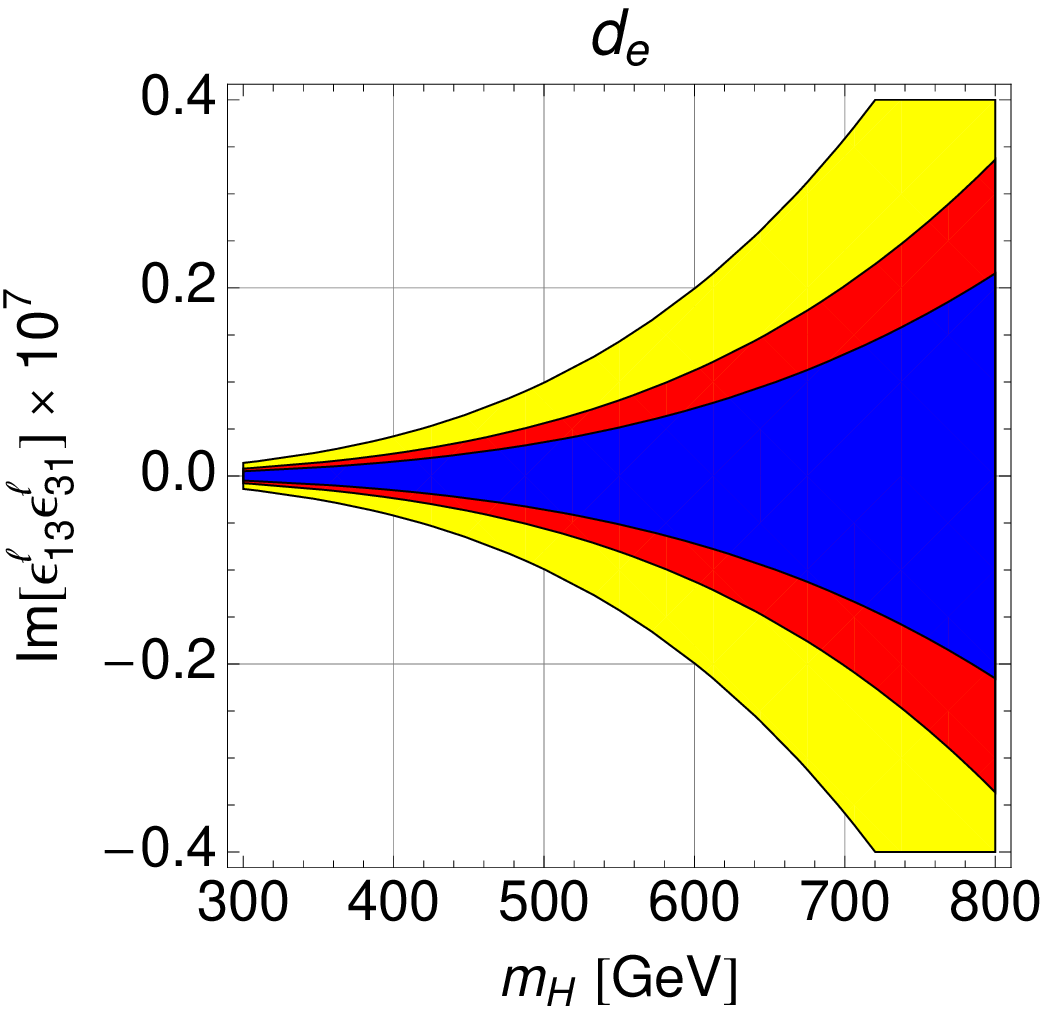}~~~
\includegraphics[width=0.4\textwidth]{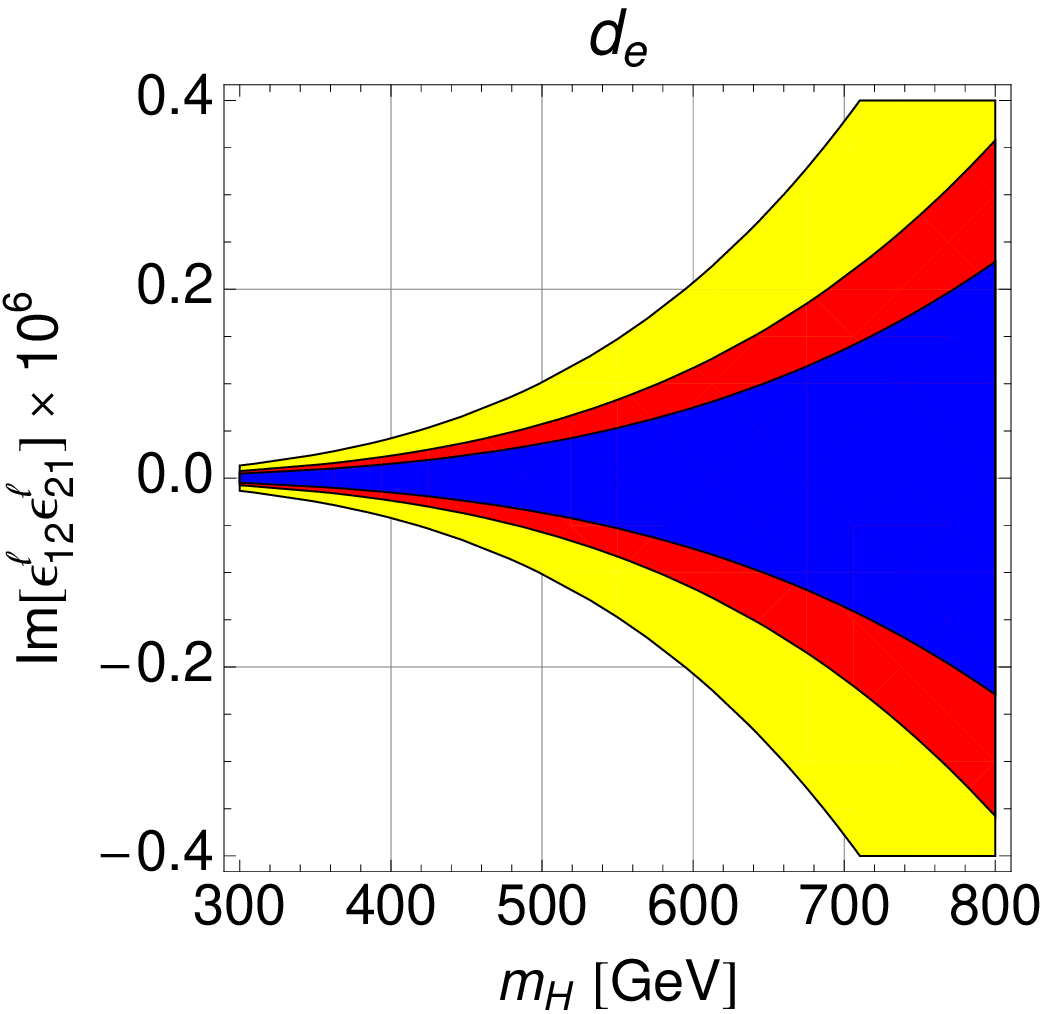}
\caption{ Allowed regions in the $\rm Im\left[ {\epsilon^\ell_{13}\epsilon^\ell_{31}} \right]$-$m_{H}$ and $\rm Im\left[ {\epsilon^\ell_{12}\epsilon^\ell_{21}} \right]$-$m_{H}$ planes from neutral Higgs contribution to ${d_{e}}$ for $\tan\beta=50$ (blue), $\tan\beta=40$ (red) and $\tan\beta=30$ (yellow). The constraints on $\rm Im\left[ {\epsilon^\ell_{11}} \right]$ are not sizable.}
\label{fig:eEDMplots}
\end{figure}

We observe that while ${d_{e}}$ enforces strong constraints on the products ${\rm Im}\left[ {\epsilon^\ell_{13}\epsilon^\ell_{31}} \right]$ and ${\rm Im}\left[ {\epsilon^\ell_{12}\epsilon^\ell_{21}} \right]$ (see Fig.~\ref{fig:eEDMplots}), ${d_{\mu}}$ and ${ d_{\tau}}$ are not capable of placing good constraints on our model parameters.
\smallskip

Similarly, following the conventions in \eq{HeffLFV}, the anomalous magnetic moments (AMMs) can be written in terms of ${c^{\ell_{i}\ell_{i}}_{R} }$ as ($\rm e > 0$)
\be
a_{\ell_i}=\, -\dfrac{4m^{2}_{\ell_{i}}}{e}\, {\rm Re}\left[ {c^{\ell_{i}\ell_{i}}_{R} }\right]   \, . 
\ee
The discrepancy between experiment and the SM prediction for the muon magnetic moment $ a_{\mu}=(g-2)/2$ is \cite{Hisano:2009ae,Passera:2004bj,Passera:2005mx,Davier:2007ua,Hagiwara:2006jt}
\be
\label{AMMbound}
\Delta a_{\mu}\, = \, a^{exp}_{\mu}-a^{SM}_{\mu} \approx \, (3 \pm 1) \times 10^{-9} \, . 
\ee
In the 2HDM of type II, the sum of the neutral and charged Higgs mediated diagrams gives the following contribution to $ a_\mu$ (for $\tan\beta=50$ and $m_{H}=500$~GeV):
\be
  a^{\rm 2HDM\, II}_{\mu} \approx \, 2.7 \times 10^{-13} \, ,
\ee
which is interfering constructively with the SM. Anyway, it can be seen that the effect is orders of magnitude smaller than the actual sensitivity and it even gets smaller for higher Higgs masses.
\medskip

\begin{figure}[t]
\centering
\includegraphics[width=0.4\textwidth]{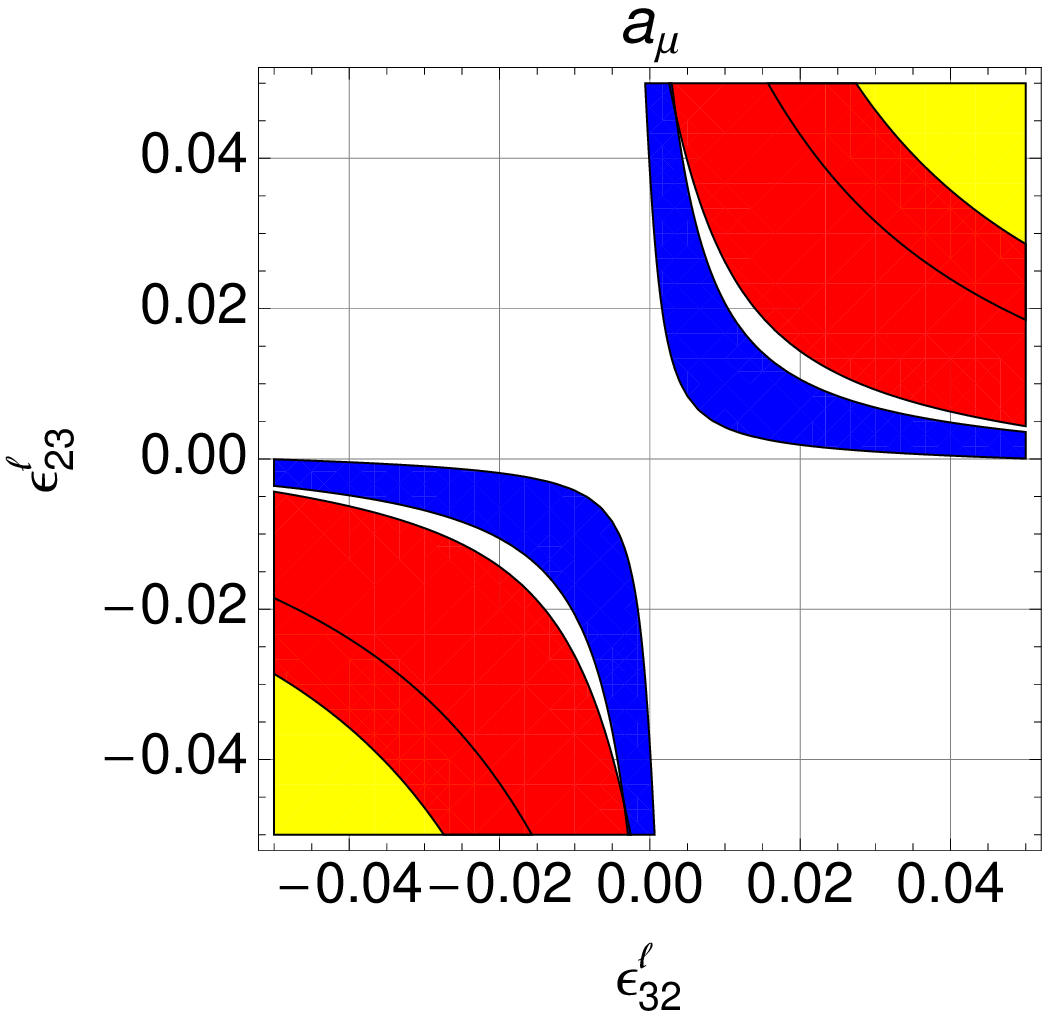}~~~
\includegraphics[width=0.42\textwidth]{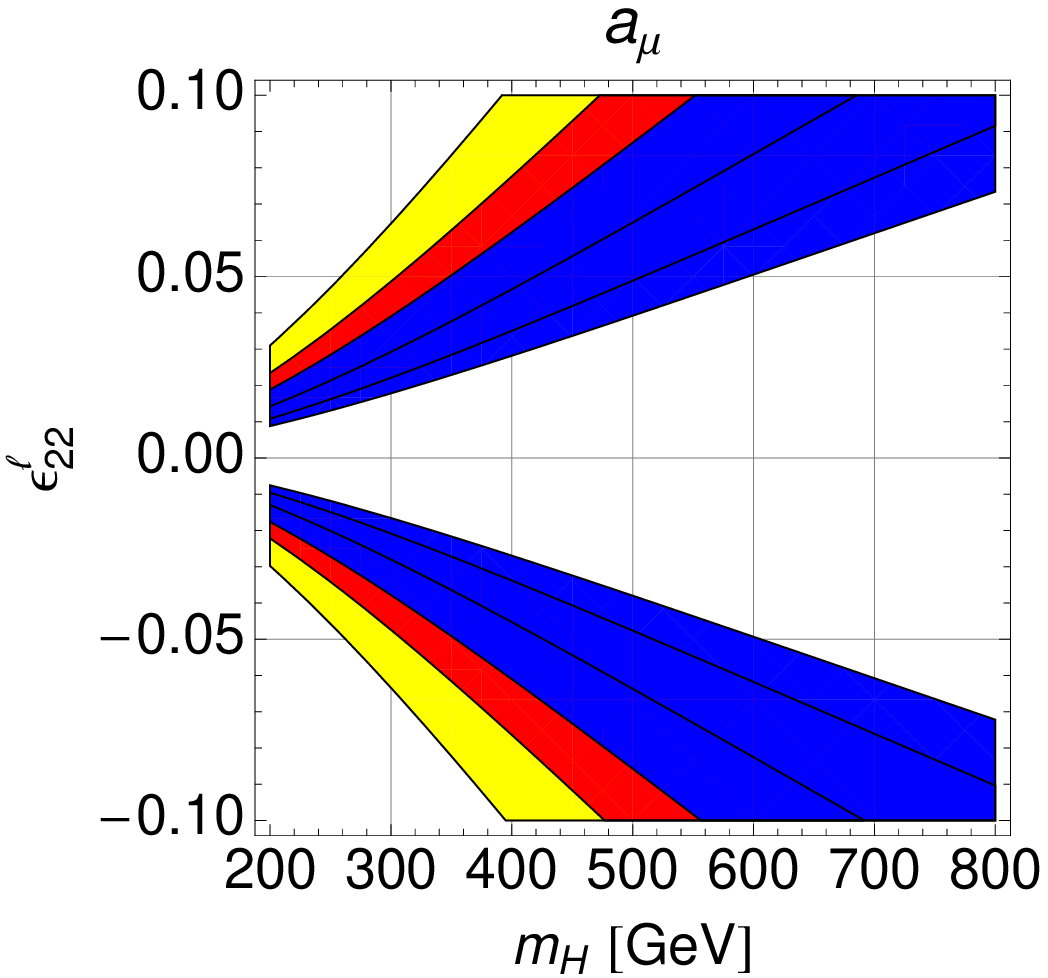}
\caption{Left: Allowed region in the $\epsilon^{\ell}_{23}$--$\epsilon^{\ell}_{32}$ plane from $\Delta a_{\mu}$ for real values of $\epsilon^{\ell}_{23}$, $\epsilon^{\ell}_{32}$ and $\tan\beta=50$, $m_{H}=700$ GeV (yellow), $m_{H}=500$ GeV (red) and $m_{H}=300$ GeV (blue). Right: Allowed region in $\epsilon^{\ell}_{22}$--$m_{H}$ plane from $\Delta a_{\mu}$ for real values of $\epsilon^{\ell}_{22}$ and $\tan\beta=50$ (blue), $\tan\beta=40$ (red) and $\tan\beta=30$ (yellow).}
\label{fig:AMMmuon}
\end{figure}

Concerning the 2HDM of type III the discrepancy between experiment and the SM prediction given in \eq{AMMbound} could be explained but only with severe fine-tuning. One would need to allow for very large values of $\epsilon^{\ell}_{22}$ which would not only violate 't Hooft's naturalness criterion but also enhance $B_s\to\mu^+\mu^-$ by orders of magnitude above the experimental limit. If one would try to explain the anomaly using $\epsilon^{\ell}_{23}$ and $\epsilon^{\ell}_{32}$ ($\epsilon^{\ell}_{12}$ and $\epsilon^{\ell}_{21}$) one would violate the bounds from $\tau^{-}\to\mu^{-}\mu^{+}\mu^{-}$ ($\mu^{-}\to e^{-}e^{+}e^{-}$ or $\mu\to e \gamma$) as illustrated in Fig.~\ref{fig:AMMmuon}.
\medskip

In conclusion, neither a type-II nor a type-III 2HDM can give a sizable effect in $a_\mu$ and both models are not capable of explaining the deviation from the SM.

\subsubsection{Electric dipole moment of the neutron}

\begin{figure}[t]
\centering
\includegraphics[width=0.41\textwidth]{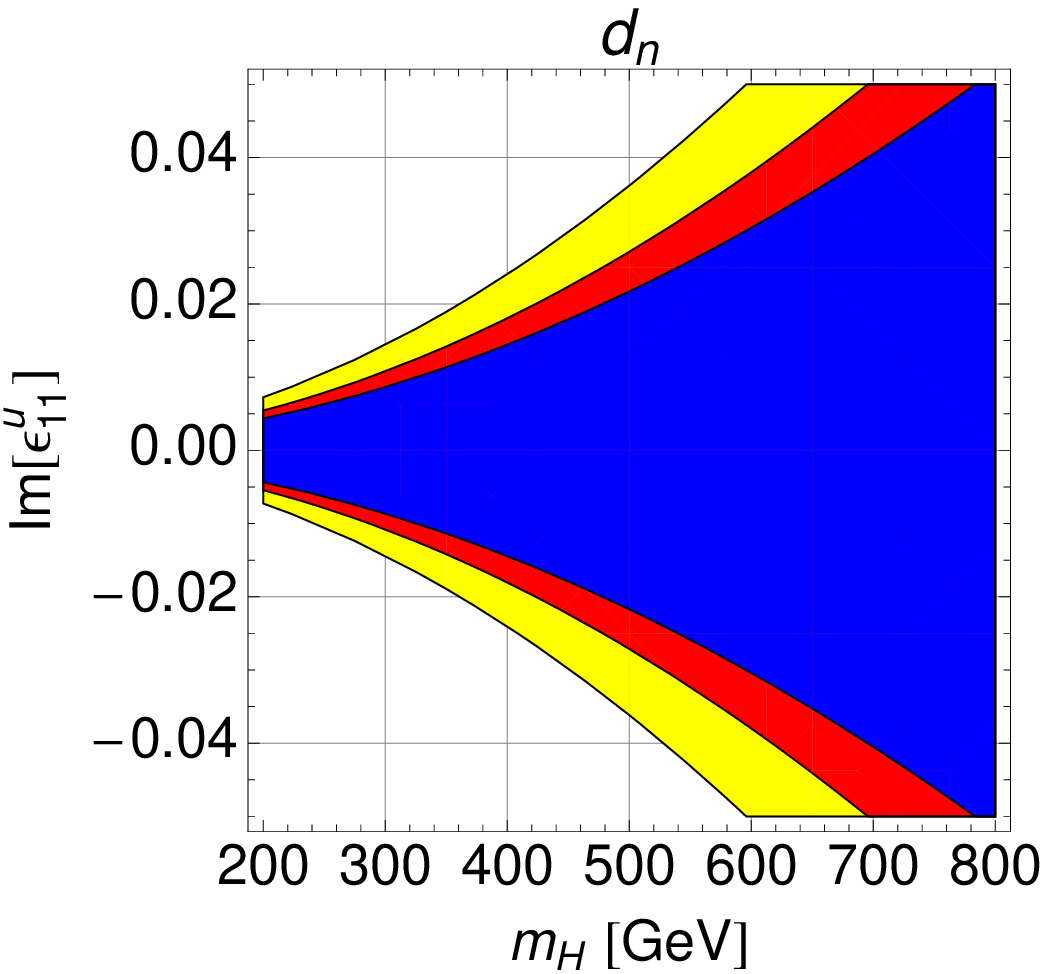}~
\includegraphics[width=0.395\textwidth]{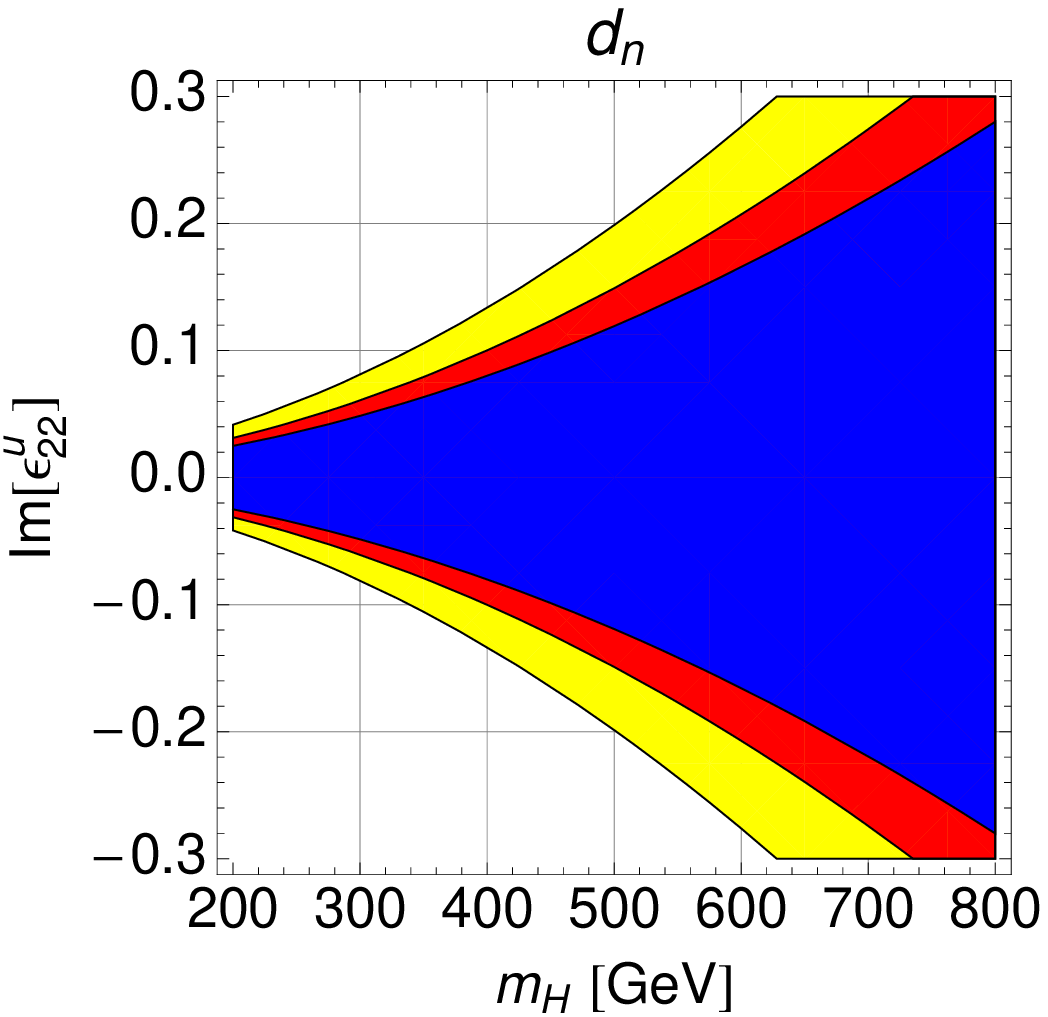}~
\caption{Allowed regions in the ${\rm Im} \left[ \epsilon^u_{11,22} \right]$--$m_{H}$ planes from the electric dipole moment of the neutron for $\tan\beta=50$ (blue), $\tan\beta=40$ (red) and $\tan\beta=30$ (yellow). We observe that $d_{n}$ can not provide good constraints on the real parts of $\epsilon^u_{11,22}$. }
\label{fig:neutronEDMb}
\end{figure} 
\begin{figure}[t]
\centering
\includegraphics[width=0.39\textwidth]{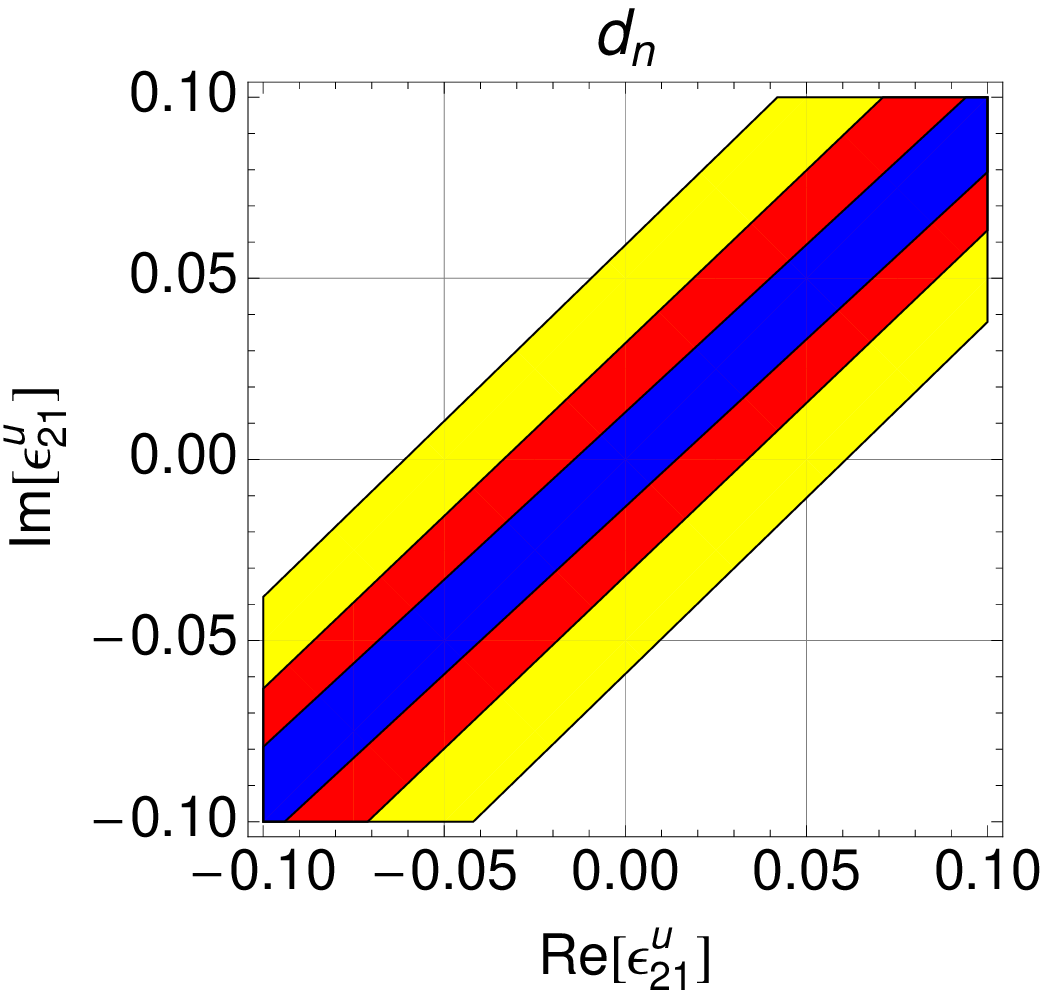}~
\includegraphics[width=0.41\textwidth]{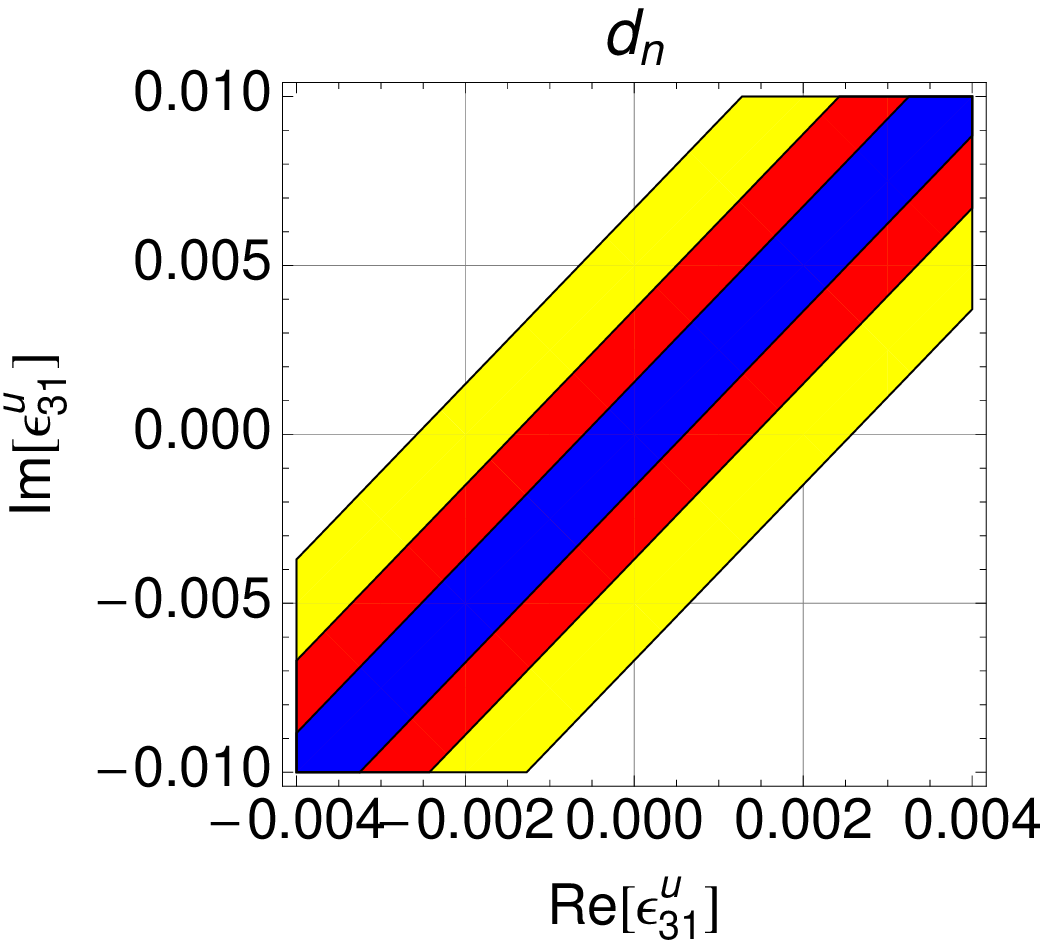}
\caption{Allowed regions in the complex $\epsilon^u_{21,31}$--planes from $d_{n}$ for $\tan\beta=50$ and $m_{H}=700$~GeV (yellow), $m_{H}=500$~GeV (red) and $m_{H}=300$~GeV (blue). We see that the absolute value of $\epsilon^u_{31}$ can only be large if it is aligned to $V_{ub}$, i.e. ${\rm Arg}[V_{ub}]={\rm Arg}[\epsilon^u_{31}] \pm \pi$ which is very important when we consider later $B\to \tau \nu$. }
\label{fig:neutronEDMb2}
\end{figure}

The neutron electric dipole moment ${ d_{n}}$ can also provide constraints on the parameters $\epsilon^{q}_{ij}$. In the SM, there is no contribution to ${ d_{n}}$ at the 1-loop level since the coefficients are real. This is also true in the type-II 2HDM.  

Using the theory estimate of Ref.~\cite{Buras:2010zm}, which is based on the QCD sum-rules calculations of Refs.~\cite{Demir:2003js,Pospelov:2000bw,Demir:2002gg,Olive:2005ru}, the neutron EDM can be written as
\begin{equation}
 {d_{n}}\, = \, (1\pm 0.5)\left[   1.4(d_{d}-0.25d_{u})+1.1 e (d^{g}_{d}+0.5d^{g}_{u})    \right]      \, ,  
 \label{neutronEDM} 
\end{equation}
where, $ d_{u}$ ($ d_{d}$) is the EDM of the up (down) quark and $ d^{g}_{u(d)}$ define the corresponding chromoelectric dipole moments which stem from the chromomagnetic dipole operator 
\begin{equation}
{ {O}}^{q_{f}q_{i}}_{R(L)}  = m_{q_{i}} \bar{q_{f}} \sigma^{\mu \nu} T^{a} P_{R(L)} q_{i} G^{a}_{\mu \nu} \, .
\end{equation}
Similar to EDMs, the (chromo) electric dipole moments of quarks are given as
\begin{equation}
\label{udEDMs}
 d^{(g)}_{q_i} =\, {2 m_{q_{i}}}\, {\rm Im}\left[ {c^{q_{i}q_{i}}_{R,(g)} } \right]   \,  .  
\end{equation}
Using the upper limit on ${d_{n}}$ (see Table~\ref{tableEDMs}) we can constrain some of $\epsilon^{u}_{ij}$ (for $\epsilon^{d}_{ij}=0$) as shown in Fig.~\ref{fig:neutronEDMb} and \ref{fig:neutronEDMb2}. These constraints are obtained for the conservative case of assuming a prefactor of $0.5$ in \eq{neutronEDM}. The explicit expressions for $ {c^{q_{i}q_{i}}_{R,(g)} }$ stemming from neutral and charged Higgs contributions to $d^{(g)}_{q_i}$ are relegated to the appendix. Note that for the neutron EDM we did not include QCD corrections.

 \section{Tree-level charged current processes}
\label{sec:charged-current}

In this section we study the constraints from processes which are mediated in the SM by a tree-level $W$ exchange and which receive additional contributions from charged Higgs exchange in 2HDMs. We study purely leptonic meson decays, semileptonic meson decays and tau lepton decays. Concerning $B$ meson decays we consider $B\to \tau \nu$, $B\to D\tau\nu$ and $B\to D^*\tau\nu$ which are, as outlined in the introduction, very interesting in the light of the observed deviation from the SM. We consider in addition $D_{(s)}\to \tau \nu$, $D_{(s)}\to \mu \nu$, $K(\pi) \to e\nu$, $K(\pi) \to \mu \nu$ and $\tau\to K(\pi) \nu$ and look for violation of lepton flavor universality via $K(\pi) \to e\nu/K(\pi)\to \mu \nu$ and $\tau\to K(\pi)\nu/K(\pi)\to\mu\nu$. Even though no deviations from the SM have been observed in these channels, they put relevant constraints on the parameter space of the type-III 2HDM.
\medskip

For purely leptonic decays of a psudoscalar meson $M$ (and also tau decays to mesons) to a lepton $\ell_j$ and a neutrino $\nu$ (which is not detected) the SM prediction is given by
\begin{equation}
{ {\cal B}_{SM}}\left[ {{M} \to \ell_{j} \nu } \right] = \frac{{{m_{{M}}}}}{{8\pi }}G_F^2 m_{\ell_{j}}^2{\tau _{{M}}}f_{{M}}^2{\left| {V_{u_f d_i}} \right|^2}{\left( {1 - \frac{{m_{\ell_{j}}^2}}{{m_{{M}}^2}}} \right)^2}\left(1+\delta^{M\ell_{j}}_{ EM}\right)\,,
\end{equation}
where $\delta^{M\ell_{j}}_{ EM}$ stands for channel dependent electromagnetic corrections (see Table~\ref{EMcorrections}), $m_M$ is the mass of the meson involved and $m_{u_f}$ ($m_{d_i}$) refers to the mass of its constituent up (down) type quark. The expression for $\tau\to M \nu$ differers by the exchange of the meson masses (life time) with the tau masses (life time) and by a factor of $1/2$ stemming from spin averaging. 
\begin{table}[t]
\centering \vspace{0.6cm}
\renewcommand{\arraystretch}{1.8}
\begin{tabular}{|c|c|c|c|}
\hline
  Ratio & Experimental value & SM prediction & $\delta^{M\ell_{j}}_{ EM}$ 
\\   \hline \hline
$  {\cal B} \left[ K \to e \nu  \right] /  {\cal B}\left[  K \to \mu \nu  \right]    $  
& $ \, \left( 2.488 \pm  0.013 \right) \times 10^{-5}$ 
& $ \, \left( 2.472 \pm  0.001 \right) \times 10^{-5}$  &  $-0.0378\pm0.0004$ \cite{Antonelli:2008jg}   \\ \hline
$   {\cal B} \left[ K \to \mu \nu  \right] /  {{ {\cal B}}}\left[  \pi \to \mu \nu  \right]    $  
& $ \, \left( 63.55 \pm  0.11 \right) \times 10^{-2} $
& $ \, \left( 63.48 \pm  1.37 \right) \times 10^{-2}   $  &$-0.0070\pm0.0018$ \cite{Antonelli:2010yf}   \\ \hline
$   {\cal B} \left[ K \to e \nu  \right] /  {{ {\cal B}}}\left[  \pi \to e \nu  \right]    $  
& $ \,\left( 1.285  \pm  0.008 \right) \times 10^{-1} $
& $ \, \left( 1.270  \pm  0.027 \right) \times 10^{-1}  $  &$-0.0070\pm0.0018$ \cite{Antonelli:2010yf}   \\ \hline
$  {\cal B} \left[ \pi \to e \nu  \right] /  {\cal B}\left[  \pi \to \mu \nu  \right]    $  
& $ \, \left( 1.230 \pm  0.004 \right) \times 10^{-4}$
& $ \,  1.234 \,  \times 10^{-4}$    & $ -3.85 \, \% $ \cite{Decker:1993py} \\ \hline
$  {\cal B} \left[ \tau \to K \nu  \right] /  {\cal B}\left[  \tau \to \pi \nu  \right]    $  
& $ \, \left( 6.46 \pm  0.10 \right) \times 10^{-2}$
&  $ \, \left( 6.56 \pm  0.16 \right) \times 10^{-2}$   &$0.0003\pm0.0044$ \cite{Banerjee:2008hg}   \\ \hline
$  {\cal B} \left[ \tau \to \pi \nu  \right] /  {\cal B}\left[  \pi \to \mu \nu  \right]    $  
& $ \, \left( 10.83 \pm  0.06 \right) \times 10^{-2}$
&  $ \,  10.87  \times 10^{-2}  $    &$+1.2 \, \%$ \cite{Decker:1993py}   \\ \hline
$  {\cal B} \left[ \tau \to K \nu  \right] /  {\cal B}\left[  K \to \mu \nu  \right]    $  
& $ \, \left( 1.102 \pm  0.016 \right) \times 10^{-2}$
&  $ \,  1.11 \times 10^{-2}   $     &$+2.0 \, \%$ \cite{Decker:1993py} \\ \hline \hline
\end{tabular}
\caption{Experimental values, SM predictions and electromagnetic corrections (in the SM) for the ratios of charged current processes. The experimental values are obtained by adding the errors of the individual branching ratios given in Ref.~\cite{Beringer:1900zz} in quadrature. The SM predictions include the uncertainties from $\delta^{M\ell_{j}}_{ EM}$ and (if involved) as well as the uncertainties due to CKM factors and decay constants. As always, we add the theory error linear to the experimental ones.}
\label{EMcorrections}
\end{table}
\medskip

NP via scalar operators can be included very easily:
\begin{equation}
{{\cal B}_{NP}}={ {\cal B}_{SM}}{\left| {1 + \frac{{m_M^2}}{\left(m_{u_f}+m_{d_i}\right){m_{\ell_j} }}\frac{ {C^{u_f d_i\,, \ell_{j}} _{R}- C^{u_f d_i\,, \ell_{j}} _{L} } }{ {C^{{ u_f d_i \,, \ell_{j} }}_{SM}} } } \right|^2}
\label{eq:BrNP}
\end{equation}
with 
\begin{equation}
{C^{{u_f d_i  \,, \ell_{j} }}_{SM}}=4 G_F V_{u_f d_i}/\sqrt{2}\,.
\end{equation}
\smallskip

All quantities in \eq{eq:BrNP} are understood to be at the meson scale $m_M$. Like for $B_s\to\mu^+\mu^-$, the SM Wilson coefficient is renormalization scale independent and the scalar Wilson coefficients evolve in the same way as the quark masses.
\smallskip

In the 2HDM III the Wilson coefficients $ C^{u_f d_i\,, \ell_{j}} _{L} $ and $ C^{u_f d_i\,, \ell_{j}} _{R}  $ are given by (neglecting terms which are not $\tan\beta$ enhanced)
\begin{equation}
\begin{gathered}
  C^{u_f d_i\,, \ell_{j}} _{R}  = -\frac{{  {{\tan }^2}\beta }}{{m_{{H^ \pm }}^2}}\left( {{V_{fi}}\frac{{{m_{{d_i}}}}}{v} - \sum\limits_{j = 1}^3 {{V_{fj}}\epsilon_{ji}^d{\mkern 1mu} } } \right){\mkern 1mu} \left( {\frac{{{m_{{\ell _j}}}}}{v} - \sum\limits_{k = 1}^3 {\epsilon_{kj}^{\ell \star}} } \right)   \, , \hfill  \\
 C^{u_f d_i\,, \ell_{j}} _{L}  = \frac{{\tan \beta }}{{m_{{H^ \pm }}^2}}\sum\limits_{j = 1}^3 {{\mkern 1mu} V_{ji}^{}\epsilon_{jf}^{\star u}{\mkern 1mu} } \left( {\frac{{{m_{{\ell _j}}}}}{v} - \sum\limits_{k = 1}^3 {\epsilon_{kj}^{\ell \star}} } \right)  \, . \hfill \\ 
\end{gathered} 
\end{equation}
Note that $C^{u_f d_i\,, \ell_{j}} _{L}  $ is only proportional to one power of $\tan\beta$ while $ C^{u_f d_i\,, \ell_{j}} _{R}  $ is proportional to $\tan^2\beta$. The Hamiltonian governing $M\to\ell_j\nu$ ($\tau\to M \nu$) and the Wilson coefficients for general scalar interactions are given in the appendix. It is important to keep in mind that, since we are dealing with lepton flavour-violating terms, we must sum over the neutrinos in the final state because the neutrino is not detected. Note that we did not include the PMNS matrix in both $C^{u_f d_i \,, \ell_{j} } _{SM} $ and $C^{u_f d_i\,, \ell_{j}} _{L,R}$ for simplifying the expressions, since it cancels in the final expression after summing over the neutrinos. 
\medskip

For semileptonic meson decays \bdtau and \bdstau, which have a three-body final state, both the SM prediction and the inclusion of NP is more complicated, as will be discussed in subsection~\ref{subsection:bdstau}.

\subsection{Tauonic charged $B$ meson decays: \btau, \bdtau and \bdstau}

As discussed in the introduction the BABAR collaboration performed an analysis of the semileptonic $B$ decays \bdtau and \bdstau using the full available data set~\cite{BaBar:2012xj,Lees:2013qea}. They find for the ratios
\begin{equation}
{\cal R}(D^{(*)})\,=\,{\cal B}(B\to D^{(*)} \tau \nu)/{\cal B}(B\to D^{(*)} \ell \nu)\,,
\end{equation}
(with $\ell=e,\mu$) the following results:
\begin{eqnarray}
{\cal R}(D)\,=\,0.440\pm0.058\pm0.042  \,,\\
{\cal R}(D^*)\,=\,0.332\pm0.024\pm0.018\,.
\end{eqnarray}
Here the first error is statistical and the second one is systematic. Comparing these measurements to the SM predictions
\begin{eqnarray}
{\cal R}_{\rm SM}(D)\,=\,0.297\pm0.017 \,, \\
{\cal R}_{\rm SM}(D^*) \,=\,0.252\pm0.003 \,,
\end{eqnarray}
we see that there is a discrepancy of 2.0\,$\sigma$ for \rd and 2.7\,$\sigma$ for \rds. For the theory predictions we used the updated results of \cite{BaBar:2012xj}, which rely on the calculations of Refs.~\cite{Kamenik:2008tj,Fajfer:2012vx} based on the results of Refs.~\cite{Korner:1987kd,Korner:1989ve,Korner:1989qb,Heiliger:1989yp,Pham:1992fr}. The measurements of both ratios ${\cal R}(D)$ and ${\cal R}(D^*)$ exceed the SM prediction, and combining them gives a $3.4\, \sigma$ deviation from the SM~\cite{BaBar:2012xj,Lees:2013qea} expectation. 
\medskip

This evidence for the violation of lepton flavour universality in \bdtau and \bdstau is further supported by the measurement of \btau by BABAR~\cite{Lees:2012ju,Aubert:2008zzb} and BELLE~\cite{Hara:2010dk}. Until recently, all measurements of \btau (the hadronic tag and the leptonic tag both from BABAR and BELLE) were significantly above the SM prediction. However, the latest BELLE result for the hadronic tag \cite{Adachi:2012mm} of ${\cal B}[B\to \tau\nu]=(0.72^{+0.27}_{-0.25}\pm0.11)\times 10^{-4}$ is in agreement with the SM prediction \cite{Charles:2004jd}: 
\begin{equation}
{\cal B}_{\rm SM}[B\to \tau\nu]=(0.796^{+0.088}_{-0.087})\times 10^{-4}\,.
\end{equation}
Averaging all measurements, one obtains the branching ratio 
\begin{equation}
{\cal B}_{exp}[B\to \tau\nu]=(1.15\pm0.23)\times 10^{-4}\,.
\end{equation}
which now disagrees with the SM prediction by $1.6\, \sigma$ using $V_{ub}$ from the global fit \cite{Charles:2004jd}.
\medskip

Combining \rd, \rds and \btau, we have evidence for violation of lepton flavor universality. Assuming that these deviations from the SM are not statistical fluctuations or underestimated theoretical or systematic uncertainties, it is interesting to ask which model of new physics can explain the measured values \cite{Fajfer:2012jt,Crivellin:2012ye,Deshpande:2012rr,He:2012zp,Celis:2012dk,Tanaka:2012nw,Bailey:2012jg,Datta:2012qk,Becirevic:2012jf,Ko:2012sv,Biancofiore:2013ki}.
\medskip

\subsubsection{\bdtau and \bdstau}
\label{subsection:bdstau}

Let us first consider  the semileptonic decays \bdtau and \bdstau. Here the Wilson coefficients $ C^{q b\,, \tau} _{R} $ and $C^{q b\,, \tau} _{L} $ affect \bdtau and \bdstau in the following way \cite{Akeroyd:2003zr,Fajfer:2012vx,Sakaki:2012ft}:
\begin{widetext}
\begin{eqnarray}
{\cal R}(D) = {\cal R}_{ SM}(D) \left(1 + 1.5 \Re\left[\frac{C_{R}^{cb\,, \tau }+C_L^{cb\,, \tau}}{C_{SM}^{cb \,, \tau}}\right]+1.0 \left|\frac{C_{R}^{cb\,, \tau}+C_L^{cb\, ,\tau}}{C_{SM}^{cb \,, \tau}}\right|^2  \right)\,, \label{RD}  \\
{\cal R}(D^*) = {\cal R}_{ SM}(D^*) \left(1 + 0.12 \Re \left[\frac{C_{R}^{cb\,, \tau}-C_L^{cb\,, \tau}}{C_{SM}^{cb \,, \tau }}\right]+0.05 \left|\frac{C_{R}^{cb\, ,\tau}-C_L^{cb\, ,\tau}}{C_{SM}^{cb \,, \tau}}\right|^2  \right)\,.
\end{eqnarray}
\end{widetext}
For our analysis we add the experimental errors in quadrature and the theoretical uncertainty linear on top of this. There are also efficiency corrections to ${\cal R}(D)$ due to the BABAR detector \cite{BaBar:2012xj} which are important in the case of large contributions from the scalar Wilson coefficients $C_{R,L}^{cb \, ,\tau}$ (i.e. if one wants to explain ${\cal R}(D)$ with destructive interference with the SM contribution). As shown in Ref.~\cite{Fajfer:2012jt}, these corrections can be effectively taken into account by multiplying the quadratic term in $C_{R,L}^{cb \, ,\tau}$ of \eq{RD} by an approximate factor of 1.5 (not included in \eq{RD}). 
\medskip

\begin{figure}[t]
\centering
\includegraphics[width=0.4\textwidth]{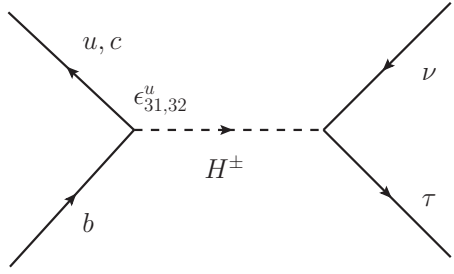}
\caption{Feynman diagram showing a charged Higgs contributing to \btau and $B\to D^{(*)}\tau\nu$ involving the flavour changing parameters $\epsilon^u_{31}$ and $\epsilon^u_{32}$ which affect \btau and $B\to D^{(*)}\tau\nu$, respectively.
\label{feynman-diagram}}
\end{figure}

\begin{figure*}[t]
\centering
\includegraphics[width=0.31\textwidth]{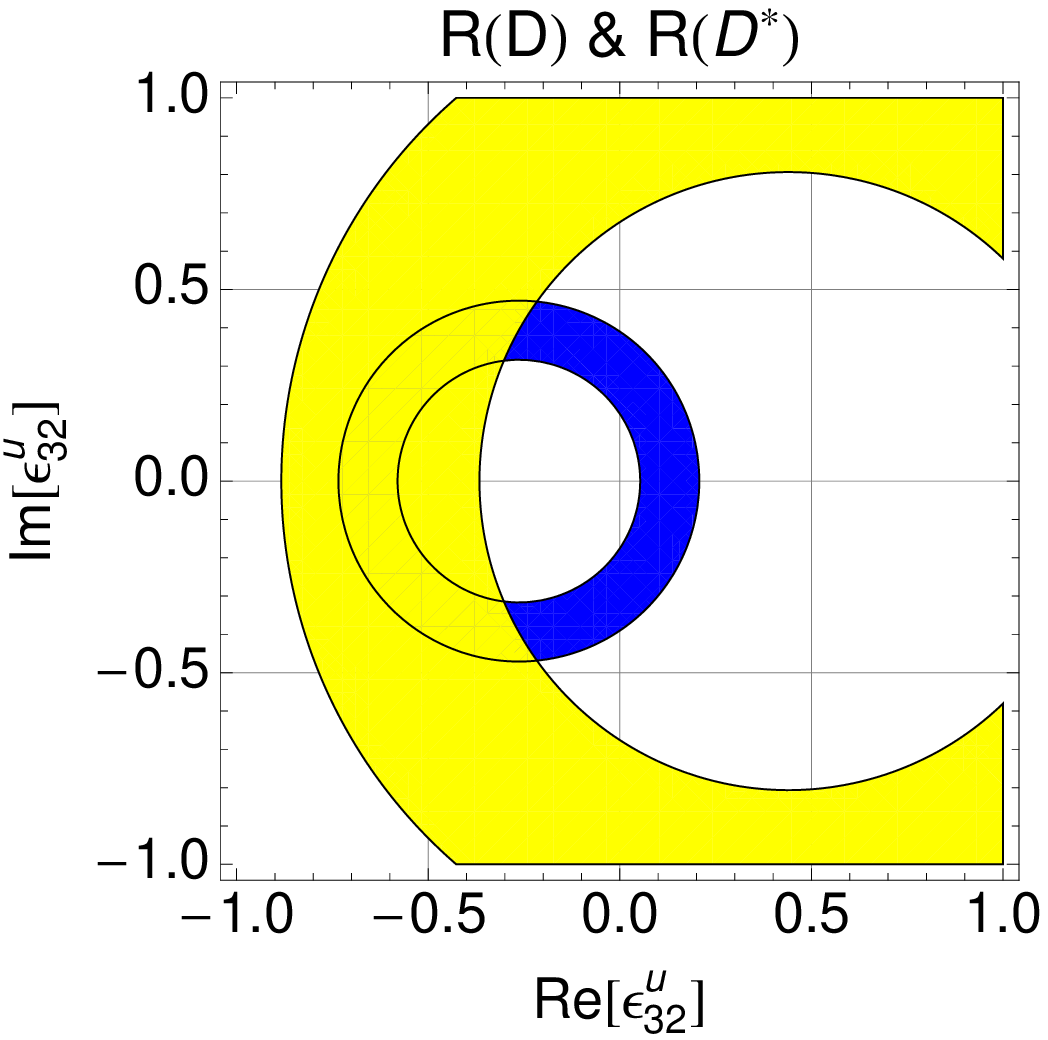}
\includegraphics[width=0.328\textwidth]{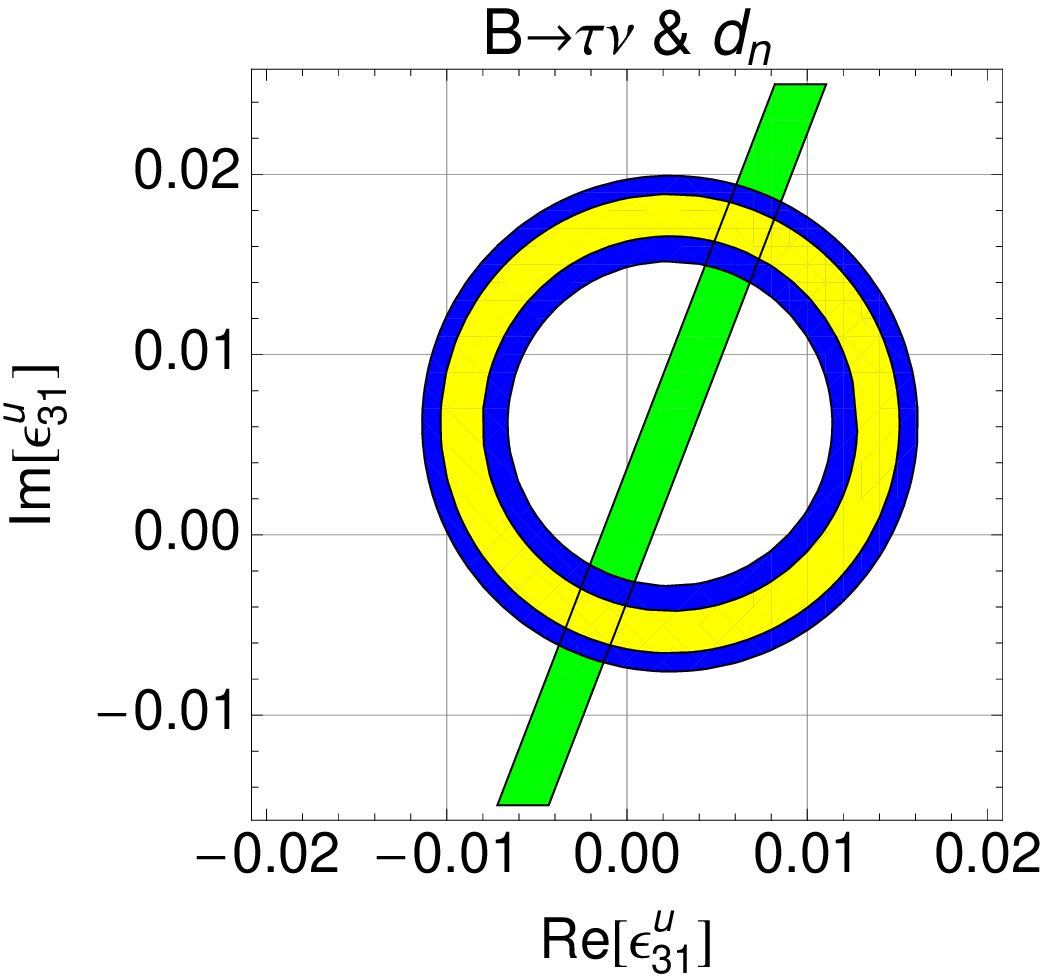}
\includegraphics[width=0.33\textwidth]{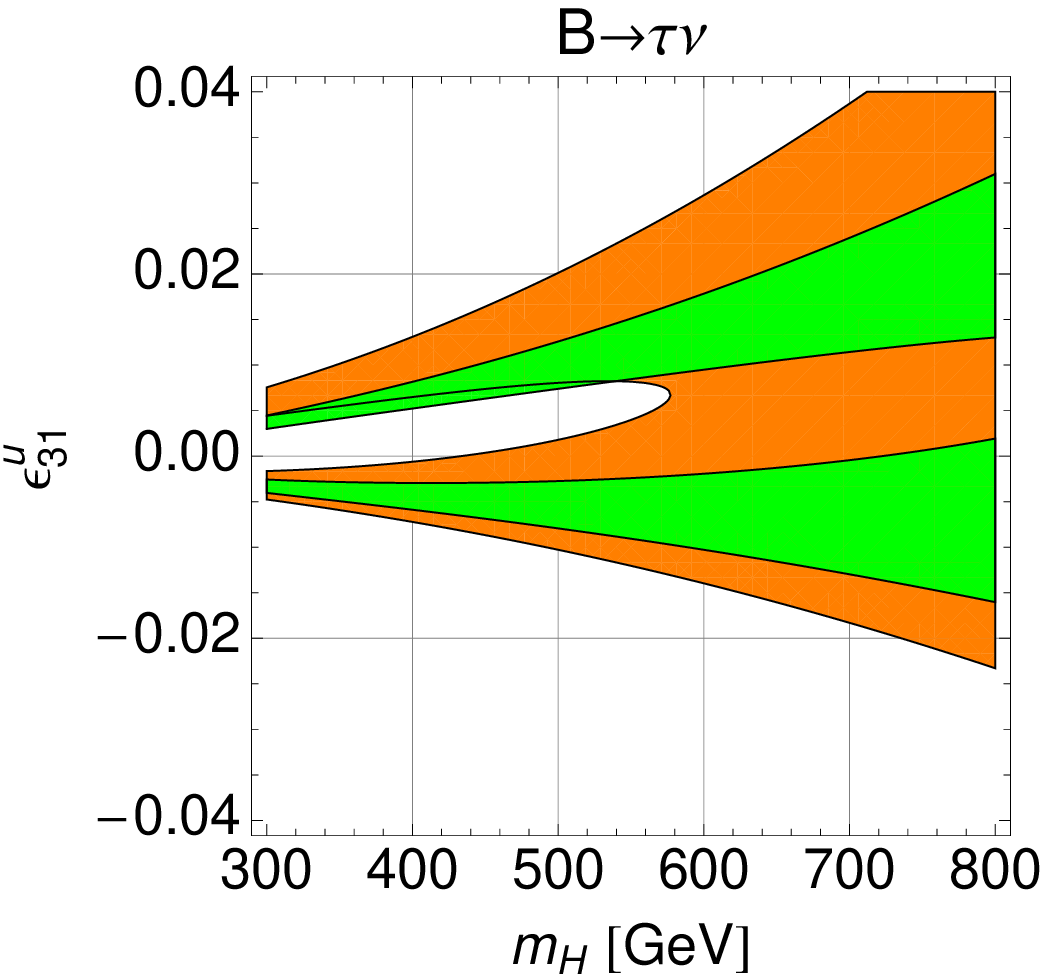}
\caption{Left: Allowed regions in the complex $\epsilon^u_{32}$--plane from \rd (blue) and \rds (yellow) for $\tan\beta=50$ and $m_H=500$~GeV. Middle:  Allowed regions in the complex $\epsilon^u_{31}$--plane combining the constraints from \btau (1 $\sigma$ (yellow) and 2 $\sigma$ (blue)) and neutron EDM (green) for $\tan\beta=50$ and $m_H=500$~GeV. Right: Allowed regions in the $m_{H}$--$\epsilon^u_{31}$ plane from \btau for real values of $\epsilon^u_{31}$ and $\tan\beta=50$~(green), $\tan\beta=30$~(orange). }
\label{2HDMIII}
\end{figure*}

\begin{figure*}[t]
\centering
\includegraphics[width=0.4\textwidth]{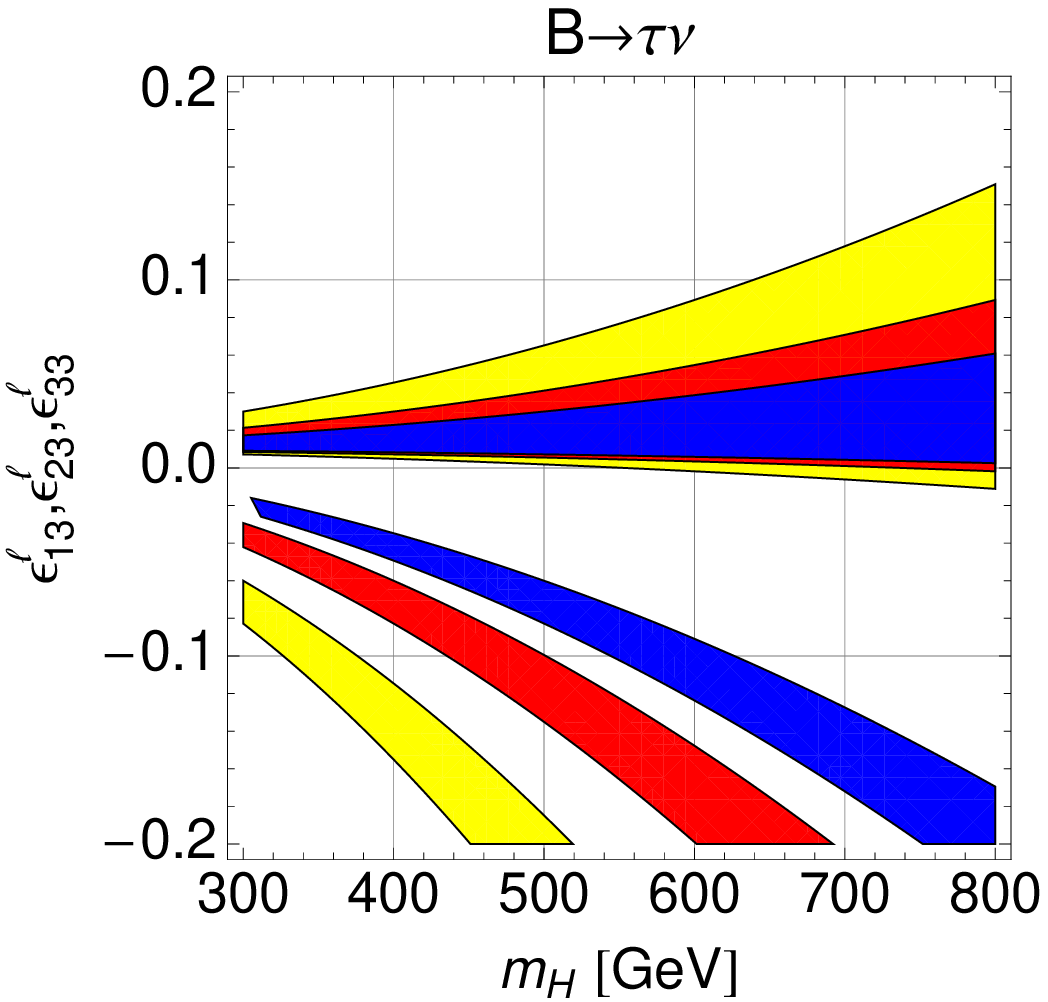}~~~~
\includegraphics[width=0.41\textwidth]{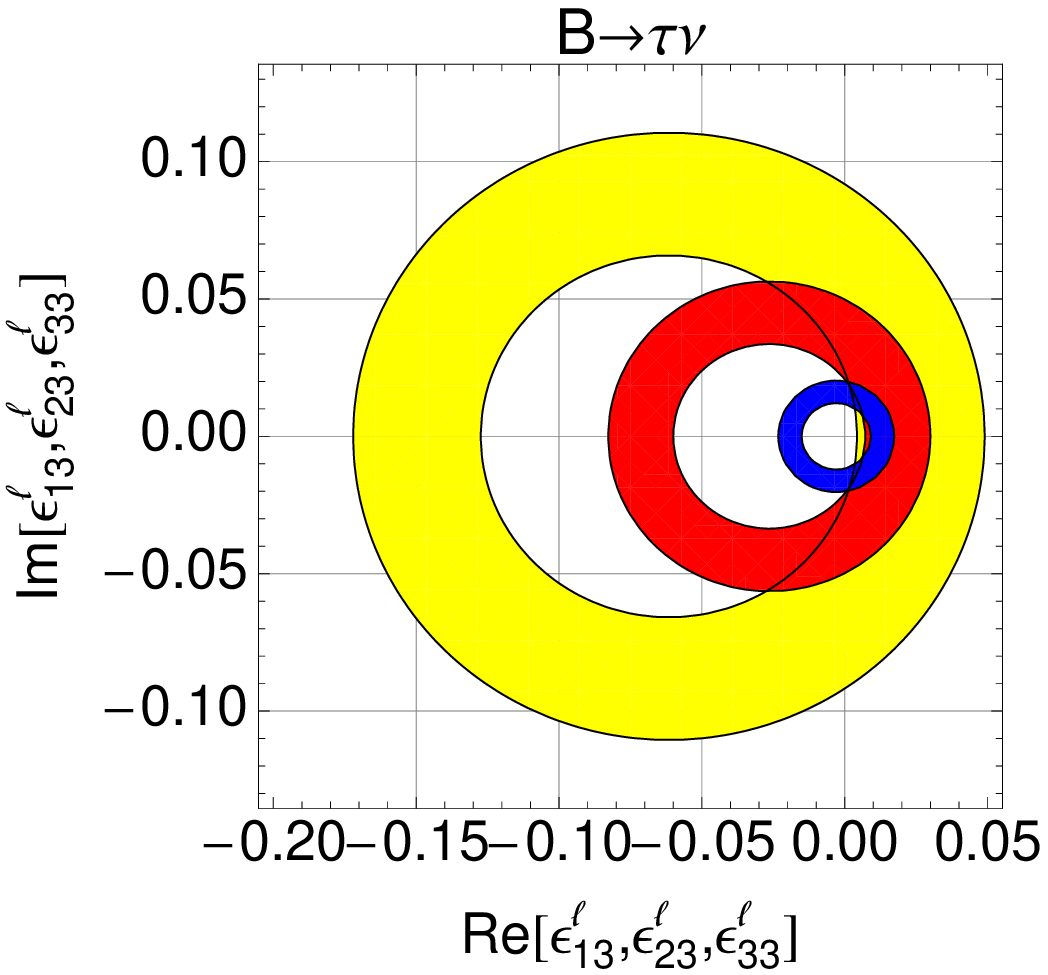}
\caption{Left: Allowed regions in the $m_{H}$--$\epsilon^\ell_{i3}$ plane from \btau for real values of $\epsilon^\ell_{i3}$ and $\tan\beta=30$~(yellow), $\tan\beta=40$~(red), $\tan\beta=50$~(blue). Right: Allowed regions in the complex $\epsilon^\ell_{13}$, $\epsilon^\ell_{23}$ and $\epsilon^\ell_{33}$--planes from \btau for $m_{H}=700$ GeV (yellow), $m_{H}=500$ GeV (red) and $m_{H}=300$ GeV (blue).}
 \label{2HDMIII-B-btau_E33}
\end{figure*}

Since $\epsilon^d_{33}$ contributes to $C_R^{cb\, ,\tau}$ (the same Wilson coefficient generated in the type-II 2HDM) it cannot simultaneously explain \rd and \rds. Therefore, we are left with $\epsilon^u_{32}$, which contributes to \bdtau and \bdstau. In the left frame of Fig.~\ref{2HDMIII} we see the allowed region in the complex $\epsilon^u_{32}$-plane, which gives the correct values for \rd and \rds within the $1\, \sigma$ uncertainties for $\tan\beta=50$ and $m_H=500$~GeV, and the middle and the right frames correspond to the allowed regions on $\epsilon^u_{31}$ from \btau.

\subsubsection{\btau}

In principle, \btau can be explained either by using $\epsilon^d_{33}$ (as in 2HDMs with MFV) or by $\epsilon^u_{31}$ (or by a combination of both of them). However, $\epsilon^d_{33}$ alone cannot explain the deviation from the SM without fine tuning, while $\epsilon^u_{31}$ is capable of doing this \cite{Crivellin:2012ye}. 
\medskip

\btau can also be used to constrain $\epsilon^\ell_{13}$, $\epsilon^\ell_{23}$ and $\epsilon^\ell_{33}$ as illustrated in Fig.~\ref{2HDMIII-B-btau_E33}. In order to obtain these constraints, we assumed that all other relevant elements ($\epsilon^d_{33}$ and $\epsilon^u_{31}$) are zero.

\begin{figure*}[t]
\centering
\includegraphics[width=0.42\textwidth]{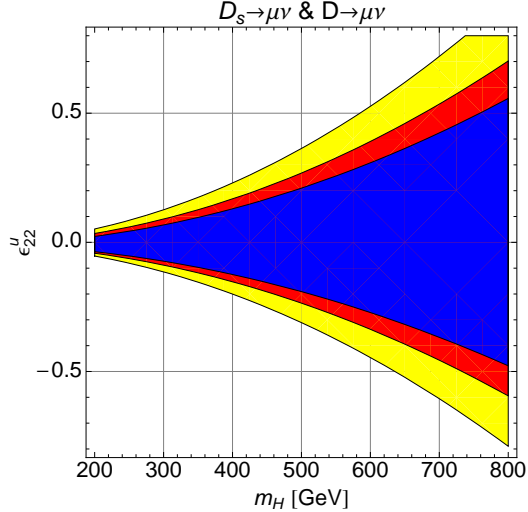}~~~
\caption{ {Left:} Allowed region in the $m_H$--$\epsilon^u_{22}$--plane (for real values of $\epsilon^u_{22}$) obtained by combining the constraints from
$ D\to\mu\nu$  and $D_s\to\mu\nu$ for $\tan\beta=30$ (yellow), $\tan\beta=40$ (red) and $\tan\beta=50$ (blue). While the upper bound on $\epsilon^u_{22}$ comes from $D_{s}\to\mu\nu$, $D \to\mu\nu$ is more constraining for negative values of $\epsilon^u_{22}$. The bound on the imaginary part of $\epsilon^u_{22}$ are very weak. The constraints from $D_{s}\to\tau\nu$ turn out to be comparable (but a bit weaker) while the ones from $D \to\tau\nu$ are weak.}
\label{D-plots}
\end{figure*}

\subsection{${D_{(s)} \to \tau \nu}$ and ${D_{(s)} \to \mu \nu}$ }

Previously, there were some indications for NP in $D_s \to\tau\nu$ \cite{Dobrescu:2008er,Alexander:2009ux,Onyisi:2009th}. However, using the new experimental values for ${\cal B}\left[D_s\to\tau\nu\right]$ (see Table~\ref{Dmutaunu}) and the improved lattice determination for the decay constant $f_{D_s}$ \cite{Simone:2011zza,Bazavov:2011aa} we find agreement between the SM predictions and experiment. Nevertheless, it is interesting to consider the constraints on the 2HDM of type III parameter space. Charged Higgs contributions to  $D_{(s)} \to \tau \nu$ and $D_{(s)} \to \mu \nu$ have been investigated in Ref.~\cite{Akeroyd:2003jb,Akeroyd:2007eh,Akeroyd:2009tn,Barranco:2013tba}. 
\begin{table}[t]
\centering \vspace{0.5cm}
\renewcommand{\arraystretch}{1.2}
\begin{tabular}{|c|c|c|}
\hline
  Process & Experimental value (bound) & SM prediction
\\   \hline \hline
~${\cal B}\left[       D_s\to\tau\nu  \right]$~  
&~ $ \, \left( 5.43\pm 0.31 \right) \times 10^{-2}$~ 
& ~$  \, \left(  5.36^{+0.54}_{-0.50}  \right) \times 10^{-2} $~ \\ \hline 
~${\cal B}\left[       D_s\to\mu\nu  \right]$~  
&~ $ \, \left( 5.90\pm 0.33 \right) \times 10^{-3}$~ 
& ~$  \, \left(  5.50^{+0.55}_{-0.52}  \right) \times 10^{-3} $~ \\ \hline 
~${\cal B}\left[       D\to\tau\nu  \right]$~  
&~ $ \, \leq 1.2  \times 10^{-3}$~ 
& ~$  \, \left(  1.10\pm 0.06  \right) \times 10^{-3} $~ \\ \hline 
~${\cal B}\left[  D \to\mu\nu  \right]$~  
&~ $ \, \left( 3.82\pm 0.33 \right) \times 10^{-4}$~ 
& ~$  \, \left(  4.15^{+0.22}_{-0.21}  \right) \times 10^{-4} $~ \\ \hline  \hline
\end{tabular}
  \caption{Experimental values (upper bounds) and SM predictions for $D_{(s)} \to \tau \nu$ and $D_{(s)} \to \mu \nu$ processes. The SM prediction for $D_{s} \to \mu \nu$ mode takes into account the EM correction effects of $+1.0 \, \%$~\cite{Alexander:2009ux,Dobrescu:2008er,Burdman:1994ip}.} 
\label{Dmutaunu}
\end{table}
\medskip 

The most important constraints on the 2HDM of type III parameter space are the ones on $\epsilon^u_{22}$ (shown in Fig.~\ref{D-plots}). $D_{(s)} \to \tau\nu$ and $D_{(s)} \to\mu \nu$ constrains ${\rm Re}\left[{\epsilon^u_{22}}\right]$ while the constraints on ${\rm Im}\left[{\epsilon^u_{22}}\right]$ are very weak. In principle, also the ratio $D_{(s)} \to \tau \nu/D_{(s)} \to \mu \nu$ could be used for constraining deviations from lepton flavor universality, but the constraints from $K(\pi) \to e \nu/K (\pi) \to \mu \nu $ and $\tau\to K  (\pi) \nu/K  (\pi)  \to\mu\nu$ turn out to be stronger.
\medskip

\begin{figure*}[t]
\centering
\includegraphics[width=0.43\textwidth]{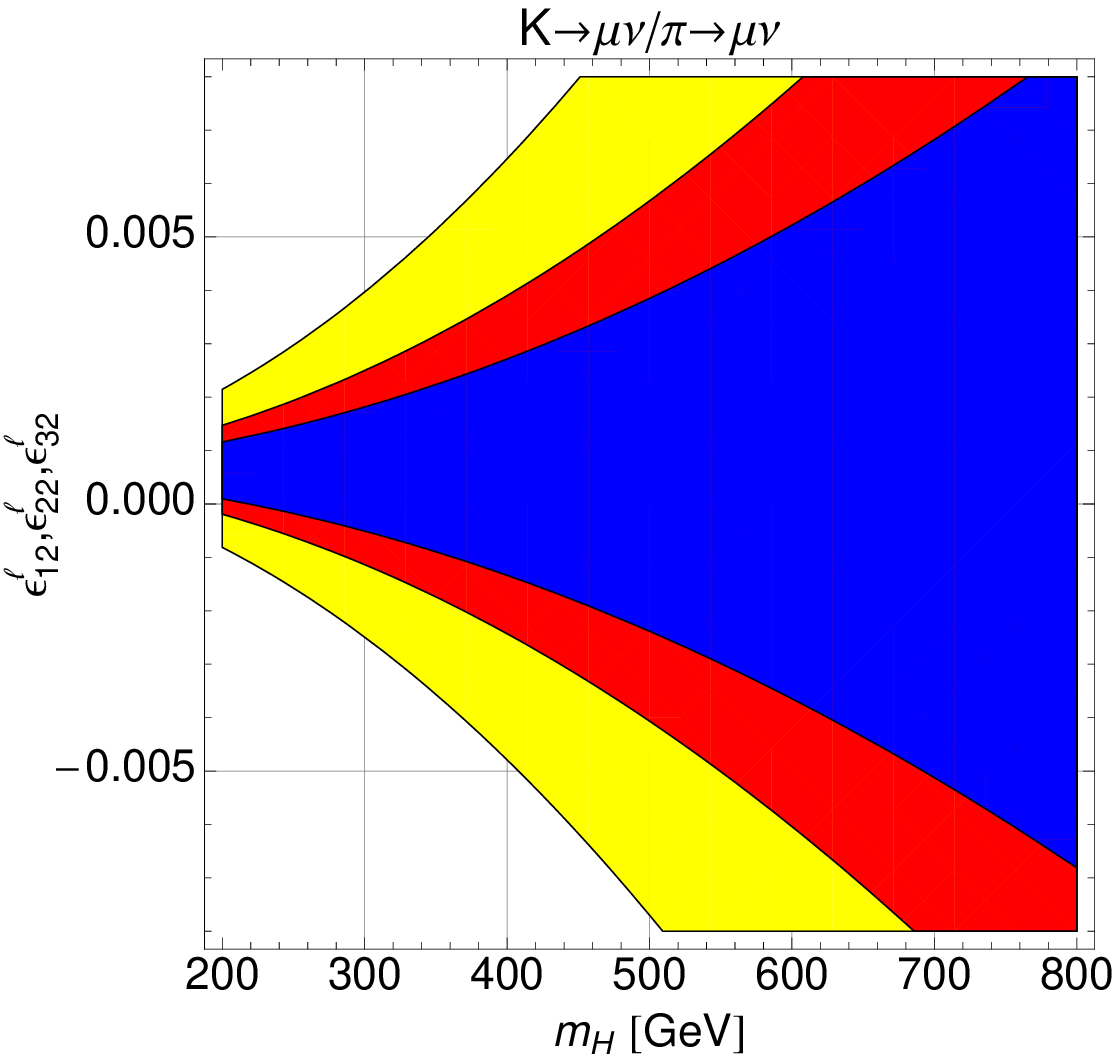}
\includegraphics[width=0.44\textwidth]{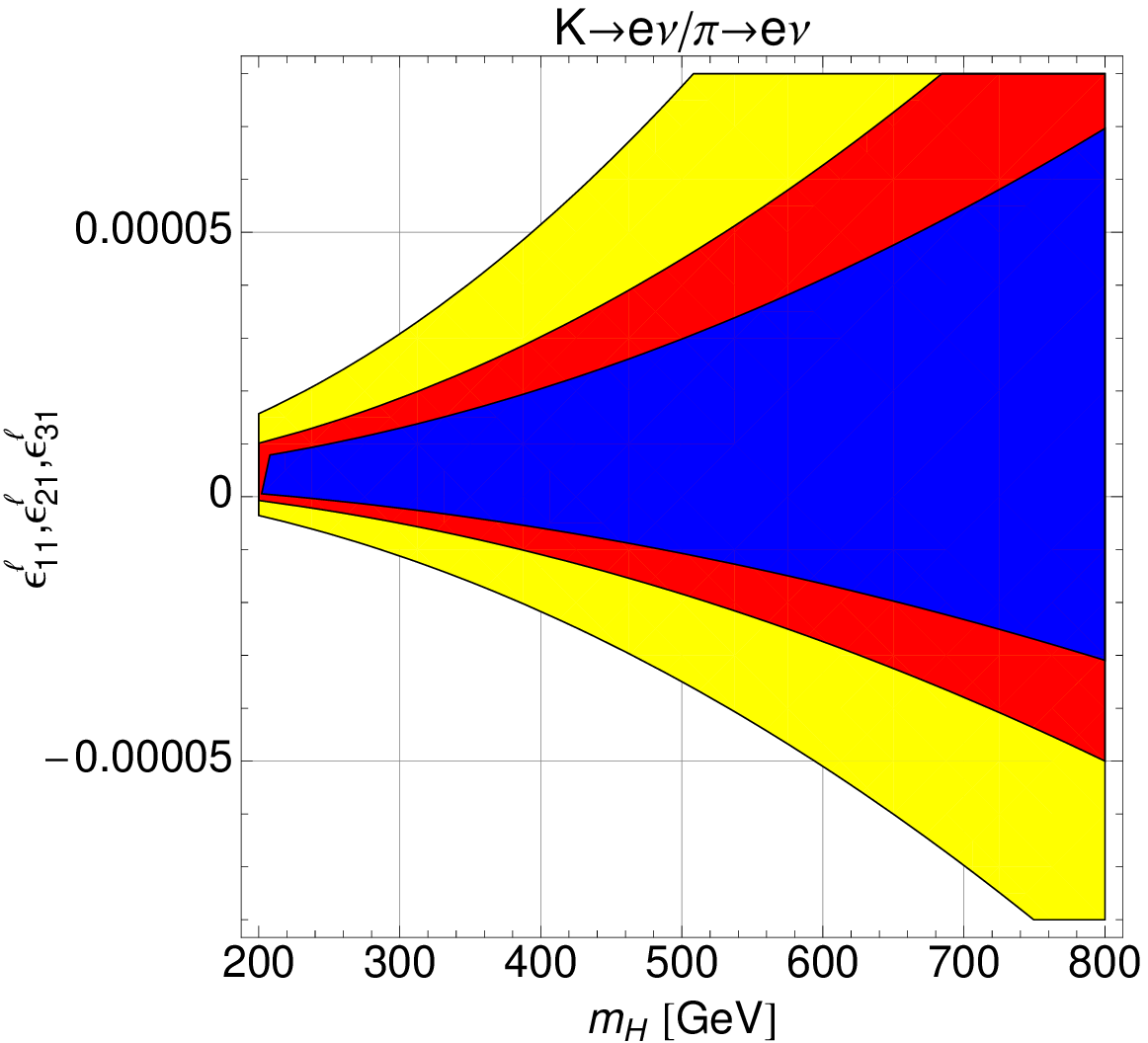}
\caption{Allowed regions in the $m_H$--$\epsilon^{\ell}_{i1,i2}$--plane from $K\to\mu\nu/\pi\to\mu\nu$ and $K\to e \nu/\pi\to e \nu$ for real values of $\epsilon^{\ell}_{i1,i2}$ and $\tan\beta=30$ (yellow), $\tan\beta=40$ (red) and $\tan\beta=50$ (blue). The constraints are weaker than the ones from $K(\pi) \to e\nu/K(\pi)\to \mu \nu$ and $\tau\to K(\pi)\nu/K(\pi)\to\mu\nu$ but cannot be avoided assuming the MFV limit ($\frac{m_{\ell_i}}{m_{\ell_j}}=\frac{\epsilon^\ell_{ii}}{\epsilon^\ell_{jj}}$).}
\label{KmuoverPimu}
\end{figure*}

\subsection{${K\to\mu\nu/\pi\to\mu\nu}$ and ${K\to e \nu/\pi\to e \nu}$ }

The ratio $ R_{K_{\ell2},\pi_{\ell2}}={\cal B}\left[ K \to \ell \nu \right]/ {\cal B}\left[ \pi \to \ell \nu \right]$ ($\ell=e,\mu$) is useful for constraining $\epsilon^{d}_{22}$, $\epsilon^{\ell}_{i1}$ and $\epsilon^{\ell}_{i2}$ because the ratio of the decay constants $f_K/f_\pi$ is known more precisely than the single decay constants \cite{Antonelli:2008jg}. 
\medskip

For obtaining the experimental values we add the errors of the individual branching ratios in quadrature and the SM values take into account the electromagnetic correction. The corresponding values are given in Table.~\ref{EMcorrections}. The errors are due to the combined uncertainties in $f_{K} /f_{\pi}$, the CKM elements and the EM corrections. We obtained the value for $V_{us}$ from $K\to\pi \ell\nu$ (which is much less sensitive to charged Higgs contributions than ${K\to\mu\nu/\pi\to\mu\nu}$) and $V_{ud}$ by exploiting CKM unitarity.
\medskip

\begin{figure*}[t]
\centering
\includegraphics[width=0.425\textwidth]{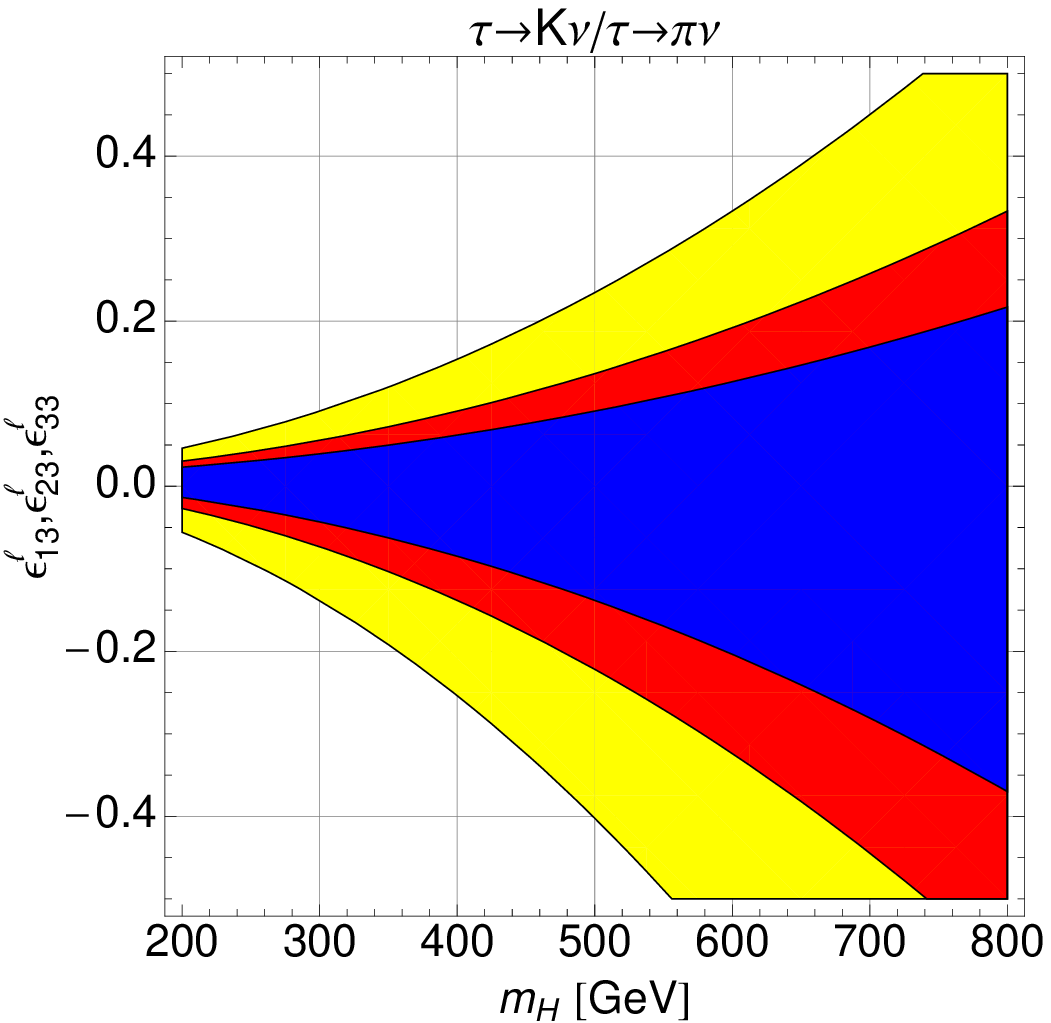}
\includegraphics[width=0.44\textwidth]{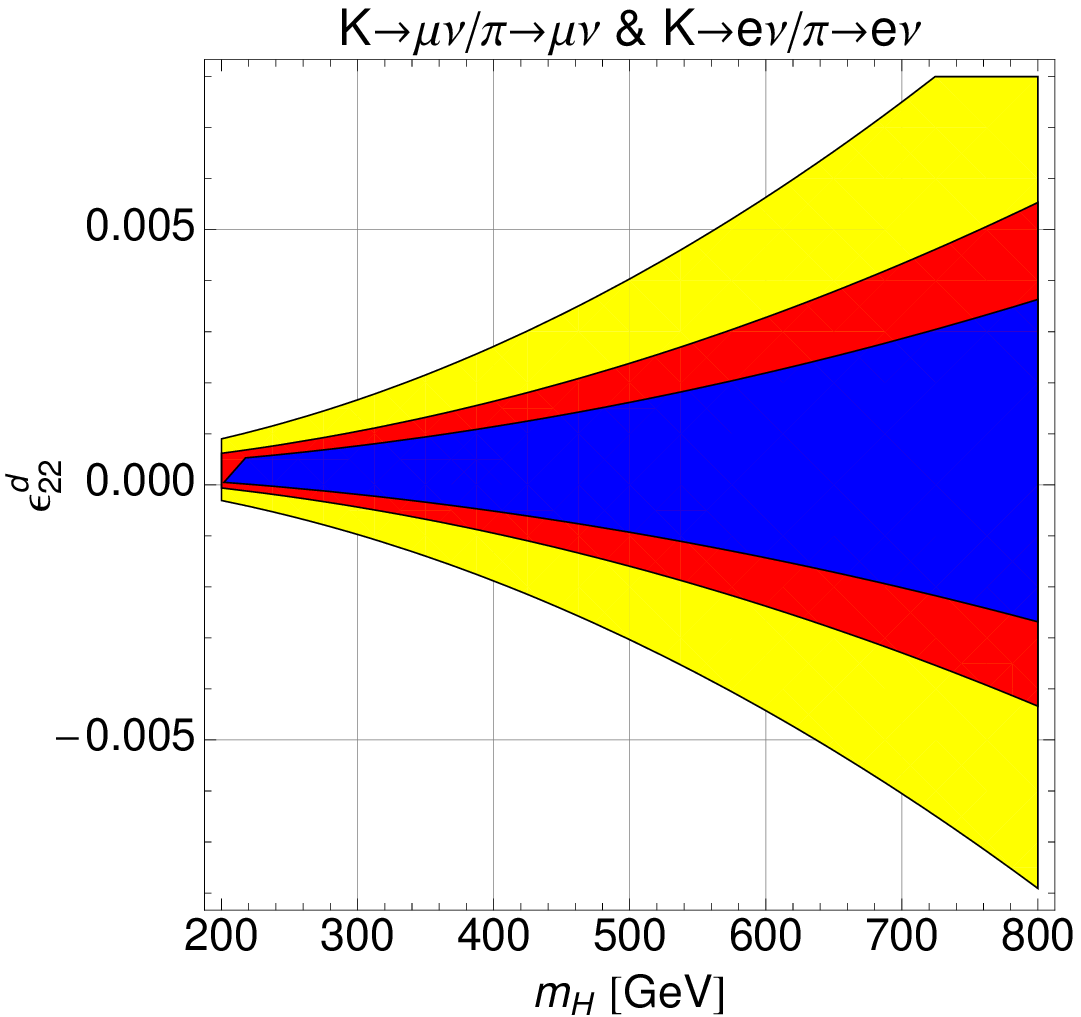}
\caption{Left: Allowed regions in the $m_H$--$\epsilon^{\ell}_{i3}$--plane from $\tau\to K\nu/\tau\to\pi\nu$. Right: Allowed regions in the $m_H$--$\epsilon^{d}_{22}$--plane obtained by combining the constraints from $K\to\mu\nu/\pi\to\mu\nu$ and $K\to e \nu/\pi\to e \nu$ for real values of $\epsilon^{d}_{22}$. In both plots $\tan\beta=30$ (yellow), $\tan\beta=40$ (red) and $\tan\beta=50$ (blue).}
\label{RKpiplot}
\end{figure*}

Fig.~\ref{RKpiplot} illustrates the allowed regions for $\epsilon^{d}_{22}$ by combining the constraints from $K\to\mu\nu/\pi\to\mu\nu$ and $K\to e \nu/\pi\to e \nu$. Like in $D_{(s)}\to\tau\nu$ and $D_{(s)}\to\mu\nu$ the constraints are on the real part of $\epsilon^{d}_{22}$ while the constraints on the imaginary part are very weak. Concerning $\epsilon^\ell_{i1}$ and $\epsilon^\ell_{i2}$ the constraints  from $K(\pi) \to e\nu/K(\pi)\to \mu \nu$ will turn out to be more stringent but the latter ones can be avoided in the limit $\frac{m_{\ell_i}}{m_{\ell_j}}=\frac{\epsilon^\ell_{ii}}{\epsilon^\ell_{jj}}$ (see Fig.~\ref{KmuoverPimu} and Fig.~\ref{KmuKeplotComb}).
\medskip

\subsection{${\tau \to K \nu/\tau \to \pi \nu}$}
The $\tau$\, is the only lepton which is heavy enough to decay into hadrons. The ratio $ { { \cal B }\left[ \tau \to K \nu \right]  }/{ { \cal B} \left[ \tau \to \pi \nu\right] } $ can be considered for putting constraints on $\epsilon^{u}_{21}$, $\epsilon^{d}_{12}$ and $\epsilon^{\ell}_{i3}$.
\medskip

The experimental and theoretical values for this ratio are given in Table.~\ref{EMcorrections}. 
We observe that the constraints from ${\bar D}^{0}\to \mu^{+}\mu^{-}$ and \dd mixing on $\epsilon^{u}_{21}$ and $K_{L}\to \mu^{+}\mu^{-}$ on $\epsilon^{d}_{12}$ are too stringent so that no sizable effects stemming from these elements are possible. Also concerning $\epsilon^{\ell}_{i3}$, as we will see in the following sections, the constraints from $\tau\to \pi \nu/ \pi \to\mu\nu$ will be stronger but again the latter ones can be avoided in the MFV limit $\frac{m_{\ell_i}}{m_{\ell_j}}=\frac{\epsilon^\ell_{ii}}{\epsilon^\ell_{jj}}$ (see Fig.~\ref{RKpiplot}).

\subsection{Tests for lepton flavour universality: ${K(\pi) \to e\nu/K(\pi)\to \mu \nu}$ and \\ ${\tau\to K(\pi)\nu/K(\pi)\to\mu\nu}$}

$K_{\ell2}\,(K \to \ell \nu)$ decays ($\ell=e, \mu$) are helicity suppressed in the SM and suffers from large theoretical uncertainties due to the decay constants. However, considering the ratio $ R_{{K_{\ell2}}}={\cal B}\left[K \to e \nu\right]/{\cal B}\left[K \to \mu \nu \right]$ the dependence on decay constants drops out. 
\medskip 

In the 2HDM of type II the charged Higgs contributions to ${K(\pi) \to e\nu/K(\pi)\to \mu \nu}$ and ${\tau\to K(\pi)\nu/K(\pi)\to\mu\nu}$ drop out. This is also true in the 2HDM of type~III (for $\epsilon^\ell_{ij}=0$ with $i\neq j$) as long as the MFV-like relation $\epsilon^\ell_{22}$/$m_\mu$=$\epsilon^\ell_{11}$/$m_e$ is not violated.

\subsubsection{$K\to e\nu/K\to \mu\nu$ and $\pi\to e\nu/\pi \to \mu\nu$}

\begin{figure}[t]
\centering
\includegraphics[width=0.44\textwidth]{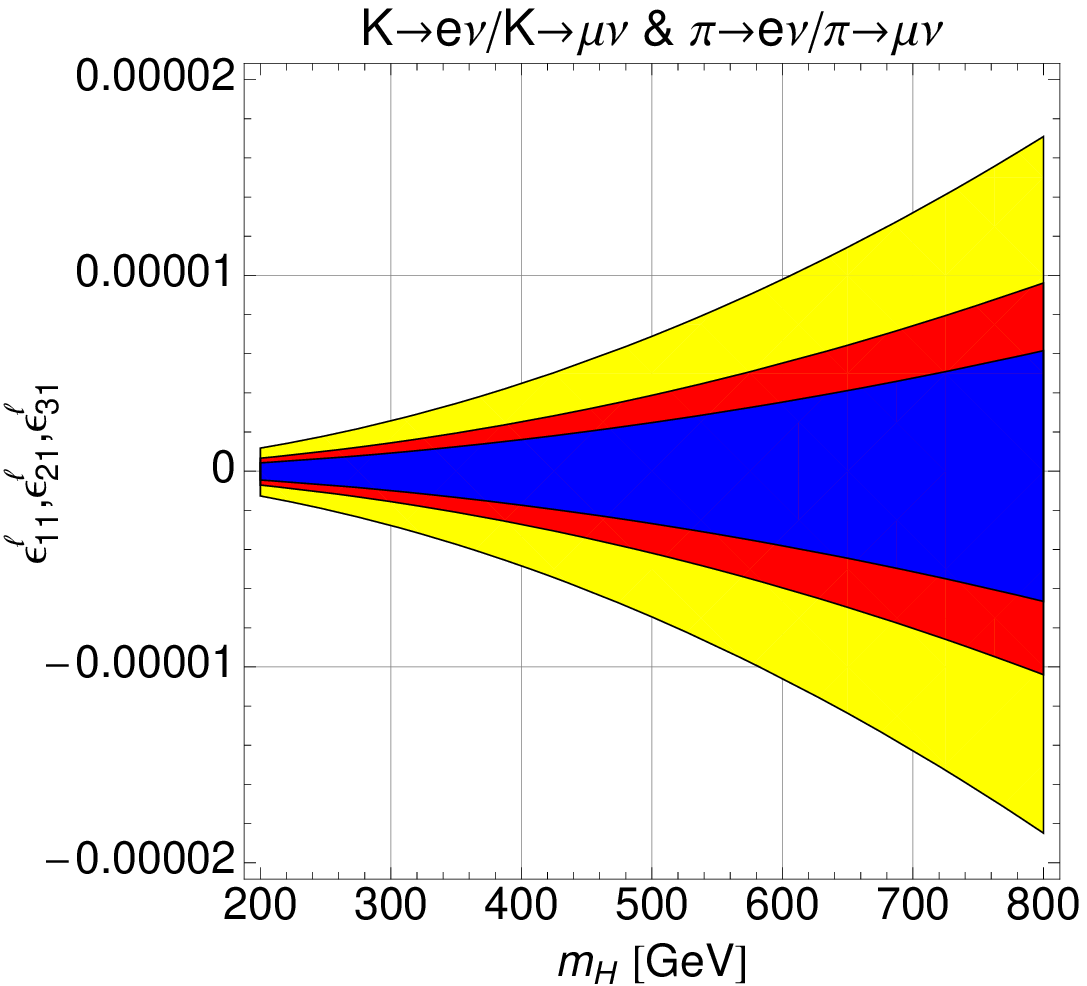}~~~~~
\includegraphics[width=0.42\textwidth]{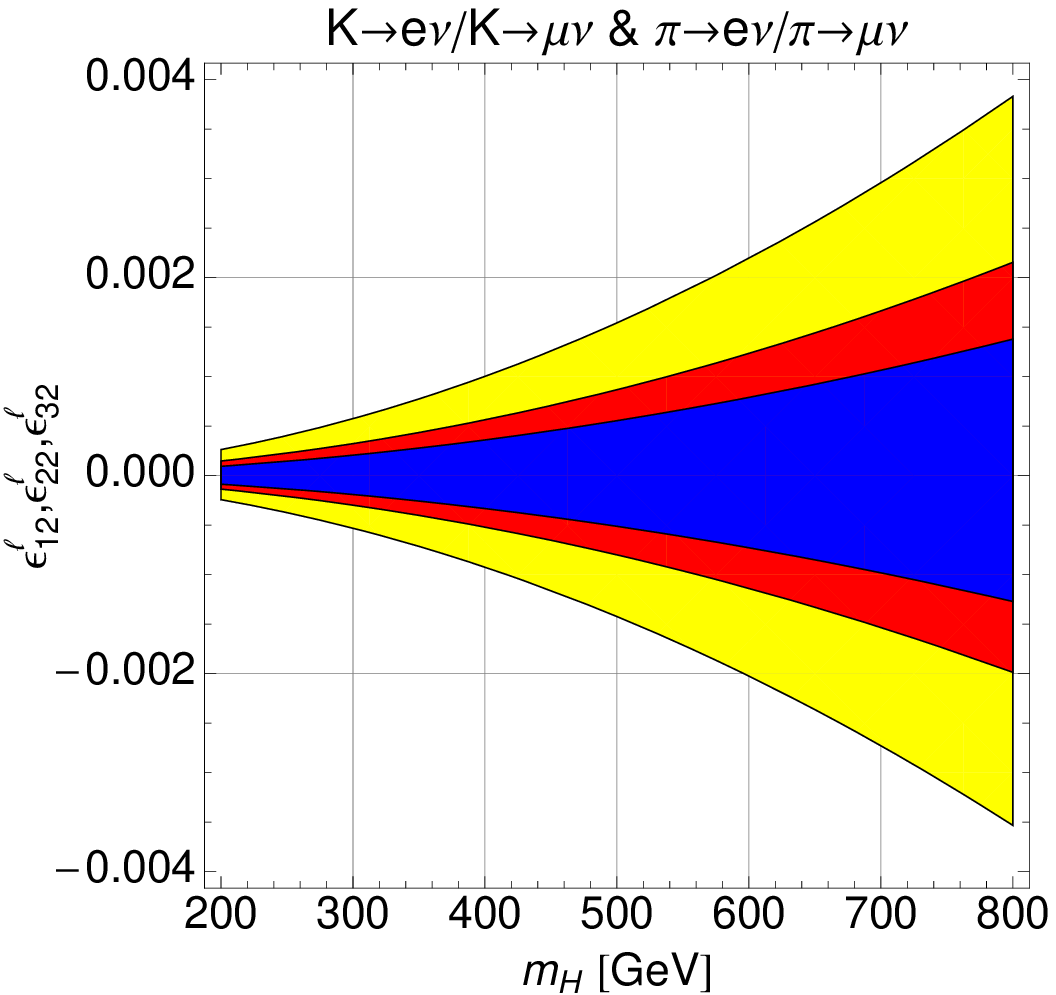}
\caption{Allowed regions in the $m_H$--$\epsilon^\ell_{ij}$--plane obtained by combining the constraints from $K\to e \nu/K\to \mu \nu$ and $\pi \to e \nu/\pi \to \mu \nu$ for real values of $\epsilon^\ell_{ij}$ and $\tan\beta=30$ (yellow), $\tan\beta=40$ (red) and $\tan\beta=50$ (blue). The constraints on $\epsilon^\ell_{i1}$ (affecting the electron coupling) are more stringent than the constraints on $\epsilon^\ell_{i2}$ (which affect the muon coupling). }
\label{KmuKeplotComb}
\end{figure}

$K\to e\nu/K\to \mu\nu$ is a very precise test of lepton flavor universality \cite{Masiero:2006ss} (see table.~\ref{EMcorrections}). Including NP entering via scalar operators modifies this ratio according to \eq{eq:BrNP}.
\medskip

We find strong constraints on $\epsilon^\ell_{i2}$ (which affect the coupling to the muon) and the constraints on $\epsilon^\ell_{i1}$ (where the coupling of the electron is involved) are even more stringent. Like for $D_{(s)} \to\tau\nu$ and  $D_{(s)} \to\mu\nu$ the constraints are much better for the real part of $\epsilon^\ell_{ij}$ than the imaginary part. Note that these constraints are obtained assuming that only one element $\epsilon^\ell_{ij}$ is non-zero. In the case $\epsilon^\ell_{22}$/$m_\mu$=$\epsilon^\ell_{11}$/$m_e$ where lepton flavor universality is restored no constraints can be obtained.
\medskip

Alternatively, the ratio $\pi\to e\nu/\pi \to \mu\nu$ can test lepton flavor universality. We find that the constraints from $\pi\to e\nu/\pi \to \mu\nu$ are comparable with the ones from $K\to e\nu/K\to \mu\nu$. Our results are illustrated in Fig.~\ref{KmuKeplotComb}.  
\medskip

\subsubsection{$\tau\to K\nu/K\to\mu\nu$ and $\tau\to \pi \nu / \pi \to\mu\nu$}

The ratios $\tau\to K\nu/K\to\mu\nu$ and $\tau\to \pi \nu / \pi \to\mu\nu$ are very similar to $K (\pi)\to e \nu/K (\pi) \to \mu \nu$: all dependences on decay constants and CKM elements drop out and they are only sensitive to NP which violates lepton-flavour universality. The corresponding experimental and the theoretical values for these ratios are given in Table.~\ref{EMcorrections}.
\smallskip

\begin{figure}[t]
\centering
\includegraphics[width=0.44\textwidth]{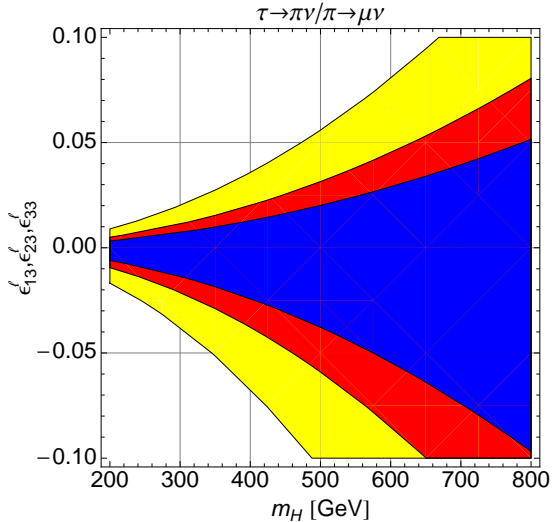}
\caption{Allowed regions in the $m_H$--$\epsilon^\ell_{i3}$--plane from $\tau\to \pi\nu/\pi\to\mu\nu$ for real values of $\epsilon^\ell_{i3}$ and $\tan\beta=30$ (yellow), $\tan\beta=40$ (red), $\tan\beta=50$ (blue). The bounds on the imaginary parts are very weak.}
\label{tauKKmuplot}
\end{figure}

We find that the constraints on $\epsilon^{\ell}_{i3}$ from $\tau\to \pi\nu/\pi\to\mu\nu$ are stronger than the ones from $\tau\to K\nu/K\to\mu\nu$ and they are shown in Fig.~\ref{tauKKmuplot}.
\medskip

\section{Upper limits and correlation for LFV processes}
\label{LFV-b-decays}

In Sec.~\ref{tree-level-constraints} we found that the neutral current lepton decays $\tau^-\to \mu^-\mu^+\mu^-$ and $\tau^-\to e^-\mu^+\mu^-$ give more stringent bounds on the elements $\epsilon^\ell_{32,23}$ and $\epsilon^\ell_{31,13}$ than the radiative decays $\tau\to \mu \gamma$ and $\tau\to e \gamma$. Also the LFV neutral meson decays $B_{s,d} \to \tau \mu$, $B_{s,d} \to \tau e$, $B_{s,d} \to \mu e$ cannot be arbitrarily large in the type-III 2HDM due to the constraints from $B_{s,d}\to\mu^+\mu^-$ and $\tau^-\to \mu^-\mu^+\mu^-$, $\tau^-\to e^-\mu^+\mu^-$, $\mu^-\to e^-e^+e^-$ (assuming again the absence of large cancellations)\footnote{see e.g. Ref.~\cite{Dedes:2002rh,Dedes:2008iw,Boubaa:2012xj} for an analysis of NP in $B_{s,d} \to \tau \mu$.}.
\medskip

Therefore, in this section we study the upper limits on $B_{s,d} \to \tau \mu$, $B_{s,d} \to \tau e$, $B_{s,d} \to \mu e$ and the correlation among $\tau^-\to \mu^-\mu^+\mu^-$, $\tau^-\to e^-\mu^+\mu^-$, $\mu^-\to e^-e^+e^-$ and $\tau\to \mu \gamma$, $\tau\to e \gamma$, $\mu\to e \gamma$ in the type-III 2HDM.
\medskip

\subsection{Neutral meson decays: $B_{s,d} \to \tau \mu$, $B_{s,d} \to \tau e$ and $B_{s,d} \to \mu e$}

In the SM (with massless neutrinos) the branching ratios for these decays vanish. Also in the 2HDM of type II these decays are not possible (even beyond tree-level). In the type-III 2HDM, these decay modes are generated in the presence of flavor-violating terms $\epsilon^\ell_{ij}$ and there exists even a tree-level neutral Higgs contribution to $B_s\to \ell_i^+\ell_j^-$ ($B_d\to \ell_i^+\ell_j^-$) if also $\epsilon^d_{23,32}\neq 0$ ($\epsilon^d_{13,31}\neq 0$).
\begin{table}[htdp]
\centering \vspace{0.8cm}
\renewcommand{\arraystretch}{1.2}
\begin{tabular}{|c| c| c| c| c|}
\hline \hline
 Observables & ~${\cal B}(B_{s} \to \mu e)$ & ~ ${\cal B}(B_{d} \to \mu e)$   &  ~${\cal B}(B_{d} \to \tau \mu)$   &  ~${\cal B}(B_{d} \to \tau e)$ 
\\   \hline \hline
Upper bounds & ~$2.0\times 10^{-7}$ \cite{Aaltonen:2009vr} & ~$6.4\times 10^{-8} $  \cite{Aaltonen:2009vr}& ~$2.2 \times 10^{-5}$ \cite{Aubert:2008cu}&  ~$2.8\times 10^{-5}$ \cite{Aubert:2008cu} \\ \hline \hline
\end{tabular}
  \caption{Upper limits (90 \% CL) on the branching ratios of the lepton flavor-violating $B$ meson decays.} 
\label{table:PDGboundsBtolAlB}
\end{table}

In the large $\tan\beta$ limit, $v\ll m_{H}$ and neglecting the smaller lepton mass, the corresponding expressions for these branching ratios take the simple form
\begin{equation}
\label{type3Boundsa}
\renewcommand{\arraystretch}{1.8}
\begin{array}{l}
{\cal B}\left[ B_{q} \to \ell^+_i \ell^-_j\right] \,  \approx \, N^q_{ij}  \left(\dfrac{\tan\beta/50}{m_{H}/500\,{\rm GeV}} \right )^{4} 2\,\left[  \left| \epsilon^{\ell}_{ji}\right|^2\left|\epsilon^{ d}_{q3}\right|^2+\left| \epsilon^{\ell}_{ij}\right|^2\left|\epsilon^{ d}_{3q}\right|^2 \right]  \,, 
\end{array}
\end{equation}
with ${q=d,s}$, $N^{q}_{ji}=N^{q}_{ij}$ and
\begin{equation}
\renewcommand{\arraystretch}{2}
\begin{array}{l}
N^s_{21} \approx  2.1 \times 10^{7} \dfrac{f_{B_s}}{0.229\, {\rm GeV}}\,, \\ 
N^d_{21} \approx  1.6 \times 10^{7} \dfrac{f_{B_d}}{0.196\, {\rm GeV}}\,, \\ 
N^s_{31,32} \approx 1.7 \times 10^{7} \dfrac{f_{B_s}}{0.229\, {\rm GeV}}\, , \\ 
N^d_{31,32} \approx 1.2 \times 10^{7}  \dfrac{f_{B_d}}{0.196\, {\rm GeV}}\,. \\ 
\end{array}
\label{type3Boundsb}
\end{equation}
Note that the expressions for the branching ratios are not symmetric in $\epsilon^\ell_{ij}$ and $\epsilon^\ell_{ji}$. Since experimentally both $B_{q} \to \ell^+_i \ell^-_j$ and $B_{q} \to \ell^-_i \ell^+_j$ are combined we compute the average
\begin{equation}
\nn {\cal B}\left[ B_{q} \to \ell_i \ell_j\right ] \,=\, \left({\cal B}\left[ B_{q} \to \ell^+_i \ell^-_j\right ]+{\cal B}\left[ B_{q} \to \ell^+_j \ell^-_i\right ]\right)/2\,.
\end{equation}

\begin{figure}[t]
\centering
\includegraphics[width=0.47\textwidth]{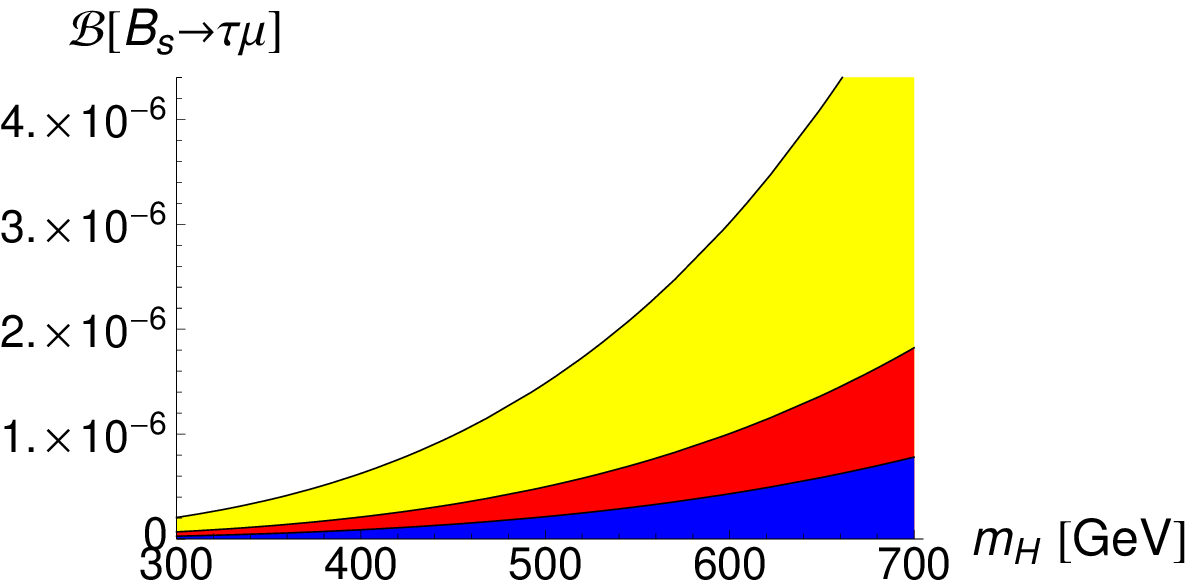}~~
\includegraphics[width=0.47\textwidth]{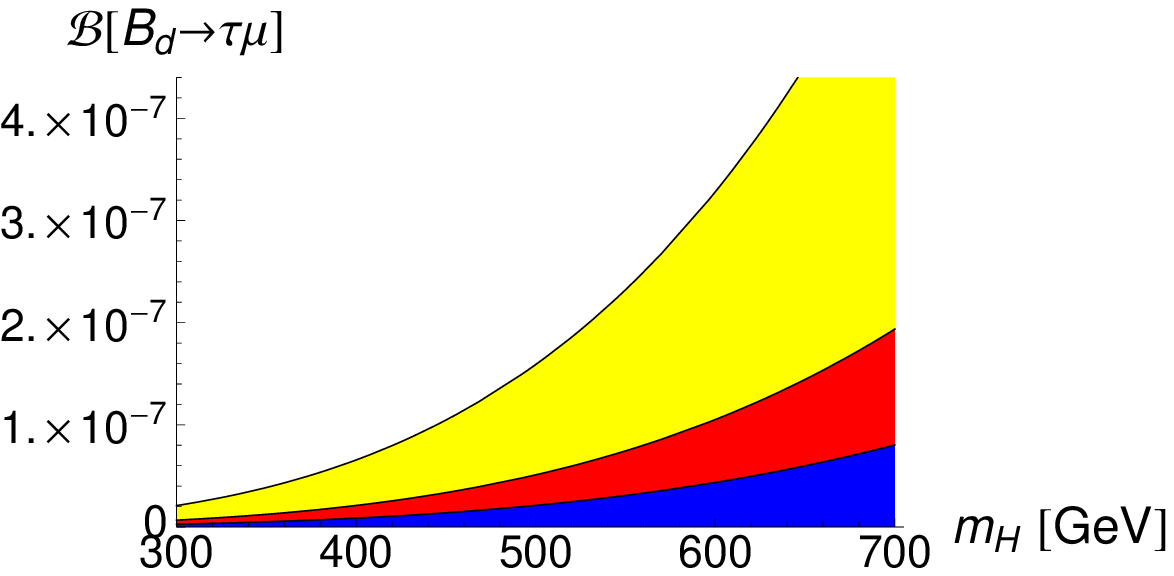}  \\ [0.2cm]
\includegraphics[width=0.47\textwidth]{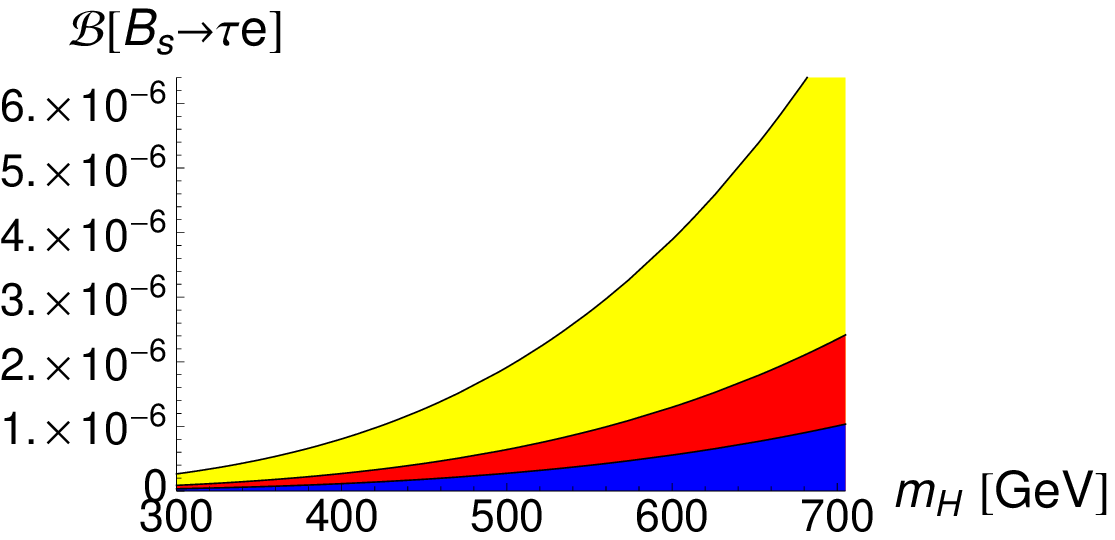}~~
\includegraphics[width=0.47\textwidth]{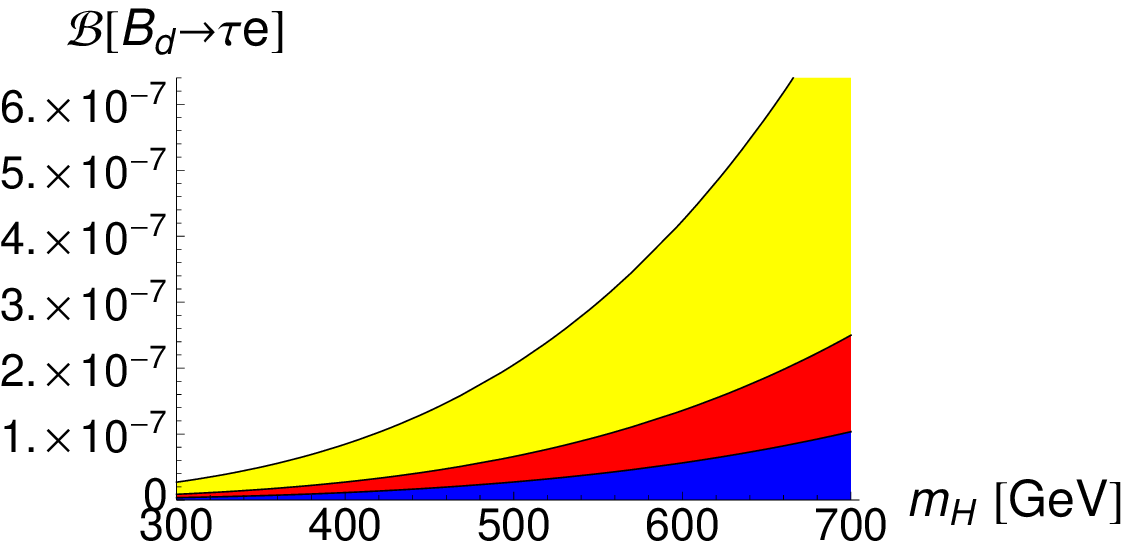} \\ [0.2cm]
\includegraphics[width=0.47\textwidth]{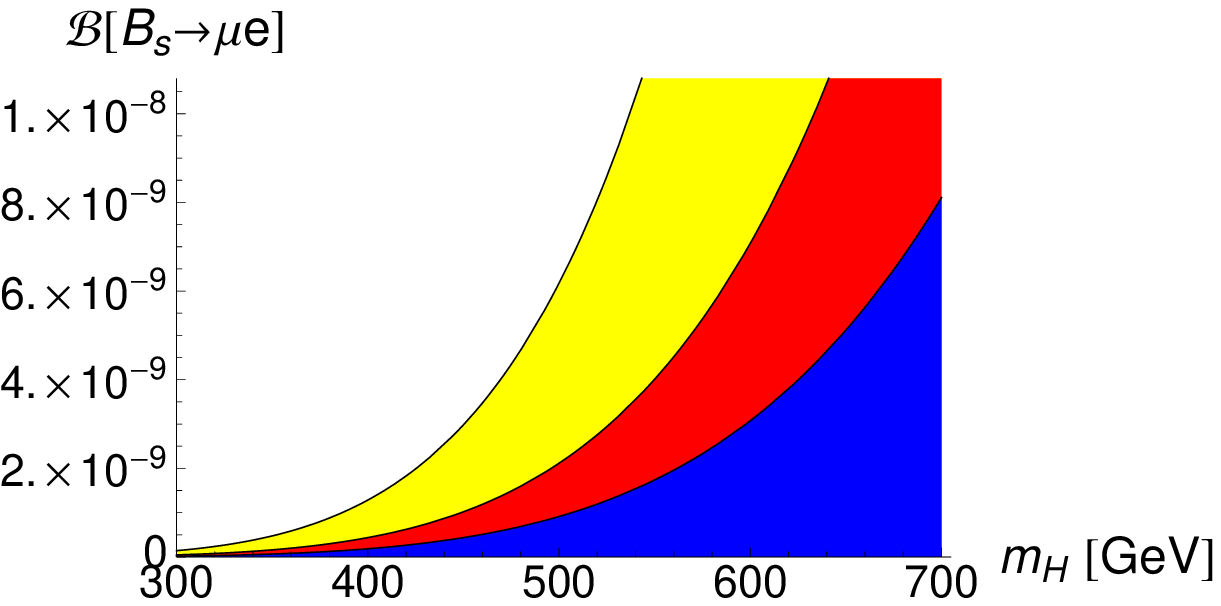}~~
\includegraphics[width=0.47\textwidth]{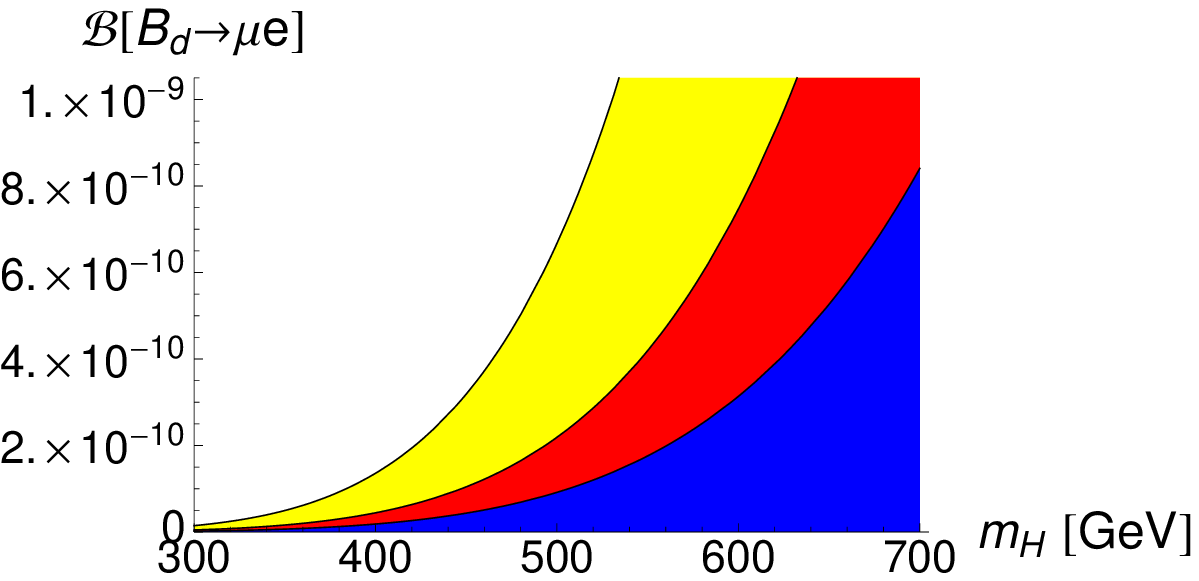}
\caption{Upper limits on the branching ratios of the lepton flavor violating $B$ meson decays as a function of $m_H$ for $\tan\beta=30$ (yellow), $\tan\beta=40$ (red) and $\tan\beta=50$ (blue). }
\label{Fig:LFVmesondecays}
\end{figure}

In order to obtain the upper limits we insert the biggest allowed values for ${\rm Abs}\left[\epsilon^{d,\ell}_{ij}\right]$. For $\epsilon^{d}_{23,32}$ ($\epsilon^{d}_{13,31}$) we use the biggest allowed absolute value compatible with the bounds from $B_{s}\to \mu^+\mu^{-}$ ($B_{d}\to \mu^+\mu^{-}$). As we can see from Fig.~\ref{fig:Bstomumu} (Fig.~\ref{fig:Bdtomumu}) the absolute value for $\epsilon^d_{32}$ ($\epsilon^d_{31}$) can be bigger than $\epsilon^d_{23}$ ($\epsilon^d_{13}$). For the leptonic parameters $\epsilon^{\ell}_{13,31}$ and $\epsilon^{\ell}_{23,32}$ we use the constraints obtained from $\tau^{-} \to \mu^{-}\mu^{+}\mu^{-}$, $\tau^{-} \to e^{-}\mu^{+}\mu^{-}$ (see Sec.~\ref{taumumumu})
\bea
\label{eq:El-limits}
\left |\epsilon^{\ell}_{31,13}  \right|  &\le&   \, 4.2 \times 10^{-3} \, \left(\dfrac{m_{H}/500\,{\rm GeV}}{\tan\beta/50} \right )^{2}  \, , \nonumber \\
\left |\epsilon^{\ell}_{32,23} \right|  &\le&   \, 3.7 \times 10^{-3}  \, \left(\dfrac{m_{H}/500\,{\rm GeV}}{\tan\beta/50} \right )^{2} \, ,
\eea
while for $\epsilon^{\ell}_{12,21}$ we use the combined constraints from $\mu^{-} \to e^{-}e^{+}e^{-}$ and from $\mu \to e \gamma$ (see Sec.~\ref{sec:muegamma}). 
\medskip

Our results are shown in Fig.~\ref{Fig:LFVmesondecays}. We see that for bigger Higgs masses larger values for the branching ratios are possible.
\medskip

\subsection{Radiative lepton decays:  $\tau\to \mu \gamma$, $\tau\to e \gamma$ and $\mu \to e \gamma$. }

In Sec.~\ref{taumumumu} and Sec.~\ref{sec:muegamma} we found that the radiative lepton decays $\tau\to \mu \gamma$ and $\tau\to e \gamma$ give less stringent bounds on the parameters $\epsilon^\ell_{23,32}$ and $\epsilon^\ell_{13,31}$ than the processes $\tau^- \to \mu^-\mu^+\mu^-$ and $\tau^- \to e^- \mu^+\mu^-$ while the constraints on $\epsilon^\ell_{12,21}$ from $\mu \to e \gamma$ are stronger than the ones from $\mu^{-} \to e^-e^+e^-$.
\medskip

\begin{figure}[t]
\centering
\includegraphics[width=0.45\textwidth]{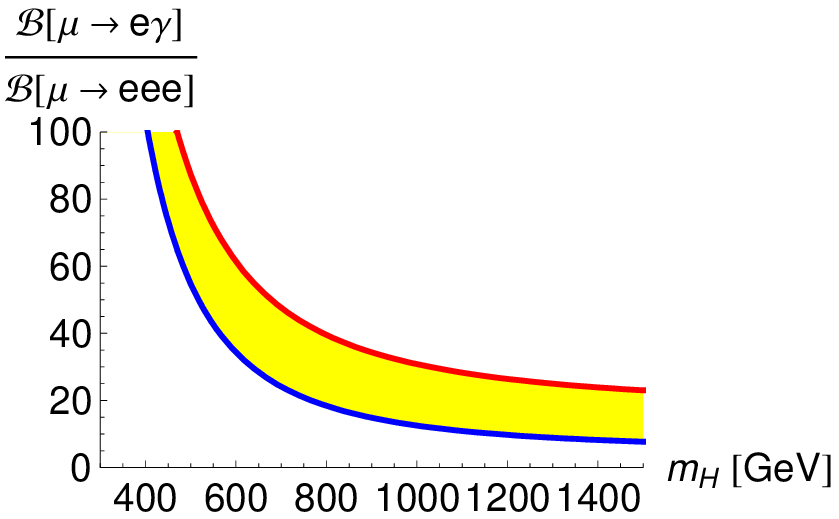}~~~
\includegraphics[width=0.45\textwidth]{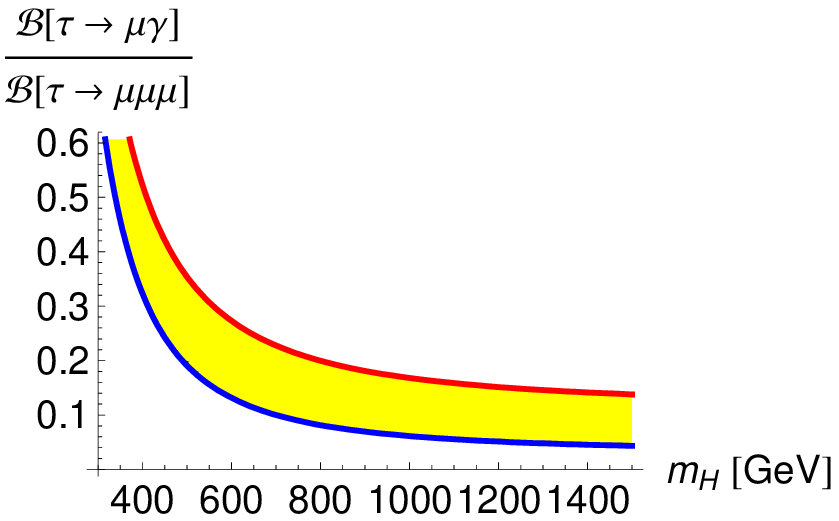}
\caption{Left: $\frac{{{\cal B} \left[ {{\mu} \to {e}\gamma } \right]}}{{{\cal B} \left[ {{\mu^{-}} \to {e^-}e^+ e^- } \right]}}$ as a function of $m_H$ assuming that only $\epsilon^\ell_{12}$ (red) or $\epsilon^\ell_{21}$ (blue) is different from zero for $\tan\beta=50$. Right: $\frac{{ {\cal B} \left[ {{\tau} \to {\mu}\gamma } \right]}}{{ {\cal B} \left[ {{\tau^{-}} \to {\mu^-}\mu^+ \mu^- } \right]}}$ as a function of $m_H$ assuming that only $\epsilon^\ell_{23}$ (red) or $\epsilon^\ell_{32}$ (blue) is different from zero for $\tan\beta=50$. \newline For scenarios in which both $\epsilon^\ell_{23}$ and $\epsilon^\ell_{32}$ ($\epsilon^\ell_{12}$ and $\epsilon^\ell_{21}$) are different from zero the 2HDM of type III predicts the ratio $\frac{{ {\cal B} \left[ {{\tau} \to {\mu}\gamma } \right]}}{{ {\cal B} \left[ {{\tau^{-}} \to {\mu^-}\mu^+ \mu^- } \right]}}$ $\left(\frac{{{\cal B} \left[ {{\mu} \to {e}\gamma } \right]}}{{{\cal B} \left[ {{\mu^{-}} \to {e^-}e^+ e^- } \right]}}\right)$ to be within the yellow region. These ratios are to a good approximation independent of $\tan\beta$ for $\tan\beta \gtrsim 20$. The behavior of $\frac{{{\cal B} \left[ {{\tau} \to {e}\gamma } \right]}}{{{\cal B} \left[ {{\tau^{-}} \to {e^-}\mu^+ \mu^- } \right]}}$ (not shown here) is very similar to the case of $3\to 2 $ transitions. }
\label{fig:mueovermueee}
\end{figure}

There are however interesting correlations between these decays in the type-III 2HDM. In the large $\tan\beta$ limit and for $v\ll m_H$ we obtain the following relation
\begin{equation}
\label{eqn:muegammaovermueee}
 \frac{{ {\cal B} \left[ {{\ell _i} \to {\ell _f}\gamma } \right]}}{{ {\cal B} \left[ {{\ell _i^-} \to {\ell _f^-}\ell_j^+ \ell_j^- } \right]}}= \frac{{\alpha _{em}^{}}}{{24\pi }}\dfrac{\left| m_{\ell_i}/v-\epsilon^\ell_{ii}\right|^{2}}{\left| m_{\ell_j}/v-\epsilon^\ell_{jj}\right|^{2 }} \frac{ \left( \left| \epsilon_{if}^\ell \right|^2 + 4 \left| \epsilon_{fi}^\ell \right|^2 \right)  }   { \left(  \left| \epsilon_{if}^\ell \right|^2 + \left| \epsilon_{fi}^\ell \right|^2  \right)  }\,.
\end{equation}
As already noted in Sec.~\ref{sec:muegamma}, we stress that this formula is only a good approximation for very heavy Higgs due to the large logarithmic term in the expression for $\ell_i\to\ell_f\gamma$ (see \eq{CiH0klitolfgamma}). Therefore, the relation in \eq{eqn:muegammaovermueee} gets modified for lighter Higgs masses as shown in Fig.~\ref{fig:mueovermueee}. We see that, as expected, for very large Higgs masses the ratios approach 

\begin{equation}
\renewcommand{\arraystretch}{1.8}
\label{muegammaovermueeeBB}
\begin{array}{l}
 \dfrac{{ {\cal B} \left[ {{\ell _i} \to {\ell _f}\gamma } \right]}}{{ {\cal B} \left[ {{\ell _i^-} \to {\ell _f^-}\ell_j^+ \ell_j^- } \right]}}= \dfrac{{\alpha _{em}^{}}}{{24\pi }}\dfrac{m_{\ell_i}^{2}}{m_{\ell_j}^{2 }}\;\; {\rm for}\;\; \epsilon_{if}^\ell\neq0\,,\\
  \dfrac{{ {\cal B} \left[ {{\ell _i} \to {\ell _f}\gamma } \right]}}{{ {\cal B} \left[ {{\ell _i^-} \to {\ell _f^-}\ell_j^+ \ell_j^- } \right]}}= \dfrac{{\alpha _{em}^{}}}{{6\pi }}\dfrac{m_{\ell_i}^{2}}{m_{\ell_j}^{2 }}\;\; {\rm for}\;\; \epsilon_{fi}^\ell\neq0\,,
 \end{array}
\end{equation}
where, we assumed that $\epsilon^{\ell}_{jj}/\epsilon^{\ell}_{ii}=m_{\ell_j}/m_{\ell_i}$ and that only one flavor changing element $\epsilon^{\ell}_{fi}$, $\epsilon^{\ell}_{if}$ is different from zero.

%
\section{Conclusions}

In this article we studied in detail the flavor phenomenology of a 2HDM with general Yukawa couplings. Motivated by the fact that the 2HDM of type~III is the decoupling limit of the MSSM we assumed a MSSM-like Higgs potential. In our analysis we proceeded in several steps:

\begin{enumerate}
	\item We gave order of magnitude constraints on the parameters $\epsilon^{q,\ell}_{ij}$ from 't~Hooft's naturalness criterion and found that all couplings except $\epsilon^{u}_{i3,3i}$ and $\epsilon^{u}_{21,22}$ should be much smaller than one.
	
	\item Considering tree-level FCNC processes we constrained the elements $\epsilon^{d}_{ij}$ ($i\neq j$) and $\epsilon^{u}_{12,21}$ from neutral meson decays to muons and from $\Delta F=2$ processes, finding that they are tiny for the values of $m_H$ and $\tan\beta$ under investigation (assuming $\epsilon^\ell_{ij}=0$). In the lepton sector the absolute values of all flavor off-diagonal elements $\epsilon^{\ell}_{ij}$ were constrained from $\tau^-\to\mu^-\mu^+\mu^-$, $\tau^-\to e^-\mu^+\mu^-$ and $\mu^-\to e^- e^+e^-$ to be very small. 
	
	\item After having found that the off-diagonal elements $\epsilon^{d}_{ij}$ must be very small due to constraints from tree-level contributions to FCNC processes we considered charged Higgs contributions to \kk, \bbs, \bbd mixing and $b\to s(d)\gamma$ arising at the one-loop level. In these contributions the so far unconstrained elements $\epsilon^{u}_{i3,3i}$ (and also $\epsilon^{u}_{22}$) enter for the first time and we found that, setting $\epsilon^{d}_{ij}=0$ (with $i\neq j$), $\epsilon^{u}_{13,23}$ should be rather small. Furthermore, the electric dipole moment of the neutron and of the charged leptons constrain $\epsilon^u_{11}$, $\epsilon^u_{22}$, $\epsilon^u_{21}$, $\epsilon^u_{31}$ and $\epsilon^\ell_{ij}$, respectively. Respecting all other constraints, no sizable effect in $a_\mu$ is possible.
	
	\item Keeping in mind the constraints from the previous steps, we considered the possible effects in charged current processes. Here we found that tests for lepton flavor universality constrain the differences $\epsilon^\ell_{ii}/m_{\ell_i}-\epsilon^\ell_{jj}/m_{\ell_j}$. Most importantly, the unconstrained elements $\epsilon^{u}_{31}$ and $\epsilon^{u}_{32}$ enter the processes \btau and $B\to D^{(*)}\tau\nu$ directly (without CKM suppression) and can remove the tension between experiment and theory prediction observed in the SM simultaneously.
	
 \item Finally we gave upper limits on the lepton flavor violating neutral $B$~meson decays in the 2HDM of type~III and correlated the radiative lepton decays to $\tau^-\to\mu^-\mu^+\mu^-$, $\tau^-\to e^-\mu^+\mu^-$ and $\mu^-\to e^- e^+e^-$.
\end{enumerate}

In Table~\ref{Table:processes1} and \ref{Table:processes2} we list all processes which have been under consideration and quote the constraints placed on the parameters $\epsilon^{q,\ell}_{ij}$ for our benchmark point $m_H=500$~GeV and $\tan\beta=50$.

\begin{table}[t]
\renewcommand{\arraystretch}{1.2}
\begin{tabular}{|l|c|}
\hline
  {\bf Observable} & {\bf Results}  \\   \hline \hline
 \multicolumn{2}{|c|}{\bf Neutral meson decays to muons}\\ \hline 
$B_{s}\to\mu^+\mu^-$ &~$\left|\epsilon _{32}^d\right| \le 3.0 \times 10^{-5}$, $\left|\epsilon _{23}^d\right| \le 1.9 \times 10^{-5}$, $\left|\epsilon _{22}^\ell \right| \le 2.0 \times 10^{-3}$   \\ \hline
$B_{d}\to\mu^+\mu^-$ & ~$\left|\epsilon _{31}^d\right| \le 1.1 \times 10^{-5}$, $\left|\epsilon _{13}^d\right| \le 9.4 \times 10^{-6}$    \\ \hline
$K_L\to\mu^+\mu^-$   &~$\left|\epsilon _{21}^d\right| \le 1.6 \times 10^{-6}$, $\left|\epsilon _{12}^d\right| \le 1.6 \times 10^{-6}$  \\ \hline
${\bar D}^0 \to\mu^+\mu^-$ &~$\left|\epsilon _{21}^u\right| \le 3.0 \times 10^{-2}$, $\left|\epsilon _{12}^u\right| \le 3.0 \times 10^{-2}$  \\ \hline\hline
\multicolumn{2}{|c|}{\bf $\Delta F=2$ processes}\\ \hline 
\bbs mixing &~ $\left|\epsilon^d_{23} \epsilon^{d\star}_{32} \right| \leq 9.2 \times 10^{-10}$,  $\left|\epsilon^u_{23}\right| \leq 0.18$, $\left|\epsilon^u_{32}\right| \leq 1.7$, $\left|\epsilon^u_{33}\right| \leq 0.7$   \\ \hline
\bbd mixing &~ $\left|\epsilon^d_{13} \epsilon^{d\star}_{31} \right| \leq 3.9 \times 10^{-11}$,  $\left|\epsilon^u_{23}\right| \leq 0.2$, $\left|\epsilon^u_{13}\right| \leq 0.04$, $\left|\epsilon^u_{31}\right| \leq 1.9$  \\ \hline
\kk mixing &~ $\left|\epsilon^d_{12} \epsilon^{d\star}_{21} \right| \leq 1.0 \times 10^{-12}$, $\left|\epsilon^u_{22}\right| \leq 0.25$, $\left|\epsilon^u_{23}\right| \leq 0.14$ \\ \hline
\dd mixing  &~ $\left|\epsilon^u_{12} \epsilon^{u\star}_{21} \right| \leq 2.0 \times 10^{-8}$, $\left|\epsilon^u_{32} \epsilon^{u\star}_{31} \right| \leq 0.02$  \\ \hline\hline
\multicolumn{2}{|c|}{\bf Radiative $B$ decays}\\ \hline 
$b\to s \gamma$&~$\left|\epsilon _{23}^u\right| \le 0.024$, $\left|\epsilon _{33}^u \right| \le 0.55$  \\ \hline
$b\to d \gamma$&~$\left|\epsilon _{13}^u\right| \le 7.0 \times 10^{-3}$  \\ \hline\hline
\multicolumn{2}{|c|}{\bf Radiative lepton decays}\\ \hline 
$\mu \to e \gamma$&~$\left|\epsilon _{12}^\ell \right| \le 1.7\times10^{-4}$, $\left|\epsilon _{21}^\ell \right| \le 2.2\times10^{-4}$, $55 \leq\frac{{ {\cal B} \left[ {{\mu} \to {e}\gamma } \right]}}{{ {\cal B} \left[ {{\mu^{-}} \to {e^-} e^+ e^- } \right]}}\leq 86$~   \\ \hline
$\tau \to e \gamma$&~ $0.19\leq\frac{{ {\cal B} \left[ {{\tau} \to {e}\gamma } \right]}}{{ {\cal B} \left[ {{\tau^{-}} \to {e^-}\mu^+ \mu^- } \right]}}\leq0.35$  \\ \hline
$\tau \to \mu \gamma$&~$0.19\leq\frac{{ {\cal B} \left[ {{\tau} \to {\mu}\gamma } \right]}}{{ {\cal B} \left[ {{\tau^{-}} \to {\mu^-}\mu^+ \mu^- } \right]}}\leq0.35$  \\ \hline\hline
\multicolumn{2}{|c|}{\bf Neural current lepton decays}\\ \hline 
$\mu^{-} \to e^{-}e^{+}e^{-}$&~$\left|\epsilon _{12,21}^\ell \right| \le 2.3\times10^{-3}$ \\ \hline
$\tau^{-} \to e^{-}\mu^{+}\mu^{-}$&~ $\left|\epsilon _{13,31}^\ell \right| \le 4.2\times10^{-3}$ \\ \hline
$\tau^{- }\to \mu^{-}\mu^{+}\mu^{-}$ &~ $\left|\epsilon _{23,32}^\ell \right| \le 3.7\times10^{-3}$ \\ \hline\hline
\end{tabular}
  \caption{Results obtained in the type-III 2HDM from various processes for $\tan\beta=50$ and $m_H=500$~GeV. } 
\label{Table:processes1}
\end{table}

\begin{table}[t]
\renewcommand{\arraystretch}{1.2}
\begin{tabular}{|l|c|}
\hline
  {\bf Observable} & {\bf Results}  \\   \hline \hline
\multicolumn{2}{|c|}{\bf Charged current processes}\\ \hline 
$B\to\tau\nu$&~$2.7 \times10^{-3} \le \left|\epsilon _{31}^u \right| \le 2.0 \times10^{-2}$,  $\left|\epsilon _{i3}^\ell \right| \le 6.0\times 10^{-2}$ \\ \hline
$B\to D\tau\nu$ \& $B\to D^{\star}\tau\nu$ &~  $0.43 \le \left|\epsilon _{32}^u \right| \le 0.74 $ \\ \hline
$D_{s}\to\tau \nu$ \& $D_{(s)}\to\mu\nu$ &~   $ \left|  {\rm Re}\left[\epsilon^{u}_{22}\right]  \right| \le 0.2$ \\ \hline
$D \to\tau\nu$&~  -- \\ \hline
$K \to \mu(e) \nu/ \pi \to \mu (e) \nu$ &~   $ \left|  {\rm Re}\left[\epsilon^{d}_{22}\right]  \right|  \le 1.0 \times 10^{-3} $ \\ \hline
$K(\pi)\to e \nu/K(\pi)\to \mu \nu$ &~  $ \left| {\rm Re}\left[\epsilon^{\ell}_{i1}\right]  \right|  \le 2.0 \times 10^{-6}$, $\left|  {\rm Re}\left[\epsilon^{\ell}_{i2}\right] \right|  \le 5.0 \times 10^{-4}$   \\ \hline
$\tau\to K(\pi)\nu/K(\pi)\to \mu \nu$~&~$-4.0 \times 10^{-2}\le {\rm Re}\left[\epsilon^{\ell}_{i3}\right] \le 2.0 \times 10^{-2}$ \\ \hline
$\tau\to K\nu/\tau \to \pi \nu$ ~&~  $\left|\epsilon^\ell_{i3}\right|\leq 0.14$ \\ \hline\hline
 \multicolumn{2}{|c|}{\bf EDMs and anomalous magnetic moments}\\ \hline
$d_e$ &   $\left | {\rm Im}\left[\epsilon^{\ell}_{12} \epsilon^{\ell}_{21} \right] \right| \le 2.5\times10^{-8}$,  $ \left| {\rm Im}\left[\epsilon^{\ell}_{13} \epsilon^{\ell}_{31} \right]  \right| \le 2.5\times10^{-9}$    \\ \hline
$d_\mu$&  ~ --\\ \hline
$d_\tau$&  ~ -- \\ \hline
$d_n$ & ~ $ \left| {\rm Im}\left[\epsilon^{u}_{11} \right]  \right| \le 2.2\times10^{-2}$, $ \left|  {\rm Im}\left[\epsilon^{u}_{22} \right] \right| \le 1.1\times10^{-1}$, ${\rm Arg}[\epsilon^u_{31}] = {\rm Arg}[V_{ub}] \pm \pi$~
\\ \hline
$a_\mu$ &  ~ {\small Deviation from the SM cannot be explained}\\ \hline\hline
 \multicolumn{2}{|c|}{\bf LVF $B$ meson decays}\\ \hline
$B_{s}\to \tau\mu$ & ~ ${\cal B}\left[B_{s}\to \tau\mu\right] \leq 2.0\times10^{-7}$ \\ \hline
$B_{s}\to \mu e$ & ~ ${\cal B}\left[B_{s}\to \mu e\right] \leq9.2\times10^{-10}$ \\ \hline
$B_{s}\to \tau e$ &  ~ ${\cal B}\left[B_{s}\to \tau e\right] \leq2.8\times10^{-7}$\\ \hline
$B_{d}\to \tau\mu$ & ~ ${\cal B}\left[B_{d}\to \tau\mu\right] \leq2.1\times10^{-8}$ \\ \hline
$B_{d}\to \mu e$ & ~ ${\cal B}\left[B_{d}\to \mu e\right] \leq9.2\times10^{-11}$ \\ \hline
$B_{d}\to \tau e$ &  ~ ${\cal B}\left[B_{d}\to \tau e\right] \leq2.8\times10^{-8}$\\ \hline \hline 
\end{tabular}
  \caption{Results obtained in the type-III 2HDM from various processes for $\tan\beta=50$ and $m_H=500$~GeV.} 
\label{Table:processes2}
\end{table}

In summary, combining the constraints from Table~\ref{Table:processes1} and~\ref{Table:processes2} the following bounds on the absolute values of the parameters ${\epsilon^q_{ij} }$ and $\epsilon^\ell_{ij}$ (for our benchmark point with $m_H=500$~GeV and $\tan\beta=50$) are obtained:
\begin{equation}
\label{EpsilonLimits}
\renewcommand{\arraystretch}{1.66}
\begin{array}{l}
\left|\epsilon _{ij}^u\right| \le {\left( {\begin{array}{*{20}{c}}
 3.4\times 10^{-4}  &~  3.0\times10^{-2} &~ 7.0\times 10^{-3}   \\
 3.0\times10^{-2} &~   1.4 \times 10^{-1}  &~   2.4\times 10^{-2}   \\
  2.0 \times 10^{-2} &~  7.4\times 10^{-1} &~  5.5\times 10^{-1} 
\end{array}} \right)_{ij}}  \\ 
\left|\epsilon _{ij}^d\right| \le {\left( {\begin{array}{*{20}{c}}
 1.3\times 10^{-4}  &~  1.6\times10^{-6} &~ 9.4\times 10^{-6}   \\
 1.6\times10^{-6} &~   2.6 \times 10^{-4}  &~   2.0\times 10^{-5}   \\
  1.1 \times 10^{-5} &~  3.0 \times 10^{-5} &~  1.4\times 10^{-2} 
\end{array}} \right)_{ij}}  \\
\left|\epsilon _{ij}^\ell \right| \le {\left( {\begin{array}{*{20}{c}}
 2.9\times 10^{-6}  &~  1.7\times10^{-4} &~ 4.2\times 10^{-3}   \\
 2.2\times10^{-4} &~   6.1 \times 10^{-4}  &~    3.7\times 10^{-3}    \\
   4.2\times 10^{-3}  &~  3.7\times 10^{-3} &~   1.0\times 10^{-2} 
\end{array}} \right)_{ij}}  \\ 
\end{array} 
\end{equation}
These bounds hold in the absence of large cancellations between different contributions. Note that in \eq{EpsilonLimits} we applied the naturalness bounds in case they were stronger than the experimental limits.
\medskip

It is interesting that \btau, \bdtau and \bdstau can be explained simultaneously in the 2HDM of type III without violating bounds from other observables and without significant fine-tuning. It remains to be seen if these tensions with the SM remain when updated experimental results and improved theory predictions will be available in the future. In order to further test the model and constrain the parameters $\epsilon^u_{32}$ ($\epsilon^u_{31}$) we propose to study $H^0,A^0\to \bar t c$ ($H^0,A^0\to \bar t u$) at the LHC.

\vspace{3mm}
{{\it{\bf Acknowledgments}} : \,}
{This work is supported by the Swiss National Science Foundation. We thank Jernej Kamenik and Ulrich Nierste for collaboration in the early stages of this work. A.C. also thanks Jernej Kamenik for useful comments on the manuscript.}

\newpage

\section{Appendix}
In this appendix, we collect the Wilson
coefficients (to the relevant precision at the matching scale) which
are needed for the calculation of $b\to
s(d) \gamma$, $\Delta F=2$ processes (i.e. neutral meson mixing), leptonic
neutral meson decays ($\Delta F=1$ processes), \btau, \bdtau, \bdstau,
$D_{(s)} \to\ell\nu_{\ell}$, $ K(\pi) \to\ell \nu_{\ell}$, $\tau \to
K(\pi) \nu$, LFV radiative lepton transitions, EDMs of charged leptons
and neutron, as well as the AMM of the muon.
In addition, we give general expressions for some
branching ratios, the explicit form of the loop functions entering our
results and summarize the input parameters used in our analysis
in tabular form.

\subsection{Loop functions}
We give the explicit form of the loop functions entering our
results. In the limit of vanishing external momentum the one and two-point
functions are defined as
\be
\renewcommand{\arraystretch}{2.0}
\begin{array}{*{20}c}
A_{0} \left(  m^{2} \right) \, =\, \dfrac{ 16\pi^{2} }{i}\mu^{4-d} \, \int{ \dfrac{d^{d}k}{ \left( 2\pi \right)^{d} } \dfrac{1}{\left(   k^{2}-m^{2} \right)  }   }   \, , \\
B_{0} \left(  m^{2}_{1},m^{2}_{2} \right) \, =\, \dfrac{ 16\pi^{2} }{i}\mu^{4-d} \, \int{ \dfrac{d^{d}k}{ \left( 2\pi \right)^{d} } \dfrac{1}{\left(   k^{2}-m^{2}_{1} \right)   \,\left(   k^{2}-m^{2}_{2} \right) }     }\, ,
\end{array}    
\ee
where $\mu$ is the renormalization scale.

The loop functions $C_{0}$ (three-point) and $D_{0}$ (four-point) 
are defined in analogy to $B_{0}$, but with
three and four propagators, respectively. Evaluating these loop
functions yields (with $d=4-2\varepsilon$) 
\be
\renewcommand{\arraystretch}{2.0}
\begin{array}{*{20}c}
A_{0} \left( m^{2} \right) \, = \,  m^{2}\, \left[ 1+
  \dfrac{1}{\varepsilon} - \gamma_{E}+\ln \left(4\pi\right) +\ln\left(
  \dfrac{\mu^{2}}{m^{2}}\right) \right]   + O\left( \varepsilon
\right) \, ,      \\
B_{0} \left(  m^{2}_{1},m^{2}_{2} \right) \, = \, 1+
\dfrac{1}{\varepsilon} - \gamma_{E}+\ln \left(4\pi\right) + \dfrac{
  m^{2}_{1} \ln\left( \dfrac{\mu^{2}}{m^{2}_{1}}\right) -   m^{2}_{2}
  \ln\left( \dfrac{\mu^{2}}{m^{2}_{2}}\right)  }{ m^{2}_{1} -
  m^{2}_{2} }   + O\left( \varepsilon \right)  \, ,  
\end{array}    
\ee

\begin{equation}
\renewcommand{\arraystretch}{2.0}
\begin{array}{*{20}c}
C_{0} \left(  m^{2}_{1},m^{2}_{2} ,m^{2}_{3} \right) \, = \,
\dfrac{B_{0} \left(  m^{2}_{1},m^{2}_{2} \right) - B_{0} \left(
  m^{2}_{1},m^{2}_{3} \right)  }{ m^{2}_{2} - m^{2}_{3}  }   \\ 
\qquad \qquad \qquad = \dfrac{  m^{2}_{1} m^{2}_{2} \ln\left(
  \dfrac{m^{2}_{1}}{m^{2}_{2}}\right)   + m^{2}_{3} m^{2}_{2}
  \ln\left( \dfrac{m^{2}_{2}}{m^{2}_{3}}\right)  +  m^{2}_{3}
  m^{2}_{1} \ln\left( \dfrac{m^{2}_{3}}{m^{2}_{1}}\right)   }{  \left(
  m^{2}_{1} - m^{2}_{2} \right) \,  \left( m^{2}_{3} - m^{2}_{1}
  \right) \,  \left( m^{2}_{2} - m^{2}_{3} \right)      }   \, , \\
D_{0} \left(  m^{2}_{1},m^{2}_{2} ,m^{2}_{3},m^{2}_{4} \right) \, = \,
\dfrac{C_{0} \left(  m^{2}_{1},m^{2}_{2},m^{2}_{3} \right) - C_{0}
  \left(  m^{2}_{1},m^{2}_{2},m^{2}_{4} \right)  }{ m^{2}_{3} -
  m^{2}_{4}  }  \, . 
\end{array} 
\end{equation}
Here, the one and the two-point loop functions $A_{0},B_{0}$ are
UV-divergent and $\varepsilon$ is the UV-regulator.

At various places also the functions $C_{2}$ and $D_{2}$ appear,
which have, compared to $C_{0}$ and $D_{0}$, an additional factor $k^2$
in the numerator of the integrand. These functions read
\begin{eqnarray}
&& C_{2} \left(  m^{2}_{1},m^{2}_{2} ,m^{2}_{3} \right) =
B_{0} \left(  m^{2}_{1},m^{2}_{2} \right) + m^{2}_{3} \, C_{0}
\left(  m^{2}_{1},m^{2}_{2} ,m^{2}_{3}\right)  \, , \nonumber \\ [0.25cm]
&& D_{2} \left(  m^{2}_{1},m^{2}_{2} ,m^{2}_{3},m^{2}_{4} \right) =
C_{0} \left(  m^{2}_{1},m^{2}_{2},m^{2}_{3} \right) + m^{2}_{4} \, D_{0}
\left(  m^{2}_{1},m^{2}_{2} ,m^{2}_{3},m^{2}_{4} \right)  \, .
\end{eqnarray}

\subsection{Radiative $ b \to s(d) \gamma$ decays}

Concering new physics contributions to $ b \to s(d) \gamma$, we work
in leading logarithmic (LL) precision in
this paper. As mentioned before, we use these processes to constrain certain
elements $\epsilon^{u}_{ij}$. For this purpose, we put the
$\epsilon^{d}_{ij}-$couplings (which are already constrained to be very
small) to zero. When also neglecting the mass of the strange quark and
further neglecting operators with mass dimension higher than six, 
we obtain the same effective Hamiltonian as in the SM, reading for $b
\to s \gamma$ (see e.g. Ref. \cite{Borzumati:1998tg}). 
\begin{equation}
{\cal{H}}_{eff}^{b \to s \gamma}= -\dfrac{4G_F }{\sqrt{2}} V_{tb}V_{ts}^\star \sum_i
C_{i} \, O_{i}\, .
\end{equation}
For $b \to d \gamma$ the CKM structure is slightly more complicated
(see e.g. Ref. \cite{Ali:1998rr}). 
In our approximation only the Wilson coefficients $C_7$ and $C_8$ of
the operators
\begin{equation}
O_7  = \dfrac{e}{{16\pi^2 }}m_b \bar{s} \sigma ^{\mu \nu } P_R b F_{\mu \nu }  \quad ; \qquad 
O_8  = \dfrac{{g_s }}{{16\pi ^2 }}m_b \bar s\sigma ^{\mu \nu } T^{a} P_R b G^{a}_{\mu \nu }  
\end{equation}
get new physics contributions. They are induced through
charged Higgs bosons propagating in the loop (neutral Higgs boson
exchange leads to power suppressed contributions which we neglect).
For $b \to s \gamma$ the new physics contributions  read 
(with $y_j=m_{u_j}^2/m_{H^+}^2$ and $\lambda_t=V_{tb} \, V_{ts}^\star$)
\bea \label{wilson3LO}
\nonumber C_{7}^{NP}&=&\frac{v^2}{\lambda_t}  \frac{1}{m_b}
\sum_{j=1}^{3}~\Gamma^{RLH^\pm \star}_{u_j d_2} \,
\Gamma^{LRH^\pm}_{u_j d_3}
\frac{C_{7,XY}^{0}(y_{j})} {m_{u_{j}}}  \\ 
&+&  \frac{v^2}{\lambda_t}   \sum_{j=1}^{3}~
\Gamma^{RLH^\pm \star}_{u_j d_2}   \, \Gamma^{RLH^\pm}_{u_j d_3}
\frac{C_{7,YY}^{0}(y_{j})} {m_{u_{j}}^{2}}  \, , \\  \nonumber 
 \nonumber C_{8}^{NP}&=&\frac{v^2}{\lambda_t}  \frac{1}{m_b}
 \sum_{j=1}^{3}~\Gamma^{RLH^\pm \star}_{u_j d_2} \,
 \Gamma^{LRH^\pm}_{u_j d_3} \frac{C_{8,XY}^{0}(y_{j})} {m_{u_{j}}}
 \\  \nonumber 
&+&  \frac{v^2}{\lambda_t}   \sum_{j=1}^{3}~
 \Gamma^{RLH^\pm \star}_{u_j d_2} \,  \Gamma^{RLH^\pm}_{u_j d_3}
 \frac{C_{8,YY}^{0}(y_{j})} {m_{u_{j}}^{2}} \, , \\  \nonumber 
\eea
while for $b \to d \gamma$ the label $d_2$ and $\lambda_t=V_{tb} \, V_{ts}^\star$ have to be
replaced by $d_1$ and $\lambda_t=V_{tb} \, V_{td}^\star$, respectively.
The functions
$C_{7,XY}^{0}$, $C_{7,YY}^{0}$, $C_{8,XY}^{0}$ and $C_{8,YY}^{0}$
were introduced in Ref. \cite{Borzumati:1998tg}; their explicit form
reads 
\bea \label{c7xyetc}
&& C_{7,XY}^{0}(y_j) = \frac{y_j}{12} \, \left[ \frac{-5y_j^2+8y_j-3+(6y_j-4)\ln
    y_j}{(y_j-1)^3} \right] \quad , \nonumber \\ 
&& C_{8,XY}^{0}(y_j) = \frac{y_j}{4} \, \left[ \frac{-y_j^2+4y_j-3-2 \ln y_j}{(y_j-1)^3}
  \right] \quad , \nonumber \\ 
&& C_{7,YY}^{0}(y_j) = \frac{y_j}{72} \, \left[
  \frac{-8y_j^3+3y_j^2+12y_j-7+(18y_j^2-12y_j)\ln y_j}{(y_j-1)^4} \right] \quad ,
\nonumber \\
&& C_{8,YY}^{0}(y_j) = \frac{y_j}{24} \, \left[ \frac{-y_j^3+6y_j^2-3y_j-2-6y_j \ln y_j}{(y_j-1)^4}
  \right] \quad .
\eea
In Eq. (\ref{wilson3LO}) we retained the contributions from
internal up- and charm-quarks, although these contributions are subleading.

\subsection{Wilson coefficients for $\Delta F=2$ processes}

The extended Higgs sector of our 2HDM of type-III also leads to 
extra contributions to $\Delta {F}=2$ processes
($B_s$, $B_d$, Kaon and $D$ mixing) which can be matched onto the effective Hamiltonian
\begin{equation}
{\cal H}^{\Delta {F}=2}_{eff}=\sum\limits_{j = 1}^5 C_j \, O_j  +
\sum\limits_{j = 1}^3 C^\prime_j \, O^\prime_j + h.c. \, , \label{Heff_DeltaF2}
\end{equation}
where the operators read in the case of $B_s$ mixing
\be
\begin{array}{llll}
O_1   \,= &\! 
 ({\bar s}_{\alpha} \gamma^{\mu} P_{L} b_{\alpha})\, 
({\bar s}_{\beta} \gamma^{\mu} P_{L} b_{\beta})     \,, 
               &  \quad 
O_2  \,= &\!
 ({\bar s}_{\alpha} P_{L} b_{\alpha})\, 
({\bar s}_{\beta}  P_{L} b_{\beta})     \,,   \\[1.002ex]
O_3  \,= &\!
 ({\bar s}_{\alpha} P_{L} b_{\beta})\, 
({\bar s}_{\beta}  P_{L} b_{\alpha})     \,, 
               &  \quad 
O_4   \,= &\!
 ({\bar s}_{\alpha} P_{L} b_{\alpha})\, 
({\bar s}_{\beta}  P_{R} b_{\beta})     \,,   \\[1.002ex]
O_5  \,= &\!
({\bar s}_{\alpha} P_{L} b_{\beta})\, 
({\bar s}_{\beta}  P_{R} b_{\alpha})     \, .
                &  \quad 
\end{array} 
\label{opbasisDF2}
\ee
$\alpha$ and $\beta$ are color indices and the primed operators can be
obtained from $O_{1,2,3}$ by interchanging $L$ and $R$. 
Similarly, the corresponding operator bases for $B_d$, Kaon and $D$
mixing follow from \eq{opbasisDF2} 
through simple adjustment of the indices.

In the following subsections we present the contributions to these
Wilson coefficients arising from: 1.) one-loop box diagrams with charged
Higgs boson exchange; 2.) tree-level contributions induced by neutral
Higgs boson exchange; 3.) box diagrams involving neutral Higgs bosons,
relevant in the case of $D$ mixing.

\subsubsection{Charged Higgs box contributions }
For definiteness, let us consider $B_s$ mixing. The corresponding
Wilson coefficients for $B_d$ and Kaon mixing follow by a simple
adjustment of the indices. We have performed our calculation in a
general $R_\xi$ gauge. The non-vanishing Wilson coefficients from pure
charged Higgs boxes are given by
\begin{equation}
\renewcommand{\arraystretch}{1.4}
\begin{array}{l}
 C_1  = \dfrac{{ - 1}}{{128\pi ^2 }}\sum\limits_{j,k = 1}^3 {\Gamma
   _{u_j d_2 }^{RL\;H^ \pm  \star} \Gamma _{u_j d_3 }^{RL\;H^ \pm  }
   \Gamma _{u_k d_2 }^{RL\;H^ \pm  \star} \Gamma _{u_k d_3 }^{RL\;H^
     \pm  } D_2 \left( {m_{u_j }^2 ,m_{u_k }^2 ,m_{H^ \pm  }^2 ,m_{H^
       \pm  }^2 } \right)} \,, \\ 
 C_2  = \dfrac{{ - 1}}{{32\pi ^2 }}\sum\limits_{j,k = 1}^3 {m_{u_j }
   m_{u_k } \Gamma _{u_j d_2 }^{LR\;H^ \pm  \star} \Gamma _{u_j d_3
   }^{RL\;H^ \pm  } \Gamma _{u_k d_2 }^{LR\;H^ \pm  \star} \Gamma
   _{u_k d_3 }^{RL\;H^ \pm  } D_0 \left( {m_{u_j }^2 ,m_{u_k }^2
     ,m_{H^ \pm  }^2 ,m_{H^ \pm  }^2 } \right)} \,, \\ 
 C_4  = \dfrac{{ - 1}}{{16\pi ^2 }}\sum\limits_{j,k = 1}^3 {m_{u_j }
   m_{u_k } \Gamma _{u_j d_2 }^{LR\;H^ \pm  \star} \Gamma _{u_j d_3
   }^{RL\;H^ \pm  } \Gamma _{u_k d_2 }^{RL\;H^ \pm  \star} \Gamma
   _{u_k d_3 }^{LR\;H^ \pm  } D_0 \left( {m_{u_j }^2 ,m_{u_k }^2
     ,m_{H^ \pm  }^2 ,m_{H^ \pm  }^2 } \right)} \,, \\ 
 C_5  = \dfrac{1}{{32\pi ^2 }}\sum\limits_{j,k = 1}^3 {\Gamma _{u_j
     d_2 }^{LR\;H^ \pm  \star} \Gamma _{u_j d_3 }^{LR\;H^ \pm  }
   \Gamma _{u_k d_2 }^{RL\;H^ \pm  \star} \Gamma _{u_k d_3 }^{RL\;H^
     \pm  } D_2 \left( {m_{u_j }^2 ,m_{u_k }^2 ,m_{H^ \pm  }^2 ,m_{H^
       \pm  }^2 } \right)} \,. 
 \label{DeltaF2charged}
 \end{array}
\end{equation}
The sum of the charged Higgs$-W^{\pm}$ and charged Higgs$-$Goldstone-boson boxes is given by
\begin{equation}
\begin{array}{l}
 C_1  = \dfrac{{g_2 ^2 }}{{32\pi ^2 }}\sum\limits_{j,k = 1}^3 \left(
 m_{u_j } m_{u_k } V_{j2}^\star V_{k3}  \Gamma _{u_j d_3}^{RL\, H^\pm  }
 \Gamma _{u_k d_2}^{RL\, H^ \pm  \star}  \right.\\
 \left.  \qquad\times \dfrac{{4M_W^2 D_0 \left( {M_W^2 ,m_{H^ \pm  }^2
       ,m_{u_j }^2 ,m_{u_k }^2 } \right)-D_2 \left( {M_W^2 ,m_{H^ \pm
       }^2 ,m_{u_j }^2 ,m_{u_k }^2 } \right)}}{{4M_W^2 }} \right)  \\ 
 C_4  = \dfrac{1}{{16\pi ^2 }}\dfrac{{g_2 ^2 }}{2}\sum\limits_{j,k =
   1}^3 \left( {V_{j3} V_{k2}^\star \Gamma _{u_j d_2}^{LR\, H^\pm  \star}
   \Gamma _{u_k d_3}^{LR\, H^\pm  } }\right.\\
\qquad\times\left.\dfrac{{C_2 \left( {\xi M_W^2 ,m_{H^ +  }^2 ,m_{u_j
      }^2 } \right) - C_2 \left( {m_{H^ +  }^2 ,m_{u_j }^2 ,m_{u_k }^2
    } \right) + m_{u_k }^2 C_0 \left( {\xi M_W^2 ,m_{H^ \pm  }^2
      ,m_{u_k }^2 } \right)}}{{M_W^2 }} \right) 
 \end{array}
\end{equation}
We stress here, that we want to use $B_s$ mixing only to constrain
certain $\epsilon^{u}_{ij}-$couplings, because the $\epsilon^{d}_{ij}-$quantities
are already contrained to be very small. We therefore only took 
systematically into account those contributions to the Wilson
coefficients which stay different from zero in the limit
$\epsilon^{d}_{ij} \to 0$.  
At first sight, the Wilson coefficient $C_4$ seems to be gauge
dependent. However, when using the unitarity of the CKM matrix
(entering the expression for $C_4$ both, explicitly and implicitly
through the $\Gamma-$quantities),
we find that the $\xi-$dependent terms are always proportional to an
element $\epsilon^d_{ij}$, which we put to zero in our analysis. Also
note that our result agrees with the one of
Ref.~\cite{Altmannshofer:2007cs}. The only difference is that we
neglected gauge dependent terms corresponding to higher dimensional
operators. The Wilson coefficients of the primed operators
can be obtained by interchanging $L$ and $R$ in the corresponding unprimed ones.
\subsubsection{{Tree-level $H_{k}^{0}$ contribution}}

The Wilson coefficients from neutral Higgs mediated tree-level
contributions to $ B_{s}$ mixing 
read:
\bea
  C_2^{H_{k}^{0}} &=& \sum\limits_{k = 1}^3  \, \dfrac{ - 1}{2 m_{H_{k}^{0}}^{2}} \,  (\Gamma _{\;d_{3} d_{2}}^{LR\,H_{k}^{0}\star})^{2}    \\ \nonumber
 C_2^{\prime\,H_{k}^{0}} &=& \sum\limits_{k = 1}^3  \, \dfrac{ - 1}{2 m_{H_{k}^{0}}^{2}} \,  (\Gamma _{\;d_{2} d_{3}}^{LR\,H_{k}^{0}})^{2}  \\ \nonumber
  C_4^{\,H_{k}^{0}} &=& \sum\limits_{k = 1}^3  \, \dfrac{ - 1}{m_{H_{k}^{0}}^{2}} \,  \Gamma _{\;d_{2} d_{3}}^{LR\,H_{k}^{0}} \,  \Gamma _{\;d_{3} d_{2}}^{LR\,H_{k}^{0}\star}
\eea
The corresponding coefficients for $B_d$, Kaon and $D$ mixing follow
by a careful adjustment of the indices.

Note that in the limit of large $\tan\beta$ and $ m_{A} \gg v$,
$C_2^{H_{k}^{0}} $ and $C_2^{\prime\,H_{k}^{0}}$ vanish and we only
get a contribution to $C_4^{H_{k}^{0}}$.

\subsubsection{ Neutral Higgs box contribution to D mixing }

The Wilson coefficients resulting from the neutral Higgs box
contribution to $D$ mixing are given as
\bea
\nonumber C_1  &=& \dfrac{{ - 1}}{{128\pi ^2 }}  \sum\limits_{j_{1},j_{2} = 1}^3 \sum\limits_{k_{1},k_{2} = 1}^3  \Gamma _{u_2 u_{j_{1}} }^{LR\;H^{0}_{k_{1}} \star}  \,  \Gamma _{u_1 u_{j_{1}} }^{LR\;H^{0}_{k_{2}} } \,  \Gamma _{u_2 u_{j_{2}} }^{LR\;H^{0}_{k_{2}} \star }  \,  \Gamma _{u_1 u_{j_{2}} }^{LR\;H^{0}_{k_{1}} } \,   
 D_2 \left( {m_{u_{j_1} }^2, m_{u_{j_2} }^2, m^{2}_{H_{k_{1}}^{0}}, m^{2}_{H_{k_{2}}^{0}} }   \right)  \,, \\  \nonumber
 \nonumber  C_2  &=& \dfrac{{ - 1}}{{32\pi ^2}}  \sum\limits_{j_{1},j_{2} = 1}^3 \sum\limits_{k_{1},k_{2} = 1}^3  { m_{u_{j_1}} m_{u_{j_2}} }    \Gamma _{u_{j_{1}} u_1 }^{LR\;H^{0}_{k_{1}} \star}   \Gamma _{u_2 u_{j_{1}} }^{LR\;H^{0}_{k_{2}} \star}  \Gamma _{u_{j_{2}} u_1 }^{LR\;H^{0}_{k_{1}} \star}   \Gamma _{u_2 u_{j_{2}} }^{LR\;H^{0}_{k_{2}} \star} \\
 &\times& D_0 \left( {m_{u_{j_1} }^2,m_{u_{j_2} }^2,m^{2}_{H_{k_{1}}^{0}},m^{2}_{H_{k_{2}}^{0}} }   \right) ,\nonumber  \\ 
 \nonumber   C_3  &=& 0 \,  , \nonumber  \\ 
 \nonumber   C_4 &=& \dfrac{{ - 1}}{{16\pi ^2 }}  \sum\limits_{j_{1},j_{2} = 1}^3 \sum\limits_{k_{1},k_{2} = 1}^3  { m_{u_{j_1}} m_{u_{j_2}}  }   \Gamma _{u_{j_{1}} u_1 }^{LR\;H^{0}_{k_{1}} \star}    \Gamma _{u_2 u_{j_{1}} }^{LR\;H^{0}_{k_{2}} \star}  \Gamma _{u_{j_{2}} u_2 }^{LR\;H^{0}_{k_{2}} }     \Gamma _{u_1 u_{j_{2}} }^{LR\;H^{0}_{k_{1}} }   \nonumber \\
  &\times& D_0 \left( {m_{u_{j_1} }^2 ,m_{u_{j_2} }^2 ,m^{2}_{H_{k_{1}}^{0}},m^{2}_{H_{k_{2}}^{0}}  } \right) , \\ \nonumber
  C_5 & =& \dfrac{{ - 1}}{{128\pi ^2 }}  \sum\limits_{j_{1},j_{2} = 1}^3 \sum\limits_{k_{1},k_{2} = 1}^3  \Gamma _{u_{j_{1}} u_{1} }^{LR\;H^{0}_{k_{1}} \star}  \,  \Gamma _{u_{j_{1}} u_{2} }^{LR\;H^{0}_{k_{2}} } \,  \Gamma _{u_1 u_{j_{2}} }^{LR\;H^{0}_{k_{1}} }  \,  \Gamma _{u_2 u_{j_{2}} }^{LR\;H^{0}_{k_{2}} \star} \, 
 D_2 \left( {m_{u_{j_1} }^2 ,m_{u_{j_2} }^2 ,m^{2}_{H_{k_{1}}^{0}},m^{2}_{H_{k_{2}}^{0}}   } \right) \, .
\eea
The indices $j_{1},j_{2}$ describe the internal up-type quarks while
$k_{1},k_{2}$ stand for neutral Higgs indices
($H^{0},h^{0},A^{0}$). Moreover, the primed Wilson coefficients can be
obtained from above by the replacement $ L\leftrightarrow R$ in the
couplings.
\subsection{Semileptonic and leptonic meson decays and tau decays: $B
  \to ({D^{(*)}})\tau \nu$, $ D_{(s)} \to \ell \nu_{\ell}$, $K(\pi)
  \to \ell \nu_{\ell} $ and $\tau\to K(\pi)\nu$ processes}

These processes are governed by the effective Hamiltonian
\begin{equation}
{\cal H}_{\rm eff}=C^{u_fd_i,\ell_j}_{\rm SM} O^{u_fd_i,\ell_j}_{\rm SM} +
C^{u_fd_i,\ell_j}_{L} O^{u_fd_i,\ell_j}_L+C^{u_fd_i,\ell_j}_R
O^{u_fd_i,\ell_j}_R+{\rm h.c.} \, ,  
\end{equation}
with the operators defined as
\begin{equation}
\renewcommand{\arraystretch}{1.4}
\begin{array}{*{20}l}
   O^{u_fd_i,\ell_j}_{\rm SM}  = \bar u_f\gamma _{\mu } P_L d_i \; \bar\ell_j \gamma _{\mu } P_L \nu\,,\\
    O^{u_fd_i,\ell_j}_{R}  = \bar u_f P_R d_i \; \bar\ell_jP_L \nu \,,\\
O^{u_f d_i,\ell_j}_{L}  = \bar u_f P_L d_i \; \bar\ell_j P_L \nu \,.   
\label{OpBasBTauNu}
\end{array}
\end{equation}
Here, for tauonic $B$ meson decays $\ell_j=\tau$, $d_i=b$ and $u_f=u$
($u_f=c$) for \btau (\bdtau and \bdstau). For $ D_{s} \to \ell_{j} \nu
$ ($ D \to \ell_{j} \nu $), $u_f=c$ and $d_i=s$ ($d$), for $\tau\to
K(\pi)\nu$, $\ell_j=\tau$, $u_f=u$ and $d_i=s$ ($d$) and for
$K(\pi)\to \ell_{j}\nu$ we have $\ell_j=\mu,e$, $u_f=u$ and $d_i=s$
($d$). The Wilson coefficients in 2HDM of type III at the matching
scale read
\begin{equation}
\renewcommand{\arraystretch}{1.8}
\begin{array}{l}
 C^{u_f d_i,\ell_{j}}_{\rm SM} = \dfrac{{ 4 G_{F}}}{\sqrt{2}} \; V_{u_f d_i}     \,,       \\
C^{u_f d_i,\ell_{j}}_{R} = \dfrac{{ -1}}{m_{H^{\pm}}^{2}} \; \Gamma_{u_f d_i}^{LR \, H^{\pm}} \;  \Gamma_{\nu\ell_{j}}^{LR \,H^{\pm} \star}  \,, \\
C^{u_f d_i,\ell_{j}}_{L} = \dfrac{{ -1}}{m_{H^{\pm}}^{2}} \; \Gamma_{u_f d_i}^{RL\,H^{\pm}} \; \Gamma_{\nu\ell_{j}}^{LR\,H^{\pm} \star}   \,.
 \end{array}
 \label{CRLBTauNu}
\end{equation}

\subsection{Lepton flavour violation (LFV): $\ell_{i} \to \ell_{f}
  \gamma$ processes}

The radiative lepton decays $\ell_{i} \to \ell_{f} \gamma$
($\ell=e,\mu$ or $\tau$) are induced by one-loop penguin diagrams
with internal neutral or charged Higgs bosons. The result for the one-loop
decay amplitude can be written as a tree-level matrix element
of the effective Hamiltonian 
\begin{equation}
\label{HeffLFV}
{\cal{H}}_{eff}=  c^{\ell_{f}\ell_{i}}_{R} \, O^{\ell_{f}\ell_{i}}_{R}
+c^{\ell_{f}\ell_{i}}_{L} \, O^{\ell_{f}\ell_{i}}_{L} \,,
\end{equation}
where $c^{\ell_{f}\ell_{i}}_{R}$ and $c^{\ell_{f}\ell_{i}}_{L}$ are
the effective Wilson coefficients of the
magnetic dipole operators
\begin{equation}
O^{\ell_{f}\ell_{i}}_{R(L)}  = m_{\ell_{i}} \bar{\ell_{f}}
\sigma_{\mu \nu} P_{R(L)} \ell_{i} F^{\mu \nu} \, .
\label{HeffLFV2}
\end{equation}
With these conventions, the branching ratio for the radiative lepton
decays $\ell_{i} \to \ell_{f} \gamma$ reads 
\begin{equation}
{ {\cal B} } \left[\ell_{i} \to \ell_{f} \gamma
  \right]\,=\,\dfrac{m_{\ell_i}^5}{4\pi \, \Gamma_{\ell_i}} \left(
|c^{\ell_{f}\ell_{i}}_{R} |^{2}+ |c^{\ell_{f}\ell_{i}}_{L} |^{2}
\right ) \,.
\label{Brmuegamma}
\end{equation}
The neutral Higgs ($H_{k}^{0}=H^{0},h^{0},A^{0}$) penguin contribution to $c^{\ell_{f}\ell_{i}}_{R}$ is given by
\begin{equation}
\renewcommand{\arraystretch}{2.0}
\begin{array}{l}
{c^{\ell_{f}\ell_{i}}_{R\,H_{k}^0}}  =   \sum\limits_{k,j = 1}^3  
 \dfrac{-e}{192 \pi^2  m^2_{H^{0}_{k}} }\, \left[
  \Gamma^{LR\,H_{k}^{0}}_{\ell_{f}    \ell_{j}}\Gamma^{LR\,H_{k}^{0}
    \star}_{\ell_{i} \ell_{j}}+\dfrac{m_{\ell_{f}}}{m_{\ell_{i}}}
  \Gamma^{LR\,H_{k}^{0} \star}_{\ell_{j} \ell_{f}}
  \Gamma^{LR\,H_{k}^{0}}_{\ell_{j} \ell_{i}}    \right. \\
~~~~~~~~~~~~~~~~~~~~~~~~~~~~~~~~~~~~ - \left.
  \dfrac{m_{\ell_{j}}}{m_{\ell_{i}}} \Gamma^{LR\,H_{k}^{0}}_{\ell_{f}
    \ell_{j}}   \Gamma^{LR\,H_{k}^{0}}_{\ell_{j} \ell_{i}}      \left(
  9+ 6 \ln\left(\dfrac{   m^{2}_{\ell_{j}}
  }{m^2_{H^{0}_{k}}}\right)  \right)      \right] \, ,
\end{array}
\label{CiH0klitolfgamma} 
\end{equation}
and ${c^{\ell_{f}\ell_{i}}_{L}}$ can be obtained from
${c^{\ell_{f}\ell_{i}}_{R}}$ by interchanging $L$ and $R$. 
Similarly, for the charged Higgs penguin contributions we find
\begin{equation}
\renewcommand{\arraystretch}{1.6}
\begin{array}{l}
  {c^{\ell_{f}\ell_{i}}_{L}}_{H^{\pm}} =   \dfrac{e}{384 \pi^2
    m^2_{H^{\pm}} }  \sum\limits_{j = 1}^3
  \Gamma^{LR\,H^{\pm}}_{\nu_{j}\ell_{i}}  \Gamma^{LR\,H^{\pm}
    \star}_{\nu_{j}\ell_{f}}  \,   ~, \\ 
    {c^{\ell_{f}\ell_{i}}_{R}}_{H^{\pm}} =
    \dfrac{m_{\ell_{f}}}{m_{\ell_{i}}}\dfrac{e}{384 \pi^2
      m^2_{H^{\pm}} }  \sum\limits_{j = 1}^3
    \Gamma^{LR\,H^{\pm}}_{\nu_{j}\ell_{i}}  \Gamma^{LR\,H^{\pm}
      \star}_{\nu_{j}\ell_{f}}  \, .  
\end{array}
\end{equation}
\subsection{Wilson coefficients for EDMs and the anomalous magnetic moment of the muon}
\subsubsection{Wilson coefficients for EDMs of charged leptons and the anomalous magnetic moment of the muon}

As in the case of the LFV processes discussed in the previous section,
we again have both neutral and charged Higgs penguin
contributions to the flavor conserving radiative transitions $\ell_{i}
\to \ell_{i} \gamma$. The corresponding effective Hamiltonian is
obtained from \eq{HeffLFV} and \eq{HeffLFV2} by identifying $\ell_f$ with $\ell_i$. 
The contribution to the effective Wilson coefficients related to
neutral Higgs bosons (propagating in the loop) reads 
\bea
  {c^{\ell_{i}\ell_{i}}_{R}}_{H_{k}^0}  &=&   \sum\limits_{k,j = 1}^3
  \dfrac{-e}{192 \pi^2  m^2_{H^{0}_{k}}
  }\, \left[    \Gamma^{LR\,H_{k}^{0} \star}_{\ell_{i} \ell_{j}}
    \Gamma^{LR\,H_{k}^{0}}_{\ell_{i} \ell_{j}} + \Gamma^{LR\,H_{k}^{0}
      \star}_{\ell_{j} \ell_{i}}   \Gamma^{LR\,H_{k}^{0}}_{\ell_{j}
      \ell_{i}}    \right. \\  \nonumber
&& - \left.   \dfrac{m_{\ell_{j}}}{m_{\ell_{i}}}
 \Gamma^{LR\,H_{k}^{0}}_{\ell_{i} \ell_{j}}
 \Gamma^{LR\,H_{k}^{0}}_{\ell_{j} \ell_{i}}      \left(   9+ 6
 \ln\left(\dfrac{  m^{2}_{\ell_{j}}  }{m^2_{H^{0}_{k}}}\right)
 \right)      \right] \, , \nonumber \\
{c^{\ell_{i}\ell_{i}}_{L}}_{H_{k}^0}  &=&
{c^{\ell_{i}\ell_{i}\star}_{R}}_{H_{k}^0}  \, , 
\eea
while the charged Higgs penguin contribution leads to the (real) coefficients
\begin{equation}
\renewcommand{\arraystretch}{1.6}
\begin{array}{l}
  {c^{\ell_{i}\ell_{i}}_{L}}_{H^{\pm}}  =    {c^{\ell_{i}\ell_{i} }_{R}}_{H^{\pm}} =  \dfrac{e}{384 \pi^2  m^2_{H^{\pm}} }  \sum\limits_{j = 1}^3 \Gamma^{LR\,H^{\pm}}_{\nu_{j}\ell_{i}}  \Gamma^{LR\,H^{\pm} \star}_{\nu_{j}\ell_{i}}  \, .
\end{array}
\end{equation}

\subsubsection{Wilson coefficients for neutron EDM}

In this section we consider 
the transitions $d \to d \gamma (g)$ and $u \to u \gamma (g)$ 
(denoted by ${d^{(g)}_{d}}$ and ${ d^{(g)}_{u}}$)
which are the building blocks for the electric dipole moment $d_n$ of
the neutron. As we are only interested in a rough estimate of $d_n$, 
we do not include QCD corrections to these building blocks. In this
approximation the latter can be described by the effective Hamiltonian 
\begin{eqnarray}
\label{Heffneutron}
{\cal H}_{eff}^{dd,uu}= &&  
c^{dd}_{R}  \,  m_{d} \, \bar{d} \sigma_{\mu \nu} P_{R} d \, F^{\mu \nu} +
c^{dd}_{L}  \,  m_{d} \, \bar{d} \sigma_{\mu \nu} P_{L} d \, F^{\mu \nu} +
\nonumber \\
&& c^{dd}_{R,g} m_{d} \, \bar{d} \sigma_{\mu \nu} P_{R} T^a d \, G^{a,\mu \nu} +
c^{dd}_{L,g}  \,  m_{d} \, \bar{d} \sigma_{\mu \nu} P_{L} T^a d \, G^{a,\mu \nu} +
(d \to u) \, .
\end{eqnarray}
The effective Wilson coefficients 
$c^{dd,uu}_{R,L}$ and $c^{dd,uu}_{R,L,g}$ again
receive neutral and charged Higgs contributions. The
neutral contributions of the Wilson coefficients (involved in
$d_d^{(g)}$) read
\bea
  {c^{d d,{H_{k}^0}}_{R}}  =   \sum\limits_{k,j = 1}^3
  \dfrac{eQ_{d}}{192 \pi^2  m^2_{H^{0}_{k}} }\, \left[
    \Gamma^{LR\,H_{k}^{0} \star}_{d d_{j}}   \Gamma^{LR\,H_{k}^{0}}_{d
      d_{j}} + \Gamma^{LR\,H_{k}^{0} \star}_{d_{j} d}
    \Gamma^{LR\,H_{k}^{0}}_{d_{j} d}    \right. \\  \nonumber
 - \left.   \dfrac{m_{d_{j}}}{m_{d}} \Gamma^{LR\,H_{k}^{0}}_{d d_{j}}
 \Gamma^{LR\,H_{k}^{0}}_{d_{j} d}      \left(   9+ 6 \ln\left(\dfrac{
   m^{2}_{d_{j}}   }{m^2_{H^{0}_{k}}} \right)  \right)      \right] \, ,
\eea
\bea
  {c^{d d,{H_{k}^0}}_{R,g}}  =   \sum\limits_{k,j = 1}^3
  \dfrac{g_{s}}{192 \pi^2  m^2_{H^{0}_{k}} }\, \left[
    \Gamma^{LR\,H_{k}^{0} \star}_{d d_{j}}   \Gamma^{LR\,H_{k}^{0}}_{d
      d_{j}} + \Gamma^{LR\,H_{k}^{0} \star}_{d_{j} d}
    \Gamma^{LR\,H_{k}^{0}}_{d_{j} d}    \right. \\  \nonumber
 - \left.   \dfrac{m_{d_{j}}}{m_{d}} \Gamma^{LR\,H_{k}^{0}}_{d d_{j}}
 \Gamma^{LR\,H_{k}^{0}}_{d_{j} d}      \left(   9+ 6 \ln\left(\dfrac{
   m^{2}_{d_{j}}   }{m^2_{H^{0}_{k}}} \right)  \right)      \right] \,  , 
\eea
and $ {c^{d d,{H_{k}^0} }_{L,(g)}} = {c^{d d,{H_{k}^0} \, \star}_{R,(g)}}$. 
The charged Higgs penguin contributions to the Wilson coefficients
(involved in $d_d^{(g)}$) read 
\begin{eqnarray}
&&  {c^{d d,{H^\pm}}_{R}}  =   \sum\limits_{j = 1}^3
  \dfrac{-e}{16 \pi^2 \, m^2_{u_j}}\, \left[
    \Gamma^{LR\,H^\pm \star}_{d u_{j}}   \Gamma^{LR\,H^\pm}_{d u_{j}}
    \, C_{7,YY}^0\left(\dfrac{m^2_{u_j}}{m^2_{H^+}}\right) 
 + \Gamma^{LR\,H^\pm \star}_{u_{j} d} \Gamma^{LR\,H^\pm}_{u_{j} d}   
    \, C_{7,YY}^0\left(\dfrac{m^2_{u_j}}{m^2_{H^+}}\right)
 \right.  \nonumber \\
&& \left. \hspace{3.5cm} + \Gamma^{LR\,H^\pm}_{d u_{j}}
 \Gamma^{LR\,H^\pm}_{u_{j} d} \dfrac{m_{u_j}}{m_d} \,
 C_{7,XY}^0\left(\dfrac{m^2_{u_j}}{m^2_{H^+}}\right) \right] \, ,
\end{eqnarray}
\begin{eqnarray}
&&  {c^{d d,{H^\pm}}_{R,g}}  =   \sum\limits_{j = 1}^3
  \dfrac{-g_s}{16 \pi^2 \, m^2_{u_j}}\, \left[
    \Gamma^{LR\,H^\pm \star}_{d u_{j}}   \Gamma^{LR\,H^\pm}_{d u_{j}}
    \, C_{8,YY}^0\left(\dfrac{m^2_{u_j}}{m^2_{H^+}}\right)
 + \Gamma^{LR\,H^\pm \star}_{u_{j} d} \Gamma^{LR\,H^\pm}_{u_{j} d}   
     \, C_{8,YY}^0\left(\dfrac{m^2_{u_j}}{m^2_{H^+}}\right)
 \right. \nonumber \\
&& \left. \hspace{3.5cm} + \Gamma^{LR\,H^\pm}_{d u_{j}}
 \Gamma^{LR\,H^\pm}_{u_{j} d} \dfrac{m_{u_j}}{m_d} \,
 C_{8,XY}^0\left(\dfrac{m^2_{u_j}}{m^2_{H^+}}\right) \right] \, ,
\end{eqnarray}
and $ {c^{d d,{H^\pm} }_{L,(g)}} = {c^{d d,{H^\pm} \, \star}_{R,(g)}}$. 

The analogous expressions for ${c^{u u,{H^\pm,H^{0}_{k}}}_{R,(g)}}$,
which are involved in the expressions of $ d^{(g)}_{u}$ are given as
\begin{eqnarray}
&&  {c^{u u,{H_k^0}}_{R}}  =   \sum\limits_{j,k = 1}^3
  \dfrac{-e \, Q_u}{16 \pi^2 \, m^2_{u_j}}\, \left[
    \Gamma^{LR\,H_k^0 \star}_{u u_{j}}   \Gamma^{LR\,H_k^0}_{u u_{j}}
    \, C_{8,YY}^0\left(\dfrac{m^2_{u_j}}{m^2_{H_k^0}}\right) 
 + \Gamma^{LR\,H_k^0 \star}_{u_{j} u} \Gamma^{LR\,H_k^0}_{u_{j} u}   
     \, C_{8,YY}^0\left(\dfrac{m^2_{u_j}}{m^2_{H_k^0}}\right)
 \right. \nonumber \\
&& \left. \hspace{3.5cm} + \Gamma^{LR\,H_k^0}_{u u_{j}}
 \Gamma^{LR\,H_k^0}_{u_{j} u} \dfrac{m_{u_j}}{m_u} \,
 C_{8,XY}^0\left(\dfrac{m^2_{u_j}}{m^2_{H_k^0}}\right) \right] \, ,
\end{eqnarray}
\begin{eqnarray}
&&  {c^{u u,{H_k^0}}_{R,g}}  =   \sum\limits_{j,k = 1}^3
  \dfrac{-g_s}{16 \pi^2 \, m^2_{u_j}}\, \left[
    \Gamma^{LR\,H_k^0 \star}_{u u_{j}}   \Gamma^{LR\,H_k^0}_{u u_{j}}
    \, C_{8,YY}^0\left(\dfrac{m^2_{u_j}}{m^2_{H_k^0}}\right) 
 + \Gamma^{LR\,H_k^0 \star}_{u_{j} u} \Gamma^{LR\,H_k^0}_{u_{j} u}   
     \, C_{8,YY}^0\left(\dfrac{m^2_{u_j}}{m^2_{H_k^0}}\right)
 \right. \nonumber \\
&& \left. \hspace{3.5cm} + \Gamma^{LR\,H_k^0}_{u u_{j}}
 \Gamma^{LR\,H_k^0}_{u_{j} u} \dfrac{m_{u_j}}{m_u} \,
 C_{8,XY}^0\left(\dfrac{m^2_{u_j}}{m^2_{H_k^0}}\right) \right] \, ,
\end{eqnarray}
\begin{eqnarray}
&&  {c^{u u,{H^\pm}}_{R}}  =   \sum\limits_{j = 1}^3
  \dfrac{-e}{1152 \pi^2 \, m^2_{H^+}}\, \left[5\,
    \Gamma^{LR\,H^\pm \star}_{u d_{j}}   \Gamma^{LR\,H^\pm}_{u d_{j}}
 + 5\, \Gamma^{LR\,H^\pm \star}_{d_{j} u} \Gamma^{LR\,H^\pm}_{d_{j} u}  
 \right.  \nonumber \\
&& \left. \hspace{3.5cm} - \Gamma^{LR\,H^\pm}_{u d_{j}}
 \Gamma^{LR\,H^\pm}_{d_{j} u} \dfrac{m_{d_j}}{m_u} \,
 12 \, \ln \left( \dfrac{m^2_{d_j}}{m^2_{H^+}}\right) \right] \, ,
\end{eqnarray}
\begin{eqnarray}
&&  {c^{u u,{H^\pm}}_{R,g}}  =   \sum\limits_{j = 1}^3
  \dfrac{g_s}{192 \pi^2 \, m^2_{H^+}}\, \left[
    \Gamma^{LR\,H^\pm \star}_{u d_{j}}   \Gamma^{LR\,H^\pm}_{u d_{j}}
 + \Gamma^{LR\,H^\pm \star}_{d_{j} u} \Gamma^{LR\,H^\pm}_{d_{j} u}   
 \right. \nonumber \\
&& \left. \hspace{3.5cm} - \Gamma^{LR\,H^\pm}_{u d_{j}}
 \Gamma^{LR\,H^\pm}_{d_{j} u} \dfrac{m_{d_j}}{m_u} \,
 \left( 9 + 6 \ln \left( \dfrac{m^2_{d_j}}{m^2_{H^+}}\right) \right) \right] \, .
\end{eqnarray}
Again, we have $ {c^{u u,{H_{k}^0} }_{L,(g)}} = {c^{u u,{H_{k}^0} \,\star}_{R,(g)}}$
and $ {c^{u u,{H^\pm} }_{L,(g)}} = {c^{u u,{H^\pm} \, \star}_{R,(g)}}$. The loop functions $C_{7,8,XY,YY}^{0}(y_{j})$ are given in 
Eq. (\ref{c7xyetc}).

\subsection{Leptonic decays of neutral mesons}

The effective Hamiltonian ${\cal H}_{eff}$ which includes the full set
of operators for the general decays $PS(\bar{q_{f}}q_{i}) \to \ell_{A}^{+}
\ell_{B}^{-}$ ($PS$ refers to the pseudo-scalar meson) reads
\begin{equation}
\renewcommand{\arraystretch}{1.6}
\begin{array}{l}
{\cal H}^{\Delta {F}=1}_{eff}=-\frac{G_{F}^{2} M_{W}^{2}}{\pi^{2}}
\left[
  C^{{q}_{f}q_{i}}_{V}O^{{q}_{f}q_{i}}_{V}+C^{{q}_{f}q_{i}}_{A}O^{{q}_{f}q_{i}}_{A}+C^{{q}_{f}q_{i}}_{S}O^{{q}_{f}q_{i}}_{S}+C^{{q}_{f}q_{i}}_{P}O^{{q}_{f}q_{i}}_{P}
  + \, {\rm primed}      \right] +{\rm h.c.}
\label{Heffsemilep}
\end{array}
\end{equation}
where the operators (together with their primed counterparts) are defined as
\be
\begin{array}{llll}
{\cal O}^{{q}_{f}q_{i}}_V \,= &\!
 (\bar{q_{f}} \gamma_\mu P_{L} q_{i})\, 
 (\bar{\ell}_B \gamma^\mu  \ell_A)\,, 
               &  \quad 
{\cal O}^{{q}_{f}q_{i}}_A \,= &\!
 (\bar{q_{f}} \gamma_\mu P_{L} q_{i})\, 
 (\bar{\ell}_B \gamma^\mu \gamma_{5}  \ell_A)\,,   \\[1.002ex]
{\cal O}^{'{q}_{f}q_{i}}_V \,= &\!
 (\bar{q_{f}} \gamma_\mu P_{R} q_{i})\, 
 (\bar{\ell}_B \gamma^\mu  \ell_A)\,, 
               &  \quad 
{\cal O}^{'{q}_{f}q_{i}}_A \,= &\!
 (\bar{q_{f}} \gamma_\mu P_{R} q_{i})\, 
 (\bar{\ell}_B \gamma^\mu \gamma_{5}  \ell_A)\,,  \\[1.002ex]
{\cal O}^{{q}_{f}q_{i}}_S \,= &\!
 (\bar{q_{f}}  P_{L} q_{i})\, 
 (\bar{\ell}_B  \ell_A)\,, 
               &  \quad 
{\cal O}^{{q}_{f}q_{i}}_P \,= &\!
 (\bar{q_{f}}  P_{L} q_{i})\, 
 (\bar{\ell}_B \gamma_{5}  \ell_A)\,, \\[1.002ex]
{\cal O}^{'{q}_{f}q_{i}}_S \,= &\!
 (\bar{q_{f}}  P_{R} q_{i})\, 
 (\bar{\ell}_B  \ell_A)\,, 
                &  \quad 
{\cal O}^{'{q}_{f}q_{i}}_P \,= &\!
 (\bar{q_{f}}  P_{R} q_{i})\, 
 (\bar{\ell}_B \gamma_{5}  \ell_A)\,.
\end{array} 
\label{opbasis}
\ee
Making use of the hadronic matrix elements 
\begin{eqnarray}
\langle 0 | \bar{q_{f}} \gamma_{\mu}\gamma_{5}q_{i} |PS \rangle&=&i f_{PS} \, p_{PS}^{\mu} \,,\\ \nonumber
\langle 0 | \bar{q_{f}} \gamma_{5}q_{i} |PS \rangle&=&-i f_{PS} \frac{M_{PS}^{2}}{(m_{q_{f}}+m_{q_{i}})} \, ,
\end{eqnarray}
one obtains the branching ratio
\begin{equation}
\begin{array}{l}
\renewcommand{\arraystretch}{2}
{\cal{B}}\left[ PS(\bar{q_{f}}q_{i}) \to \ell_{A}^{+} \ell_{B}^{-}
  \right] = \dfrac{G_{F}^{4} M_{W}^{4} }{32 \pi^{5}}  \,
f\left(x_{A}^{2},x_{B}^{2}\right) \,  M_{PS} \,  f_{PS}^{2} \,
\left(m_{l_{A}}+m_{l_{B}}\right)^{2}   \,   \tau_{PS}   \\
 ~~~~~~~~~~~ \times \left\{       \left|  \dfrac{M_{PS}^{2}
  \left(C^{{q}_{f}q_{i}}_{P}-C^{'{q}_{f}q_{i}}_{P}\right)}{\left(m_{q_{f}}+m_{q_{i}}\right)\left(m_{l_{A}}+m_{l_{B}}\right)} 
-\left(C^{{q}_{f}q_{i}}_{A}-C^{'{q}_{f}q_{i}}_{A}\right)  \right|^{2} \times
\left[1-(x_{A}-x_{B})^{2}\right]  \right.   \\
~~~~~~~~~~+ \left. \left|  \dfrac{M_{PS}^{2} \left(C^{{q}_{f}q_{i}}_{S}-C^{'{q}_{f}q_{i}}_{S}\right)}{(m_{q_{f}}+m_{q_{i}})(m_{l_{A}}+m_{l_{B}})} 
+\dfrac{(m_{l_{A}}-m_{l_{B}})}{(m_{l_{A}}+m_{l_{B}})}\left(C^{{q}_{f}q_{i}}_{V}-C^{'{q}_{f}q_{i}}_{V}\right)
\right| ^{2} \times \left[1-(x_{A}+x_{B})^{2}\right]    \right\} \,,
\label{BRmostgen}
\end{array}
\end{equation}
where the function $f(x_{i},x_{j})$ and the ratio $x_{i}$ are defined
as \cite{Chankowski:2000ng}
\be
\nonumber f(x_{i},x_{j})=\sqrt{1-2 (x_{i}+x_{j})+(x_{i}-x_{j})^{2}}~, ~~x_{i}=\frac{m_{\ell_{i}}}{M_{PS}} \, .
\ee
%
\subsubsection{Wilson coefficients}
\begin{itemize}
 
\item {\it {Tree level neutral Higgs contributions to
    $PS(\bar{q_{f}}q_{i}) \to \ell_{A}^+ \ell_{B}^-$} in the 2HDM of
  type III}

The non-vanishing Wilson coefficients of the operators in \eq{Heffsemilep}
induced through tree-level neutral Higgs ($
H_{k}^{0}=H^{0},h^{0},A^{0}$) exchange read
\begin{equation}
\renewcommand{\arraystretch}{1.6}
\begin{array}{l}
  C_S^{{q}_{f}q_{i}}= \dfrac{{ \pi^{2}}}{2 G_{F}^{2}
    M_{W}^{2}}\sum\limits_{k = 1}^3 \dfrac{1}{m^{2}_{H_{k}^{0}}}\,
  \left( \Gamma _{\ell_{B} \ell_{A}}^{LR\, H_{k}^{0}}  + \Gamma
  _{\ell_{B} \ell_{A}}^{RL \, H_{k}^{0}}          \right) \,  \Gamma
  _{q_{f} q_{i}}^{RL\, H_{k}^{0}}        \\

  C_P^{{q}_{f}q_{i}}  = \dfrac{{ \pi^{2}}}{2 G_{F}^{2}
    M_{W}^{2}}\sum\limits_{k = 1}^3 \dfrac{1}{m^{2}_{H_{k}^{0}}}\,
  \left( \Gamma _{\ell_{B} \ell_{A}}^{LR\, H_{k}^{0}}  - \Gamma
  _{\ell_{B} \ell_{A}}^{RL \, H_{k}^{0}}          \right) \,  \Gamma
  _{q_{f} q_{i}}^{RL\, H_{k}^{0}}        \\

  C_S^{'{q}_{f}q_{i}}= \dfrac{{ \pi^{2}}}{2 G_{F}^{2}
    M_{W}^{2}}\sum\limits_{k = 1}^3 \dfrac{1}{m^{2}_{H_{k}^{0}}}\,
  \left( \Gamma _{\ell_{B} \ell_{A}}^{LR\, H_{k}^{0}}  + \Gamma
  _{\ell_{B} \ell_{A}}^{RL \, H_{k}^{0}}          \right) \,  \Gamma
  _{q_{f} q_{i}}^{LR\, H_{k}^{0}}        \\

   C_P^{'{q}_{f}q_{i}} = \dfrac{{ \pi^{2}}}{2 G_{F}^{2}
     M_{W}^{2}}\sum\limits_{k = 1}^3 \dfrac{1}{m^{2}_{H_{k}^{0}}}\,
   \left( \Gamma _{\ell_{B} \ell_{A}}^{LR\, H_{k}^{0}}  - \Gamma
   _{\ell_{B} \ell_{A}}^{RL \, H_{k}^{0}}          \right) \,  \Gamma
   _{q_{f} q_{i}}^{LR\, H_{k}^{0}}           \, .    

 \end{array}
\end{equation}

\item {\it {Loop-induced charged Higgs contributions to $B_{s}\to \mu^{+} \mu^{-}$ in the 2HDM of type II}  }

As mentioned earlier, we also include in our analysis the 2HDM of type II loop-induced charged Higgs contributions to $B_{s}\to \mu^{+} \mu^{-}$ from Ref.~\cite{Logan:2000iv}:
\bea
  C^{bs}_S &=& C^{bs}_P = -\dfrac{ m_{b} \, V^{*}_{tb}V_{ts}}{2} \dfrac{m_{\mu}}{2 M^{2}_{W}}  \, \tan^{2}\beta \, \dfrac{\log \left( m^{2}_{H}/m^{2}_{t}  \right) }{  m^{2}_{H}/m^{2}_{t} -1}     \,   , 
\eea
where $m_{b}$ and $m_t$ are understood to be running masses evaluated at the matching scale.

\end{itemize}
\subsection{Flavour-changing lepton decays}
The general expressions for the branching ratios of $\tau^{-} \to e^{-}\mu^{+}\mu^{-}$ and $\tau^{-}\to \mu^{-}\mu^{+}\mu^{-}$ have the form
\begin{equation}
\renewcommand{\arraystretch}{2.4}
\begin{array}{l}
{\cal B} \left[ {\tau^{-}  \to e^{-}\mu^{+} \mu^{-} } \right] = \dfrac{{m_\tau ^5}}{{12{{\left( {8\pi } \right)}^3}{\Gamma _\tau }}}
\left( {{{\left| {\dfrac{{ \Gamma _{\tau e}^{LR\, H_k^0\star }\Gamma _{\mu \mu }^{LR\,H_k^0}}}{{m_{H_k^0}^2}}} \right|}^2} + {{\left| {\dfrac{{\Gamma _{e\tau }^{LR \,H_k^0\star}\Gamma _{\mu \mu }^{LR\, H_k^0}}}{{m_{H_k^0}^2}}} \right|}^2} }\right.\\
\qquad\qquad\qquad\qquad\qquad\qquad \;\;\left.{
+ {{\left| {\dfrac{{\Gamma _{\tau e}^{LR\, H_k^0}\Gamma _{\mu \mu }^{LR\, H_k^0}}}{{m_{H_k^0}^2}}} \right|}^2} + {{\left| {\dfrac{{\Gamma _{e\tau }^{LR\, H_k^0}\Gamma _{\mu \mu }^{LR\, H_k^0}}}{{m_{H_k^0}^2}}} \right|}^2}} \right)\\
{\cal B} \left[ {\tau^{-}  \to \mu^{-} \mu^{+} \mu^{-} } \right] = \dfrac{{m_\tau ^5}}{{12{{\left( {8\pi } \right)}^3}{\Gamma _\tau }}}\dfrac{1}{2}\left( {2 {{{\left| {\dfrac{{\Gamma _{\tau \mu }^{LR\,H_k^0\star}\Gamma _{\mu \mu }^{LR\, H_k^0}}}{{m_{H_k^0}^2}}} \right|}^2} + 2{{\left| {\dfrac{{\Gamma _{\mu \tau }^{LR\, H_k^0\star}\Gamma _{\mu \mu }^{LR\,H_k^0}}}{{m_{H_k^0}^2}}} \right|}^2}}  }\right.\\
\qquad\qquad\qquad \qquad\qquad\qquad\;\;\;\; \left.{+ {{\left| {\dfrac{{\Gamma _{\tau \mu }^{LR\, H_k^0}\Gamma _{\mu \mu }^{LR\, H_k^0}}}{{m_{H_k^0}^2}}} \right|}^2} + {{\left| {\dfrac{{\Gamma _{\mu \tau }^{LR\, H_k^0}\Gamma _{\mu \mu }^{LR\,H_k^0}}}{{m_{H_k^0}^2}}} \right|}^2}} \right) \, .
\end{array}
\label{taumumumuFull}
\end{equation}
Note that the (not explicitly denoted) sum over the Higgses must be
performed before taking the various absolute values in \eq{taumumumuFull}. 
\subsection{Input parameters}
In this section we list our input parameters in tabular form.
\begin{table}[h]
\label{tab:input1-2} {
\centering \vspace{0.8cm}

\begin{tabular}{|c|c|}
 \hline
 Parameter & Value~(GeV) \\   \hline \hline
${\overline m}_{u}$(2 GeV)& $0.00219\pm 0.00015$   \cite{Colangelo:2010et}  \\ \hline
${\overline m}_{d}$(2 GeV)& $0.00467\pm 0.00020$   \cite{Colangelo:2010et}   \\ \hline
${\overline m}_{s}$(2 GeV)& $0.095\pm 0.006$  \cite{Colangelo:2010et} \\ \hline
${\overline m}_{c}(m_{c})$& $1.28\pm 0.04$ \cite{Blossier:2010cr}  \\ \hline
${\overline m}_{b}(m_{b})$& $4.243\pm 0.043$  \cite{Amhis:2012bh}  \\ \hline
${\overline m}_{t}(m_{t})$& $165.80 \pm 0.54\pm0.72$  \cite{Charles:2004jd} \\ \hline \hline
\end{tabular}
~~~~~~~~~~~~~~~~
\begin{tabular}{|c|c|}
 \hline
 Parameter & Value \\   \hline \hline
$M_{W}$&$80.40$~GeV    \\ \hline
$M_{Z}$&$91.19$~GeV   \\ \hline
${\alpha_{s}(M_{Z})}$&$0.119$      \\ \hline
$G_{F}$&$1.16637\times10^{-5}$~\text{GeV}$^{-2}$  \\ \hline
${\alpha_{em}}^{-1}$&$137$   \\ \hline
$v$&$174.10$~GeV  \\ \hline  \hline 
\end{tabular}}
\caption{Left: Input values for the quark masses used in our article. In the numerical analysis, we used the NNLO expressions in $\alpha_s$ for the running (see for example Ref.~\cite{Buras:1998raa}) in order to obtain the quark-mass values at higher scales. Right: Electroweak parameters and the strong coupling constant used in our analysis. Concerning the running of $\alpha_s$ we used NNLO expressions (given for example in Ref.~\cite{Beringer:1900zz}).}
\end{table}
\begin{table}[h]
\label{tab:input3-4} {
\centering \vspace{0.8cm}

\begin{tabular}{|c|c|}
 \hline
 Parameter & Value  \\   \hline \hline
${f_{B_{s}}}/{f_{B}}$ &$1.221\pm0.010\pm0.033$  \cite{Charles:2004jd}   \\ \hline
$f_{D}$ &$218.9\pm11.3$ MeV     \cite{Bazavov:2011aa}  \\ \hline
$f_{D_{s}}$ &$249\pm2\pm5$ MeV     \cite{Charles:2004jd}  \\ \hline
${f_{D_{s}}}/{f_{D}}$ &$1.188\pm0.025 $  \cite{Bazavov:2011aa}    \\ \hline
$f_{K}$ &$156.3\pm0.3\pm1.9$ MeV     \cite{Charles:2004jd}   \\ \hline
${f_{K}}/{f_{\pi}}$ &$1.193\pm0.005$   \cite{Colangelo:2010et}   \\ \hline \hline
\end{tabular}
~~~~~~~~~~~~~~~~
\begin{tabular}{|c|c|}
 \hline
 ~Meson massses~& Values (GeV) ~\\   \hline \hline
$m_{B^{\pm}(B^{0})}$& $5.279$  \\ \hline
$m_{B_{s}}$& $5.367$    \\ \hline
$m_{D^{\pm}(D^{0})}$& $1.870$~($1.865$)    \\ \hline
$m_{D_{s}}$& $1.969$    \\ \hline
$m_{K^{\pm}(K^{0})}$& $0.494$~($0.498$)    \\ \hline
$m_{\pi^{\pm}(\pi^{0})}$& $0.140$~($0.135$)    \\ \hline \hline
\end{tabular}
}
\caption{Left: Values for decay constants of Ref.~\cite{Charles:2004jd} obtained by averaging the lattice results of Ref.~\cite{Blossier:2009bx,Bernard:2007ps,Follana:2007uv,Aubin:2008ie,Beane:2006kx,Durr:2010hr,Bernard:2002pc,Bazavov:2009ii,Davies:2010ip,Dimopoulos:2011gx,Gamiz:2009ku,McNeile:2011ng,Albertus:2010nm,Bazavov:2011aa}. Right: Meson masses according to the particle data group (see online update of Ref.~\cite{Beringer:1900zz}).
}
\end{table}

\newpage

\bibliographystyle{hieeetr}

\bibliography{2HDM3}

\begin{thebibliography}{100}

\bibitem{Lee:1973iz}
T.~Lee, ``{A Theory of Spontaneous T Violation},'' {\em Phys.Rev.}, vol.~D8,
  pp.~1226--1239, 1973.

\bibitem{Gunion:1989we}
J.~F. Gunion, H.~E. Haber, G.~L. Kane, and S.~Dawson, ``{THE HIGGS HUNTER'S
  GUIDE},'' {\em Front.Phys.}, vol.~80, pp.~1--448, 2000.

\bibitem{Branco:2011iw}
G.~Branco, P.~Ferreira, L.~Lavoura, M.~Rebelo, M.~Sher, {\em et~al.}, ``{Theory
  and phenomenology of two-Higgs-doublet models},'' {\em Phys.Rept.}, vol.~516,
  pp.~1--102, 2012, 1106.0034.

\bibitem{Kim:1986ax}
J.~E. Kim, ``{Light Pseudoscalars, Particle Physics and Cosmology},'' {\em
  Phys.Rept.}, vol.~150, pp.~1--177, 1987.

\bibitem{Peccei:1977hh}
R.~Peccei and H.~R. Quinn, ``{CP Conservation in the Presence of Instantons},''
  {\em Phys.Rev.Lett.}, vol.~38, pp.~1440--1443, 1977.

\bibitem{Trodden:1998qg}
M.~Trodden, ``{Electroweak baryogenesis: A Brief review},'' 1998,
  hep-ph/9805252.

\bibitem{Haber:1984rc}
H.~E. Haber and G.~L. Kane, ``{The Search for Supersymmetry: Probing Physics
  Beyond the Standard Model},'' {\em Phys.Rept.}, vol.~117, pp.~75--263, 1985.

\bibitem{BaBar:2012xj}
B.~Aubert {\em et~al.}, ``{Evidence for an excess of $B \to D^{(*)} \tau \nu$
  decays},'' 2012, 1205.5442.

\bibitem{Lees:2013qea}
J.~Lees {\em et~al.}, ``{Measurement of an Excess of $B \to D^{(*)} \tau \nu$
  Decays and Implications for Charged Higgs Bosons},'' 2013, 1303.0571.

\bibitem{Lees:2012ju}
J.~Lees {\em et~al.}, ``{Evidence of $B\to \tau \nu$ decays with hadronic $B$
  tags},'' 2012, 1207.0698.

\bibitem{Aubert:2008zzb}
B.~Aubert {\em et~al.}, ``{A Search for $B^{+} \to \ell^{+} \nu_{\ell}$
  Recoiling Against $B^{-}\to D^{0} \ell^{-} {\bar\nu} X$ },'' {\em Phys.Rev.},
  vol.~D81, p.~051101, 2010.

\bibitem{Hara:2010dk}
K.~Hara {\em et~al.}, ``{Evidence for $B^- \to \tau^- \bar{\nu}$ with a
  Semileptonic Tagging Method},'' {\em Phys.Rev.}, vol.~D82, p.~071101, 2010,
  1006.4201.

\bibitem{Adachi:2012mm}
I.~Adachi {\em et~al.}, ``{Measurement of $B^- \to \tau^- \bar{\nu}_\tau$ with
  a Hadronic Tagging Method Using the Full Data Sample of Belle},'' 2012,
  1208.4678.

\bibitem{Charles:2004jd}
J.~Charles {\em et~al.}, ``{CP violation and the CKM matrix: Assessing the
  impact of the asymmetric $B$ factories},'' {\em Eur.Phys.J.}, vol.~C41,
  pp.~1--131, 2005, hep-ph/0406184.

\bibitem{MFV}
G.~D'Ambrosio, G.~F. Giudice, G.~Isidori, and A.~Strumia, ``{Minimal flavour
  violation: An effective field theory approach},'' {\em Nucl. Phys.},
  vol.~B645, pp.~155--187, 2002, hep-ph/0207036.

\bibitem{Crivellin:2012ye}
A.~Crivellin, C.~Greub, and A.~Kokulu, ``{Explaining $B\to D\tau\nu$, $B\to
  D^*\tau\nu$ and $B\to \tau\nu$ in a 2HDM of type III},'' {\em Phys.Rev.},
  vol.~D86, p.~054014, 2012, 1206.2634.

\bibitem{Buras:2010mh}
A.~J. Buras, M.~V. Carlucci, S.~Gori, and G.~Isidori, ``{Higgs-mediated FCNCs:
  Natural Flavour Conservation vs. Minimal Flavour Violation},'' {\em JHEP},
  vol.~1010, p.~009, 2010, 1005.5310.

\bibitem{Blankenburg:2011ca}
G.~Blankenburg and G.~Isidori, ``{ $B \to \tau \nu$ in multi-Higgs models with
  MFV},'' 2011, 1107.1216.

\bibitem{Hamzaoui:1998nu}
C.~Hamzaoui, M.~Pospelov, and M.~Toharia, ``{Higgs mediated FCNC in
  supersymmetric models with large tan Beta},'' {\em Phys.Rev.}, vol.~D59,
  p.~095005, 1999, hep-ph/9807350.

\bibitem{Babu:1999hn}
K.~S. Babu and C.~F. Kolda, ``{Higgs mediated $B^0 \to \mu^{+} \mu^{-}$ in
  minimal supersymmetry},'' {\em Phys. Rev. Lett.}, vol.~84, pp.~228--231,
  2000, hep-ph/9909476.

\bibitem{Carena:1999py}
M.~S. Carena, D.~Garcia, U.~Nierste, and C.~E. Wagner, ``{Effective Lagrangian
  for the $\bar{t} b H^{+}$ interaction in the MSSM and charged Higgs
  phenomenology},'' {\em Nucl.Phys.}, vol.~B577, pp.~88--120, 2000,
  hep-ph/9912516.

\bibitem{Isidori:2001fv}
G.~Isidori and A.~Retico, ``{Scalar flavor changing neutral currents in the
  large tan beta limit},'' {\em JHEP}, vol.~0111, p.~001, 2001, hep-ph/0110121.

\bibitem{Buras:2002vd}
A.~J. Buras, P.~H. Chankowski, J.~Rosiek, and L.~Slawianowska, ``{ $\Delta
  M_{d,s}, B^0_{d,s} \to \mu^{+} \mu^{-}$ and $B \to X_{s} \gamma$ in
  supersymmetry at large $\tan\beta$},'' {\em Nucl. Phys.}, vol.~B659, p.~3,
  2003, hep-ph/0210145.

\bibitem{Hofer:2009xb}
L.~Hofer, U.~Nierste, and D.~Scherer, ``{Resummation of $\tan(\beta)$-enhanced
  supersymmetric loop corrections beyond the decoupling limit},'' {\em JHEP},
  vol.~0910, p.~081, 2009, 0907.5408.

\bibitem{Isidori:2002qe}
G.~Isidori and A.~Retico, ``{$B_{s,d} \to \ell^{+} \ell^{-}$ and $K_{L} \to
  \ell^{+} \ell^{-}$ in SUSY models with nonminimal sources of flavor
  mixing},'' {\em JHEP}, vol.~0209, p.~063, 2002, hep-ph/0208159.

\bibitem{Crivellin:2010er}
A.~Crivellin, ``{Effective Higgs Vertices in the generic MSSM},'' {\em Phys.
  Rev.}, vol.~D83, p.~056001, 2011, 1012.4840.

\bibitem{Noth:2010jy}
D.~Noth and M.~Spira, ``{Supersymmetric Higgs Yukawa Couplings to Bottom Quarks
  at next-to-next-to-leading Order},'' {\em JHEP}, vol.~06, p.~084, 2011,
  1001.1935.

\bibitem{Crivellin:2012zz}
A.~Crivellin and C.~Greub, ``{Two-loop SQCD corrections to Higgs-quark-quark
  couplings in the generic MSSM},'' 2012, 1210.7453.

\bibitem{Okada:1990vk}
Y.~Okada, M.~Yamaguchi, and T.~Yanagida, ``{Upper bound of the lightest Higgs
  boson mass in the minimal supersymmetric standard model},'' {\em
  Prog.Theor.Phys.}, vol.~85, pp.~1--6, 1991.

\bibitem{Haber:1990aw}
H.~E. Haber and R.~Hempfling, ``{Can the mass of the lightest Higgs boson of
  the minimal supersymmetric model be larger than $m_Z$?},'' {\em
  Phys.Rev.Lett.}, vol.~66, pp.~1815--1818, 1991.

\bibitem{Ellis:1990nz}
J.~R. Ellis, G.~Ridolfi, and F.~Zwirner, ``{Radiative corrections to the masses
  of supersymmetric Higgs bosons},'' {\em Phys.Lett.}, vol.~B257, pp.~83--91,
  1991.

\bibitem{Brignole:1991wp}
A.~Brignole, ``{Radiative corrections to the supersymmetric charged Higgs boson
  mass},'' {\em Phys.Lett.}, vol.~B277, pp.~313--323, 1992.

\bibitem{Chankowski:1992er}
P.~H. Chankowski, S.~Pokorski, and J.~Rosiek, ``{Complete on-shell
  renormalization scheme for the minimal supersymmetric Higgs sector},'' {\em
  Nucl.Phys.}, vol.~B423, pp.~437--496, 1994, hep-ph/9303309.

\bibitem{Dabelstein:1994hb}
A.~Dabelstein, ``{The One loop renormalization of the MSSM Higgs sector and its
  application to the neutral scalar Higgs masses},'' {\em Z.Phys.}, vol.~C67,
  pp.~495--512, 1995, hep-ph/9409375.

\bibitem{Pilaftsis:1999qt}
A.~Pilaftsis and C.~E. Wagner, ``{Higgs bosons in the minimal supersymmetric
  standard model with explicit CP violation},'' {\em Nucl.Phys.}, vol.~B553,
  pp.~3--42, 1999, hep-ph/9902371.

\bibitem{Carena:2000yi}
M.~S. Carena, J.~R. Ellis, A.~Pilaftsis, and C.~Wagner, ``{Renormalization
  group improved effective potential for the MSSM Higgs sector with explicit CP
  violation},'' {\em Nucl.Phys.}, vol.~B586, pp.~92--140, 2000, hep-ph/0003180.

\bibitem{Freitas:2007dp}
A.~Freitas, E.~Gasser, and U.~Haisch, ``{Supersymmetric large $\tan(\beta)$
  corrections to $\Delta_{M_{d,s}}$ and $B_{d, s} \to \mu^{+} \mu^{-}$
  revisited},'' {\em Phys.Rev.}, vol.~D76, p.~014016, 2007, hep-ph/0702267.

\bibitem{Gorbahn:2009pp}
M.~Gorbahn, S.~Jager, U.~Nierste, and S.~Trine, ``{The supersymmetric Higgs
  sector and $B-\bar{B}$ mixing for large $\tan\beta$ },'' {\em Phys.Rev.},
  vol.~D84, p.~034030, 2011, 0901.2065.

\bibitem{Crivellin:2011jt}
A.~Crivellin, L.~Hofer, and J.~Rosiek, ``{Complete resummation of
  chirally-enhanced loop-effects in the MSSM with non-minimal sources of
  flavor-violation},'' 2011, 1103.4272.

\bibitem{Rosiek:1995kg}
J.~Rosiek, ``{Complete set of Feynman rules for the MSSM: Erratum},'' 1995,
  hep-ph/9511250.

\bibitem{Glashow:1976nt}
S.~L. Glashow and S.~Weinberg, ``{Natural Conservation Laws for Neutral
  Currents},'' {\em Phys.Rev.}, vol.~D15, p.~1958, 1977.

\bibitem{Miki:2002nz}
T.~Miki {\em et~al.}, ``{Effects of charged Higgs boson and QCD corrections in
  ${\bar B} \to D \tau {\bar \nu_{\tau}}$ },'' pp.~116--124, 2002,
  hep-ph/0210051.

\bibitem{WahabElKaffas:2007xd}
A.~Wahab El~Kaffas, P.~Osland, and O.~M. Ogreid, ``{Constraining the
  Two-Higgs-Doublet-Model parameter space},'' {\em Phys.Rev.}, vol.~D76,
  p.~095001, 2007, 0706.2997.

\bibitem{Deschamps:2009rh}
O.~Deschamps, S.~Descotes-Genon, S.~Monteil, V.~Niess, S.~T'Jampens, {\em
  et~al.}, ``{The Two Higgs Doublet of Type II facing flavour physics data},''
  {\em Phys.Rev.}, vol.~D82, p.~073012, 2010, 0907.5135.

\bibitem{Bertolini:1990if}
S.~Bertolini, F.~Borzumati, A.~Masiero, and G.~Ridolfi, ``{Effects of
  supergravity induced electroweak breaking on rare $B$ decays and mixings},''
  {\em Nucl.Phys.}, vol.~B353, pp.~591--649, 1991.

\bibitem{Ciuchini:1997xe}
M.~Ciuchini, G.~Degrassi, P.~Gambino, and G.~Giudice, ``{Next-to-leading QCD
  corrections to $B \to X_{s} \gamma$: Standard model and two Higgs doublet
  model},'' {\em Nucl.Phys.}, vol.~B527, pp.~21--43, 1998, hep-ph/9710335.

\bibitem{Borzumati:1998tg}
F.~Borzumati and C.~Greub, ``{2HDMs predictions for ${\bar B} \to X_{s} \gamma$
  in NLO QCD},'' {\em Phys.Rev.}, vol.~D58, p.~074004, 1998, hep-ph/9802391.

\bibitem{Misiak:2006zs}
M.~Misiak, H.~Asatrian, K.~Bieri, M.~Czakon, A.~Czarnecki, {\em et~al.},
  ``{Estimate of ${\cal B}({\bar B} \to X_{s} \gamma) $ at $O(\alpha^{2}_{s})$
  },'' {\em Phys.Rev.Lett.}, vol.~98, p.~022002, 2007, hep-ph/0609232.

\bibitem{Hermann:2012fc}
T.~Hermann, M.~Misiak, and M.~Steinhauser, ``{$\bar{B}\to X_s \gamma$ in the
  Two Higgs Doublet Model up to Next-to-Next-to-Leading Order in QCD},'' 2012,
  1208.2788.

\bibitem{He:1988tf}
X.~He, T.~Nguyen, and R.~Volkas, ``{B MESON RARE DECAYS IN TWO HIGGS DOUBLETS
  MODELS},'' {\em Phys.Rev.}, vol.~D38, p.~814, 1988.

\bibitem{Skiba:1992mg}
W.~Skiba and J.~Kalinowski, ``{$B_s \to \tau^{+} \tau^{-}$ decay in a two Higgs
  doublet model},'' {\em Nucl.Phys.}, vol.~B404, pp.~3--19, 1993.

\bibitem{Logan:2000iv}
H.~E. Logan and U.~Nierste, ``{ $B_{s,d} \to \ell^+ \ell^-$ in a two Higgs
  doublet model},'' {\em Nucl.Phys.}, vol.~B586, pp.~39--55, 2000,
  hep-ph/0004139.

\bibitem{Hou:1992sy}
W.-S. Hou, ``{Enhanced charged Higgs boson effects in $B^{-} \to \tau
  {\bar\nu}, \mu {\bar\nu}$, and $b \to \tau {\bar\nu} X$ },'' {\em Phys.Rev.},
  vol.~D48, pp.~2342--2344, 1993.

\bibitem{Akeroyd:2003zr}
A.~Akeroyd and S.~Recksiegel, ``{The Effect of $H^{\pm}$ on $B^{\pm} \to
  \tau^{\pm} \nu_{\tau}$ and $B^{\pm} \to \mu^{\pm} \nu_{\mu}$ },'' {\em
  J.Phys.G}, vol.~G29, pp.~2311--2317, 2003, hep-ph/0306037.

\bibitem{Fajfer:2012vx}
S.~Fajfer, J.~F. Kamenik, and I.~Nisandzic, ``{On the $B \to D^* \tau \nu_\tau$
  Sensitivity to New Physics},'' 2012, 1203.2654.

\bibitem{Tanaka:1994ay}
M.~Tanaka, ``{Charged Higgs effects on exclusive semitauonic $B$ decays},''
  {\em Z.Phys.}, vol.~C67, pp.~321--326, 1995, hep-ph/9411405.

\bibitem{Nierste:2008qe}
U.~Nierste, S.~Trine, and S.~Westhoff, ``{Charged-Higgs effects in a new $B \to
  D \tau \nu$ differential decay distribution},'' {\em Phys.Rev.}, vol.~D78,
  p.~015006, 2008, 0801.4938.

\bibitem{Akeroyd:2003jb}
A.~Akeroyd, ``{Effect of $H^{\pm}$ on $D^{+}_{s} \to \mu \nu$ and $D^{+}_{s}
  \to \tau \nu$ },'' {\em Prog.Theor.Phys.}, vol.~111, pp.~295--299, 2004,
  hep-ph/0308260.

\bibitem{Akeroyd:2007eh}
A.~Akeroyd and C.~H. Chen, ``{Effect of $H^{\pm}$ on $B \to \tau \nu_{\tau}$
  and $D^{+}_{s} \to \tau,\mu \nu$ },'' {\em Phys.Rev.}, vol.~D75, p.~075004,
  2007, hep-ph/0701078.

\bibitem{Akeroyd:2009tn}
A.~Akeroyd and F.~Mahmoudi, ``{Constraints on charged Higgs bosons from
  $D^{+}_{s} \to \mu \nu$ and $D^{+}_{s} \to \tau \nu$ },'' {\em JHEP},
  vol.~0904, p.~121, 2009, 0902.2393.

\bibitem{Antonelli:2008jg}
M.~Antonelli {\em et~al.}, ``{Precision tests of the Standard Model with
  leptonic and semileptonic kaon decays},'' 2008, 0801.1817.

\bibitem{CMS}
S.~Chatrchyan {\em et~al.}, ``{Search for neutral Higgs bosons decaying to tau
  pairs in pp collisions at ${\sqrt s}=7$ TeV},'' {\em Phys.Lett.}, vol.~B713,
  pp.~68--90, 2012, 1202.4083.

\bibitem{Pich:2009sp}
A.~Pich and P.~Tuzon, ``{Yukawa Alignment in the Two-Higgs-Doublet Model},''
  {\em Phys.Rev.}, vol.~D80, p.~091702, 2009, 0908.1554.

\bibitem{Jung:2010ik}
M.~Jung, A.~Pich, and P.~Tuzon, ``{Charged-Higgs phenomenology in the Aligned
  two-Higgs-doublet model},'' {\em JHEP}, vol.~1011, p.~003, 2010, 1006.0470.

\bibitem{Mahmoudi:2009zx}
F.~Mahmoudi and O.~Stal, ``{Flavor constraints on the two-Higgs-doublet model
  with general Yukawa couplings},'' {\em Phys.Rev.}, vol.~D81, p.~035016, 2010,
  0907.1791.

\bibitem{Iltan:2001rp}
E.~Iltan, ``{Electric dipole moments of charged leptons and lepton flavor
  violating interactions in the general two Higgs doublet model},'' {\em
  Phys.Rev.}, vol.~D64, p.~013013, 2001, hep-ph/0101017.

\bibitem{Cheng:1987rs}
T.~Cheng and M.~Sher, ``{Mass Matrix Ansatz and Flavor Nonconservation in
  Models with Multiple Higgs Doublets},'' {\em Phys.Rev.}, vol.~D35, p.~3484,
  1987.

\bibitem{Aaij:2012ct}
R.~Aaij {\em et~al.}, ``{First evidence for the decay $B_s \to \mu^{+}\mu^{-} $
  },'' 2012, 1211.2674.

\bibitem{Buras:2012ru}
A.~J. Buras, J.~Girrbach, D.~Guadagnoli, and G.~Isidori, ``{On the Standard
  Model prediction for $BR(B_{s,d} \to \mu^{+} \mu^{-} )$ },'' 2012, 1208.0934.

\bibitem{Isidori:2003ts}
G.~Isidori and R.~Unterdorfer, ``{On the short distance constraints from
  $K_{L,S} \to \mu^{+} \mu^{-}$ },'' {\em JHEP}, vol.~0401, p.~009, 2004,
  hep-ph/0311084.

\bibitem{Beringer:1900zz}
J.~Beringer {\em et~al.}, ``{Review of Particle Physics (RPP)},'' {\em
  Phys.Rev.}, vol.~D86, p.~010001, 2012.

\bibitem{InamiLim81}
T.~Inami and C.~Lim {\em Prog. Theor. Phys}, vol.~65, p.~297, 1981.

\bibitem{Ciuchini:1997bw}
M.~Ciuchini, E.~Franco, V.~Lubicz, G.~Martinelli, I.~Scimemi, {\em et~al.},
  ``{Next-to-leading order QCD corrections to $\Delta F = 2$ effective
  Hamiltonians},'' {\em Nucl.Phys.}, vol.~B523, pp.~501--525, 1998,
  hep-ph/9711402.

\bibitem{Buras:2000if}
A.~J. Buras, M.~Misiak, and J.~Urban, ``{Two loop QCD anomalous dimensions of
  flavor changing four quark operators within and beyond the standard model},''
  {\em Nucl.Phys.}, vol.~B586, pp.~397--426, 2000, hep-ph/0005183.

\bibitem{Bona:2007vi}
M.~Bona {\em et~al.}, ``{Model-independent constraints on $\Delta$ F=2
  operators and the scale of new physics},'' {\em JHEP}, vol.~0803, p.~049,
  2008, 0707.0636.

\bibitem{Becirevic:2001jj}
D.~Becirevic, M.~Ciuchini, E.~Franco, V.~Gimenez, G.~Martinelli, {\em et~al.},
  ``{$B_d - \bar{B}_d$ mixing and the $B_d \to J/\psi K_s$ asymmetry in general
  SUSY models},'' {\em Nucl.Phys.}, vol.~B634, pp.~105--119, 2002,
  hep-ph/0112303.

\bibitem{Ciuchini:1998ix}
M.~Ciuchini, V.~Lubicz, L.~Conti, A.~Vladikas, A.~Donini, {\em et~al.},
  ``{Delta M(K) and epsilon(K) in SUSY at the next-to-leading order},'' {\em
  JHEP}, vol.~9810, p.~008, 1998, hep-ph/9808328.

\bibitem{Lubicz:2008am}
V.~Lubicz and C.~Tarantino, ``{Flavour physics and Lattice QCD: Averages of
  lattice inputs for the Unitarity Triangle Analysis},'' {\em Nuovo Cim.},
  vol.~B123, pp.~674--688, 2008, 0807.4605.

\bibitem{Becirevic:2001yv}
D.~Becirevic, V.~Gimenez, G.~Martinelli, M.~Papinutto, and J.~Reyes,
  ``{Combined relativistic and static analysis for all Delta B=2 operators},''
  {\em Nucl.Phys.Proc.Suppl.}, vol.~106, pp.~385--387, 2002, hep-lat/0110117.

\bibitem{Ciuchini:2000de}
M.~Ciuchini, G.~D'Agostini, E.~Franco, V.~Lubicz, G.~Martinelli, {\em et~al.},
  ``{2000 CKM triangle analysis: A Critical review with updated experimental
  inputs and theoretical parameters},'' {\em JHEP}, vol.~0107, p.~013, 2001,
  hep-ph/0012308.

\bibitem{Lenz:2010gu}
A.~Lenz, U.~Nierste, J.~Charles, S.~Descotes-Genon, A.~Jantsch, {\em et~al.},
  ``{Anatomy of New Physics in $B - \bar{B}$ mixing},'' {\em Phys.Rev.},
  vol.~D83, p.~036004, 2011, 1008.1593.

\bibitem{UTfit:2012zm}
M.~Bona {\em et~al.}, ``{The UTfit Collaboration Average of D meson mixing
  data: Spring 2012},'' 2012, 1206.6245.

\bibitem{Hayasaka:2010np}
K.~Hayasaka, K.~Inami, Y.~Miyazaki, K.~Arinstein, V.~Aulchenko, {\em et~al.},
  ``{Search for Lepton Flavor Violating Tau Decays into Three Leptons with 719
  Million Produced $\tau^{+}\tau^{-}$ Pairs},'' {\em Phys.Lett.}, vol.~B687,
  pp.~139--143, 2010, 1001.3221.

\bibitem{Bellgardt:1987du}
U.~Bellgardt {\em et~al.}, ``{Search for the Decay $\mu^+ \to e^+ e^+ e^-$},''
  {\em Nucl.Phys.}, vol.~B299, p.~1, 1988.

\bibitem{StoneICHEP12}
S.~Stone, ``Planery talk at the international conference on high energy physics
  (ichep 2012), melbourne, australia,'' {\em Planery talk at the International
  Conference on High Energy Physics (ICHEP 2012), Melbourne, Australia,}, July
  4-11th, 2012.

\bibitem{Collaboration:2012fk}
B.~Collaboration, ``Precision measurement of the $b \to x_s \gamma$ photon
  energy spectrum, branching fraction, and direct cp asymmetry $a_{CP}(b \to
  x_{s+d}\gamma)$,'' 07 2012, 1207.2690v1,1207.5772.

\bibitem{Lees:2012wg}
J.~Lees {\em et~al.}, ``{Exclusive Measurements of $b \to s \gamma$ Transition
  Rate and Photon Energy Spectrum},'' 2012, 1207.2520.

\bibitem{Amhis:2012bh}
Y.~Amhis {\em et~al.}, ``{Averages of $b$-hadron, $c$-hadron, and $\tau$-lepton
  properties as of early 2012},'' 2012, 1207.1158.

\bibitem{Benzke:2010js}
M.~Benzke, S.~J. Lee, M.~Neubert, and G.~Paz, ``{Factorization at Subleading
  Power and Irreducible Uncertainties in $\bar B\to X_s\gamma$ Decay},'' {\em
  JHEP}, vol.~1008, p.~099, 2010, 1003.5012.

\bibitem{Hurth:2010tk}
T.~Hurth and M.~Nakao, ``{Radiative and Electroweak Penguin Decays of B
  Mesons},'' {\em Ann.Rev.Nucl.Part.Sci.}, vol.~60, pp.~645--677, 2010,
  1005.1224.

\bibitem{delAmoSanchez:2010ae}
P.~del Amo~Sanchez {\em et~al.}, ``{Study of $B \to X \gamma$ decays and
  determination of $|V_{td}/V_{ts}|$ },'' {\em Phys.Rev.}, vol.~D82, p.~051101,
  2010, 1005.4087.

\bibitem{Wang:2011sn}
W.~Wang, ``{ $b \to s \gamma$ and $b \to d \gamma$ (B factories) },'' 2011,
  1102.1925.

\bibitem{Crivellin:2011ba}
A.~Crivellin and L.~Mercolli, ``{ $B \to X_d \gamma$ and constraints on new
  physics},'' {\em Phys.Rev.}, vol.~D84, p.~114005, 2011, 1106.5499.

\bibitem{Ali:1998rr}
A.~Ali, H.~Asatrian, and C.~Greub, ``{Inclusive decay rate for $B \to X_{d}
  \gamma$ in next-to-leading logarithmic order and CP asymmetry in the standard
  model},'' {\em Phys.Lett.}, vol.~B429, pp.~87--98, 1998, hep-ph/9803314.

\bibitem{Hurth:2003dk}
T.~Hurth, E.~Lunghi, and W.~Porod, ``{Untagged ${\bar B} \to X_{s+d} \gamma$ CP
  asymmetry as a probe for new physics},'' {\em Nucl.Phys.}, vol.~B704,
  pp.~56--74, 2005, hep-ph/0312260.

\bibitem{Adam:2013mnn}
J.~Adam {\em et~al.}, ``{New constraint on the existence of the $\mu^+ \to e^+
  \gamma$ decay},'' 2013, 1303.0754.

\bibitem{Aubert:2009ag}
B.~Aubert {\em et~al.}, ``{Searches for Lepton Flavor Violation in the Decays
  $\tau^{\pm} \to e^{\pm} \gamma$ and $\tau^{\pm} \to \mu^{\pm} \gamma$ },''
  {\em Phys.Rev.Lett.}, vol.~104, p.~021802, 2010, 0908.2381.

\bibitem{Hayasaka:2007vc}
K.~Hayasaka {\em et~al.}, ``{New search for $\tau \to \mu \gamma$ and $\tau \to
  e \gamma$ decays at Belle},'' {\em Phys.Lett.}, vol.~B666, pp.~16--22, 2008,
  0705.0650.

\bibitem{Hudson:2011zz}
J.~Hudson, D.~Kara, I.~Smallman, B.~Sauer, M.~Tarbutt, {\em et~al.},
  ``{Improved measurement of the shape of the electron},'' {\em Nature},
  vol.~473, pp.~493--496, 2011.

\bibitem{Bennett:2008dy}
G.~Bennett {\em et~al.}, ``{An Improved Limit on the Muon Electric Dipole
  Moment},'' {\em Phys.Rev.}, vol.~D80, p.~052008, 2009, 0811.1207.

\bibitem{Inami:2002ah}
K.~Inami {\em et~al.}, ``{Search for the electric dipole moment of the tau
  lepton},'' {\em Phys.Lett.}, vol.~B551, pp.~16--26, 2003, hep-ex/0210066.

\bibitem{Baker:2006ts}
C.~Baker, D.~Doyle, P.~Geltenbort, K.~Green, M.~van~der Grinten, {\em et~al.},
  ``{An Improved experimental limit on the electric dipole moment of the
  neutron},'' {\em Phys.Rev.Lett.}, vol.~97, p.~131801, 2006, hep-ex/0602020.

\bibitem{Hisano:2009ae}
J.~Hisano, M.~Nagai, P.~Paradisi, and Y.~Shimizu, ``{Waiting for $\mu \to e
  \gamma$ from the MEG experiment},'' {\em JHEP}, vol.~0912, p.~030, 2009,
  0904.2080.

\bibitem{Passera:2004bj}
M.~Passera, ``{The Standard model prediction of the muon anomalous magnetic
  moment},'' {\em J.Phys.}, vol.~G31, pp.~R75--R94, 2005, hep-ph/0411168.

\bibitem{Passera:2005mx}
M.~Passera, ``{Status of the standard model prediction of the muon $g-2$ },''
  {\em Nucl.Phys.Proc.Suppl.}, vol.~155, pp.~365--368, 2006, hep-ph/0509372.

\bibitem{Davier:2007ua}
M.~Davier, ``{The Hadronic contribution to $(g-2)_{\mu}$ },'' {\em
  Nucl.Phys.Proc.Suppl.}, vol.~169, pp.~288--296, 2007, hep-ph/0701163.

\bibitem{Hagiwara:2006jt}
K.~Hagiwara, A.~Martin, D.~Nomura, and T.~Teubner, ``{Improved predictions for
  $g-2$ of the muon and $\alpha_{QED} (M_{Z}^{2}) $ },'' {\em Phys.Lett.},
  vol.~B649, pp.~173--179, 2007, hep-ph/0611102.

\bibitem{Buras:2010zm}
A.~J. Buras, G.~Isidori, and P.~Paradisi, ``{EDMs versus CPV in $B_{s,d}$
  mixing in two Higgs doublet models with MFV},'' {\em Phys.Lett.}, vol.~B694,
  pp.~402--409, 2011, 1007.5291.

\bibitem{Demir:2003js}
D.~A. Demir, O.~Lebedev, K.~A. Olive, M.~Pospelov, and A.~Ritz, ``{Electric
  dipole moments in the MSSM at large tan beta},'' {\em Nucl.Phys.}, vol.~B680,
  pp.~339--374, 2004, hep-ph/0311314.

\bibitem{Pospelov:2000bw}
M.~Pospelov and A.~Ritz, ``{Neutron EDM from electric and chromoelectric dipole
  moments of quarks},'' {\em Phys.Rev.}, vol.~D63, p.~073015, 2001,
  hep-ph/0010037.

\bibitem{Demir:2002gg}
D.~A. Demir, M.~Pospelov, and A.~Ritz, ``{Hadronic EDMs, the Weinberg operator,
  and light gluinos},'' {\em Phys.Rev.}, vol.~D67, p.~015007, 2003,
  hep-ph/0208257.

\bibitem{Olive:2005ru}
K.~A. Olive, M.~Pospelov, A.~Ritz, and Y.~Santoso, ``{CP-odd phase correlations
  and electric dipole moments},'' {\em Phys.Rev.}, vol.~D72, p.~075001, 2005,
  hep-ph/0506106.

\bibitem{Antonelli:2010yf}
M.~Antonelli, V.~Cirigliano, G.~Isidori, F.~Mescia, M.~Moulson, {\em et~al.},
  ``{An Evaluation of $|V_{us}|$ and precise tests of the Standard Model from
  world data on leptonic and semileptonic kaon decays},'' {\em Eur.Phys.J.},
  vol.~C69, pp.~399--424, 2010, 1005.2323.

\bibitem{Decker:1993py}
R.~Decker and M.~Finkemeier, ``{Radiative corrections to the decay $\tau \to
  \pi (K) \nu_{\tau}$ },'' {\em Phys.Lett.}, vol.~B316, pp.~403--406, 1993,
  hep-ph/9307372.

\bibitem{Banerjee:2008hg}
S.~Banerjee, ``{Lepton Universality, $|V_{us}|$ and search for second class
  current in tau decays},'' 2008, 0811.1429.

\bibitem{Kamenik:2008tj}
J.~F. Kamenik and F.~Mescia, ``{ $B \to D \tau \nu$ Branching Ratios:
  Opportunity for Lattice QCD and Hadron Colliders},'' {\em Phys.Rev.},
  vol.~D78, p.~014003, 2008, 0802.3790.

\bibitem{Korner:1987kd}
J.~Korner and G.~Schuler, ``{Exclusive Semileptonic Decays of Bottom Mesons in
  the Spectator Quark Model},'' {\em Z.Phys.}, vol.~C38, p.~511, 1988.

\bibitem{Korner:1989ve}
J.~Korner and G.~Schuler, ``{LEPTON MASS EFFECTS IN SEMILEPTONIC B MESON
  DECAYS},'' {\em Phys.Lett.}, vol.~B231, p.~306, 1989.

\bibitem{Korner:1989qb}
J.~Korner and G.~Schuler, ``{Exclusive Semileptonic Heavy Meson Decays
  Including Lepton Mass Effects},'' {\em Z.Phys.}, vol.~C46, p.~93, 1990.

\bibitem{Heiliger:1989yp}
P.~Heiliger and L.~Sehgal, ``{SEMILEPTONIC DECAYS OF B MESONS INTO $\tau
  \nu_{\tau}$ },'' {\em Phys.Lett.}, vol.~B229, p.~409, 1989.

\bibitem{Pham:1992fr}
X.-Y. Pham, ``{New method for determination of the $D \to {\bar K^{*}} e \nu$
  axial form-factors without resort to angular distributions},'' {\em
  Phys.Rev.}, vol.~D46, p.~R1909, 1992.

\bibitem{Fajfer:2012jt}
S.~Fajfer, J.~F. Kamenik, I.~Nisandzic, and J.~Zupan, ``{Implications of lepton
  flavor universality violations in $B$ decays},'' 2012, 1206.1872.

\bibitem{Deshpande:2012rr}
N.~Deshpande and A.~Menon, ``{Hints of R-parity violation in $B$ decays into
  $\tau\nu$},'' 2012, 1208.4134.

\bibitem{He:2012zp}
X.-G. He and G.~Valencia, ``{B decays with $\tau$-leptons in non-universal
  left-right models},'' 2012, 1211.0348.

\bibitem{Celis:2012dk}
A.~Celis, M.~Jung, X.-Q. Li, and A.~Pich, ``{Sensitivity to charged scalars in
  $B\to D^{(*)}\tau\nu_\tau$ and $B\to\tau\nu_\tau$ decays},'' 2012, 1210.8443.

\bibitem{Tanaka:2012nw}
M.~Tanaka and R.~Watanabe, ``{New physics in the weak interaction of $\bar B\to
  D^{(*)}\tau\bar\nu$},'' 2012, 1212.1878.

\bibitem{Bailey:2012jg}
J.~A. Bailey, A.~Bazavov, C.~Bernard, C.~Bouchard, C.~DeTar, {\em et~al.},
  ``{Refining new-physics searches in $B \to D \tau \nu$ decay with lattice
  QCD},'' {\em Phys.Rev.Lett.}, vol.~109, p.~071802, 2012, 1206.4992.

\bibitem{Datta:2012qk}
A.~Datta, M.~Duraisamy, and D.~Ghosh, ``{Diagnosing New Physics in $b \to c \,
  \tau \, \nu_\tau$ decays in the light of the recent BaBar result},'' {\em
  Phys.Rev.}, vol.~D86, p.~034027, 2012, 1206.3760.

\bibitem{Becirevic:2012jf}
D.~Becirevic, N.~Kosnik, and A.~Tayduganov, ``{$\bar B\to D\tau\bar \nu_\tau$
  vs. $\bar B\to D\mu\bar \nu_\mu$},'' {\em Phys.Lett.}, vol.~B716,
  pp.~208--213, 2012, 1206.4977.

\bibitem{Ko:2012sv}
P.~Ko, Y.~Omura, and C.~Yu, ``{$B \to D^{(*)} \tau \nu$ and $B \to \tau \nu$ in
  chiral $U(1)'$ models with flavored multi Higgs doublets},'' 2012, 1212.4607.

\bibitem{Biancofiore:2013ki}
P.~Biancofiore, P.~Colangelo, and F.~De~Fazio, ``{On the anomalous enhancement
  observed in $B \to D^{(*)} \, \tau \, {\bar \nu}_\tau $ decays},'' 2013,
  1302.1042.

\bibitem{Sakaki:2012ft}
Y.~Sakaki and H.~Tanaka, ``{Constraints of the Charged Scalar Effects Using the
  Forward-Backward Asymmetry on $B\to D^{(*)}\tau\bar{\nu_{\tau}}$},'' 2012,
  1205.4908.

\bibitem{Dobrescu:2008er}
B.~A. Dobrescu and A.~S. Kronfeld, ``{Accumulating evidence for nonstandard
  leptonic decays of $D_s$ mesons},'' {\em Phys.Rev.Lett.}, vol.~100,
  p.~241802, 2008, 0803.0512.

\bibitem{Alexander:2009ux}
J.~Alexander {\em et~al.}, ``{Measurement of $B({D_s^+ \to \ell^+ \nu})$ and
  the Decay Constant $f_{D_s^+}$ From 600 $/pb^{-1}$ of $e^\pm$ Annihilation
  Data Near 4170 MeV},'' {\em Phys.Rev.}, vol.~D79, p.~052001, 2009, 0901.1216.

\bibitem{Onyisi:2009th}
P.~Onyisi {\em et~al.}, ``{Improved Measurement of Absolute Branching Fraction
  of $D^{+}_{s} \to \tau \nu_{\tau}$ },'' {\em Phys.Rev.}, vol.~D79, p.~052002,
  2009, 0901.1147.

\bibitem{Simone:2011zza}
J.~N. Simone, ``{The decay constants $f_{D_s}$ and $f_{D^+}$ form lattice
  QCD},'' {\em Int.J.Mod.Phys.Conf.Ser.}, vol.~02, pp.~92--96, 2011.

\bibitem{Bazavov:2011aa}
A.~Bazavov {\em et~al.}, ``{$B$- and $D$-meson decay constants from
  three-flavor lattice QCD},'' {\em Phys.Rev.}, vol.~D85, p.~114506, 2012,
  1112.3051.

\bibitem{Barranco:2013tba}
V.~G.~M. J.~Barranco, D.~Delepine and L.~Lopez-Lozano, ``{Constraining New
  Physics with $D$ meson decays},'' 2013, 1303.3896.

\bibitem{Burdman:1994ip}
G.~Burdman, J.~T. Goldman, and D.~Wyler, ``{Radiative leptonic decays of heavy
  mesons},'' {\em Phys.Rev.}, vol.~D51, pp.~111--117, 1995, hep-ph/9405425.

\bibitem{Masiero:2006ss}
A.~Masiero and P.~Paradisi, ``{The B - TAU FCNC connection in SUSY Unified
  Theories},'' {\em J.Phys.Conf.Ser.}, vol.~53, pp.~248--254, 2006,
  hep-ph/0609262.

\bibitem{Dedes:2002rh}
A.~Dedes, J.~R. Ellis, and M.~Raidal, ``{Higgs mediated $B^0_{s,d} \to \mu
  \tau$, $e \tau$ and $\tau \to 3 \mu$, $e \mu \mu$ decays in supersymmetric
  seesaw models},'' {\em Phys.Lett.}, vol.~B549, pp.~159--169, 2002,
  hep-ph/0209207.

\bibitem{Dedes:2008iw}
A.~Dedes, J.~Rosiek, and P.~Tanedo, ``{Complete One-Loop MSSM Predictions for
  $B^0 \to \ell^+ \ell^{'-}$ at the Tevatron and LHC},'' {\em Phys.Rev.},
  vol.~D79, p.~055006, 2009, 0812.4320.

\bibitem{Boubaa:2012xj}
D.~Boubaa, A.~Datta, M.~Duraisamy, and S.~Khalil, ``{Predictions for $B \to
  \tau \bar{\mu} + \mu \bar{\tau} $},'' 2012, 1211.5168.

\bibitem{Aaltonen:2009vr}
T.~Aaltonen {\em et~al.}, ``{Search for the Decays $B^0_{(s)} \to e^+ \mu^-$
  and $B^0_{(s)}\to e^+ e^-$ in CDF Run II},'' {\em Phys.Rev.Lett.}, vol.~102,
  p.~201801, 2009, 0901.3803.

\bibitem{Aubert:2008cu}
B.~Aubert {\em et~al.}, ``{Searches for the decays $B^0 \to \ell^\pm \tau^\mp$
  and $B^{+} \to \ell^{+} \nu$ $(\ell=e, \mu^{})$ using hadronic tag
  reconstruction},'' {\em Phys.Rev.}, vol.~D77, p.~091104, 2008, 0801.0697.

\bibitem{Altmannshofer:2007cs}
W.~Altmannshofer, A.~J. Buras, and D.~Guadagnoli, ``{The MFV limit of the MSSM
  for low $\tan(\beta)$: Meson mixings revisited},'' {\em JHEP}, vol.~0711,
  p.~065, 2007, hep-ph/0703200.

\bibitem{Chankowski:2000ng}
P.~H. Chankowski and L.~Slawianowska, ``{ $B^0_{s,d} \to \mu^+ \mu^-$ decay in
  the MSSM},'' {\em Phys.Rev.}, vol.~D63, p.~054012, 2001, hep-ph/0008046.

\bibitem{Colangelo:2010et}
G.~Colangelo, S.~Durr, A.~Juttner, L.~Lellouch, H.~Leutwyler, {\em et~al.},
  ``{Review of lattice results concerning low energy particle physics},'' {\em
  Eur.Phys.J.}, vol.~C71, p.~1695, 2011, 1011.4408.

\bibitem{Blossier:2010cr}
B.~Blossier {\em et~al.}, ``{Average up/down, strange and charm quark masses
  with Nf=2 twisted mass lattice QCD},'' {\em Phys.Rev.}, vol.~D82, p.~114513,
  2010, 1010.3659.

\bibitem{Buras:1998raa}
A.~J. Buras, ``{Weak Hamiltonian, CP violation and rare decays},''
  pp.~281--539, 1998, hep-ph/9806471.

\bibitem{Blossier:2009bx}
B.~Blossier {\em et~al.}, ``{Pseudoscalar decay constants of kaon and D-mesons
  from $N_f=2$ twisted mass Lattice QCD},'' {\em JHEP}, vol.~0907, p.~043,
  2009, 0904.0954.

\bibitem{Bernard:2007ps}
C.~Bernard, C.~E. DeTar, L.~Levkova, S.~Gottlieb, U.~Heller, {\em et~al.},
  ``{Status of the MILC light pseudoscalar meson project},'' {\em PoS},
  vol.~LAT2007, p.~090, 2007, 0710.1118.

\bibitem{Follana:2007uv}
E.~Follana, C.~Davies, G.~Lepage, and J.~Shigemitsu, ``{High Precision
  determination of the pi, K, D and D(s) decay constants from lattice QCD},''
  {\em Phys.Rev.Lett.}, vol.~100, p.~062002, 2008, 0706.1726.

\bibitem{Aubin:2008ie}
C.~Aubin, J.~Laiho, and R.~S. Van~de Water, ``{Light pseudoscalar meson masses
  and decay constants from mixed action lattice QCD},'' {\em PoS},
  vol.~LATTICE2008, p.~105, 2008, 0810.4328.

\bibitem{Beane:2006kx}
S.~Beane, P.~Bedaque, K.~Orginos, and M.~Savage, ``{$f_{K}/f_{\pi}$ in Full QCD
  with Domain Wall Valence Quarks},'' {\em Phys.Rev.}, vol.~D75, p.~094501,
  2007, hep-lat/0606023.

\bibitem{Durr:2010hr}
S.~Durr, Z.~Fodor, C.~Hoelbling, S.~Katz, S.~Krieg, {\em et~al.}, ``{The ratio
  $F_{K}/F_{\pi}$ in QCD},'' {\em Phys.Rev.}, vol.~D81, p.~054507, 2010,
  1001.4692.

\bibitem{Bernard:2002pc}
C.~Bernard {\em et~al.}, ``{Lattice calculation of heavy light decay constants
  with two flavors of dynamical quarks},'' {\em Phys.Rev.}, vol.~D66,
  p.~094501, 2002, hep-lat/0206016.

\bibitem{Bazavov:2009ii}
A.~Bazavov {\em et~al.}, ``{The $D_s$ and $D^+$ Leptonic Decay Constants from
  Lattice QCD},'' {\em PoS}, vol.~LAT2009, p.~249, 2009, 0912.5221.

\bibitem{Davies:2010ip}
C.~Davies, C.~McNeile, E.~Follana, G.~Lepage, H.~Na, {\em et~al.}, ``{Update:
  Precision $D_s$ decay constant from full lattice QCD using very fine
  lattices},'' {\em Phys.Rev.}, vol.~D82, p.~114504, 2010, 1008.4018.

\bibitem{Dimopoulos:2011gx}
P.~Dimopoulos {\em et~al.}, ``{Lattice QCD determination of $m_b$, $f_B$ and
  $f_{B_s}$ with twisted mass Wilson fermions},'' {\em JHEP}, vol.~1201,
  p.~046, 2012, 1107.1441.

\bibitem{Gamiz:2009ku}
E.~Gamiz, C.~T. Davies, G.~P. Lepage, J.~Shigemitsu, and M.~Wingate, ``{Neutral
  $B$ Meson Mixing in Unquenched Lattice QCD},'' {\em Phys.Rev.}, vol.~D80,
  p.~014503, 2009, 0902.1815.

\bibitem{McNeile:2011ng}
C.~McNeile, C.~Davies, E.~Follana, K.~Hornbostel, and G.~Lepage,
  ``{High-Precision $f_{B_s}$ and HQET from Relativistic Lattice QCD},'' {\em
  Phys.Rev.}, vol.~D85, p.~031503, 2012, 1110.4510.

\bibitem{Albertus:2010nm}
C.~Albertus, Y.~Aoki, P.~Boyle, N.~Christ, T.~Dumitrescu, {\em et~al.},
  ``{Neutral B-meson mixing from unquenched lattice QCD with domain-wall light
  quarks and static b-quarks},'' {\em Phys.Rev.}, vol.~D82, p.~014505, 2010,
  1001.2023.

\end{thebibliography}

\end{document}